# BLACK-TO-WHITE HOLE SCENARIO: FOUNDATION AND EVAPORATION

PIERRE MARTIN-DUSSAUD

SUPERVISORS:
Carlo Rovelli
Simone Speziale

RAPPORTEURS:
Aurélien Barrau
Seth Major

EXAMINERS:
Francesca Vidotto
Alejandro Perez

To my beloved family
and my unforgettable friends

# SHORT ABSTRACT

The physics of the beginning of the 20th century experienced two major conceptual revolutions that changed the way we see the world. General relativity, on the one hand, describes space-time on a large scale; quantum mechanics, on the other hand, deals with the microscopic behaviour of matter. Since then, physicists have been looking for a theory of quantum gravity, which would bring the two languages together. It is expected that such a theory profoundly changes our understanding of black holes, these extremely dense astrophysical objects, long remained in the shade of calculations, and now observed in droves in the sky. Quantum theory has already shown, as a first approximation, that a black hole slowly evaporates. Quantum gravity also predicts that such an object could metamorphose into a white hole, its time-reverse. In this thesis, we study the foundations of such a scenario and propose a mathematical model that includes the phenomenon of evaporation.

*Version française*

La physique du début du xx^e^ siècle a connu deux grandes révolutions conceptuelles qui ont bouleversé notre façon de voir le monde. La relativité générale, d'une part, décrit l'espace-temps à grande échelle ; la mécanique quantique, d'autre part, traite du comportement microscopique de la matière. Depuis lors, les physiciens sont en quête d'une théorie de la gravité quantique, qui réunirait les deux langages. Il est attendu qu'une telle théorie modifie profondément notre compréhension des trous noirs, ces astres d'une densité extrême, longtemps restés dans l'ombre des calculs, et désormais observés en nombre dans le ciel. La théorie quantique a déjà montré, en première approximation, qu'un trou noir s'évapore lentement. La gravité quantique prédit par ailleurs qu'un tel astre pourrait se métamorphoser en trou blanc, son symétrique temporel. Dans cette thèse, nous étudions les fondements d'un tel scénario et proposons un modèle mathématique qui inclut le phénomène d'évaporation.

# LONG ABSTRACT

The physics of the beginning of the 20th century experienced two major conceptual revolutions that changed the way we see the world. General relativity, on the one hand, describes space-time on a large scale; quantum mechanics, on the other hand, deals with the microscopic behaviour of matter. Since then, physicists have been looking for a theory of quantum gravity, which would bring the two languages together.

Among other possible approaches, loop quantum gravity emerged in the early 1990s, applying a canonical quantification method to general relativity. As one of its main feature, this theory presents a discrete image of space. A bit later, the theory of spin-foams came to complete the picture, by pursuing a covariant approach to the problem.

After 30 years of important theoretical developments, it has become urgent to put these theories to the test. However, the task turns out to be difficult, as the physical regime of quantum gravity has so far escaped experimental reach. Besides primordial cosmology, black holes seem to be the best candidates for highlighting quantum gravitational effects. After a long time remained in the shade of calculations, these extremely dense astrophysical objects are now observed in droves in the sky.

As early as 1974, calculations of quantum field theory in curved space showed, to the surprise of physicists, that black holes evaporate very slowly. In other words, black holes emit thermal radiation, so that their mass gradually decreases. This major discovery has also raised the information paradox, which has troubled theorists for almost 50 years.

However, if the predictions of quantum gravity are to be believed, a bouncing phenomenon could occur even before evaporation begins. The central singularity, predicted by general relativity, should thus be smoothed out and give way to a continuous transition from the black hole to a white hole. Although never observed until now, white holes are exact solutions of the Einstein equations, corresponding to the time-reverse of black holes. Physically, their horizon delimits an area of space-time from which all matter is expelled.

According to this black-to-white hole scenario, black holes would be collapsed stars, on the verge of bouncing back. Their coming explosion would only be a matter of time. This hypothetical phenomenon has already been the subject of analyses and calculations. It has thus been shown that the characteristic bouncing time could be longer than the characteristic evaporation time. The goal of this thesis was therefore to review the foundations of this scenario and to propose a substantial modification of its mathematical model to take into account the progressive effects of the evaporation.

Evaporation is therefore brought back to the fore. It dominates the first phase of the time evolution of black holes. Their mass gradually decreases until reaching the Planck scale. Then, the quantum transition process occurs and causes the metamorphosis from the black hole to the white hole. This time, the white hole is just a long-lived remnant. Seen from the outside, it is a particle of planckian size, interacting weakly, but its immense interior volume contains the famous information that was feared to be lost. It will finally take a much longer time for the white hole to explode than it took for the black hole to evaporate.

*Version française*

La physique du début du xx^e^ siècle a connu deux grandes révolutions conceptuelles qui ont bouleversé notre façon de voir le monde. La relativité générale, d'une part, décrit l'espace-temps à grande échelle ; la mécanique quantique, d'autre part, traite du comportement microscopique de la matière. Depuis lors, les physiciens sont en quête d'une théorie de la gravité quantique, qui réunirait les deux langages.

Parmi les différentes approches possibles, la gravité quantique à boucle a émergé au début des années 1990, en appliquant à la relativité générale, une méthode canonique de quantification. Entre autres caractères, cette théorie présente une image discrète de l'espace. Un peu plus tard, la théorie des mousses de spin est venue compléter le tableau, en poursuivant une approche covariante du problème.

Après 30 ans de développements théoriques importants, il est devenu urgent de mettre ces théories au banc d'essai. La tâche se révèle néanmoins ardue, tant le régime physique de la gravité quantique échappe jusqu'à présent aux expériences. Outre la cosmologie primordiale, les trous noirs semblent être les meilleurs candidats pour mettre en évidence des effets quantiques gravitationnels. Longtemps restés dans l'ombre des calculs, ces astres d'une densité extrême sont désormais observés en nombre dans le ciel.

Dès 1974, des calculs de théorie quantique des champs en espace courbe ont montré, à la surprise des physiciens, que les trous noirs s'évaporent très lentement. Autrement dit, les trous noirs émettent un rayonnement thermique, si bien que leur masse diminue progressivement. Cette découverte majeure n'est pas sans difficulté, comme en témoigne le paradoxe de l'information, qui agite les théoriciens depuis bientôt 50 ans.

Cependant, si l'on en croit les prédictions de la gravité quantique, un phénomène de rebond pourrait survenir avant même que l'évaporation commence. La singularité centrale, prédite par la relativité générale, devrait ainsi laisser place à une transition continue du trou noir vers un trou blanc. Bien que n'ayant été jusqu'à présent jamais observés, les trous blancs sont des solutions exactes des équations d'Einstein, correspondant au renversement temporel des trous noirs. Physiquement, leur horizon délimite une zone de l'espace-temps de laquelle toute matière se trouve expulsée.

Selon ce scénario du trou noir-et-blanc, les trous noirs seraient en fait des étoiles effondrées, en train de rebondir sur elle-même. Leur explosion prochaine ne serait donc qu'une question de temps. Ce phénomène hypothétique a déjà fait l'objet d'analyses et de calculs. Il a ainsi été montré que le temps caractéristique de rebond pourrait être finalement plus long que le temps caractéristique d'évaporation. L'objet de cette thèse a donc été de revoir les fondements de ce scénario, et de proposer une modification substantielle de son modèle mathématique afin de prendre en compte les effets progressifs de l'évaporation.

L'évaporation est donc remise au premier plan. C'est elle qui domine la première phase de l'évolution temporelle des trous noirs. La masse de ceux-ci diminuent progressivement jusqu'à atteindre l'échelle de Planck. C'est alors qu'intervient le processus quantique de transition qui provoque la métamorphose du trou noir en trou blanc. Cette fois, le trou blanc n'est plus qu'un vestige rémanent. Vu de l'extérieur, c'est une particule de taille planckienne, interagissant faiblement, mais son immense volume intérieur contient la fameuse information que l'on croyait perdue. Il lui faudra finalement pour exploser beaucoup plus de temps qu'il en a fallu au trou noir pour s'évaporer.

# PUBLICATIONS

Some ideas and figures have appeared previously in the following publications:

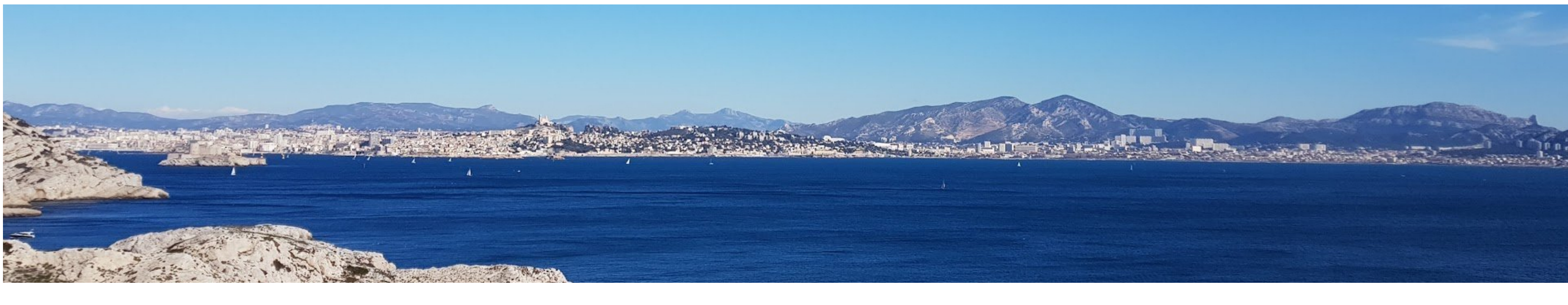

# ACKNOWLEDGEMENTS

Marseilles was founded circa 600 BCE. The new colony was celebrated with the marriage of Protis, a greek settler from Phocaea, and Gyptis, the daughter of the chief of the local Ligurian tribe. It is my turn to raise the ceremonial cup to the people, who have brought their sweat and stones to the edification of this PhD thesis.

**Carlo Rovelli,** who has been a constant source of inspiration.

**Simone Speziale,** for his twisted intellectual stimulation.

**Alejandro Perez,** for his enthusiasm and his humour.

**The Italian Connection,** Elena De Paoli, Giorgio Sarno, Tommaso De Lorenzo, Valeria Gelardi, Pietro Dona, Francesco Gozzini, Giovanni Rabuffo, Nicolas Pedreschi, Francesca Melozzi, Riccardo Catalano...

**The other journeymen,** Thibaut Josset, Marios Christodoulou, Fabio D'Ambrosio, Daniel Martinez, Lautaro Amadei, Eduardo Velásquez, Andrea Di Biagio, Josephine Thomas, Adrien Kuntz, Hongguang Liu, Andrea Calcinari, Farshid Soltani...

**The Basic Research Community for Physics,** and in particular, Alex Thomas, Adrian Fernandez Cid, Carlos Zapata-Carratala, Federico Zalamea, Flavio Del Santo, Isha Kotecha, Jan Glowacki, Jan-Hendrik Treude, Jérémy Attard, Johannes Kleiner, Robin Lorenz, Robin Reuben, Vaclav Zatloukal...

**All the researchers,** who have contributed in a way or another: Abhay Ashtekar, Andrew Hamilton, Aurélien Barrau, Bianca Dittrich, Eugenio Bianchi, Francesca Vidotto, Hal M. Haggard, James M. Bardeen, John Baez, Jorma Louko, Laurent Freidel, Philipp Höhn, Seth Major...

**All the amazing teachers and mentors,** who have taught me so much: the astronomy club Orion43, Georges Paturel, Olivier Bordellès, Dominique Roux, Françoise Wotkiewicz, Alain Audras, Hubert Guillon, Roger Mansuy, Julien Cubizolles, Yves Duval, Ulrich Sauvage, Henning Samtleben, Marc Magro, Pascal Degiovanni, François Gieres, Sébastien Tanzilli, Olivier Alibart, Benoit Vicedo...

**The Marseillais,** of birth or circumstance, who have made life more colourful these last few years: Lucas Trottmann, Adrian Moral Saiz, Brian Alejandro Castro Agudelo, Ewen Corre, Vladimir

Laletin, Violaine Chevrier, Tristan Martine, Gérald Souchet, Bernard Borgialli, Jean-Luc Mélenchon, Johan Laflotte, Lucas Matray, Théo Challande, Lucas Variol, Victor Chollat-Namy, Serge Long, Antoine Pinaut, Christina Lasserre, Roméo Bigué, Hugo Maney, Théo Maillet, Timothée Verdier, Raphaël Cohen, Karim Khelfaoui, Jordan Zbili, Maxime Logerot, William et Maëva, Rémi et Cathy, Gisèle Tappero, Julian Augé, Ivan Vidulic, Ante Ivcevic, Alice Billon, Céline Carru, Filippo Pasquali, my students, le théâtre du Carré Rond (Dominique Lamour et Michel Adjriou), Garance Coiffure, Pizza Charly, Lan Thai, O'Bidul (Fabrice Bazin), L'Art de la Fromagerie (Michel Lando), Le Fournil de l'Étoile, la Boulangerie de Quartier, La Brasserie Communale, la Baleine, Les Variétés...

# FOREWORD

> *La mer s'avance insensiblement et sans bruit, rien ne semble se casser, rien ne bouge, l'eau est si loin, on l'entend à peine... Pourtant elle finit par entourer la substance rétive, celle-ci peu à peu devient une presqu'île, puis une île, puis un îlot, qui finit par être submergé à son tour, comme s'il s'était finalement dissous dans l'océan s'étendant à perte de vue...*[1]
>
> — A. Grothendieck in [86] p. 502.

Three years of PhD in Marseilles have taught me to sail over the large sea of loop quantum gravity. From the early canonical formulation of general relativity to the state-of-the-art spin-foams of the EPRL model, a patient work has made me familiar with the basic techniques that have been invented so far to quantize gravity in a background-independent way.

Meanwhile, I acquainted myself with the world of scientific research, and I would like to say a word about it, before delving into physics itself. However noble science may seem, we shall not forget that it is a human enterprise, embedded in the flow of historical contingencies. This PhD manuscript has been written in the context of the COVID-19 pandemic, the first global anguish since WWII, causing the lockdown of more than half of the world population. This time of break has been a time to draw a few human lessons, and I would like to take this opportunity to make some comments on the world of research. As I am just finishing a PhD, my eyes have not yet been accustomed enough not to notice a few disturbing aspects of this world.

The two last centuries have seen an extreme fragmentation of research domains. It probably pertains to a more general division of work, taking place in a competitive global economy. There was a need for experts, a single person gathering an enormous amount of information about a very specific subject. Along this line, I have heard many times that a good PhD student should focus on a very specific question, follow a very narrow path and stick to it. Yet, I have often observed that this approach leads many, either to dogmatism or to disgust. Such advice is dispensed with benevolence, to face two major torments, to publish and to find a job. It is despairing to see how

1 Translation: *The sea advances imperceptibly and without noise, nothing seems to break, nothing moves, the water is so far you hardly hear it... Yet it ends up surrounding the retive substance, little by little, it becomes a peninsula, then an island, then an islet, which ends up being submerged in turn, as if it had finally dissolved in the ocean stretching as far as the eye can see...* In this excerpt, Grothendieck explains his method for solving mathematical problems, as opposed to the 'chisel and hammer' (brute force) method.

much the publicly declared noble intentions of scientists are privately confessed to being overtaken by such petty considerations.

Publishing has become the obsession of any serious researcher. From an anthropological point of view, it has really become the human practice that defines what science is about. But the exponential growth of articles production should not obscure the fact that papers, although cited, are not read. To be honest, they are often hardly readable. As a fetish, they help to convince oneself that *something* has been achieved. As an economic item, they fulfil a purely bibliometric function, showing that their authors are still in the game, that they still *produce*. You are congratulated whenever you publish, whatever you publish. Citations are brandished as certifications that the path is right and that research is making progress. Meanwhile, teaching activities or pedagogical papers are despised for not being original. Innovation has been erected as a new totem, but in practice, it is often just noisy merchandising of old ideas, if not fencing. In short, we publish faster than we think.

This trend is pushed forward by the few historical publishers who dominate the market. They demand and pretend to publish only breakthroughs and highly-original research. It is just one more lie of their economical model, which rests on the free-work of researchers to produce and review articles, and the paywall for them to access the published works. Despite general disapproval of this blatant despoliation of public money, the initiatives to fight it are few, and publishing in *Nature* or *Science* is still a dream for many of us.

In this competitive atmosphere, the specialisation of science has created clans, jealous of their expertise and positions. In that respect, a PhD is an initiatory ritual to enter a closed community, to learn its jargon, to manage its keywords, to hear its gossip and silent grudges too. The allegiance is not the result of a careful pondering of scientific arguments, but rather a bet of trust between human beings.

Finding a job has become a necessary threat for young researchers to comply with the existing system. It justifies some blindspots of our behaviours, like the intense use of air travel, contrasting with the usual diligence of scientists to engage against climate change. What's more, the efficiency of this job market is by far questionable, as so many days of work are wasted into paper-work, filling up dozens of applications, writing hyperbolic cover letters, and fashionable research statements.

My portrait of the economy of science production is certainly partial and somehow exaggerated. But I think its exaggeration goes together with an excess of truth. It is not a lesson of moral, as I am far from being myself exempt from contradictions. But I keep thinking things may be different.

In the last few decades, the fast development of the Internet network has made available to everyone an unprecedented amount of highly specialised knowledge. The paradigm of production and consumption of scientific knowledge has switched from rarity to overabundance. This calls for the rise of wide-ranging physicists, working at the intersection of fields, seeking for structural understanding, conceptual clarity and synthesis more than fashionable bells and whistles. Slow science.

During this PhD, my everyday motivation was nourished by a personal feng shui that seeks constant harmony between conceptual and technical approaches. I wish the future will let me the chance to pursue in my way this millennial quest for understanding Nature.

Marseille, June 2020.

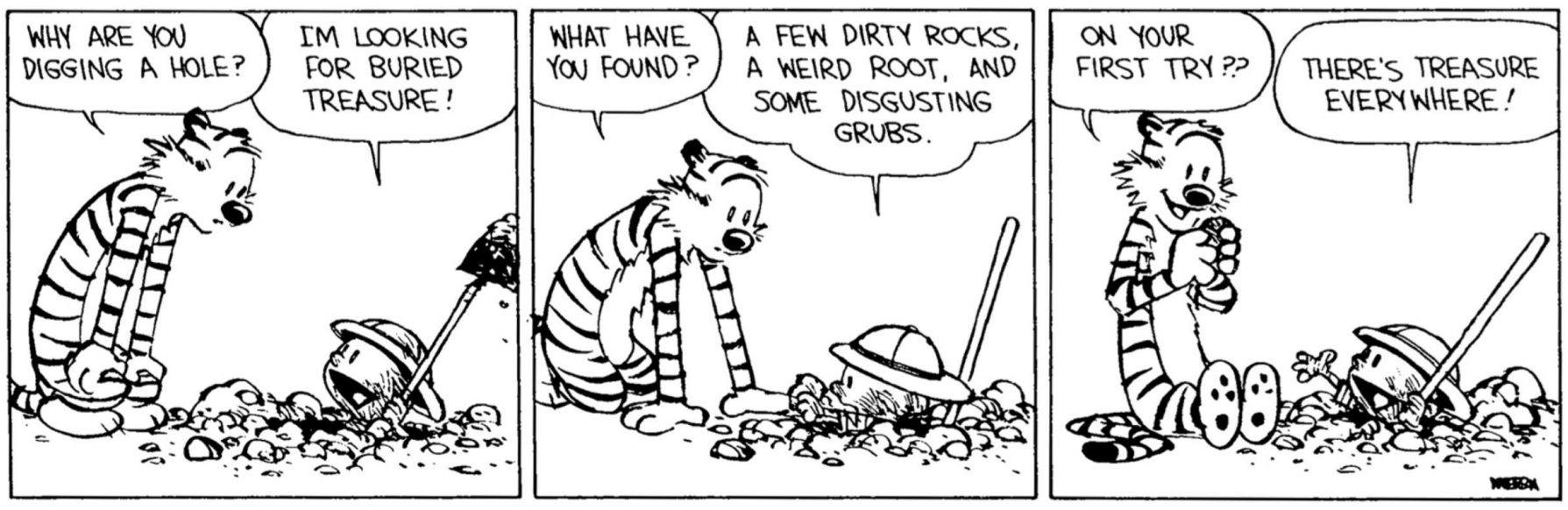

# CONTENTS

# LIST OF FIGURES

# ACRONYMS

CGHS Callan-Giddings-Harvey-Strominger

CSCO Complete Set of Commuting Observables

EPRL Engle-Pereira-Rovelli-Livine

GGV Gel'fand-Graev-Vilenkin

GMS Gel'fand-Minlos-Shapiro

GR General Relativity

LHS Left Hand Side

LQC Loop Quantum Cosmology

LQG Loop Quantum Gravity

QFT Quantum Field Theory

QISS Quantum Information Structure of Space-time

QM Quantum Mechanics

RHS Right Hand Side

RQM Relational Quantum Mechanics

Contrary to a wide-spread academic convention of mysterious origin, we do not capitalise 'figure'. There is no reason to scream in the middle of a perfectly calm sentence. Similarly, we do not capitalise proper adjectives, like 'euclidean', 'galilean', 'cartesian', 'newtonian', 'lagrangian', 'riemannian', 'planckian', 'lorentzian' as their copyrights are outdated and now in public domain.

# NOTATIONS

| | |
|---|---|
| $\stackrel{\text{def}}{=}$ | Equal by definition |
| $\propto$ | Proportional to |
| $\sim$ | Of the order of, scaling as |
| $\cong$ | Isomorphic to |
| $\approx$ | Approximately equal to |
| iff | If and only if |
| s | Second |
| m | Meter |
| kg | Kilogram |
| J | Joule |
| K | Kelvin |
| $G$ | Newton constant, $G \approx 6.7 \times 10^{-11}$ m³ kg⁻¹ s⁻² |
| $\hbar$ | reduced Planck constant, $\hbar \approx 1.1 \times 10^{-34}$ J s |
| $k_B$ | Boltzmann constant, $k_B \approx 1.4 \times 10^{-23}$ J K⁻¹ |
| $c$ | Speed of light in vacuum, $c \approx 3.0 \times 10^{8}$ m s⁻¹ |
| $\gamma$ | Immirzi parameter |
| $l_P$ | Planck length, $l_P \approx 1.6 \times 10^{-35}$ m |
| $\rho_P$ | Planck density, $\rho_P \approx 5.2 \times 10^{96}$ kg m⁻³ |
| $\tau_H$ | Hawking evaporation time |
| $\tau_U$ | Age of the universe, $\tau_U \approx 1.4 \times 10^{10}$ years $\approx 4.4 \times 10^{17}$ s |
| $M_\odot$ | Solar mass, $M_\odot \approx 2.0 \times 10^{30}$ kg |
| $\det M$ | Determinant of $M$ |
| $\operatorname{Tr} M$ | Trace of $M$ |
| $\operatorname{Sp}(M)$ | Spectrum of $M$ |
| $M^\dagger$ | Hermitian conjugate of $M$ |
| $M^T$ | Transpose of $M$ |
| $\epsilon_{ijk}$ | Levi-Civita symbol |
| $\binom{p}{n}$ | Binomial coefficient $\frac{n!}{p!(n-p)!}$ |
| $]a, b[$ | Open interval from $a$ to $b$, sometimes denoted $(a, b)$ |
| $\boldsymbol{z}$ | $(z_0, z_1) \in \mathbb{C}^2$ |
| $\bar{z}$, $z^*$ | Complex conjugate of $z$ |
| $d\Omega^2$ | The usual metric on the 2-sphere, $d\Omega^2 = d\theta^2 + \sin^2\theta \, d\phi^2$ |
| $\Box$ | d'Alembert operator $\Box = \nabla_a \nabla^a$ |
| $W$ | Lambert $W$-function |

# INTRODUCTION

Greek temples all share the same basic architecture. The inner chamber, called the *naos*, shelters the deity. In the 247 pages of this thesis, the naos is located chapter 15. In the unfortunate event of an earthquake, it is the room more likely to resist. The tourist in a rush may well go directly to it, although his eyes may not have had time to accommodate to the obscurity of the place. He may not be able to appreciate the colours of the paintings, nor the subtle delicateness of the burning incenses.

Our deity is a planckian black-to-white hole. It is so small that you would not even notice its presence if it were sitting in front of you. However, their presence in the pantheon might be required for the harmony of the universe. Arguments from quantum gravity suggest it could be the legitimate heir of black holes. As such, they might release the information that their parents had trapped, and their potential abundance might balance the missing dark matter.

But let's be honest, the deity is never really in the naos. Only a statue of it can be found. In our case, you will find a *model* of a black-to-white hole. However blessed it may be, the materiality of the statue does not guarantee the actual existence of the deity. Similarly, our model is not a proof of physical existence, however close it may be to solve the semi-classical Einstein equations. It is a special kind of temple that we are entering in, where doubt is prefered to faith.

It has taken a hundred years for black holes to finally show their dark face after they were unconsciously described by Schwarzschild in 1916. On 10 April 2019, the team of the Event Horizon Telescope revealed to the world the first image ever taken of a black hole. It has come as the crowning achievement of a series of evidence, accumulated since the 1970s, of the existence of such obscure objects.

But now that their existence is unanimously acknowledged, physicists want to know how they die. Hawking has shown convincingly that black holes evaporate, emitting thermal radiation which shrinks them very slowly. About the final stages of the evaporation, myths are many. If only quantum gravity was an accomplished science, the answer would probably be revealed. But such a Grail has not been found yet, and we shall content ourselves with preliminary incantations, like Quantum Field Theory (QFT) in curved space-time. Here, we defend a scenario suggesting the ultimate metamorphosis of black holes into long-lived white holes. If not yet proven, we are convinced that such a scenario deserves consideration, if not ritual sacrifice.

Before reaching chapter 15, the visitor is advised to spend time in the preceding chapters of part II, the *pronaos*, that is the porch, where to get prepared to meet the deity. The way is paved, step by step, to introduce ex-votos and ritual spells: conformal diagrams (a sacred penrosian pictorial art of general relativity), semi-classical Einstein equations (an iterative approach to quantum gravity), the information-loss paradox (the main scholastic controversy of the last few decades), white holes (the forgotten time-reverse of black holes).

The *crepidoma* is the multilevel platform over which the temple is erected. Like heavy blocks of marbles, each chapter of part I is a relief for the overall structure. It lays a solid mathematical ground for a sound physical building. It is a necessary mathematical propaedeutic, introducing a propitiatory language to ease later physical discussions. It essentially focuses on the two main groups of interest for quantum gravity, SU(2) and $SL_2(\mathbb{C})$, going through their representation theory, recoupling theory, and harmonic analysis. The crepidoma takes time to be built, and no temple is possible without it, but once it is done, it remains almost unnoticed to physicists, who walk upon it quietly.

The naos is not the ultimate goal of our visit. After the cult statue is met, there comes a time for prayers, gathered in part III. In a sense, it is still unfinished work. The beautiful formalism of spin-foams and coherent states is invoked in the hope of a miracle: the computation of the amplitude of the black-to-white transition. Such a miracle has not occurred yet, but the prayers are going on, and faith is increasing.

The core of the temple is surrounded by a hallway of columns, the *peristasis*. It offers shade and rest to visitors. The temple sits upon the acropolis of theoretical physics, rising in the cradle of philosophy. In the two chapters of part IV, the reader is invited to contemplate the landscape. The meditation starts with the concept of *relationality*: Nature is seen through the eyes of an observer. The notion is subtle and it sheds light upon the meaning of a quantum amplitude in a background independent theory of space-time. The second chapter clarifies the meaning of the related notion of *(non)-locality*.

There are many temples dedicated to black holes. Many have fallen into ruins, but most have contributed, to some extent, to the collective pursuit of knowledge. In this thesis, I have tried to clarify the mathematical description of the evaporating black-to-white hole, with special attention to related foundational issues. If the journey has been enlightening for me, I hope this manuscript will be useful to others.

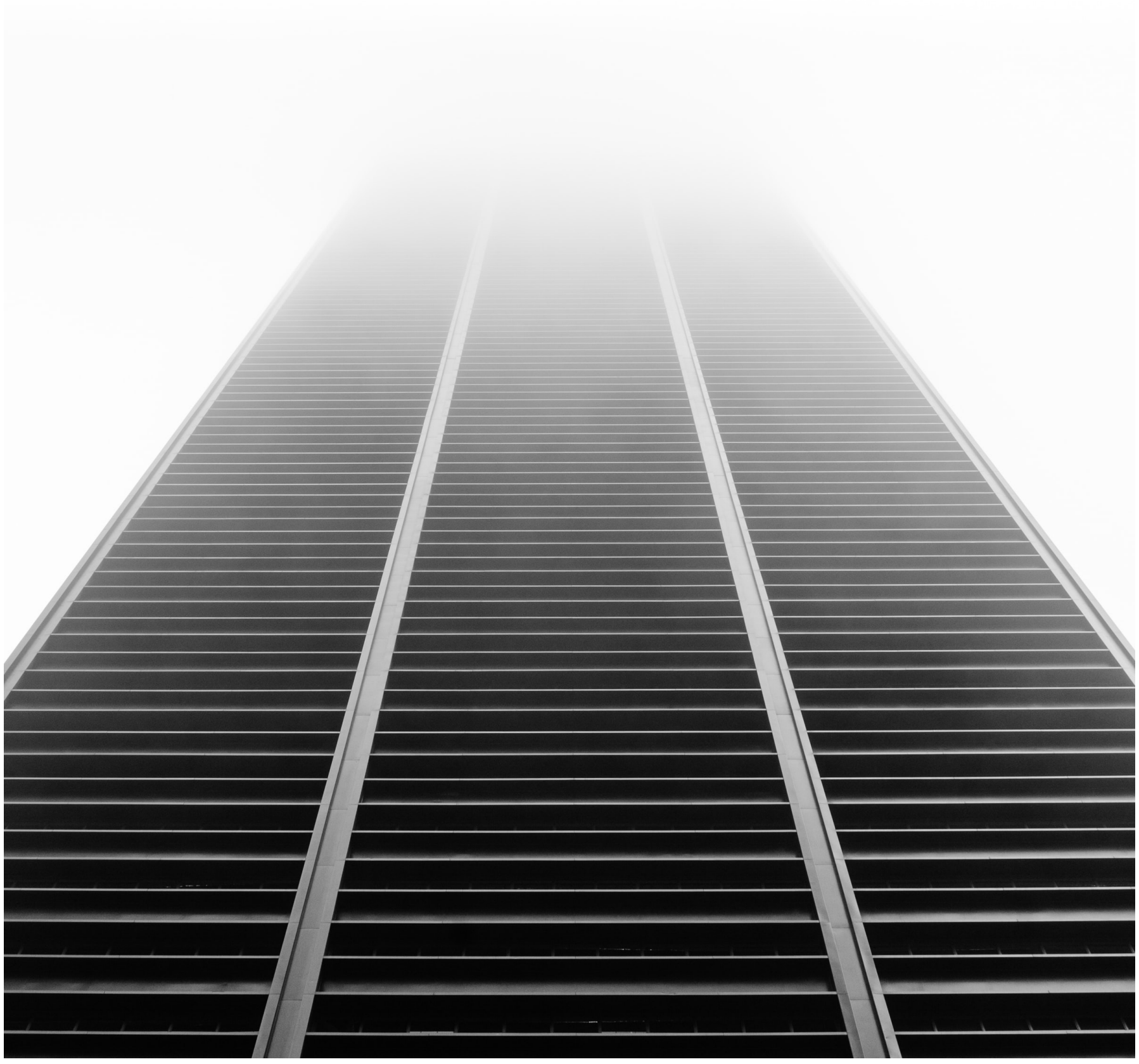

Part I

# MATHEMATICAL PROPAEDEUTIC

The central role of group theory in physics has been largely revealed in the modern theories of the 20th century. Quantum gravity makes no exception. The two main groups of interest for quantum gravity are $SL_2(\mathbb{C})$ and its subgroup SU(2). This may seem natural since $SL_2(\mathbb{C})$ is, in some sense, the 'quantum version' of the restricted Lorentz group $SO^+(3,1)$, which is an important symmetry group of Minkowski spacetime, and similarly, SU(2) is the 'quantum version' of the group of space rotations; but the reason why these groups come out in quantum gravity is actually more subtle (see chapter 16).

Even though many monographs exist devoted to this theory, the different tools needed (e.g. representation theory, harmonic analysis, recoupling theory...) are often dispersed in different books, with different conventions and notations. This was the initial motivation for the compilation of the review '*A primer of group theory for Loop Quantum Gravity and spin-foams*' [122]. It served three main purposes:

1. A concise introduction for students to the essential mathematical tools of Loop Quantum Gravity (LQG). It bridges a gap between the level of students at the end of a master programme, and the minimum level required to start doing research in LQG.
2. A convenient compendium for researchers. Instead of having each formula in a different heavy, old book, the most useful ones were gathered in a short toolbox.
3. A translational hub between the conventions of the main references. For many notions, each author tends to use their own notations, which makes it difficult to switch easily from one reference to another. The review shows explicitly how they relate to one another.

This first part is a reworking of the review [122]. However, to keep it concise, this thesis has been deprived of the latter translational role. Also, there are fewer reminders and proofs. Nevertheless, it has been completed with three chapters (1, 3, 6) absent of [122].

Although most of the technical content is not new, the overall compilation is. This part offers mathematical support for the three following parts, although it is clear that not all of it will be used later. It could have been an appendix, but appendices are botched, whereas introductory chapters are well-groomed. Appendices are never read, while this part may be worth it. To a large extent, it is interesting per se.

The plan is the following:

Ch. 1 discusses the foundational notion of equality.

Ch. 2 wraps up all the basics of SU(2) and $SL_2(\mathbb{C})$, the two Lie groups of main interest for quantum gravity.

Ch. 3 lights up some geometrical aspects of spheres.

Ch. 4 catalogues various possible realisations SU(2)-irreps used in the literature.

Ch. 5 condenses the main results of the recoupling theory of SU(2).

Ch. 6 shows how functions over SU(2) can be decomposed in harmonics.

Ch. 7 renders all the flourish of the representations of $SL_2(\mathbb{C})$.

Ch. 8 attempts to generalise recoupling theory to $SL_2(\mathbb{C})$.

# 1 EQUALITY

Since its axiomatic formalisation, notably carried out by Zermelo and Fraenkel, modern mathematics is based on the notion of *set*. According to the usual bourbakian reconstruction, sets are endowed with *structures*, which turn them into *spaces*. However, this way does not permit to recover exactly the intuitive notion of *mathematical object*. Indeed, different spaces can describe the same mathematical object. In fact, different descriptions of the same object are related by *isomorphisms*, that are bijections which preserve the structures.

A simple example is provided by *the circle*. For physicists, there is no doubt about what a circle is. The mathematicians, who are both praised and mocked for giving precise definitions, do surely have a good definition of a circle. However, a fair sample of mathematicians will not give you one, but two definitions of a circle. The real mathematicians define it as a submanifold of the plane $\mathbb{R}^2$:

$$S^1 \stackrel{\text{def}}{=} \left\{(x,y) \in \mathbb{R}^2 \mid x^2 + y^2 = 1\right\}. \tag{1}$$

The complex mathematicians define it as a subgroup of $\mathbb{C}$:

$$U(1) \stackrel{\text{def}}{=} \{\lambda \in \mathbb{C} \mid |\lambda| = 1\}, \tag{2}$$

$S^1$ and $U(1)$ are different sets. They are even different spaces, as the former is a differentiable manifold, and the latter a group. Nevertheless, both deserve the name of 'circle' and they can be regarded as the 'same thing' through the following isomorphism

$$f : \begin{cases} S^1 & \to & U(1) \\ (x,y) & \mapsto & x + iy \end{cases} \tag{3}$$

It is an isomorphism, as it enables to translate the group and the manifold structures from one set to another. It is much more than being only a bijection. A bijection is quite a weak requirement, as it only preserves the cardinal of sets, so that the circle is also in bijection for instance with the disk. Mathematicians have fancy names to distinguish all the kinds of isomorphism:

- Bijection, between sets;
- Homeomorphism, between topological spaces;
- Isometry, between metric spaces;
- Homomorphism, between groups;

- Diffeomorphism, between differentiable manifolds.

In all cases, we denote

$$A \cong B \tag{4}$$

to signify the existence of an isomorphism between $A$ and $B$, which kind of isomorphism should be clear from the context.

Since $S^1$ and $U(1)$ share all the same structures, it is tempting to say that we should regard them as really the same object, and write $S^1 = U(1)$. This idea is reminiscent of Leibniz definition of equality:

> $x = y$, *if and only if,* $x$ *and* $y$ *have all the same properties.*

Unfortunately, his definition is too fuzzy to be useful, as $x$ and $y$ can *never* share *all* the same properties, just because, for instance, '$x$' and '$y$' are not written alike. The domain of properties has to be restricted to get a consistent definition of equality. As a result, even in mathematics, the meaning of the symbol '$=$' is not as sharp as people usually believe, it is always some kind of $\approx$. In our case, a mathematician refuses to write $S^1 = U(1)$ as long as he refuses to write $\mathbb{R}^2 = \mathbb{C}$. But the difference between '$=$' and '$\cong$' is inessential. When many structures are shared, it is fine to write simply $A = B$.

Our brain is a champion for performing this kind of identification, but a computer trying to do the same would often run into 'typing' errors. Category theory provides an appropriate language for speaking about these identifications. On the contrary, the theory of sets suffers from a certain formalist rigidity due to its requirement to define all objects based on the notion of set.

Some other people would probably say that $S^1$ and $U(1)$ are actually two incarnations of a third object, which is, *really*, the circle. In this view, the circle belongs to the platonic world of Ideas, and only contingent witnesses of it are seen *in real life*. Such a discussion would have been a delight for medieval scholasticism, but is useless for proving theorems about circles.

Physicists are usually more flexible with notations and definitions, as long as 'it is clear what it means'. As Feynman says

> *We cannot define anything precisely. If we attempt to, we get into that paralysis of thought that comes to philosophers, who sit opposite each other, one saying to the other, 'You don't know what you are talking about!'. The second one says, 'What do you mean by know? What do you mean by talking? What do you mean by you?' ([65], lecture 8)*

My opinion is more qualified. The flexibility with definitions and notations can be a strength when it makes us agile to juggle with concepts, but it is a weakness when it blurs the beauty of details. It took me a while to understand this simple lesson, maybe because it is not often said explicitly. So I thought someone would appreciate to read it here someday.

# 2 WARMUP

As we wanted this thesis to be almost self-contained, this chapter is a melting pot of the basic algebraic mathematical tools that will be later used extensively. It also fixes many of the notations. If you already feel warmed up, you would do well skipping this chapter. If you have never seen these notions in your life, you would do better to first learn them with an introductory book. Good ones are for instance Knapp [113], Hall [92], and Bernard-Laszlo-Renard [30].

## 2.1 BASICS OF $SL_2(\mathbb{C})$

$\mathcal{M}_2(\mathbb{C})$ is the algebra of $2 \times 2$ complex matrices. It is an *algebra* because it is a vector space (with addition of matrices) endowed with a bilinear product (the usual matrix product).

$GL_2(\mathbb{C})$ is a *linear group* defined by

$$GL_2(\mathbb{C}) \stackrel{\text{def}}{=} \{M \in \mathcal{M}_2(\mathbb{C}) \mid \det M \neq 0\}. \tag{5}$$

It is a 4-dimensional complex Lie group.

$\mathfrak{gl}_2(\mathbb{C})$ is the Lie algebra of $GL_2(\mathbb{C})$. It is actually isomorphic to $\mathcal{M}_2(\mathbb{C})$, when it is endowed with the Lie product $[M, N] = MN - NM$.

$SL_2(\mathbb{C})$ is a *special linear group*, defined by

$$SL_2(\mathbb{C}) \stackrel{\text{def}}{=} \{M \in GL_2(\mathbb{C}) \mid \det M = 1\}. \tag{6}$$

It is a 3-dimensional complex Lie sub-group of $GL_2(\mathbb{C})$. Topologically, $SL_2(\mathbb{C})$ is not compact but it is simply connected.

$\mathfrak{sl}_2(\mathbb{C})$ is the Lie algebra of $SL_2(\mathbb{C})$. One can show that

$$\mathfrak{sl}_2(\mathbb{C}) = \{M \in \mathcal{M}_2(\mathbb{C}) \mid \operatorname{Tr} M = 0\}. \tag{7}$$

It is a 3-dimensional complex Lie sub-algebra of $\mathfrak{gl}_2(\mathbb{C})$.

$\sigma_1, \sigma_2, \sigma_3$ are the *Pauli matrices*, defined by

$$\sigma_1 \stackrel{\text{def}}{=} \begin{pmatrix} 0 & 1 \\ 1 & 0 \end{pmatrix}, \quad \sigma_2 \stackrel{\text{def}}{=} \begin{pmatrix} 0 & -i \\ i & 0 \end{pmatrix}, \quad \sigma_3 \stackrel{\text{def}}{=} \begin{pmatrix} 1 & 0 \\ 0 & -1 \end{pmatrix}. \tag{8}$$

They form a basis of $\mathfrak{sl}_2(\mathbb{C})$. Interestingly, they satisfy

$$[\sigma_i, \sigma_j] = 2i\, \epsilon_{ijk}\, \sigma_k. \tag{9}$$

*Here and everywhere else, Einstein notation is understood over repeated indices.*

With the identity matrix $\mathbb{1}$, the Pauli matrices also provide a basis of the *complex* vector space $\mathcal{M}_2(\mathbb{C})$: any $a \in \mathcal{M}_2(\mathbb{C})$ can be written uniquely

$$a = a_0 \mathbb{1} + \sum_{k=1}^{3} a_k \sigma_k \quad \text{with} \quad a_0, a_1, a_2, a_3 \in \mathbb{C}. \tag{10}$$

Note that in this basis, the determinant reads

$$\det a = a_0^2 - a_1^2 - a_2^2 - a_3^2. \tag{11}$$

$H_2(\mathbb{C})$ is the *real* vector space of $2 \times 2$ *hermitian* matrices, defined by

$$H_2(\mathbb{C}) \stackrel{\text{def}}{=} \{M \in \mathcal{M}_2(\mathbb{C}) \mid M^\dagger = M\}. \tag{12}$$

A basis is also given by the Pauli matrices: any $h \in H_2(\mathbb{C})$ can be written uniquely as

$$h = h_0 \mathbb{1} + \sum_{k=1}^{3} h_k \sigma_k \quad \text{with} \quad h_0, h_1, h_2, h_3 \in \mathbb{R}. \tag{13}$$

$H_2^{++}(\mathbb{C})$, the set of $2 \times 2$ hermitian *positive-definite* matrices, is

$$H_2^{++}(\mathbb{C}) \stackrel{\text{def}}{=} \{M \in H_2(\mathbb{C}) \mid \forall \lambda \in Sp(H), \quad \lambda > 0\}, \tag{14}$$

with $Sp(H)$, the *spectrum* of H, i. e. the set of its eigenvalues.

## 2.2 SPACETIME SYMMETRIES

$\mathbb{M}$ is the spacetime of special relativity, called *Minkowski spacetime*. Mathematically, it is the vector space $\mathbb{R}^4$, endowed with a *lorentzian inner product*, whose signature is either $(-,+,+,+)$ (general relativists convention) or $(+,-,-,-)$ (particle physicists convention).

$\mathbb{P}$ is the group of all isometries (distance-preserving transformations) of $\mathbb{M}$, called the *Poincaré group* (or sometimes the *inhomogeneous Lorentz group*).

O(3,1) is the linear subgroup of isometries that leave the origin fixed, called the *Lorentz group* (or sometimes the *homogeneous Lorentz group*), and sometimes also denoted O(1,3). The Poincaré group $\mathbb{P}$ can be decomposed as a semi-direct product $\mathbb{P} = O(3,1) \ltimes \mathbb{R}^4$. O(3,1) is composed of four connected components related to each other by the operators of parity (space-reversal) and time-reversal.

$SO^+(3,1)$ is the identity component of O(3,1). It forms a subgroup made of transformations that preserves the orientation and the direction of time. It is called the *proper orthochronous Lorentz group*, or the *restricted Lorentz group*.

As a real vector space, Minkowski spacetime $\mathbb{M}$ is isomorphic to $H_2(\mathbb{C})$, with the map

$$X = (t, x, y, z) \mapsto h = t\mathbb{1} + x\sigma_1 + y\sigma_2 + z\sigma_3 = \begin{pmatrix} t+z & x-iy \\ x+iy & t-z \end{pmatrix}. \quad (15)$$

The inverse map is given by

$$h \mapsto X = \frac{1}{2}(\mathrm{Tr}\, h, \mathrm{Tr}\, h\sigma_1, \mathrm{Tr}\, h\sigma_2, \mathrm{Tr}\, h\sigma_3), \quad (16)$$

and the pseudo-scalar product (with convention $(-,+,+,+)$)

$$\begin{aligned} X \cdot X' &= -tt' + xx' + yy' + zz' \\ &= \frac{1}{4}\mathrm{Tr}\left(hh' - h\sigma_1 h'\sigma_1 - h\sigma_2 h'\sigma_2 - h\sigma_3 h'\sigma_3\right) \end{aligned} \quad (17)$$

Note that the pseudo-norm of $\mathbb{M}$ is mapped to the determinant over $H_2(\mathbb{C})$:

$$X \cdot X = -\det h. \quad (18)$$

From the latter property, we see that the action of $a \in SL_2(\mathbb{C})$ upon $h \in H_2(\mathbb{C})$, given by

$$h \mapsto aha^\dagger, \quad (19)$$

defines a linear isometry on $\mathbb{M}$. Thus, it defines a homomorphism between $SO^+(3,1)$ and $SL_2(\mathbb{C})$, and it is easy to show the following isomorphism of groups

$$SL_2(\mathbb{C})/Z_2 \cong SO^+(3,1). \quad (20)$$

with $Z_2 = \{\mathbb{1}, -\mathbb{1}\}$, the 2-element group. $SL_2(\mathbb{C})$ is said to be the double cover, or the universal cover, of $SO^+(3,1)$. For that reason it is sometimes called the *Lorentz spin group*. This gives a first glimpse on the role of $SL_2(\mathbb{C})$ in fundamental physics.

## 2.3 SUB-GROUPS OF $SL_2(\mathbb{C})$

There are many sub-groups of $SL_2(\mathbb{C})$. We describe below the main ones. The figure 1 shows the relations of inclusion between them.

$SU(2)$, the unitary special group, is defined by:

$$SU(2) \stackrel{\mathrm{def}}{=} \left\{u \in SL_2(\mathbb{C}) \mid u^\dagger u = \mathbb{1}\right\}. \quad (21)$$

Any $u \in SU(2)$ can be uniquely written as

$$u = u_0 e + i\sum_{k=1}^{3} u_k\sigma_k \quad \text{with} \quad u_0, u_1, u_2, u_3 \in \mathbb{R}$$
$$\text{and} \quad \sum_{k=0}^{3} u_k^2 = 1. \quad (22)$$

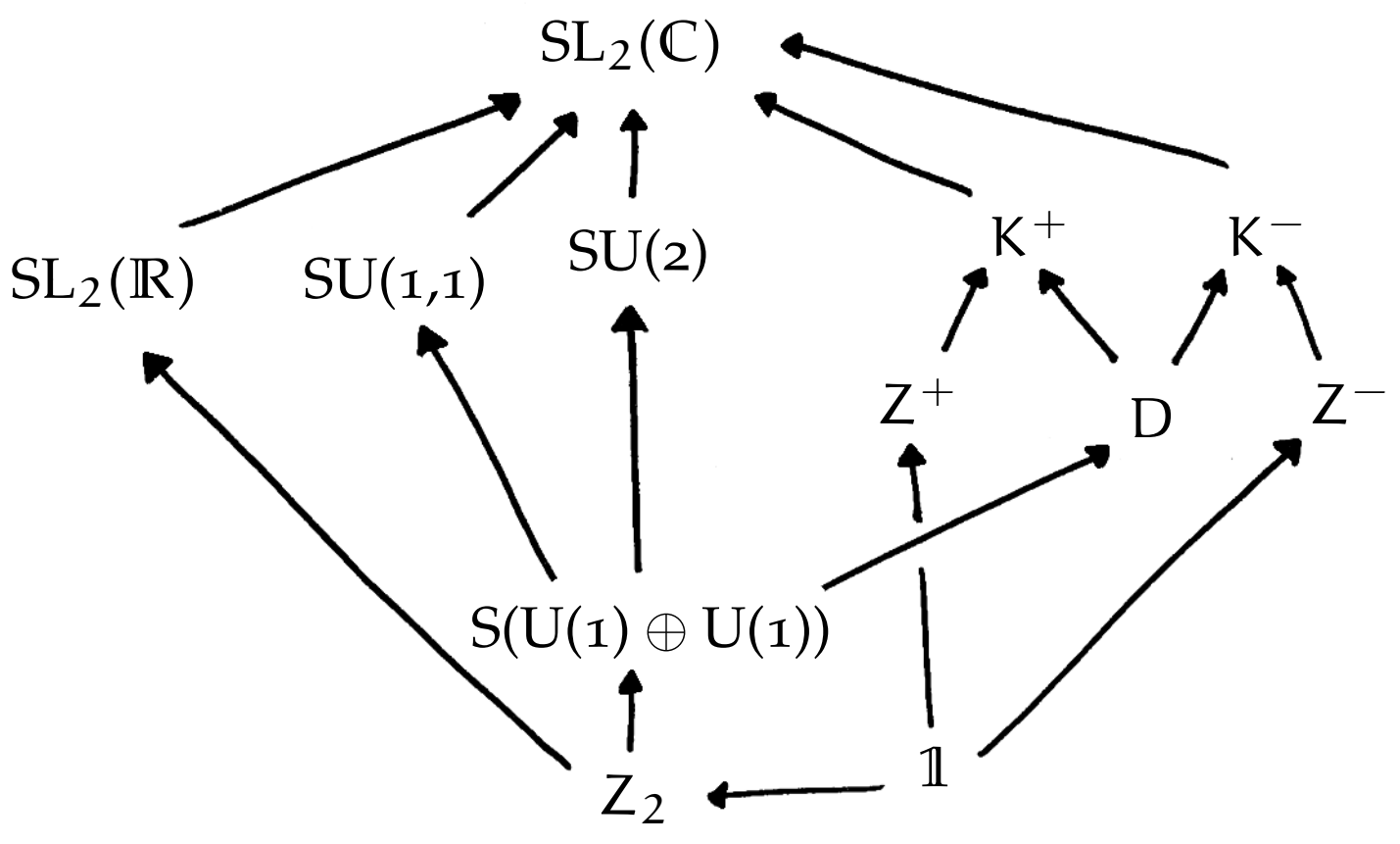

Figure 1: This graph represents the relations of inclusions between the subgroups of $SL_2(\mathbb{C})$.

Through the isomorphism (15) and the action (19), the definition (21) enables us to see SU(2) as the stabiliser (also called *little group* or *isotropy group*) of the unit time vector $(1,0,0,0)$. Physically, it means that SU(2) only acts over the space, and not in the time direction. Choosing another time direction, related to $(1,0,0,0)$ by a boost $\Lambda$, would have defined another stabilizer, isomorphic to SU(2), which makes physicists sometimes talk of *a* SU(2), as if there were several.

SU(1,1) is defined by:

$$\mathrm{SU(1,1)} \stackrel{\text{def}}{=} \left\{ v \in SL_2(\mathbb{C}) \mid v^\dagger \sigma_3 v = \sigma_3 \right\}. \tag{23}$$

Any $v \in$ SU(1,1) can be uniquely written as:

$$v = v_0 e + v_1 \sigma_1 + v_2 \sigma_2 + i v_3 \sigma_3 \quad \text{with} \quad v_0, v_1, v_2, v_3 \in \mathbb{R} \quad \text{and} \quad v_0^2 - v_1^2 - v_2^2 + v_3^2 = 1. \tag{24}$$

Similarly to the SU(2) case, SU(1,1) can be understood by its action in Minkoswki spacetime as the stabiliser of $(0,0,0,1)$.

$SL_2(\mathbb{R})$, the real linear special group, is defined by

$$SL_2(\mathbb{R}) \stackrel{\text{def}}{=} \{ a \in \mathcal{M}_2(\mathbb{R}) \mid \det a = 1 \}, \tag{25}$$

and interestingly it is also

$$SL_2(\mathbb{R}) = \left\{ a \in SL_2(\mathbb{C}) \mid a^\dagger \sigma_2 a = \sigma_2 \right\}. \tag{26}$$

Any $a \in$ SU(1,1) can be uniquely written as:

$$a = a_0 e + a_1 \sigma_1 + i a_2 \sigma_2 + a_3 \sigma_3 \quad \text{with} \quad a_0, a_1, a_2, a_3 \in \mathbb{R} \quad \text{and} \quad a_0^2 - a_1^2 + a_2^2 - a_3^2 = 1. \tag{27}$$

Again $SL_2(\mathbb{R})$ can be understood by its action in Minkoswki spacetime as the stabiliser of $(0,0,1,0)$.

$K_+$ AND $K_-$, the upper and lower triangular groups, are defined by:

$$\begin{aligned} K_+ &\stackrel{\text{def}}{=} \left\{ \begin{pmatrix} \lambda^{-1} & \mu \\ 0 & \lambda \end{pmatrix} \mid \lambda \in \mathbb{C}^* \quad \text{and} \quad \mu \in \mathbb{C} \right\} \\ K_- &\stackrel{\text{def}}{=} \left\{ \begin{pmatrix} \lambda^{-1} & 0 \\ \mu & \lambda \end{pmatrix} \mid \lambda \in \mathbb{C}^* \quad \text{and} \quad \mu \in \mathbb{C} \right\}. \end{aligned} \tag{28}$$

They are also called the *Borel sub-groups* or the *parabolic sub-groups*.

$Z_+$ AND $Z_-$, are defined by

$$\begin{aligned} Z_+ &\stackrel{\text{def}}{=} \left\{ \begin{pmatrix} 1 & z \\ 0 & 1 \end{pmatrix} \mid z \in \mathbb{C} \right\} \\ Z_- &\stackrel{\text{def}}{=} \left\{ \begin{pmatrix} 1 & 0 \\ z & 1 \end{pmatrix} \mid z \in \mathbb{C} \right\}. \end{aligned} \tag{29}$$

$D$, the diagonal group is defined by

$$D \stackrel{\text{def}}{=} \left\{ \begin{pmatrix} \delta & 0 \\ 0 & \delta^{-1} \end{pmatrix} \mid \delta \in \mathbb{C}^* \right\}. \tag{30}$$

$S(U(1) \oplus U(1))$, defined by

$$S(U(1) \oplus U(1)) \stackrel{\text{def}}{=} \left\{ \begin{pmatrix} e^{i\theta} & 0 \\ 0 & e^{-i\theta} \end{pmatrix} \mid \theta \in \mathbb{R} \right\}. \tag{31}$$

It is the maximal torus (i. e. the biggest compact, connected, abelian Lie subgroup) of $SU(2)$. We have obviously

$$S(U(1) \oplus U(1)) \cong U(1). \tag{32}$$

$Z_2$, the center, defined as the subset of $SL_2(\mathbb{C})$ which commute with all $SL_2(\mathbb{C})$, is shown to be

$$Z_2 = \{\mathbb{1}, -\mathbb{1}\}. \tag{33}$$

Since it is a normal subgroup (as any center of any group), the quotient $SL_2(\mathbb{C})/Z_2$ is also a group, which can be shown to be isomorphic to the restricted Lorentz group $SO^+(3,1)$, as was already said in equation (20).

*A subgroup* $H \subset G$ *is said to be normal if* $ghg^{-1} \in H$, *for all* $g \in G$ *and* $h \in H$.

## 2.4 DECOMPOSITION OF $SL_2(\mathbb{C})$

The structural properties of a matrix group can be grasped through the study of its decompositions. We are going to present four different decompositions of $SL_2(\mathbb{C})$.

■ **Polar decomposition.** For all $M \in GL_2(\mathbb{C})$, there exists a unique unitary matrix $U \in U(2)$ and a unique positive-definite hermitian matrix $H \in H_2^{++}(\mathbb{C})$ such that:

$$M = HU. \tag{34}$$

Remarks:

1. The order does not matter, and the theorem would also be true with $M = UH$.
2. If $M \in SL_2(\mathbb{C})$, then $U \in SU(2)$ and $\det H = 1$.
3. It is called 'polar' because it is a generalisation of the polar decomposition of complex numbers $z = re^{i\theta}$. It can be generalised further to any $GL_n(\mathbb{C})$.

■ **Cartan decomposition.** For all $g \in SL_2(\mathbb{C})$, there exists $u, v \in SU(2)$ and $r \in \mathbb{R}_+$ such that:

$$g = u\, e^{r\sigma_3/2}\, v^{-1}. \tag{35}$$

Remarks:

1. The number $r$ is called *the rapidity of the boost along the axis* $z$.
2. This theorem can be generalized to the case of $SL_n(\mathbb{C})$.
3. The rapidity $r$ is uniquely determined but $u$ and $v$ are not. The other possible choices are $(u\, e^{i\theta\sigma_3}, v\, e^{i\theta\sigma_3})$, with $\theta \in \mathbb{R}$.
4. The polar decomposition of $SL_2(\mathbb{C})$ is a particular case of the Cartan decomposition where $v^{-1} = \mathbb{1}$. This requirement makes it unique.

■ **Gauss decomposition.** Let $g \in SL_2(\mathbb{C})$ such that $g_{22} \neq 0$. There exists a unique triplet $(z_+, d, z_-) \in Z_+ \times D \times Z_-$ such that

$$g = z_+ d z_- \tag{36}$$

■ **Iwasawa decomposition.** For any matrix $M \in SL_2(\mathbb{C})$, there exists a unique triplet $(Z, D, U) \in Z_+ \times D_{\mathbb{R}_+} \times SU(2)$ such that

$$M = ZDU = \begin{pmatrix} 1 & z \\ 0 & 1 \end{pmatrix} \begin{pmatrix} \lambda^{-1} & 0 \\ 0 & \lambda \end{pmatrix} \begin{pmatrix} \alpha & -\beta^* \\ \beta & \alpha^* \end{pmatrix} \tag{37}$$

with $(z, \lambda, \alpha, \beta) \in \mathbb{C} \times \mathbb{R}_+^* \times \mathbb{C}^2$.

## 2.5 BASICS OF SU(2)

Let us now focus on the special unitary subgroup

$$\mathrm{SU}(2) \stackrel{\text{def}}{=} \left\{ u \in \mathrm{SL}_2(\mathbb{C}) \mid u^\dagger u = \mathbb{1} \right\}. \tag{38}$$

It is a 3-dimensional real Lie subgroup of the 6-dimensional real Lie group $\mathrm{SL}_2(\mathbb{C})$. Any $u \in \mathrm{SU}(2)$ can be uniquely written as

$$u = \begin{pmatrix} \alpha & -\beta^* \\ \beta & \alpha^* \end{pmatrix} \quad \text{with} \quad (\alpha, \beta) \in \mathbb{C}^2, |\alpha|^2 + |\beta|^2 = 1, \tag{39}$$

or equivalently

$$u = \begin{pmatrix} a + ib & -c + id \\ c + id & a - ib \end{pmatrix} \quad \text{with} \quad (a, b, c, d) \in \mathbb{R}^4 \quad \text{and} \quad a^2 + b^2 + c^2 + d^2 = 1. \tag{40}$$

The latter expression shows that SU(2) is diffeomorphic to $S^3$, the unit sphere of $\mathbb{R}^4$. Therefore it is connected, simply connected and compact. The center of SU(2) is $Z_2$ and the quotient $\mathrm{SU}(2)/Z_2$ is a group, which happens to be isomorphic to SO(3) (see section 2.6). The real Lie algebra of SU(2) is

$$\mathfrak{su}(2) = \left\{ M \in \mathcal{M}_2(\mathbb{C}) \mid M^\dagger = -M \text{ and } \operatorname{Tr} M = 0 \right\}. \tag{41}$$

It is a real vector space, of which a basis is given by $(i\sigma_1, i\sigma_2, i\sigma_3)$. Since SU(2) is a compact Lie group, any element of SU(2) can be written (non uniquely) as the exponential of an element of the associated Lie algebra $\mathfrak{su}(2)$ (it is a general theorem for compact Lie groups).

■ **Exponential decomposition.** If $u \in \mathrm{SU}(2)$, there exists a (non-unique) $\vec{\alpha} \in \mathbb{R}^3$ such that

$$u = e^{i\vec{\alpha}\cdot\vec{\sigma}} = \cos\|\vec{\alpha}\|\mathbb{1} + i\sin\|\vec{\alpha}\|\frac{\vec{\alpha}\cdot\vec{\sigma}}{\|\vec{\alpha}\|}. \tag{42}$$

■ **Euler angles decomposition.** For all $u \in \mathrm{SU}(2)$, there exists $\alpha, \beta, \gamma \in \mathbb{R}$ (called Euler angles) such that:

$$u = e^{-\frac{i\alpha}{2}\sigma_3} e^{-\frac{i\beta}{2}\sigma_2} e^{-\frac{i\gamma}{2}\sigma_3} \tag{43}$$

The choice can be made unique by restricting the domain of definition of the angles, with for instance $\alpha \in ]-2\pi, 2\pi[$, $\beta \in [0, \pi]$ and $\gamma \in [|\alpha|, 4\pi - |\alpha|[$. Explicitly the Right Hand Side (RHS) is

*We use the French notation for open intervals, that is $]a, b[$ instead of the more familiar $(a, b)$ on the other side of the pond.*

$$\begin{pmatrix} e^{-\frac{i(\alpha+\gamma)}{2}}\cos\beta/2 & -e^{\frac{i(\gamma-\alpha)}{2}}\sin\beta/2 \\ e^{-\frac{i(\gamma-\alpha)}{2}}\sin\beta/2 & e^{\frac{i(\alpha+\gamma)}{2}}\cos\beta/2 \end{pmatrix}. \tag{44}$$

## 2.6 THE ROTATIONS SO(3)

As we have said in section 2.3, the action of SU(2) over Minkowski spacetime, given by equation (19), preserves the time direction. Then, SU(2) acts on the spatial dimensions as the group of rotations over the euclidean space $\mathbb{R}^3$. The group of rotations of $\mathbb{R}^3$ is

$$SO(3) \stackrel{\text{def}}{=} \left\{ M \in \mathcal{M}_3(\mathbb{R}) \mid M^T M = \mathbb{1} \text{ and } \det M = 1 \right\}. \tag{45}$$

It is a Lie group whose Lie algebra is

$$\mathfrak{so}(3) = \left\{ M \in \mathcal{M}_3(\mathbb{R}) \mid M^T + M = 0 \text{ and } \operatorname{Tr} M = 0 \right\}. \tag{46}$$

We can show the following isomorphism of Lie algebra

$$\mathfrak{so}(3) \cong \mathfrak{su}(2). \tag{47}$$

Besides we have the following group isomorphism:

$$SO(3) \cong SU(2)/Z_2. \tag{48}$$

This can be seen with the map $g$, which sends the SU(2) matrix

$$u = \begin{pmatrix} \alpha & -\beta^* \\ \beta & \alpha^* \end{pmatrix} \tag{49}$$

to the SO(3) matrix

$$g(u) = \begin{pmatrix} \frac{1}{2}(\alpha^2 + \alpha^{*2} - \beta^2 - \beta^{*2}) & \frac{i}{2}(\alpha^2 - \alpha^{*2} - \beta^2 + \beta^{*2}) & \alpha\beta^* + \alpha^*\beta \\ \frac{i}{2}(-\alpha^2 + \alpha^{*2} - \beta^2 + \beta^{*2}) & \frac{1}{2}(\alpha^2 + \alpha^{*2} + \beta^2 + \beta^{*2}) & i(-\alpha\beta^* + \alpha^*\beta) \\ -\alpha\beta - \alpha^*\beta^* & i(-\alpha\beta + \alpha^*\beta^*) & \alpha\alpha^* - \beta\beta^* \end{pmatrix}. \tag{50}$$

It is a 2-to-1 onto homomorphism from SU(2) to SO(3). It satisfies notably $g(u) = g(-u)$. Topologically, SO(3) is homeomorphic to the sphere $S^3$ with the antipodal points being identified. It is connected, but not simply connected. The action of the homomorphism (50) over the Euler decomposition (43), shows that any rotation $r \in SO(3)$, can be decomposed as

$$\begin{aligned} r = r_z(\alpha) r_y(\beta) r_z(\gamma) \\ \text{with } r_z(\phi) = \begin{pmatrix} \cos\phi & \sin\phi & 0 \\ -\sin\phi & \cos\phi & 0 \\ 0 & 0 & 1 \end{pmatrix} \\ \text{and} \quad r_y(\beta) = \begin{pmatrix} \cos\beta & 0 & \sin\beta \\ 0 & 1 & 0 \\ -\sin\beta & 0 & \cos\beta \end{pmatrix}. \end{aligned} \tag{51}$$

where $(\alpha, \beta, \gamma)$ are (any choice of) Euler angles of (any choice of) one of the two antecedents of $r$ by $g$. The unicity of the decomposition can be obtained for instance with the restriction $\alpha \in ]-\pi, \pi[$, $\beta \in [0, \pi]$ and $\gamma \in [|\alpha|, 2\pi - |\alpha|[$.

## 2.7 MEASURE AND INTEGRATION

HAAR MEASURE. A Borel set in SU(2) is any subset of SU(2) obtained from open sets through countable union, countable intersection, or taking the complement. All Borel sets form an algebra called the Borel algebra $\mathcal{B}(\mathrm{SU}(2))$. A Borel SU(2)-measure $\mu$ is a non-negative function over $\mathcal{B}(\mathrm{SU}(2))$ for which $\mu(\emptyset) = 0$, and which is countable additive (the measure of a disjoint union is the sum of the measures of each set). A Borel measure is said to be quasi-regular if it is both

1. Outer regular: $\mu(S) = \inf\{\mu(U) \mid S \subseteq U, U \text{ open}\}$;
2. Inner regular: $\mu(S) = \sup\{\mu(K) \mid K \subseteq S, K \text{ compact}\}$.

It can be shown that there exists a unique quasi-regular Borel measure $\mu$ over SU(2) which is

1. Invariant: $\mu(u) = \mu(gu) = \mu(ug)$;
2. Normalised: $\mu(\mathrm{SU}(2)) = 1$.

It is called the (two-sided normalised) *Haar measure* of SU(2). The Haar measure enables the definition of integrals of functions $f$ over SU(2):

$$\int_{\mathrm{SU}(2)} f(u)\, d\mu(u) \quad \text{also denoted} \quad \int_{\mathrm{SU}(2)} f(u)\, du. \tag{52}$$

THE HILBERT SPACE $L^2(\mathrm{SU}(2))$. The space of complex functions over SU(2) satisfying

$$\int_{\mathrm{SU}(2)} |f(u)|^2 du < \infty, \tag{53}$$

is denoted $L^2(\mathrm{SU}(2))$. It is an infinite-dimensional Hilbert space with the scalar product

$$(f_1, f_2) \stackrel{\text{def}}{=} \int_{\mathrm{SU}(2)} f_1^*(u) f_2(u) du. \tag{54}$$

MEASURE OVER $\mathrm{SL}_2(\mathbb{C})$. In GGV ([75] pp. 214-215), an invariant measure $da$ over $\mathrm{SL}_2(\mathbb{C})$ is defined, so that for any $g \in \mathrm{SL}_2(\mathbb{C})$, we have

$$da = d(ga) = d(ag) = d\left(a^{-1}\right). \tag{55}$$

It is given explicitly by

$$da = \left(\frac{i}{2}\right)^3 |a_{12}|^{-2} da_{11} d\overline{a_{11}}\, da_{12} d\overline{a_{12}}\, da_{22} d\overline{a_{22}}. \tag{56}$$

In Rühl [164], the invariant measure is given in terms of the coefficients in the decomposition (10), by

$$da = \frac{1}{\pi^4} \delta\left(a_0 - \sum_{k=1}^{3} a_k^2 - 1\right) da_0 d\overline{a_0}\, da_1 d\overline{a_1}\, da_2 d\overline{a_2}\, da_3 d\overline{a_3}. \tag{57}$$

It is normalised so that the induced measure over SU(2) is the same Haar measure defined in the previous section. In Rühl ([164] p. 285), it is shown that using the Cartan decomposition $a = ue^{r\sigma_3/2}v^{-1}$, we have:

$$d\mu(a) = \frac{1}{4\pi}\sinh^2 r\, dr\, du\, dv. \tag{58}$$

## 2.8 REPRESENTATIONS

REPRESENTATION OF GROUPS. A good way to understand the structural properties of a group is to look for its action on vector spaces. By 'action', we mean specifically a linear action that preserves the group product: it is called a *representation*. In physics, notably in quantum mechanics, we often focus on representations over Hilbert spaces.

Let G be a locally compact group, and $GL(\mathcal{H})$ the group of bounded linear operators over a Hilbert space $\mathcal{H}$ that admit a bounded inverse. A (bounded continuous) *representation* $\rho$ of G over $\mathcal{H}$ is a homomorphism $\rho : G \to GL(\mathcal{H})$, such that the action map $G \times \mathcal{H} \to \mathcal{H}$ is continuous. In the case of a finite-dimensional Hilbert space $\mathcal{H}$, $GL(\mathcal{H})$ is just the space of invertible linear maps, and a representation is any linear action of G over $\mathcal{H}$. It is called *unitary* if it preserves the scalar product.

REPRESENTATION OF LIE ALGEBRAS. There are also representations of Lie algebras, which are linear actions preserving the Lie bracket. Any representation of a Lie group defines, by differentiation, a representation of its Lie algebra. Precisely, if $\rho : G \to GL(\mathcal{H})$ is a representation of G, the *differential of* $\rho$, is the linear map $D\rho : \mathfrak{g} \to \mathfrak{gl}(\mathcal{H})$ defined for all $X \in \mathfrak{g}$ by:

$$(D\rho)(X) \stackrel{\text{def}}{=} \frac{d}{dt}\rho(e^{tX})\bigg|_{t=0}. \tag{59}$$

Moreover, for all $X \in \mathfrak{g}$,

$$\rho(e^X) = e^{D\rho(X)}. \tag{60}$$

One can show that:

1. If $F \subset \mathcal{H}$ is stable for $\rho$, then F is also stable for $D\rho$.
2. If $D\rho$ is irreducible, then $\rho$ is also irreducible.
3. If G is connected, the converses of (1) and (2) are also true.

Conversely, given a Lie algebra $\mathfrak{g}$, there is no unique Lie group associated to it, but there is a unique simply connected one G, which is obtained by exponentiation of $\mathfrak{g}$. Then, given any morphism of Lie algebra $\phi$, there exists a morphism of a Lie group $\rho$ such that $\phi = D\rho$. Thus, a representation of $\mathfrak{g}$ will infer a representation on each of its associated Lie groups.

IRREP. A representation is *irreducible* if it admits no closed stable subspace other than $\{0\}$ and $\mathcal{H}$. For brevity, we commonly say 'irrep' instead of 'irreducible representation'. They can be seen as the building blocks of the other representations. From two representations, one can build others using notably the direct sum and the tensor product. If $V$ and $W$ are two vector spaces of representation of a group $G$ and its algebra $\mathfrak{g}$, we define a representation over the direct sum $V \oplus W$ by

$$\begin{aligned} \forall g \in G, &\quad g \cdot (v + w) = g \cdot v + g \cdot w \\ \forall X \in \mathfrak{g}, &\quad X \cdot (v + w) = X \cdot v + X \cdot w. \end{aligned} \tag{61}$$

We also define a representation over the tensor product $V \otimes W$

$$\begin{aligned} \forall g \in G, &\quad g \cdot (v \otimes w) = (g \cdot v) \otimes (g \cdot w) \\ \forall X \in \mathfrak{g}, &\quad X \cdot (v \otimes w) = (X \cdot v) \otimes w + v \otimes (X \cdot w). \end{aligned} \tag{62}$$

## 2.9 INTERTWINERS

If $V$ and $W$ are two vector spaces of representation of a group $G$ and its algebra $\mathfrak{g}$, an *intertwiner* (or *equivariant map* or *intertwining operator*) is a linear map $T : V \to W$ satisfying:

$$T(g \cdot v) = g \cdot T(v). \tag{63}$$

The space of intertwiners, denoted $\mathrm{Hom}_G(V, W)$, is a subspace of the vector space of linear maps $\mathrm{Hom}(V, W)$. A useful result is the following isomorphism

*In the language of category theory, an intertwiner is nothing but a natural transformation between two functors, each functor being a representation of the group.*

$$\mathrm{Hom}_G(V, W) \cong \mathrm{Inv}_G(V \otimes W^*), \tag{64}$$

where $W^*$ is the dual space of $W$ and

$$\mathrm{Inv}_G(E) \overset{\text{def}}{=} \{\psi \in E \mid \forall g \in G,\ g \cdot \psi = \psi\}. \tag{65}$$

Two representations are *equivalent* if there is an invertible intertwiner between them. An invertible intertwiner is a way to *identify* two representations, as if there were only a change of notation between them. It is common to alleviate the notations by making the intertwiner implicit, and using instead the symbol of congruence '$\cong$', which should be understood as 'equal from the perspective of the group representation'. For instance, anticipating on section 4.2, we denote $|jm\rangle \cong v_{j-m}$ instead of $|jm\rangle = T(v_{j-m})$, and similarly for operators, we write $J_+ \cong e$, rather than $J_+ = T \circ e \circ T^{-1}$. Thus, two equivalent representations will often be presented as two *realisations* of the same representation. The symbol $\cong$ is not a strict equality '$=$' in the mathematical sense since it only identifies *some* of the structures on the two sides of the equation. See chapter 1 for a discussion about what being equal or isomorphic means.

SCHUR'S LEMMA. If $T : V \to W$ is an intertwiner between two finite irreps of G, then either $T = 0$, or T is bijective. Moreover, if the irreps are unitary and T is bijective, then for any other bijective intertwiner $T'$ there exists $\lambda \in \mathbb{C}$ such that $T' = \lambda T$. This lemma is useful notably to prove the following and very important theorem.

PETER-WEYL'S THEOREM. An important case is when the group G is compact (e.g. SU(2), but not $SL_2(\mathbb{C})$). In this case, we have the following properties:

1. Any complex finite representation of G can be endowed with a hermitian product which makes the representation unitary.
2. Any unitary irrep of G is finite-dimensional.
3. Any unitary representation can be decomposed into a direct sum of irreps.

Theses results justify notably that focusing on unitary irreps of SU(2), as we do in chapter 4, is sufficient to describe all possible finite or unitary representations of SU(2). Finally, the compactness of G enables to define the space of square-integrable functions $L^2(G)$ with the Haar measure, and then

4. The linear span of all matrix coefficients of all finite unitary irreps of G is dense in $L^2(G)$.

A proof can be found in Knapp ([113] pp. 17-20).

## 2.10 INDUCED REPRESENTATIONS

There is a well-known method to build a representation of group, *induced* from a representation of one of its subgroups. We present below two possible formal definitions of the method (see the book [127] for details). In section 7.3, we will apply the method to construct the principal series of irreps of $SL_2(\mathbb{C})$.

Consider a group G and K one of its subgroup. Say $\rho$ is a representation of K over a vector space V. We are going to build a representation of G using $\rho$. We first define a vector space $\mathcal{H}$, then a group homomorphism $U : G \to GL(\mathcal{H})$. There are two equivalent ways to proceed.

1. Let $\mathcal{H}$ be the vector space of functions $f : G \to V$ such that

$$\forall g \in G, \quad \forall k \in K, \quad f(gk) = \rho(k) f(g). \tag{66}$$

For all $g \in G$, we define the linear map $U(g) : \mathcal{H} \to \mathcal{H}$ by

$$\forall f \in \mathcal{H}, \quad \forall x \in G, \quad U(g)f(x) = f(g^{-1}x). \tag{67}$$

2. Denote the quotient $M \stackrel{\text{def}}{=} G/K$. Let $P(M, K)$ be a K-principal bundle over M. Denote $P \times_\rho V$ the associated vector bundle of base M. It has V as fibre. Let

$$\mathcal{H} = \Gamma\left(P \times_\rho V\right), \tag{68}$$

the set of sections of $P \times_\rho V$. For all $g \in G$, we define the linear map $U(g)$ by

$$\forall f \in \mathcal{H}, \quad \forall x \in G/K, \quad (U(g)f)(x) = f(g^{-1}x). \tag{69}$$

In both cases, $(U, \mathcal{H})$ is the representation of G *induced* from the representation $(\rho, V)$ of the subgroup K.

As an example, consider the trivial subgroup $\{e\}$ of a Lie group G, and its trivial representation over $\mathbb{C}$. The induced representation is then given by the Hilbert space $L^2(G)$, endowed with a left-invariant (resp. right-invariant) measure, and the linear action $g \cdot f(h) = f(g^{-1}h)$ (resp. $g \cdot f(h) = f(hg)$). It is also called the left (resp. right) regular representation.

# 3 MUSIC OF THE SPHERES

In this chapter, we present three ways to think about spheres and we introduce two essential bundles over the sphere. All these geometrical tools are omnipresent in theoretical physics and especially in quantum gravity.

## 3.1 VARIATIONS UPON A SPHERE

In chapter 1, we have seen how subtle the definition of a circle may be. Here, we do it again with three equivalent descriptions of the sphere:

1. The submanifold $S^2$;
2. The Riemann sphere $\bar{\mathbb{C}}$;
3. The complex projective line $\mathbb{C}P^1$.

### 3.1.1 *The sphere $S^2$*

Define the sphere $S^2$ as

$$S^2 \overset{\text{def}}{=} \left\{(x, y, z) \in \mathbb{R}^3 \mid x^2 + y^2 + z^2 = 1\right\}. \tag{70}$$

It is a topological space with the induced topology of $\mathbb{R}^3$, meaning the open subsets of $S^2$ are the intersection of $S^2$ with the open sets of $\mathbb{R}^3$. $S^2$ is also a 2-dimensional differentiable manifold. It can be parametrised with the spherical coordinates $(\theta, \phi)$ as

$$S^2 = \{(\sin\theta\cos\phi, \sin\theta\sin\phi, \cos\theta) \mid \theta \in [0, \pi] \text{ and } \phi \in [0, 2\pi[\}. \tag{71}$$

$S^2$ is endowed with a metric induced from the euclidian metric of $\mathbb{R}^3$

$$ds^2 = dx\, dy\, dz. \tag{72}$$

In spherical coordinates, it reads

$$ds^2 = \sin\theta\, d\theta\, d\phi. \tag{73}$$

### 3.1.2 *The Riemann sphere*

A topological space is locally compact if any two different points always admit disjoint compact neighbourhoods. Such a space X can be compactified by adding a single point to it. The resulting compact space is denoted $\bar{X}$, and called the *Alexandroff extension*. The Alexandroff extension of $\mathbb{C}$ is the *Riemann sphere*, denoted $\bar{\mathbb{C}} = \mathbb{C} \cup \{\infty\}$. It is a Riemann surface, that is a uni-dimensional complex manifold, and it is diffeomorphic to $S^2$:

$$\bar{\mathbb{C}} \cong S^2. \tag{74}$$

Different diffeomorphisms are used in the literature. Here we give the *stereographic projection from the south pole*, from $S^2$ to $\bar{\mathbb{C}}$, that reads,

$$(x, y, z) \mapsto \zeta = \frac{-x + iy}{1 + z}, \tag{75}$$

or, in spherical coordinates,

$$(\theta, \phi) \mapsto \zeta = -\tan\frac{\theta}{2} e^{-i\phi}. \tag{76}$$

The inverse is

$$\zeta \mapsto \begin{pmatrix} x \\ y \\ z \end{pmatrix} = \frac{1}{1 + |\zeta|^2} \begin{pmatrix} -\zeta - \zeta^* \\ i(\zeta^* - \zeta) \\ 1 - |\zeta|^2 \end{pmatrix}. \tag{77}$$

The metric now takes the form

$$ds^2 = \frac{4 d\zeta\, d\zeta^*}{(1 + |\zeta|^2)^2}, \tag{78}$$

with $d\zeta = d\,\mathrm{Re}(\zeta) + i\, d\,\mathrm{Im}(\zeta)$. The metric enables to measure lengths and areas. $\bar{\mathbb{C}}$ is actually a *Kähler manifold*, so that its metric can be locally written as the second derivative of a potential, in our case, $F(\zeta) = \log\left(1 + |\zeta|^2\right)$, and

$$ds^2 = 4 \frac{\partial}{\partial \zeta} \left( \frac{\partial F}{\partial \zeta^*} \right) d\zeta d\zeta^*. \tag{79}$$

The areas are measured with the symplectic 2-form

$$\omega = 2i \frac{d\zeta \wedge d\zeta^*}{(1 + |\zeta|^2)^2}. \tag{80}$$

### 3.1.3 *The complex projective line*

The set $\mathbb{C}^2$ is a complex vector space. A *vector line* d is a uni-dimensional linear subspace of $\mathbb{C}^2$. The *complex projective line* $\mathbb{C}P^1$ is the set of vector lines of $\mathbb{C}^2$.

A more sophisticated way of saying the same thing consists in defining an equivalent relation between two points $(z_0, z_1)$ and $(w_0, w_1)$, each in $\mathbb{C}^2$, such as

$$(z_0, z_1) \sim (w_0, w_1) \quad \Leftrightarrow \quad \exists \lambda \in \mathbb{C}, \quad (z_0, z_1) = (\lambda w_0, \lambda w_1). \tag{81}$$

$\mathbb{C}P^1$ is then defined as the quotient space:

$$\mathbb{C}P^1 \stackrel{\text{def}}{=} \mathbb{C}^2/\sim. \tag{82}$$

If $(x_0, x_1) \in \mathbb{C}^2 \setminus \{0\}$, the vector line $d$ passing through it is denoted $[x_0 : x_1]$, and $(x_0, x_1)$ are called the *homogeneous coordinates* of $d$. The surjective function,

$$p : \begin{cases} \mathbb{C}^2 \setminus \{0\} & \to & \mathbb{C}P^1 \\ (x_0, x_1) & \mapsto & [x_0 : x_1] \end{cases} \tag{83}$$

induces a topology on $\mathbb{C}P^1$ ($U \subset \mathbb{C}P^1$ is an open set if, and only if, $p^{-1}(U)$ is an open set), so that $p$ is continuous. In fact, $\mathbb{C}P^1$ is a complex manifold, with the open cover given by

$$\begin{aligned} U_0 &\stackrel{\text{def}}{=} \{[x_0 : x_1] \mid x_0 \neq 0\} \\ U_1 &\stackrel{\text{def}}{=} \{[x_0 : x_1] \mid x_1 \neq 0\} \end{aligned} \tag{84}$$

The *north pole* is $N \stackrel{\text{def}}{=} [1 : 0] \in U_0$ and the *south pole* is $S \stackrel{\text{def}}{=} [0 : 1] \in U_1$. In fact, $U_0 = \mathbb{C}P^1 \setminus \{S\}$ and $U_1 = \mathbb{C}P^1 \setminus \{N\}$. The north map is

$$\zeta_0 : \begin{cases} U_0 & \to & \mathbb{C} \\ [x_0 : x_1] & \mapsto & \frac{x_1}{x_0}. \end{cases} \tag{85}$$

The south map $\zeta_1$ is defined similarly. It is well-defined whatever the choice of homogeneous coordinates. $\zeta_0$ and $\zeta_1$ are homeomomorphisms and $\zeta_0 \circ \zeta_1^{-1}(z) = \frac{1}{z}$ is holomorphic. Thus $\mathbb{C}P^1$ is diffeomorphic to $\bar{\mathbb{C}}$ (and thus to $S^2$)

$$\mathbb{C}P^1 \cong \bar{\mathbb{C}}. \tag{86}$$

## 3.2 TAUTOLOGICAL BUNDLE $\mathcal{O}(-1)$

We now introduce to an essential fibre bundle which plays a crucial role in the representation of $SL_2(\mathbb{C})$, as described in chapter 7.

The set $\mathcal{O}(-1)$ is the disjoint union of vector lines of $\mathbb{C}^2$:

$$\mathcal{O}(-1) \stackrel{\text{def}}{=} \bigsqcup_{d \in \mathbb{C}P^1} d = \bigcup_{d \in \mathbb{C}P^1} \{(d, x) \mid x \in d\}. \tag{87}$$

It is a subset of $\mathbb{C}P^1 \times \mathbb{C}^2$. Note the difference with the simple union and the set of vector lines:

$$\bigcup_{d \in \mathbb{C}P^1} d = \mathbb{C}^2 \quad \text{and} \quad \bigcup_{d \in \mathbb{C}P^1} \{d\} = \mathbb{C}P^1. \tag{88}$$

Notations are subtle. The specificity of $\mathcal{O}(-1)$ is its structure of holomorphic line bundle:

$$\mathbb{C} \to \mathcal{O}(-1) \to \mathbb{C}P^1, \tag{89}$$

with the usual notation 'fibre $\to$ bundle $\to$ base'. It is called the *tautological bundle* because the fibre over a point $d \in \mathbb{C}P^1$ is $d$ itself.

DEFINITION. The projection over the base space is

$$\pi : \begin{cases} \mathcal{O}(-1) & \to & \mathbb{C}P^1 \\ (d, x) & \mapsto & d. \end{cases} \tag{90}$$

The fibre over $d \in \mathbb{C}P^1$ is $\pi^{-1}(d) \cong \mathbb{C}$. $(U_0, U_1)$ forms an open covering of $\mathbb{C}P^1$ so that $(\pi^{-1}(U_0), \pi^{-1}(U_1))$ is an open covering of $\mathcal{O}(-1)$. In fact, $\mathcal{O}(-1)$ is locally trivialised on this covering by

$$\begin{aligned} t_0 &: \begin{cases} U_0 \times \mathbb{C} & \to & \pi^{-1}(U_0) \\ (d, z) & \mapsto & (d, (z, \zeta_0(d)z)) \end{cases} \\ t_1 &: \begin{cases} U_1 \times \mathbb{C} & \to & \pi^{-1}(U_1) \\ (d, z) & \mapsto & (d, (\zeta_1(d)z, z)) \end{cases} \end{aligned} \tag{91}$$

As trivialisations, $t_0$ and $t_1$ are diffeomorphisms satisfying $\pi \circ t(d, z) = d$. They are related by a transition function

$$t_{10} : \begin{cases} U_0 \cap U_1 & \to & \mathbb{C} \\ d & \mapsto & \zeta_0(d) \end{cases} \tag{92}$$

which satisfies

$$t_0(d, x) = t_1(d, t_{10}(d)x). \tag{93}$$

All these properties define a line bundle. The manifold $\mathcal{O}(-1)$ and $\mathbb{C}P^1$ are complex, and the trivialisations and the transition functions are holomorphic, so that the $\mathcal{O}(-1)$ is a holomorphic line bundle.

NOTATION. The transition function $t_{10}$ is nothing but $\zeta_0$ restricted to $U_0 \cap U_1$. In fact, for any $k \in \mathbb{Z}$, one can define a holomorphic line bundle $\mathcal{O}(k)$ over $\mathbb{C}P^1$, for which the transition function $t_{10}$ is the restriction of $\zeta_0^{-k}$. When $k = -1$, we get the tautological bundle $\mathcal{O}(-1)$. When $k = 0$, we get the trivial bundle $\mathcal{O}(0) = \mathbb{C}P^1 \times \mathbb{C}$. It is a theorem that any holomorphic line bundle over $\mathbb{C}P^1$ is one of the $\mathcal{O}(k)$. Over other Riemann surfaces, there are generally many more line bundles, sometimes even a continuous family of them. A hard theorem by Grothendieck states that any holomorphic vector bundle over $\mathbb{C}P^1$ is a direct sum of some $\mathcal{O}(k)$ [102].

SECTION. A global section of $\mathcal{O}(-1)$ is a continuous map $s : \mathbb{C}P^1 \to \mathcal{O}(-1)$ satisfying $\pi \circ s = \mathrm{Id}_{\mathbb{C}P^1}$. Actually, the only global holomorphic section of $\mathcal{O}(-1)$ is the null section ($s_0 : d \mapsto (d, (0, 0))$). So, it is more interesting to look at local sections, defined over open sets of $\mathbb{C}P^1$. The set of local sections has the mathematical structure of a sheaf.

Let's consider sections over $U_0$. Any such section $s : U_0 \to \mathcal{O}(-1)$ can be written as $s(d) = (d, f(d))$, with $f : U_0 \to \mathbb{C}^2$ a continuous function satisfying $f(d) \in d$. Then, the map $\sigma \overset{\text{def}}{=} f \circ \zeta_0^{-1} : \mathbb{C} \to \mathbb{C}^2$ is a continuous function satisfying $\sigma_1(z) = z\,\sigma_0(z)$, such that $\sigma_0$ does not vanish. Thus, any section over $U_0$ is uniquely characterised by a continuous function $\sigma_0 : \mathbb{C} \to \mathbb{C} \setminus \{0\}$, and conversely any such function defines a section.

INTEGRATION. $\mathcal{O}(-1)$ is a real differentiable manifold of dimension 4. The image of a section $s(U_0) \subset \mathcal{O}(-1)$ is a 2-dimensional real submanifold. A 2-form $\alpha$ defined over $\mathcal{O}(-1)$ can be integrated over $s(U_0)$:

$$\int_{s(U_0)} \alpha \tag{94}$$

Such a 2-form $\alpha$ is itself a section of the vector bundle $\Lambda^2 T^*\mathcal{O}(-1)$ of base space $\mathcal{O}(-1)$. In this bundle, a basis of sections over $\pi^{-1}(U_0)$ is given by

$$dz_0 \wedge dz_1,\ dz_0 \wedge d\bar{z}_0,\ dz_0 \wedge d\bar{z}_1,\ d\bar{z}_0 \wedge dz_1,\ dz_1 \wedge d\bar{z}_1, d\bar{z}_0 \wedge d\bar{z}_1,$$

denoted $D_i(z_0, z_1)$ with $i \in \{1, ..., 6\}$. So $\alpha$ can be written in this basis and

$$\int_{s(U_0)} \alpha = \int_{\mathbb{C}} \alpha_i(\sigma_0(z), \sigma_1(z)) D_i(\sigma_0(z), \sigma_1(z)). \tag{95}$$

A 2-form $\alpha$ is homogeneous of degree 0 if it is constant on each fibre, i. e.

$$\forall \lambda \in \mathbb{C}, \quad \alpha(z_0, z_1) = \alpha(\lambda z_0, \lambda z_1). \tag{96}$$

In such a case, the integral does not depend on the chosen section:

$$\int_{s(U_0)} \alpha = \int_{\mathbb{C}} \alpha_i(1, z) D_i(1, z). \tag{97}$$

This fact will be used in section 7.3 to define the principal series of representations of $SL_2(\mathbb{C})$.

## 3.3 HOPF FIBRATION

In the preceding section, we have seen the tautological bundle over $\mathbb{C}P^1$. We now describe another useful bundle over $\mathbb{C}P^1$, the *Hopf fibration*:

$$S^1 \to S^3 \to S^2. \tag{98}$$

To illustrate the central role of the Hopf fibration in physics, we refer the reader to the beautiful review paper [187].

$\mathbb{C}^2$ and $\mathfrak{su}(2)$ are respectively the fundamental and the adjoint representations of SU(2). The fundamental representation is simply given by the multiplication of a matrix by a vector, and the adjoint representation is given by

$$\forall u \in SU(2),\ \forall x \in \mathfrak{su}(2), \quad u \cdot x = u x u^{-1}. \tag{99}$$

The *Hopf map*

$$p : \begin{cases} \mathbb{C}^2 & \to & \mathfrak{su}(2) \\ \mathbf{z} & \mapsto & \sum_{k=1}^{3} (\mathbf{z}^\dagger \sigma_k \mathbf{z})\ i\sigma_k \end{cases} \tag{100}$$

*We use the shorthand $\mathbf{z} = (z_0, z_1)$, and $a\,\mathbf{z}$ denotes the matrix multiplication of $a$ with $\begin{pmatrix} z_0 \\ z_1 \end{pmatrix}$.*

is an intertwiner of the representations of SU(2). It satifies notably

$$\mathbf{z}\mathbf{z}^\dagger = \frac{1}{2}\left(\|p(\mathbf{z})\|\mathbb{1} + p(\mathbf{z}) \cdot \vec{\sigma}\right). \tag{101}$$

Using the standard isomorphism between $\mathfrak{su}(2)$ and $\mathbb{R}^3$ (see section 2.5), the Hopf map reads

$$p : \begin{cases} \mathbb{C}^2 & \to & \mathbb{R}^3 \\ (z, w) & \mapsto & (zw^* + wz^*, i(zw^* - wz^*), |z|^2 - |w|^2) \end{cases} \tag{102}$$

Then using the standard isomorphism between $\mathbb{C}^2$ and $\mathbb{R}^4$, it reads

$$p : \begin{cases} \mathbb{R}^4 & \to & \mathbb{R}^3 \\ (x, y, z, w) & \mapsto & (2(xz + yw), 2(yz - xw), x^2 + y^2 - z^2 - w^2) \end{cases} \tag{103}$$

The sphere $S^3$ is defined as

$$S^3 \stackrel{\text{def}}{=} \left\{(x, y, z, t) \in \mathbb{R}^4 \mid x^2 + y^2 + z^2 + t^4 = 1\right\}. \tag{104}$$

When we restrict $p$ to $S^3$, we get the *Hopf projection*

$$\mathfrak{p} : S^3 \to S^2. \tag{105}$$

In terms of Euler angles and spherical coordinates, it reads

$$\mathfrak{p} : \begin{cases} S^3 & \to & S^2 \\ (\theta, \phi, \psi) & \mapsto & (\theta, \phi) \end{cases} \tag{106}$$

Using the fact $S^2 \cong \mathbb{C}P^1$ and the following identification

$$S^3 \cong \left\{(z_0, z_1) \in \mathbb{C}^2 \mid |z_0|^2 + |z_1|^2 = 1\right\}, \tag{107}$$

we can show that the Hopf projection reads

$$\mathfrak{p} : \begin{cases} S^3 & \to & \mathbb{C}P^1 \\ (z_0, z_1) & \mapsto & [z_0 : z_1] \end{cases} \tag{108}$$

So the Hopf projection is a restriction of the surjective map (83) that sends a point of $\mathbb{C}^2$ to the complex line to which it belongs. Then, for any $d \in \mathbb{C}P^1$, we have

$$\mathfrak{p}^{-1}(d) \cong U(1). \tag{109}$$

$\mathfrak{p}^{-1}(U_0)$ and $\mathfrak{p}^{-1}(U_1)$ form an open cover of $S^3$. In fact, $S^3$ is locally trivialised over each of them through:

$$\tau_0 : \begin{cases} U_0 \times U(1) & \to & \mathfrak{p}^{-1}(U_0) \\ (d, \lambda) & \mapsto & \left( \frac{\lambda}{\sqrt{1+|\zeta_0(d)|^2}}, \frac{\zeta_0(d)\lambda}{\sqrt{1+|\zeta_0(d)|^2}} \right) \end{cases}$$
$$\tau_1 : \begin{cases} U_1 \times U(1) & \to & \mathfrak{p}^{-1}(U_1) \\ (d, \lambda) & \mapsto & \left( \frac{\zeta_1(d)\lambda}{\sqrt{1+|\zeta_1(d)|^2}}, \frac{\lambda}{\sqrt{1+|\zeta_1(d)|^2}} \right) \end{cases} \tag{110}$$

$\tau_0$ and $\tau_1$ are indeed diffeomorphisms satisfying $\mathfrak{p} \circ \tau(d, \lambda) = d$. The transition function is

$$\tau_{10} : \begin{cases} U_0 \cap U_1 & \to & \mathbb{C} \\ d & \mapsto & \arg \zeta_0(d) \end{cases} \tag{111}$$

So, it satisfies

$$\tau_0(d, \lambda) = \tau_1(d, \tau_{10}(d)\lambda). \tag{112}$$

All these properties, define indeed the structure of a fibre bundle. $S^3$ is actually a U(1)-principal bundle over $\mathbb{C}P^1$. Using standard isomorphisms, we get the nice staking of spheres:

$$S^1 \to S^3 \to S^2. \tag{113}$$

Finally, the Hopf fibration can also be seen as a re-branding of the quotient

$$S^2 \cong SU(2)/U(1). \tag{114}$$

This is the consequence of a more general theorem. If G is a Lie group and H a Lie subgroup, then there exists a unique structure of smooth manifold on G/H such that the quotient map $\pi : G \to G/H$ defines a fibre bundle of fibre H.

# 4 REPRESENTATION THEORY OF SU(2)

Motivated by their omnipresence in quantum physics, we are going to study the representations of SU(2) over finite-dimensional Hilbert spaces. Note that real representations of SU(2) also exists (see [109]) but they are ignored by physicists. Due to Peter-Weyl's theorem, a complex finite representation of a compact group can be decomposed into a direct sum of irreps. So we will focus on irreps of SU(2).

## 4.1 IRREPS OF SU(2)

To start with, it is important to notice the following one-to-one correspondence between sets of finite-dimensional representations:

1. Holomorphic representations[1] of $SL_2(\mathbb{C})$,
2. Representations of SU(2),
3. Representations of $\mathfrak{su}(2)$,
4. $\mathbb{C}$-linear representations[2] of $\mathfrak{sl}_2(\mathbb{C})$.

It is a particular case of the so-called *Weyl's unitary trick*. Concretely, we go from one sets to another through:

$(1) \Rightarrow (2)$ Restriction of the action of $SL_2(\mathbb{C})$ to its subgroup SU(2).

$(2) \Rightarrow (3)$ Differentiation as shown in equation (59).

$(3) \Rightarrow (4)$ Using $\mathfrak{sl}_2(\mathbb{C}) \cong \mathfrak{su}(2) \oplus i\,\mathfrak{su}(2)$.

$(4) \Rightarrow (1)$ Exponentiation as shown in equation (60).

Importantly, this correspondence preserves invariant subspaces and equivalences of representations. In particular, it means that it is now sufficient to our purpose to find all the $\mathbb{C}$-linear irreps of $\mathfrak{sl}_2(\mathbb{C})$.

THEOREM. *For all $n \in \mathbb{N}$, there exists a $n$-dimensional $\mathbb{C}$-linear irrep of $\mathfrak{sl}_2(\mathbb{C})$, unique up to equivalence.*

1 Here, 'holomorphic' means that the map defined by the representation over the vector space is holomorphic.
2 '$\mathbb{C}$-linear' means that we regard $\mathfrak{sl}_2(\mathbb{C})$ as a complex (and not real) vector space. We will care about the $\mathbb{R}$-linear representations of $\mathfrak{sl}_2(\mathbb{C})$ in section 7.1.

A proof can be found in [30]. The $(n+1)$-dimensional irrep is fully characterised by the action of the elements $h, e, f \in \mathfrak{sl}_2(\mathbb{C})$, defined by

$$\begin{aligned} h &\overset{\text{def}}{=} \sigma_3 = \begin{pmatrix} 1 & 0 \\ 0 & -1 \end{pmatrix} \\ e &\overset{\text{def}}{=} \frac{\sigma_1 + i\sigma_2}{2} = \begin{pmatrix} 0 & 1 \\ 0 & 0 \end{pmatrix} \\ f &\overset{\text{def}}{=} \frac{\sigma_1 - i\sigma_2}{2} = \begin{pmatrix} 0 & 0 \\ 1 & 0 \end{pmatrix}, \end{aligned} \tag{115}$$

which satisfy the commuting relations

$$[h, e] = 2e \qquad [h, f] = -2f \qquad [e, f] = h.$$

Their action over a basis $(v_i)$, $i \in \{0, ..., n\}$, is given by

$$h \cdot v_k = 2(j-k)v_k, \quad e \cdot v_k = k(n-k+1)v_{k-1}, \quad f \cdot v_k = v_{k+1}. \tag{116}$$

The 3-dimensional complex vector space $\mathfrak{sl}_2(\mathbb{C})$ can also be seen as a 6-dimensional real vector space, which has $\mathfrak{su}(2)$ as its subspace. Thus, by restriction of the action of $\mathfrak{sl}_2(\mathbb{C})$ to $\mathfrak{su}(2)$, the previously found $\mathbb{C}$-linear irreps of $\mathfrak{sl}_2(\mathbb{C})$, define also irreps of $\mathfrak{su}(2)$. Finally, by exponentiating with (42), we find all irreps of SU(2) over complex vector spaces.

## 4.2 ANGULAR MOMENTUM REALISATION

In physics textbooks, the representations of $\mathfrak{sl}_2(\mathbb{C})$ are indexed by half-integers, called *spins*. To each spin $j \in \mathbb{N}/2$ is associated a Hilbert space $\mathcal{Q}_j$ of dimension $2j+1$. The canonical basis, also called the *magnetic basis*, is composed of the vectors (or 'kets' in the Dirac language) denoted

$$|j, m\rangle \quad \text{with} \quad m \in \{-j, -j+1, ..., 0, ..., j-1, j\}. \tag{117}$$

It is made orthonormal by choosing the scalar product that satisfies

$$\langle j, m | j, n\rangle = \delta_{mn}. \tag{118}$$

We now define the *angular momentum observables* $J_i \overset{\text{def}}{=} \frac{1}{2}\sigma_i$, sometimes called simply *generators of* SU(2) or *generators of rotations*. Notice that the $J_i$ are elements of $i\,\mathfrak{su}(2)$, and not of $\mathfrak{su}(2)$, as observables are required to be hermitian. They satisfy

*In some textbooks, the generators are defined as $J_i \overset{def}{=} \frac{\hbar}{2}\sigma_i$, which has the dimension of an angular momentum. Consider that we are working in Planck units.*

$$\left[J_i, J_j\right] = i\epsilon_{ijk}J_k. \tag{119}$$

We then define their linear action over $\mathcal{Q}_j$ by

$$\begin{aligned} J_1 \left|j,m\right\rangle &= \frac{1}{2}\sqrt{(j-m)(j+m+1)}\left|j,m+1\right\rangle \\ &\qquad + \frac{1}{2}\sqrt{(j+m)(j-m+1)}\left|j,m-1\right\rangle , \\ J_2 \left|j,m\right\rangle &= \frac{1}{2i}\sqrt{(j-m)(j+m+1)}\left|j,m+1\right\rangle \\ &\qquad - \frac{1}{2i}\sqrt{(j+m)(j-m+1)}\left|j,m-1\right\rangle , \\ J_3 \left|j,m\right\rangle &= m\left|j,m\right\rangle . \end{aligned} \tag{120}$$

It is somehow simpler to remember the action of the *ladder operators* $J_\pm \overset{\text{def}}{=} J_1 \pm iJ_2$,

$$\begin{aligned} J_+ \left|j,m\right\rangle &= \sqrt{(j-m)(j+m+1)}\left|j,m+1\right\rangle , \\ J_- \left|j,m\right\rangle &= \sqrt{(j+m)(j-m+1)}\left|j,m-1\right\rangle . \end{aligned} \tag{121}$$

*Sometimes the action of $J_+$ over $|jm\rangle$ is written with a constant phase $e^{i\delta}$. It defines an equivalent representation, but the choice of $\delta = 0$ (called the Condon-Shortley convention, from [48]) is the most widespread.*

The action of the generators $J_i$ over $\mathcal{Q}_j$ defines a $(2j+1)$-dimensional irrep of SU(2), called the *spin-*$j$ *representation*. This is shown by exhibiting the following equivalence with the irreps defined in the previous section:

$$|jm\rangle \cong v_{j-m} \qquad J_3 \cong h/2 \qquad J_+ \cong e \qquad J_- \cong f. \tag{122}$$

Finally notice that $|jm\rangle$ is also an eigenvector of the *total angular momentum* $\vec{J}^2 \overset{\text{def}}{=} J_1^2 + J_2^2 + J_3^2$:

$$\vec{J}^2\,|jm\rangle = j(j+1)\,|jm\rangle . \tag{123}$$

In fact, the $|jm\rangle$ form the unique orthonormal basis that diagonalises simultaneously the commuting operators $J_3$ and $\vec{J}^2$. We say that $J_3$ and $\vec{J}^2$ form a *Complete Set of Commuting Observables (CSCO)*. From a mathematical perspective, notice also that $\vec{J}^2$ is not an element of the algebra $i\,\mathfrak{su}(2)$, but an element of the *universal enveloping algebra* $\mathcal{U}(i\,\mathfrak{su}(2))$ whose action can be easily computed by successive action of $\mathfrak{su}(2)$. Since $\vec{J}^2$ has the property to be a quadratic element that commutes with all of $\mathcal{U}(i\,\mathfrak{su}(2))$, it is called the *Casimir operator* of $\mathcal{U}(i\,\mathfrak{su}(2))$.

WIGNER MATRIX. The exponentiation of the action of the generators of SU(2) defines a linear action of the group SU(2) (see eq. (60)). The *Wigner matrix* $D^j(g)$ represents the action of $g \in$ SU(2) in the $|j,m\rangle$ basis. It is thus a square matrix of size $2j+1$, whose coefficients are the functions

$$D^j_{mn}(g) \overset{\text{def}}{=} \langle j,m|g|j,n\rangle . \tag{124}$$

★ NOTA BENE. One should be aware of a small ambiguity in the notation '$\langle j,m|g|j,n\rangle$' that arises when $g$ is a matrix that belongs simultaneously to SU(2) and to $\mathfrak{su}(2)$. Then it should be said explicitly if one considers the group action or the algebra action when computing $\langle j,m|g|j,n\rangle$, because it gives a

different result. This ambiguity comes from the fact that physicists do not usually write explicitly whether they consider the group representation $\rho$, or its differential $D\rho$. Mathematicians would write $\langle j, m|\rho(g)|j, n\rangle$, or $\langle j, m|D\rho(g)|j, n\rangle$. From equation (42), if $g = e^a \in SU(2) \cap \mathfrak{su}(2)$, with $a \in \mathfrak{su}(2)$, then $\rho(g) = e^{D\rho(a)}$, but $\rho(g) \neq D\rho(e^a) = D\rho(g)$. In the definition of the Wigner matrix above, it is the group action which is considered.

From Shur's lemma, it can be shown that the functions $D^j_{mn}$ form an orthogonal family of $L^2(SU(2))$:

$$\int_{SU(2)} dg\, \overline{D^{j'}_{m'n'}(g)} D^j_{mn}(g) = \frac{1}{2j+1} \delta_{jj'} \delta_{mm'} \delta_{nn'}. \tag{125}$$

In fact, the Peter-Weyl theorem even asserts that the functions $D^j_{mn}$ form a basis of $L^2(SU(2))$, i. e. any function $f \in L^2(SU(2))$ can be written

$$f(g) = \sum_{j \in \mathbb{N}/2} \sum_{m=-j}^{j} \sum_{n=-j}^{j} f^j_{mn} D^j_{mn}(g), \tag{126}$$

with coefficients $f^j_{mn} \in \mathbb{C}$. It implies notably an equivalence between the following Hilbert spaces

$$L^2(SU(2)) \cong \bigoplus_{j \in \mathbb{N}/2} (\mathcal{Q}_j \otimes \mathcal{Q}^*_j). \tag{127}$$

The equivalence is not per se a surprise, since all Hilbert spaces of the same dimension are isomorphic, but more interesting is the specific form of the isomorphism, i. e. $D^j_{mn} \cong |j, n\rangle \otimes \langle j, m|$. We are going now to derive explicit expressions for computing $D^j_{mn}(g)$, but we first need to introduce another realisation of the spin-j irreps.

### 4.3 HOMOGENEOUS REALISATION

Let $\mathbb{C}_{2j}[z_0, z_1]$ be the vector space of polynomials of two complex variables, homogeneous of degree $2j \in \mathbb{N}$. If $P(z_0, z_1) \in \mathbb{C}_{2j}[z_0, z_1]$, it can be written as

$$P(z_0, z_1) = \sum_{k=0}^{2j} a_k z_0^k z_1^{2j-k}, \tag{128}$$

with coefficients $a_0, ..., a_{2j} \in \mathbb{C}$.The action of SU(2) given by

*We would get equivalent realisations by defining the action with $P(a^{-1}\mathbf{z})$, $P(a^\dagger \mathbf{z})$ or $P(\mathbf{z}a)$. In fact $P(\mathbf{z}a) = P(a^T\mathbf{z})$.*

$$g \cdot P(\mathbf{z}) = P(g^T \mathbf{z}) \tag{129}$$

defines a $(2j+1)$-dimensional group representation. It induces the following action of the generators

$$J_+ \cong z_0 \frac{\partial}{\partial z_1} \qquad J_- \cong z_1 \frac{\partial}{\partial z_0} \qquad J_3 \cong \frac{1}{2}\left( z_0 \frac{\partial}{\partial z_0} - z_1 \frac{\partial}{\partial z_1} \right). \tag{130}$$

This representation is equivalent to the spin-j irrep through the correspondence:

$$|j, m\rangle \cong \left(\frac{(2j)!}{(j+m)!(j-m)!}\right)^{1/2} z_0^{j+m} z_1^{j-m}, \tag{131}$$

The RHS is sometimes denoted with Dirac notations $\langle z_0 z_1 | jm\rangle$. We have thus found another realisation of the spin-j irrep, called the *homogeneous realisation*. It is very convenient to derive an explicit expression for the Wigner matrix coefficients.

WIGNER MATRIX FORMULA.

$$\begin{aligned} D^j_{mn}(g) = &\left(\frac{(j+m)!(j-m)!}{(j+n)!(j-n)!}\right)^{1/2} \\ &\times \sum_k \binom{j+n}{k}\binom{j-n}{j+m-k} g_{11}^k g_{21}^{j+n-k} g_{12}^{j+m-k} g_{22}^{k-m-n}. \end{aligned} \tag{132}$$

The sum is done over the integers $k \in \{\max(0, m+n), ..., \min(j+m, j+n)\}$. A proof is found in [122]. From this, we show that

$$\overline{D^j_{mn}(u)} = (-1)^{m-n} D^j_{-m,-n}(u). \tag{133}$$

EULER ANGLES EXPRESSION. Wigner proposed also another explicit expression for his matrix, in terms of the Euler angles. If $u \in SU(2)$, and $\alpha, \beta, \gamma \in \mathbb{R}^3$ are the Euler angles of $u$, such that $u = e^{-\frac{i\alpha}{2}\sigma_3} e^{-\frac{i\beta}{2}\sigma_2} e^{-\frac{i\gamma}{2}\sigma_3}$, then

$$D^j_{m'm}(u) = e^{-i(\alpha m' + \gamma m)} d^j_{m'm}(\beta)$$

with the *reduced Wigner matrix*

$$\begin{aligned} d^j_{m'm}(\beta) = &\left(\frac{(j+m')!(j-m')!}{(j+m)!(j-m)!}\right)^{\frac{1}{2}} \\ &\times \sum_{k=\max(0,m'+m)}^{\min(j+m',j+m)} (-1)^{m'+j-k} \binom{j+m}{k}\binom{j-m}{j-k+m'} \\ &\times \left(\cos\frac{\beta}{2}\right)^{2k-m-m'} \left(\sin\frac{\beta}{2}\right)^{m+m'+2j-2k}. \end{aligned} \tag{134}$$

It is already implemented in the Wolfram Language with the command

$$\texttt{WignerD}[\{j, m, n\}, \alpha, \beta, \gamma] = e^{i(\alpha m + \gamma n)} d^j_{mn}(-\beta). \tag{135}$$

## 4.4 PROJECTIVE REALISATION

The spin-j irrep can also be realised over $\mathbb{C}_{2j}[z]$, the vector space of complex polynomials of one variable $z$ of degree at most $2j$. This realisation is obtained from the $\mathbb{C}_{2j}[z_0, z_1]$ realisation by the map:

$$\begin{cases} \mathbb{C}_{2j}[z_0, z_1] & \to \mathbb{C}_{2j}[z] \\ P(z_0, z_1) & \mapsto P(z, 1) \end{cases} \tag{136}$$

This map is constructed from a projection from $\mathbb{C}^2$ to $\mathbb{C}$, hence the name 'projective' that we give to this realisation. Sometimes it is also named the 'holomorphic' realisation. From this we deduce the action of SU(2)

$$a \cdot f(z) = (a_{12}z + a_{22})^{2j} f\left(\frac{a_{11}z + a_{21}}{a_{12}z + a_{22}}\right), \tag{137}$$

and of the generators

$$J_+ \cong -z^2 \frac{d}{dz} + 2jz \quad J_3 \cong z\frac{d}{dz} - j \quad J_- \cong \frac{d}{dz}. \tag{138}$$

The canonical basis becomes

$$|j, m\rangle \cong \sqrt{\frac{(2j)!}{(j+m)!(j-m)!}} z^{j+m}. \tag{139}$$

We can give the following explicit expression for the scalar product that makes the canonical basis orthonormal:

$$\langle f|g\rangle \overset{\text{def}}{=} \frac{i}{2} \frac{2j+1}{\pi} \int_{\mathbb{C}} \overline{f(z)} g(z) \frac{dz d\bar{z}}{(1+|z|^2)^{2j+2}}. \tag{140}$$

## 4.5 SPINORIAL REALISATION

We now introduce a last realisation of the spin-j irreps, which relies on notations developed by Penrose [143]. It was found useful for *twistor theory* [144], and later in quantum gravity for the so-called *twisted geometries* [69, 118]. The notions are a kind of gymnastics that needs some time to be learnt, but finally bears fruit in the long run.

ABSTRACT INDICES. We are going to use the clever *conventions of abstract indices* of Penrose ([143] pp. 68-115). To start with, we need a set of 'abstract indices' $\mathcal{L}$, that is to say a countable set of symbols. We use for instance capital letters:

$$\mathcal{L} \overset{\text{def}}{=} \{A, B, ..., Z, A_0, ..., Z_0, A_1, ...\}. \tag{141}$$

Then we denote $\mathfrak{S}^\bullet \overset{\text{def}}{=} \mathbb{C}^2$, and for any abstract index $A \in \mathcal{L}$, $\mathfrak{S}^A \overset{\text{def}}{=} \mathfrak{S}^\bullet \times \{A\}$. Obviously $\mathfrak{S}^A$ is isomorphic to $\mathbb{C}^2$ as a complex vector space. An element of $\mathfrak{S}^A$ will be typically denoted $z^A = (z, A) \in \mathfrak{S}^A$.

The abstract index $A$ serves as a marker to 'type' the vector $z \in \mathbb{C}^2$ (thus $z^A \neq z^B$). This notation is very efficient to deal with several copies of the same space (here $\mathbb{C}^2$), like in tensor theory.

The vector space of linear forms from $\mathbb{C}^2$ to $\mathbb{C}$, is called the dual space, and denoted $\mathfrak{S}_\bullet$. Similarly, we denote $\mathfrak{S}_A \stackrel{\text{def}}{=} \mathfrak{S}_\bullet \times \{A\}$, which is trivially isomorphic to the dual space of $\mathfrak{S}^A$. Its elements, called covectors, are denoted with an abstract lower capital index, $z_A$. Then the evaluation of a covector $y_A = (y, A)$ on a vector $z^A = (z, A)$ (called a 'contraction') is denoted $y_A z^A = y(z) \in \mathbb{C}$ (the order does not matter $y_A z^A = z^A y_A$).

SPINORS. Consider the space of *formal* (commutative and associative) finite sums of *formal* (commutative and associative) products of elements, one from each $\mathfrak{S}^{A_1}, ..., \mathfrak{S}^{A_p}, \mathfrak{S}_{B_1}, ..., \mathfrak{S}_{B_q}$. A typical element can be written:

$$t^{A_1 ... A_p}{}_{B_1 ... B_q} = \sum_{i=1}^{m} z_{1,i}{}^{A_1} ... z_{p,i}{}^{A_p}\, y^{1,i}{}_{B_1} ... y^{q,i}{}_{B_q}. \qquad (142)$$

Then impose the rules

1. (Homogeneity)

$$\forall \alpha \in \mathbb{C}, \quad (\alpha z_1{}^{A_1})\, z_2{}^{A_2} ... z_p{}^{A_p} = z_1{}^{A_1}\, (\alpha z_2{}^{A_2}) ... z_p{}^{A_p} \qquad (143)$$

2. (Distributivity)

$$(z_1{}^{A_1} + z_2{}^{A_2})\, z_3^{A_3} ... z_p{}^{A_p} = z_1{}^{A_1}\, z_3{}^{A_3} ... z_p{}^{A_p} + z_2{}^{A_2}\, z_3{}^{A_3} ... z_p{}^{A_p}. \qquad (144)$$

The resulting space is a vector space denoted $\mathfrak{S}^{A_1 ... A_p}_{B_1 ... B_q}$. Its elements are called *spinors* of type $(p, q)$, and its dimension is $2^{p+q}$.

★ NOTA BENE. The spinor space $\mathfrak{S}^{A_1 ... A_p}_{B_1 ... B_q}$ is isomorphic, but not equal, to $\mathfrak{S}^{A_1} \otimes ... \otimes \mathfrak{S}^{A_p} \otimes \mathfrak{S}_{B_1} \otimes ... \otimes \mathfrak{S}_{B_q}$. The difference is the commutativity of the product. For instance the formal product of $\mathfrak{S}^{AB}$ is commutative (by assumption) in the sense that, for $z^A \in \mathfrak{S}^A$ and $y^B \in \mathfrak{S}^B$, $z^A y^B = y^B z^A$, whereas the tensor product is not, $z^A \otimes y^B \neq y^B \otimes z^A$, simply because $z^A \otimes y^B \in \mathfrak{S}^A \otimes \mathfrak{S}^B$ and $y^B \otimes z^A \in \mathfrak{S}^B \otimes \mathfrak{S}^A$ do not belong to the same set. The usual tensor product $\otimes$ imposes an arbitrary ordering between the vectors of each space, while the abstract indices notation is a way to allow the commutation of vectors at the price of constantly keeping track of the vector space to which they belong with a label. If you have to correct an exam, either you keep the pile of copies in a rigid and arbitrary defined order, or you ask the students to write their name on their copy, so that it does not really matter if the copies are mixed up while you fell in the stairs.

The spinor space is endowed with a bunch of basic operations defined by a set of rules. It would be utterly non-pedagogical to state these rules in the most general case. On the contrary, they are very intuitive for simple examples and generalise without ambiguity for higher-order spinors.

1. (Index substitution) If $z^A = (z, A) \in \mathfrak{S}^A$, we denote $z^B = (z, B) \in \mathfrak{S}^B$. Thus $z^A \neq z^B$.

2. (Index permutation) If $t^{AB} = \sum_i z_i^A y_i^B \in \mathfrak{S}^{AB}$, we denote $t^{BA} = \sum_i z_i^B y_i^A = \sum_i y_i^A z_i^B \in \mathfrak{S}^{AB}$.

*$S_n$ denotes the group of permutation of $\{1, ..., n\}$.*

3. (Symmetrisation) $t^{(AB)} \stackrel{\text{def}}{=} \frac{1}{2}(t^{AB} + t^{BA})$ or generally

$$z^{(A_1 \ldots A_n)} \stackrel{\text{def}}{=} \frac{1}{n!} \sum_{\sigma \in S_n} z^{A_{\sigma(1)} \ldots A_{\sigma(n)}}. \tag{145}$$

4. (Anti-symmetrisation) $t^{[AB]} \stackrel{\text{def}}{=} \frac{1}{2}(t^{AB} - t^{BA})$ or generally

$$z^{[A_1 \ldots A_n]} \stackrel{\text{def}}{=} \frac{1}{n!} \sum_{\sigma \in S_n} \epsilon_\sigma z^{A_{\sigma(1)} \ldots A_{\sigma(n)}}, \tag{146}$$

with $\epsilon_\sigma$ the signature of the permutation $\sigma$.

5. (Contraction) If $t^A_B = \sum_i z_i^A y_B^i \in \mathfrak{S}^A_B$, then $t^A_A = \sum_i z_i^A y_A^i \in \mathbb{C}$.

INDEX DUALISATION. We denote the canonical basis of $\mathbb{C}^2$:

$$e_0 \stackrel{\text{def}}{=} \begin{pmatrix} 1 \\ 0 \end{pmatrix} \qquad e_1 \stackrel{\text{def}}{=} \begin{pmatrix} 0 \\ 1 \end{pmatrix}. \tag{147}$$

It is easy to show that there exists a unique normalised skew-symmetric spinor of type $(0, 2)$. It is denoted $\epsilon_{AB}$, and satisfies by definition:

$$\epsilon_{AB} = -\epsilon_{BA}, \qquad \epsilon_{AB} e_0^A e_1^B = 1. \tag{148}$$

$\epsilon_{AB}$ corresponds over $\mathbb{C}^2$ to the unique 2-form $\epsilon$ normalised by the condition $\epsilon(e_0, e_1) = 1$, which is nothing but the determinant over $\mathbb{C}^2$. For two vectors $z = (z_0, z_1)$ and $y = (y_0, y_1)$, we show easily that:

$$\epsilon_{AB} z^A y^B = z_0 y_1 - z_1 y_0. \tag{149}$$

*The convention $z_A = \epsilon_{AB} z^B$ is also encountered (it changes a sign), but we prefer the ace of heart choice.*

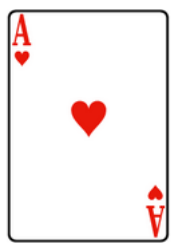

Interestingly, $\epsilon_{AB}$ defines a canonical mapping between $\mathfrak{S}^A$ and $\mathfrak{S}_A$, given by

$$z^B \mapsto z_B = z^A \epsilon_{AB}. \tag{150}$$

It is called *index dualisation*. The covectors of the dual space $\mathfrak{S}_\bullet$ can also be described by a pair of components in the dual basis. The index dualisation can then be expressed in components as $((z_0, z_1), A) \mapsto ((-z_1, z_0), A)$. Similarly to the usual Dirac notation $|z\rangle = (z_0, z_1)$, a notation is sometimes introduced for the dual $[z| = (-z_1, z_0)$. With this choice, the RHS of (149) reads $[z|y\rangle$.

CONJUGATION. The conjugate of $z \in \mathbb{C}$ is denoted $\bar{z}$ or $z^*$. We define the conjugation over $\mathfrak{S}^A$ by $\overline{z^A} = \overline{(z, A)} \stackrel{\text{def}}{=} (\bar{z}, \dot{A}) = \bar{z}^{\dot{A}} \in \mathfrak{S}^{\dot{A}}$. Thus we have introduced a new set of abstract indices, the dotted indices:

$$\dot{\mathcal{L}} \stackrel{\text{def}}{=} \{\dot{A}, \dot{B}, ..., \dot{Z}, \dot{A}_0, ..., \dot{Z}_0, \dot{A}_1, ...\}. \tag{151}$$

We impose moreover that $\overline{\bar{z}^{\dot{A}}} = z^A$, i.e. $\ddot{A} = A$, so that the conjugation is an involution. Importantly, we regard the set $\mathcal{L}$ and $\dot{\mathcal{L}}$ as incompatible classes of abstract indices, meaning that we forbid index substitution between them two. In other words, dotted and undotted indices commute: for any $t^{A\dot{B}} \in \mathfrak{S}^{A\dot{B}}$, we have $t^{A\dot{B}} = t^{\dot{B}A}$.

★ NOTA BENE. One way to formalise this 'incompatibility' between dotted and undotted indices would be to define rather $z^A = (z, A, 0)$ and $z^{\dot{A}} = (z, A, 1)$. Thus the index substitution $z^A = (z, A, 0) \mapsto z^A = (z, B, 0) = z^B$ clearly does not enable to translate from a dotted to an undotted index. Only the complex conjugation can through $z^A = (z, A, 0) \mapsto (\bar{z}, A, 1) = z^{\dot{A}}$.

INNER PRODUCT. We define the map J by:

$$J\begin{pmatrix} z_0 \\ z_1 \end{pmatrix} = \begin{pmatrix} -\overline{z_1} \\ \overline{z_0} \end{pmatrix}. \tag{152}$$

Using the previously introduced generalised Dirac notation, we read $J\,|z\rangle = |z]$. Since $J^2 = -1$, the map J behaves over $\mathbb{C}^2$ very much as the imaginary number i behaves over $\mathbb{C}$. For this reason the map J is said to define a *complex structure* over $\mathbb{C}^2$. A combination of $\epsilon_{AB}$ and J defines an inner product over $\mathfrak{S}^A$:

$$-\epsilon_{AB}(Jz)^A y^B = \overline{z_0}y_0 + \overline{z_1}y_1. \tag{153}$$

In generalised Dirac notations, we read $-[Jz|y\rangle = \langle z|y\rangle$, which is consistent with the usual Dirac notation for the scalar product.

REPRESENTATION. The vector space $\mathcal{M}_2(\mathbb{C})$ is isomorphic to $\mathfrak{S}^A_B$, through the isomorphism that associates to any $t \in \mathcal{M}_2(\mathbb{C})$ the unique spinor $t^A{}_B$ such that:

$$\forall z \in \mathbb{C}^2, \quad (tz)^A = t^A{}_B\, z^B. \tag{154}$$

Then the groups $SL_2(\mathbb{C})$ and SU(2) can be represented over $\mathfrak{S}^{A_1...A_p}$ such as:

$$u \cdot z^{A_1...A_p} = u^{A_1}{}_{B_1}...u^{A_p}{}_{B_p}\, z^{B_1...B_p}. \tag{155}$$

Yet this representation is not irreducible, since it is stable over the subspace of completely symmetric spinors $\mathfrak{S}^{(A_1...A_p)}$. Thus $SL_2(\mathbb{C})$ and SU(2) can be represented on the vector space $\mathfrak{S}^{(A_1...A_p)}$ of dimension $p+1$. A basis is given by

$$e_{i_1}^{(A_1}...e_{i_p}^{A_p)} \quad \text{with} \quad i_1, ..., i_p \in \{0, 1\}, \tag{156}$$

which can also be denoted as

$$e_0^{(A_1}...e_0^{A_m}e_1^{A_{m+1}}...e_1^{A_p)} \quad \text{with} \quad m \in \{0, ..., p\}. \tag{157}$$

This representation is irreducible and equivalent to the spin-($j = \frac{p}{2}$) irrep through the intertwiner:

$$e_0^{(A_1} ... e_0^{A_m} e_1^{A_{m+1}} ... e_1^{A_p)} \cong z_0^m z_1^{p-m}. \tag{158}$$

# 5 RECOUPLING THEORY OF SU(2)

The SU(2) irreps provide the fundamental building blocks of quantum space-time. From a mathematical perspective, irreps are the fundamental bricks from which other representations are built. Indeed any finite representation of SU(2) is *completely reducible*, i. e. it can be written as a direct sum of irreps. In particular, a tensor product of irreps can be *decomposed* into a direct sum of irreps, i.e. there exists a bijective intertwiner that maps the tensor product to a direct sum of irreps. Such an intertwiner is sometimes called a '*coupling tensor*' (see Moussouris [130] pp. 10-11). This naming comes from quantum physics: when two systems *couple* (i.e. interact), the total system is described by states of the tensor product of the Hilbert spaces of the subsystems. Notice that there may exist several coupling tensors between a tensor product and its decomposition into a sum of irreps. It is precisely the goal of '*recoupling theory*' to describe these coupling tensors and to understand how one can translate from one decomposition to another.

## 5.1 CLEBSCH-GORDAN COEFFICIENTS

Given $\mathcal{Q}_{j_1}$ and $\mathcal{Q}_{j_2}$, two irreps of SU(2), the tensor representation is defined over $\mathcal{Q}_{j_1} \otimes \mathcal{Q}_{j_2}$. The canonical basis of $\mathcal{Q}_{j_1} \otimes \mathcal{Q}_{j_2}$ is given by the elements

$$|j_1 m_1; j_2 m_2\rangle \stackrel{\text{def}}{=} |j_1, m_1\rangle \otimes |j_2, m_2\rangle \tag{159}$$

where $m_1$ and $m_2$ belong to the usual range of magnetic indices. Interestingly, this basis is the unique orthonormal basis that diagonalises simultaneously the commuting operators:

$$J_3 \otimes \mathbb{1}, \quad \mathbb{1} \otimes J_3, \quad \vec{J}^2 \otimes \mathbb{1}, \quad \mathbb{1} \otimes \vec{J}^2. \tag{160}$$

Another CSCO on $\mathcal{Q}_{j_1} \otimes \mathcal{Q}_{j_2}$ is given by

$$J_3 \otimes \mathbb{1} + \mathbb{1} \otimes J_3, \quad \left(\vec{J} \otimes \mathbb{1} + \mathbb{1} \otimes \vec{J}\right)^2, \quad \vec{J}^2 \otimes \mathbb{1}, \quad \mathbb{1} \otimes \vec{J}^2. \tag{161}$$

$\left(\vec{J} \otimes \mathbb{1} + \mathbb{1} \otimes \vec{J}\right)^2$ *is a shorthand for* $(J_1 \otimes \mathbb{1} + \mathbb{1} \otimes J_1)^2 + (J_2 \otimes \mathbb{1} + \mathbb{1} \otimes J_2)^2 + (J_3 \otimes \mathbb{1} + \mathbb{1} \otimes J_3)^2$

Therefore, there exists an orthonormal basis that diagonalises them simultaneously. It is denoted

$$|j_1 j_2; k, n\rangle \quad \text{with} \quad k \in \{|j_1 - j_2|, ..., j_1 + j_2\} \quad \text{and} \quad n \in \{-k, ..., k\}, \tag{162}$$

and characterised by the action of the operators:

$$\begin{aligned}(J_3 \otimes \mathbb{1} + \mathbb{1} \otimes J_3)\,|j_1 j_2; k; n\rangle &= n\,|j_1 j_2; k; n\rangle \\ \left(\vec{J} \otimes \mathbb{1} + \mathbb{1} \otimes \vec{J}\right)^2 |j_1 j_2; k; n\rangle &= k(k+1)\,|j_1 j_2; k; n\rangle \\ \vec{J}^2 \otimes \mathbb{1}\,|j_1 j_2; k; n\rangle &= j_1(j_1+1)\,|j_1 j_2; k; n\rangle \\ \mathbb{1} \otimes \vec{J}^2\,|j_1 j_2; k; n\rangle &= j_2(j_2+1)\,|j_1 j_2; k; n\rangle\,. \end{aligned} \tag{163}$$

This result proves that $\mathcal{Q}_{j_1} \otimes \mathcal{Q}_{j_2}$ can be decomposed into a direct sum of irreps, namely we have the following equivalence of representations

$$\mathcal{Q}_{j_1} \otimes \mathcal{Q}_{j_2} \cong \bigoplus_{k=|j_1-j_2|}^{j_1+j_2} \mathcal{Q}_k. \tag{164}$$

The equivalence is given by the bijective intertwiner

$$\iota \begin{cases} \mathcal{Q}_{j_1} \otimes \mathcal{Q}_{j_2} \to \bigoplus_{k=|j_1-j_2|}^{j_1+j_2} \mathcal{Q}_k \\ |j_1, j_2; km\rangle \mapsto |km\rangle\,. \end{cases} \tag{165}$$

We define the *Clebsch-Gordan coefficients* by the scalar product

$$C^{jm}_{j_1 m_1 j_2 m_2} \overset{\text{def}}{=} \langle j_1 m_1; j_2 m_2 | j_1 j_2; jm\rangle \tag{166}$$

so that

$$|j_1 j_2; j, m\rangle = \sum_{m_1=-j_1}^{j_1} \sum_{m_2=-j_2}^{j_2} C^{jm}_{j_1 m_1 j_2 m_2}\,|j_1 m_1 j_2 m_2\rangle\,. \tag{167}$$

The Clebsh-Gordan coefficients can be seen as the matrix coefficients of the intertwiner $\iota$ in the canonical bases.

REMARKS

1. Due to the Condon-Shortley convention for the SU(2)-action, we have $C^{jm}_{j_1 m_1 j_2 m_2} \in \mathbb{R}$.

2. The coefficients $C^{jm}_{j_1 m_1 j_2 m_2}$ are well-defined and non-zero, only if the following *Clebsch-Gordan inequality* (aka *triangle inequality*) is satisfied
$$|j_1 - j_2| \leqslant j \leqslant j_1 + j_2. \tag{168}$$
Otherwise, we choose by convention, that $C^{jm}_{j_1 m_1 j_2 m_2} = 0$.

3. If $m \neq m_1 + m_1$, then $C^{jm}_{j_1 m_1 j_2 m_2} = 0$.

4. Since the $|jm\rangle$ form an orthonormal basis, we have the following 'orthogonality relation'
$$\sum_{m_1=-j_1}^{j_1} \sum_{m_2=-j_2}^{j_2} C^{jm}_{j_1 m_1 j_2 m_2} C^{j'm'}_{j_1 m_1 j_2 m_2} = \delta_{j,j'}\delta_{m,m'}. \tag{169}$$

5. Another consequence is the decomposition of products of Wigner matrices into sums, like

$$D^{j_1}_{m_1 n_1}(g) D^{j_2}_{m_2 n_2}(g) = \sum_{j \in \mathbb{N}/2} \sum_{m=-j}^{j} \sum_{m'=-j}^{j} C^{jm}_{j_1 m_1 j_2 m_2} C^{jm'}_{j_1 n_1 j_2 n_2} D^{j}_{mm'}(g). \quad (170)$$

The Clebsch-Gordan coefficients are numbers, but their definition is quite implicit. Hopefully, we have explicit formulas to compute them!

EXPLICIT FORMULA.

*Other similar expressions can be found in Varshalovich ([189] p. 238), notably in terms of the hyper-geometrical function ${}_3F_2$.*

$$C^{jm}_{j_1 m_1 j_2 m_2} = \delta_{m, m_1+m_2} \sqrt{2j+1} \times \sqrt{\frac{(j+m)!(j-m)!(-j+j_1+j_2)!(j-j_1+j_2)!(j+j_1-j_2)!}{(j+j_1+j_2+1)!(j_1+m_1)!(j_1-m_1)!(j_2+m_2)!(j_2-m_2)!}} \times \sum_k \frac{(-1)^{k+j_2+m_2}(j+j_2+m_1-k)!(j_1-m_1+k)!}{(j-j_1+j_2-k)!(j+m-k)!k!(k+j_1-j_2-m)!} \quad (171)$$

In the Wolfram Language, they are implemented as

$$C^{j_3 m_3}_{j_1 m_1 j_2 m_2} = \texttt{ClebschGordan}[\{j_1, m_1\}, \{j_2, m_2\}, \{j_3, m_3\}]. \quad (172)$$

## 5.2 INVARIANT SUBSPACE

Generally speaking, a tensor product of $n$ irreps can be decomposed into a direct sum

$$\bigotimes_{i=1}^{n} \mathcal{Q}_{j_i} \cong \bigoplus_{k=0}^{J} \Big( \underbrace{\mathcal{Q}_k \oplus \ldots \oplus \mathcal{Q}_k}_{d_k \text{ times}} \Big), \quad (173)$$

where $J \stackrel{\text{def}}{=} \sum_i j_i$ and $d_k$ is the degeneracy of the irrep $\mathcal{Q}_k$. Here, 'decomposing' means 'finding a bijective intertwiner between the two spaces'. Concretely, such a decomposition is obtained by applying successively the decomposition of a product of only two irreps, as given by equation (164). It is usual to denote the operator $(J_i)_k \stackrel{\text{def}}{=} \mathbb{1} \otimes \ldots \otimes J_i \otimes \ldots \otimes \mathbb{1}$ corresponding to $J_i$ acting on the $k^{\text{th}}$ Hilbert space of the product. The three components $(J_1)_k$, $(J_2)_k$, $(J_3)_k$ form the vectorial operator $\vec{J}_k$.

We define the SU(2)-invariant subspace as

$$\mathrm{Inv}_{\mathrm{SU}(2)} \left( \bigotimes_{i=1}^{n} \mathcal{Q}_{j_i} \right) \stackrel{\text{def}}{=} \left\{ \psi \in \bigotimes_{i=1}^{n} \mathcal{Q}_{j_i} \mid \forall g \in \mathrm{SU}(2),\ g \cdot \psi = \psi \right\}. \quad (174)$$

It can also be characterized by the action of the algebra:

$$\mathrm{Inv}_{\mathrm{SU}(2)} \left( \bigotimes_{i=1}^{n} \mathcal{Q}_{j_i} \right) = \left\{ \psi \in \bigotimes_{i=1}^{n} \mathcal{Q}_{j_i} \mid \forall s \in \mathfrak{su}(2),\ s \cdot \psi = 0 \right\}. \quad (175)$$

From this, it is easy to see that

$$\mathrm{Inv}_{SU(2)}\left(\bigotimes_{i=1}^{n}\mathcal{Q}_{j_i}\right)\cong\underbrace{\mathcal{Q}_0\oplus\ldots\oplus\mathcal{Q}_0}_{d_0\ \text{times}},\tag{176}$$

where $\mathcal{Q}_0 \cong \mathbb{C}$ is the trivial representation. Interestingly, we also have the following isomorphism:

$$\mathrm{Inv}_{SU(2)}\left(\bigotimes_{i=1}^{n}\mathcal{Q}_{j_i}\right)\cong\mathrm{Hom}_{SU(2)}\left(\bigotimes_{i=1}^{n}\mathcal{Q}_{j_i},\mathcal{Q}_0\right),\tag{177}$$

where the RHS is the vector space of SU(2)-intertwiners between $\bigotimes_{i=1}^{n}\mathcal{Q}_{j_i}$ and $\mathcal{Q}_0$. It is a particular case of equation (64).

ORTHOGONAL PROJECTOR. By definition, the orthogonal projector

$$P:\bigotimes_{i=1}^{n}\mathcal{Q}_{j_i}\to\mathrm{Inv}_{SU(2)}\left(\bigotimes_{i=1}^{n}\mathcal{Q}_{j_i}\right)\tag{178}$$

satisfies

$$P^2=P\quad\text{and}\quad P^\dagger=P.\tag{179}$$

It is easy to show that

$$P=\int_{SU(2)}du\bigotimes_{k=1}^{n}D^{j_k}(u).\tag{180}$$

If $|\iota\rangle$ is an orthonormal basis of $\mathrm{Inv}_{SU(2)}\left(\bigotimes_{i=1}^{n}\mathcal{Q}_{j_i}\right)$, then $P$ can also be written as

$$P=\sum_{\iota}|\iota\rangle\langle\iota|.\tag{181}$$

## 5.3 WIGNER'S $3jm$-SYMBOL

We can decompose $\mathcal{Q}_{j_1}\otimes\mathcal{Q}_{j_2}\otimes\mathcal{Q}_{j_3}$ into a direct sum by applying equation (164) twice, and first on the left tensor product:

$$\mathcal{Q}_{j_1}\otimes\mathcal{Q}_{j_2}\otimes\mathcal{Q}_{j_3}\to\left(\bigoplus_{j_{12}}\mathcal{Q}_{j_{12}}\right)\otimes\mathcal{Q}_{j_3}\to\bigoplus_{j_{12}=|j_1-j_2|}^{j_1+j_2}\ \bigoplus_{k=|j_{12}-j_3|}^{j_{12}+j_3}\mathcal{Q}_k\tag{182}$$

Thus we construct an orthonormal basis of $\mathcal{Q}_{j_1}\otimes\mathcal{Q}_{j_2}\otimes\mathcal{Q}_{j_3}$ given by the states

$$\begin{aligned}&|(j_1j_2)j_3;j_{12}kn\rangle\\&=\sum_{m_1,m_2,m_3,m_{12}}C^{j_{12}m_{12}}_{j_1m_1j_2m_2}C^{kn}_{j_{12}m_{12}j_3m_3}\bigotimes_{i=1}^{3}|j_i,m_i\rangle,\\&\qquad\text{with } j_{12}\in\{|j_1-j_2|,\ldots,j_1+j_2\}\\&\qquad\text{and } k\in\{|j_{12}-j_3|,\ldots,j_{12}+j_3\}\\&\qquad\qquad\text{and } n\in\{-k,\ldots,k\}.\end{aligned}\tag{183}$$

If the Clebsch-Gordan condition (168) is satisfied, then one can be show that

$$\mathrm{Inv}_{SU(2)}\left(\mathcal{Q}_{j_1}\otimes\mathcal{Q}_{j_2}\otimes\mathcal{Q}_{j_3}\right) = \mathrm{Span}\{|(j_1j_2)j_3; j_3 00\rangle\}, \tag{184}$$

so that $\mathrm{Inv}_{SU(2)}(\mathcal{Q}_{j_1}\otimes\mathcal{Q}_{j_2}\otimes\mathcal{Q}_{j_3})$ is one dimensional. Otherwise

$$\mathrm{Inv}_{SU(2)}(\mathcal{Q}_{j_1}\otimes\mathcal{Q}_{j_2}\otimes\mathcal{Q}_{j_3}) = \{0\}. \tag{185}$$

Now supposing that the condition is satisfied, there exists a unique unit vector in $\mathrm{Inv}_{SU(2)}(\mathcal{Q}_{j_1}\otimes\mathcal{Q}_{j_2}\otimes\mathcal{Q}_{j_3})$,

$$|0\rangle = \sum_{m_1,m_2,m_3}\begin{pmatrix} j_1 & j_2 & j_3\\ m_1 & m_2 & m_3\end{pmatrix}\bigotimes_{k=1}^{3}|j_k, m_k\rangle, \tag{186}$$

such that the coefficients $\begin{pmatrix} j_1 & j_2 & j_3\\ m_1 & m_2 & m_3\end{pmatrix}$ are real and satisfy the symmetry properties

$$\begin{pmatrix} j_1 & j_2 & j_3\\ m_1 & m_2 & m_3\end{pmatrix} = \begin{pmatrix} j_3 & j_1 & j_2\\ m_3 & m_1 & m_2\end{pmatrix} = \begin{pmatrix} j_2 & j_3 & j_1\\ m_2 & m_3 & m_1\end{pmatrix}. \tag{187}$$

The coefficients $\begin{pmatrix} j_1 & j_2 & j_3\\ m_1 & m_2 & m_3\end{pmatrix}$ are called the *Wigner's* 3jm-*symbol* and are related to the Clebsch-Gordan coefficients by

$$\begin{pmatrix} j_1 & j_2 & j_3\\ m_1 & m_2 & m_3\end{pmatrix} = \frac{(-1)^{j_1-j_2-m_3}}{\sqrt{2j_3+1}}C^{j_3,-m_3}_{j_1m_1j_2m_2}. \tag{188}$$

In Mathematica, they are given by

$$\texttt{ThreeJSymbol}[\{j_1,m_1\},\{j_2,m_2\},\{j_3,m_3\}] = \begin{pmatrix} j_1 & j_2 & j_3\\ m_1 & m_2 & m_3\end{pmatrix}. \tag{189}$$

REMARKS.

1. These symbols satisfy nice symmetry properties such as:

$$\begin{aligned}\begin{pmatrix} j_1 & j_2 & j_3\\ m_1 & m_2 & m_3\end{pmatrix} &= (-1)^{j_1+j_2+j_3}\begin{pmatrix} j_2 & j_1 & j_3\\ m_2 & m_1 & m_3\end{pmatrix}\\ &= (-1)^{j_1+j_2+j_3}\begin{pmatrix} j_1 & j_2 & j_3\\ -m_1 & -m_2 & -m_3\end{pmatrix}.\end{aligned} \tag{190}$$

2. The orthogonality relation (169) becomes

$$\sum_{jm}(2j+1)\begin{pmatrix} j_1 & j_2 & j\\ m_1 & m_2 & m\end{pmatrix}\begin{pmatrix} j_1 & j_2 & j\\ m_1' & m_2' & m\end{pmatrix} = \delta_{m_1m_1'}\delta_{m_2m_2'} \tag{191}$$

$$\sum_{m_1m_2}(2j+1)\begin{pmatrix} j_1 & j_2 & j\\ m_1 & m_2 & m\end{pmatrix}\begin{pmatrix} j_1 & j_2 & j'\\ m_1 & m_2 & m'\end{pmatrix} = \delta_{jj'}\delta_{mm'} \tag{192}$$

3. Equating (180) and (181) in the magnetic basis, we find

$$\int_{SU(2)} D^{j_1}_{m_1 n_1}(u) D^{j_2}_{m_2 n_2}(u) D^{j_3}_{m_3 n_3}(u) du = \begin{pmatrix} j_1 & j_2 & j_3 \\ m_1 & m_2 & m_3 \end{pmatrix} \begin{pmatrix} j_1 & j_2 & j_3 \\ n_1 & n_2 & n_3 \end{pmatrix}. \quad (193)$$

## 5.4 WIGNER'S $4jm$-SYMBOL

Similarly to the previous section, we can decompose $\mathcal{Q}_{j_1} \otimes \mathcal{Q}_{j_2} \otimes \mathcal{Q}_{j_3} \otimes \mathcal{Q}_{j_4}$ into a direct sum by applying equation (164) successively. We get

$$\mathcal{Q}_{j_1} \otimes \mathcal{Q}_{j_2} \otimes \mathcal{Q}_{j_3} \otimes \mathcal{Q}_{j_4} \cong \bigoplus_{j_{12}=|j_1-j_2|}^{j_1+j_2} \bigoplus_{k=|j_{12}-j_3|}^{j_{12}+j_3} \bigoplus_{l=|k-j_4|}^{k+j_4} \mathcal{Q}_l \quad (194)$$

In particular, we can see that

$$\mathrm{Inv}_{SU(2)}\left(\bigotimes_{i=1}^{4} \mathcal{Q}_{j_i}\right) \cong \underbrace{\mathcal{Q}_0 \oplus \ldots \oplus \mathcal{Q}_0}_{d_0 \text{ times}} \quad (195)$$

with $d_0 = \max(|j_1 - j_2|, |j_3 - j_4|) - \min(j_1 + j_2, j_3 + j_4)$. An orthonormal basis of $\mathrm{Inv}_{SU(2)}\left(\bigotimes_{i=1}^{4} \mathcal{Q}_{j_i}\right)$ is given by

$$|j\rangle_{12} = \sqrt{2j+1} \sum_{\substack{m_1, m_2, \\ m_3, m_4}} \begin{pmatrix} j_1 & j_2 & j_3 & j_4 \\ m_1 & m_2 & m_3 & m_4 \end{pmatrix}^{(j)} \bigotimes_{k=1}^{4} |j_k, m_k\rangle, \quad (196)$$

with $j \in \{\max(|j_1 - j_2|, |j_3 - j_4|), \ldots, \min(j_1 + j_2, j_3 + j_4)\}$, and

$$\begin{pmatrix} j_1 & j_2 & j_3 & j_4 \\ m_1 & m_2 & m_3 & m_4 \end{pmatrix}^{(j)} \stackrel{\text{def}}{=} \sum_{m} (-1)^{j-m} \begin{pmatrix} j_1 & j_2 & j \\ m_1 & m_2 & m \end{pmatrix} \begin{pmatrix} j & j_3 & j_4 \\ -m & m_3 & m_4 \end{pmatrix} \quad (197)$$

This basis has the interesting property that it diagonalises $(\vec{J}_1 + \vec{J}_2)^2$:

$$(\vec{J}_1 + \vec{J}_2)^2 \, |j\rangle_{12} = j(j+1) \, |j\rangle_{12}, \quad (198)$$

and this explains the notation '12' in index. The $4jm$-symbol also satisfy orthogonality relations:

$$\sum_{m_1, m_2, m_3} \begin{pmatrix} j_1 & j_2 & j_3 & j_4 \\ m_1 & m_2 & m_3 & m_4 \end{pmatrix}^{(j_{12})} \begin{pmatrix} j_1 & j_2 & j_3 & l_4 \\ m_1 & m_2 & m_3 & n_4 \end{pmatrix}^{(l_{12})} = \frac{\delta_{j_{12} l_{12}}}{2j_{12}+1} \frac{\delta_{j_4 l_4} \delta_{m_4 n_4}}{2j_4+1}. \quad (199)$$

Finally we can show, similarly to equation (193), that

$$\int_{SU(2)} D^{j_1}_{m_1 n_1}(u) D^{j_2}_{m_2 n_2}(u) D^{j_3}_{m_3 n_3}(u) D^{j_4}_{m_4 n_4}(u)\, du$$
$$= \sum_j (2j+1) \begin{pmatrix} j_1 & j_2 & j_3 & j_4 \\ m_1 & m_2 & m_3 & m_4 \end{pmatrix}^{(j)} \begin{pmatrix} j_1 & j_2 & j_3 & j_4 \\ n_1 & n_2 & n_3 & n_4 \end{pmatrix}^{(j)} . \tag{200}$$

## 5.5 wigner's 6j-symbol

In the previous section, we have exhibited an orthonormal basis for $\mathrm{Inv}_{SU(2)}\left(\bigotimes_{i=1}^4 \mathcal{Q}_{j_i}\right)$. It is built from one possible decomposition of $\bigotimes_{i=1}^4 \mathcal{Q}_{j_i}$ into irreps. Another possible decomposition leads to another basis

$$|j\rangle_{23} = \sum_{\substack{m_1, m_2, \\ m_3, m_4}} \sqrt{2j+1} \begin{pmatrix} j_4 & j_1 & j_2 & j_3 \\ m_4 & m_1 & m_2 & m_3 \end{pmatrix}^{(j)} \bigotimes_{i=1}^{4} |j_i, m_i\rangle , \tag{201}$$

that diagonalises $(\vec{J}_2 + \vec{J}_3)^2$. The change of basis is given by

$$_{12}\langle j|k\rangle_{23}$$
$$= \sqrt{2j+1}\sqrt{2k+1}(-1)^{j_1+j_2+j_3-j_4-2j-2k} \begin{Bmatrix} j_1 & j_2 & j \\ j_3 & j_4 & k \end{Bmatrix} \tag{202}$$

where we have defined a new symbol:

$$\begin{Bmatrix} j_1 & j_2 & j_3 \\ j_4 & j_5 & j_6 \end{Bmatrix} \stackrel{\text{def}}{=} \sum_{m_1,\dots,m_6} (-1)^{\sum_{i=1}^6 (j_i - m_i)}$$
$$\times \begin{pmatrix} j_1 & j_2 & j_3 \\ -m_1 & -m_2 & -m_3 \end{pmatrix} \begin{pmatrix} j_1 & j_5 & j_6 \\ m_1 & -m_5 & m_6 \end{pmatrix}$$
$$\times \begin{pmatrix} j_4 & j_2 & j_6 \\ m_4 & m_2 & -m_6 \end{pmatrix} \begin{pmatrix} j_3 & j_4 & j_5 \\ m_3 & -m_4 & m_5 \end{pmatrix} \tag{203}$$

In the Wolfram Language, it is returned by the function

$$\texttt{SixJSymbol}[\{j_1, j_2, j_3\}, \{j_4, j_5, j_6\}] = \begin{Bmatrix} j_1 & j_2 & j_3 \\ j_4 & j_5 & j_6 \end{Bmatrix} . \tag{204}$$

This symbol satisfies the symmetries

$$\begin{Bmatrix} j_1 & j_2 & j_3 \\ j_4 & j_5 & j_6 \end{Bmatrix} = \begin{Bmatrix} j_2 & j_1 & j_3 \\ j_5 & j_4 & j_6 \end{Bmatrix} = \begin{Bmatrix} j_3 & j_2 & j_1 \\ j_6 & j_5 & j_4 \end{Bmatrix} = \begin{Bmatrix} j_4 & j_2 & j_3 \\ j_1 & j_5 & j_6 \end{Bmatrix} . \tag{205}$$

Similarly one can define the symbols $9j$ and $15j$.

The $\{6j\}$-symbol appeared in quantum gravity when Ponzano and Regge realised that the $\{6j\}$-symbol approximate the action of general relativity in the semi-classical limit [150]. More precisely, they have shown that

$$\begin{Bmatrix} j_1 & j_2 & j_3 \\ j_4 & j_5 & j_6 \end{Bmatrix} \underset{j_i \to \infty}{\sim} \frac{1}{\sqrt{12\pi V}} \cos\left(S(j_i) + \pi/4\right), \tag{206}$$

where $V$ is the volume of a tetrahedron whose edges have a length of $j_i + 1/2$, and $S(j_i)$ is the so-called *Regge action*, which is a discrete 3-dimensional version of the Einstein-Hilbert action. This result was a important source of inspiration for later development of spin-foams.

## 5.6 GRAPHICAL CALCULUS

The recoupling theory of SU(2) can be nicely implemented graphically. The underlying philosophy of it is to take advantage of the two dimensions offered by our sheets of paper and our black-boards to literally *draw* our calculations, rather than restricting oneself to the usual one-dimensional lines of calculations. If done properly, the method can help to understand the structure of analytical expressions and make computations faster. Of course, the first principle of graphical calculus is that there should be a one-to-one correspondence between analytical expressions and diagrams. There exist many conventions for this correspondence in the literature, so we have chosen one that seems to be quite popular [166], and which is described in detail by Varshalovich ([189], Chap. 11).

DEFINITIONS. The basic object of this graphical calculus is the 3-valent node, that represents the Wigner's $3jm$ symbol:

$$\begin{pmatrix} j_1 & j_2 & j_3 \\ m_1 & m_2 & m_3 \end{pmatrix} =$$

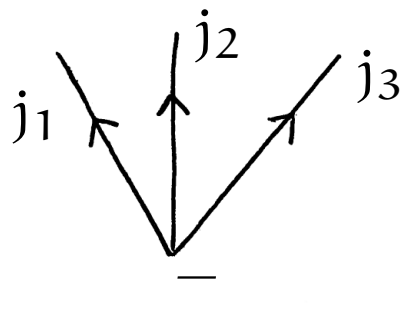

$=$

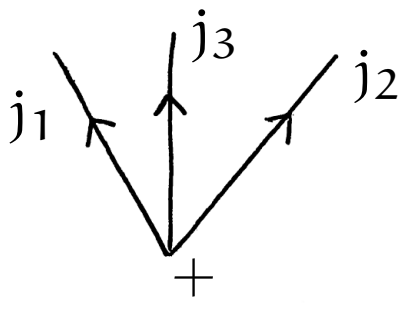

. (207)

Remarks:

1. The signs $+/-$ on the nodes indicate the sense of rotation (anticlockwise/clockwise) in which the spins must be read. To alleviate notations we decide not to write them in the following by choosing conventionally that the default sign of the nodes is minus, if not otherwise specified.
2. The arrows on the wires will be used below to define the operation of summation.

3. Everywhere, we implicitly assume that the Clebsch-Gordan inequalities are satisfied.
4. The magnetic indices are implicit in the diagram, which creates no ambiguity, as long as we associate the $m_i$ to the spin $j_i$.
5. The symmetry properties (187) are naturally implemented on the diagram, which also guarantees the one-to-one correspondence between the analytical expression and the diagram.
6. Only the topology of the diagram matters, which means that all topological deformations are allowed.

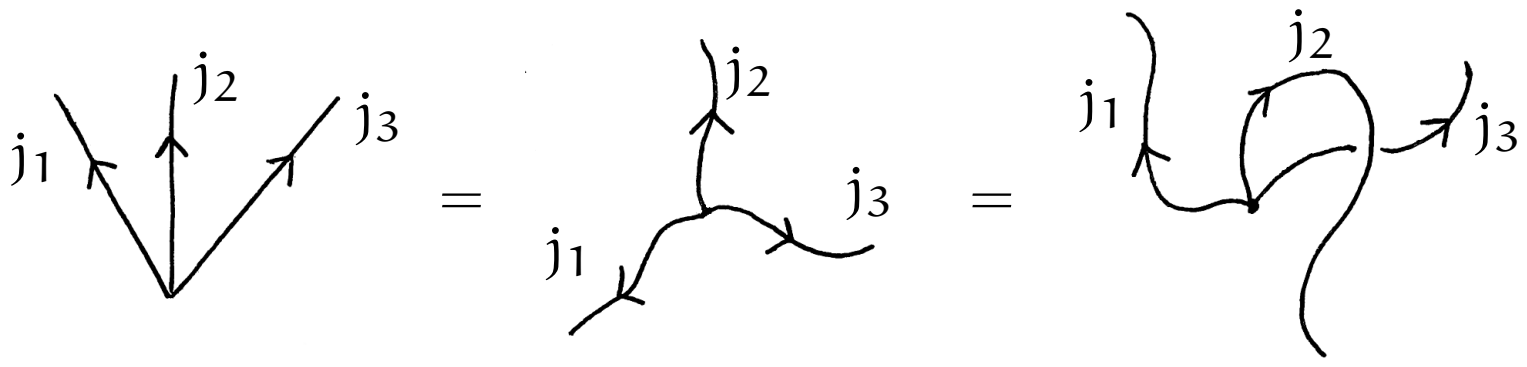

(208)

This principle of topological equivalence is a strong principle of graphical calculus, that will hold for any other diagram constructed later.

Then we can define graphically the two basic operations of algebra: multiplication and summation. Multiplication is implemented simply by juxtaposition of two diagrams:

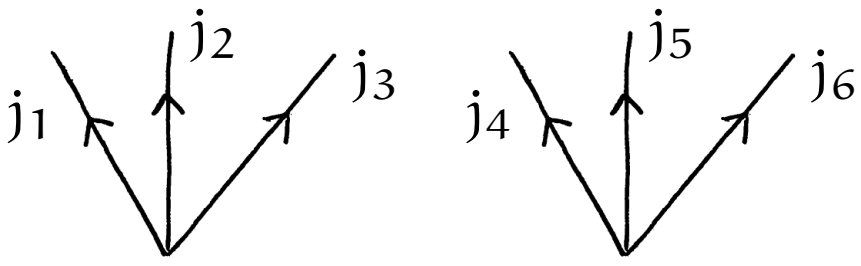

$$= \begin{pmatrix} j_1 & j_2 & j_3 \\ m_1 & m_2 & m_3 \end{pmatrix} \begin{pmatrix} j_4 & j_5 & j_6 \\ m_4 & m_5 & m_6 \end{pmatrix} \quad (209)$$

To define summation, we shall first tell more about the orientation of external wires. As you may have noticed, the arrows on the wires are all outgoing. Now we define also the ingoing orientation with the general rule that inverting the orientation of an external line $(jm)$ amounts analytically to transforming $m$ to $-m$ and multiplying the overall expression by a factor $(-1)^{j-m}$. For instance

*The inversion of the orientation can be seen as the contraction with the 'metric tensor' $\epsilon_{mm'} = (-1)^{j-m}\delta_{m,-m'}$.*

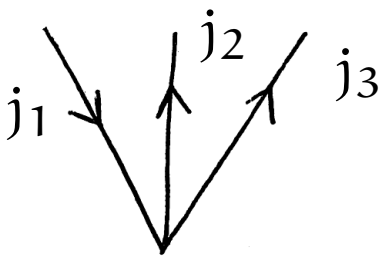

$$= (-1)^{j_1-m_1} \begin{pmatrix} j_1 & j_2 & j_3 \\ -m_1 & m_2 & m_3 \end{pmatrix}. \quad (210)$$

The summation over a magnetic index m (from $-j$ to $j$) is now represented by gluing two external wires with the same spin $j$ and magnetic index $m$, but of opposite directions, like:

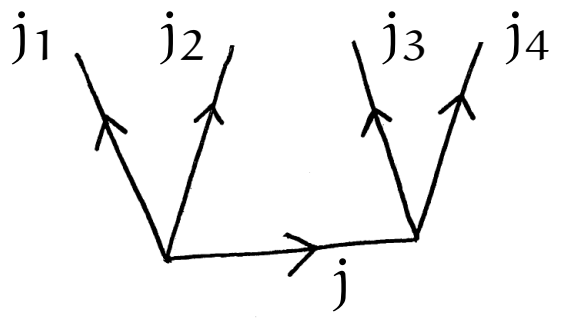

$$= \sum_{m=-j}^{j}$$

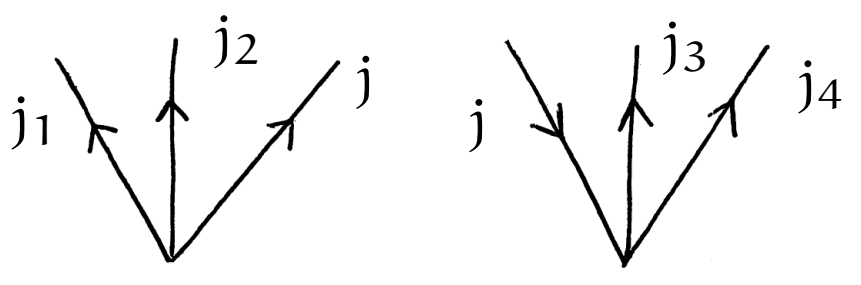

(211)

On the RHS, we recognise the definition of the $4jm$-symbol, so that

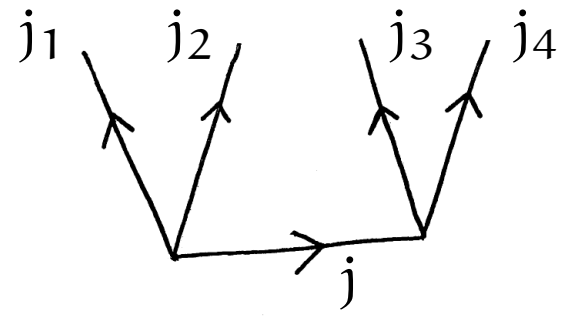

$$= \begin{pmatrix} j_1 & j_2 & j_3 & j_4 \\ m_1 & m_2 & m_3 & m_4 \end{pmatrix}^{(j)} \tag{212}$$

The line between two nodes, whose magnetic index is summed over, is called an *internal* line, in opposition to external lines, which have a free hand. The rule of inversion for internal lines is a bit different than for external ones, as it just gives a global phase:

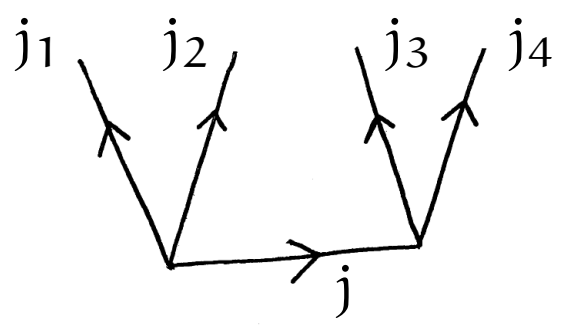

$$= (-1)^{2j}$$

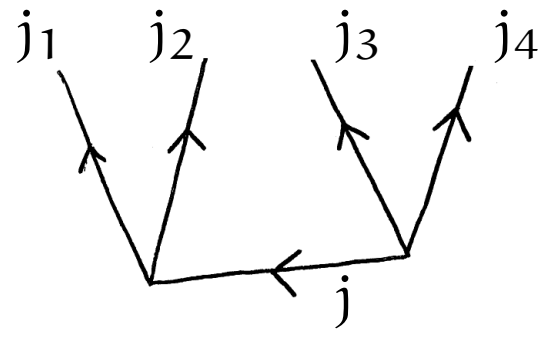

.

A powerful aspect of graphical calculus comes from the representation of the Kronecker delta with a single line

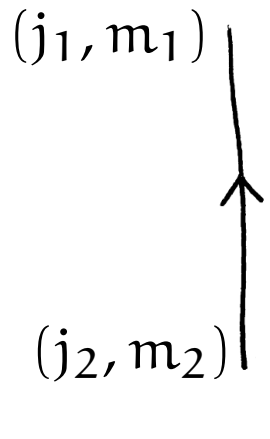

$$= \delta_{j_1 j_2} \delta_{m_1 m_2}$$

or

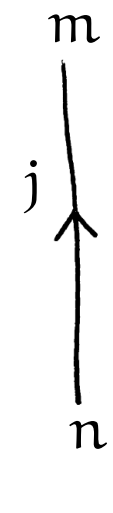

$$= \delta_{mn} \, . \tag{213}$$

We can apply the rule of summation to compute its trace:

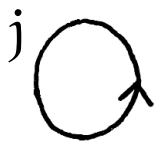

$$= 2j + 1. \tag{214}$$

This kind of diagrams with no external lines encode numbers. All magnetic indices are summed over, so that it is only a function of the internal spins $j_i$, what we can call an *invariant function*. On the contrary, diagrams with external lines encode tensors with one index per free-hand. Also, the orthogonality relation (192) now reads

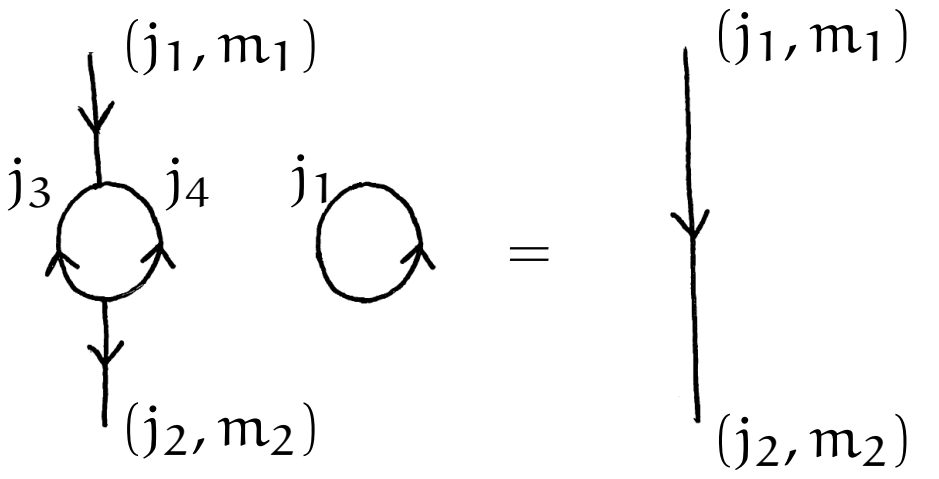

(215)

It gives a way to remove internal loops from diagrams.

LEMMAS. From all the rules described above, the following lemma can already be checked.

1. Reversing all external lines has no effect:

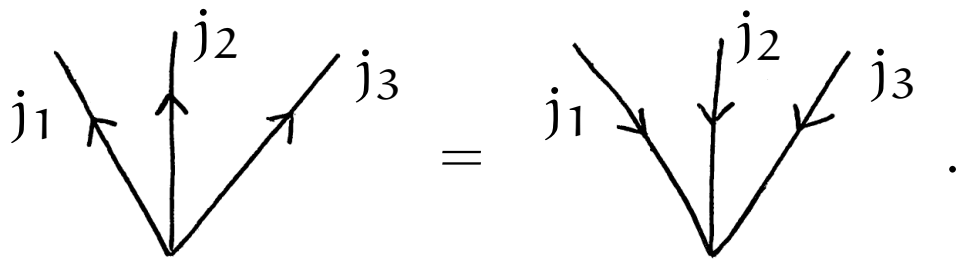

(216)

2. Changing the sign of the node gives a phase

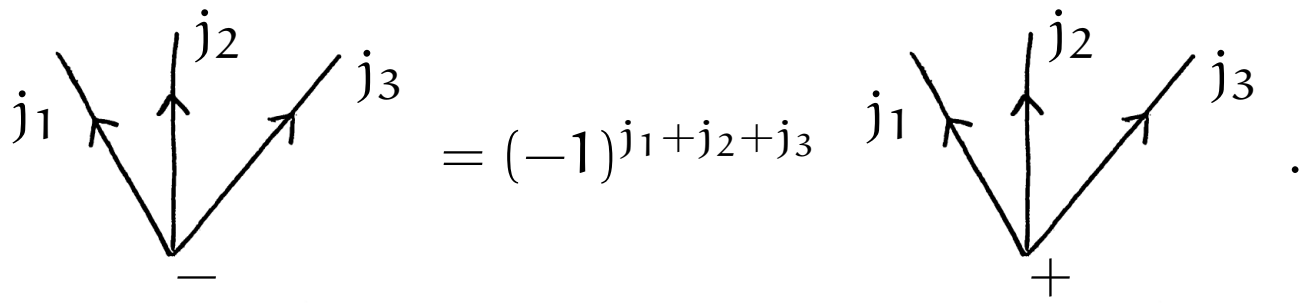

(217)

3. The evaluation of the so-called Θ-graph:

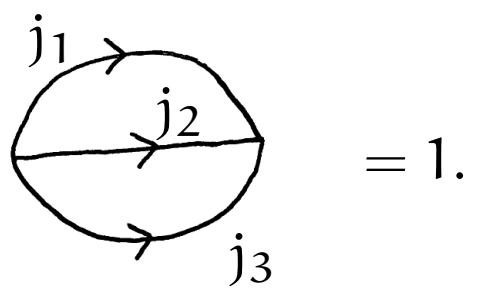

(218)

4. Similarly, equation (199) implies

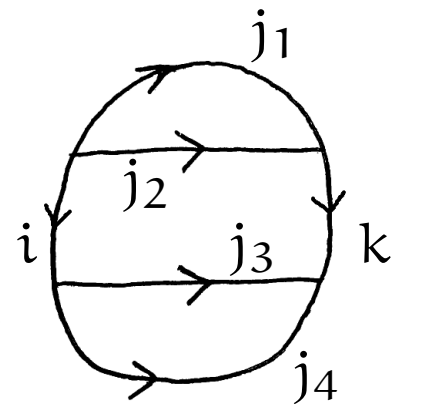

$$= \frac{\delta_{i,k}}{d_i}. \tag{219}$$

INVARIANT FUNCTIONS. One nice thing about this graphical calculus is that it makes it easy to represent and to remember the Wigner 6j-symbol:

$$\begin{Bmatrix} j_1 & j_2 & j_3 \\ j_4 & j_5 & j_6 \end{Bmatrix} =$$ 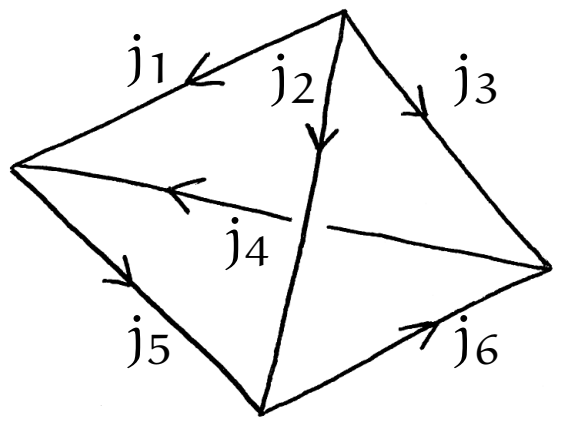 (220)

As we can see, the 6j-symbol looks like a tetrahedron. We can define other invariant functions in the same spirit, like 9j-symbol:

$$\begin{Bmatrix} j_1 & j_2 & j_3 \\ j_4 & j_5 & j_6 \\ j_7 & j_8 & j_9 \end{Bmatrix} =$$ 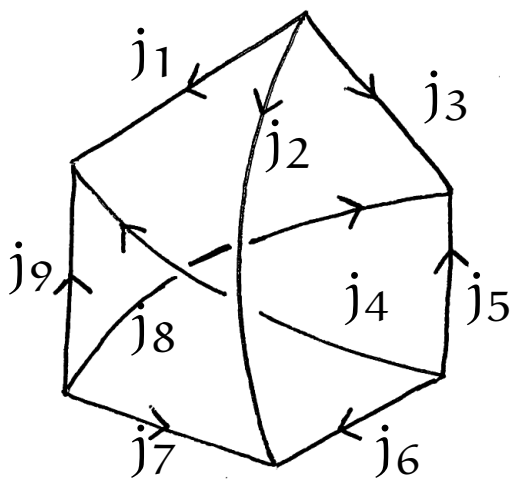 (221)

Notice that we could have also defined the 9j-symbol to be rather

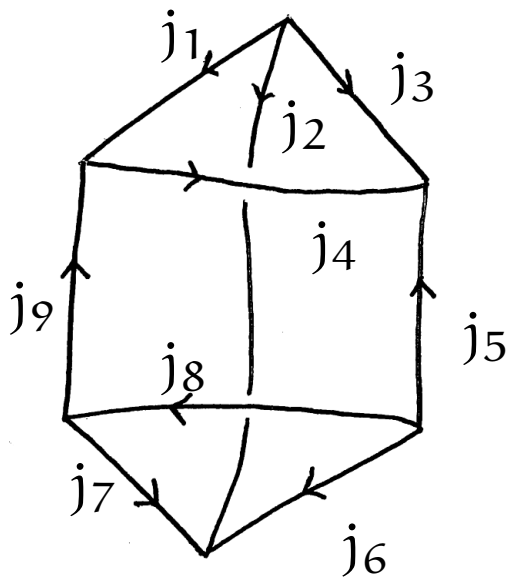 (222)

but this one can be actually rewritten as a product of two 6j-symbols. Such a decomposition cannot be done with the 9j-symbol of equation (221), so that it is said 'irreducible'. We also have the 15j-symbol:

*For more details on the 9j-symbol, see [59] pp. 100-114.*

$$\begin{Bmatrix} j_1 & j_2 & j_{11} \\ j_4 & j_5 & j_{15} \\ j_7 & j_3 & j_{14} \\ j_9 & j_6 & j_{13} \\ j_8 & j_{10} & j_{12} \end{Bmatrix} =$$ 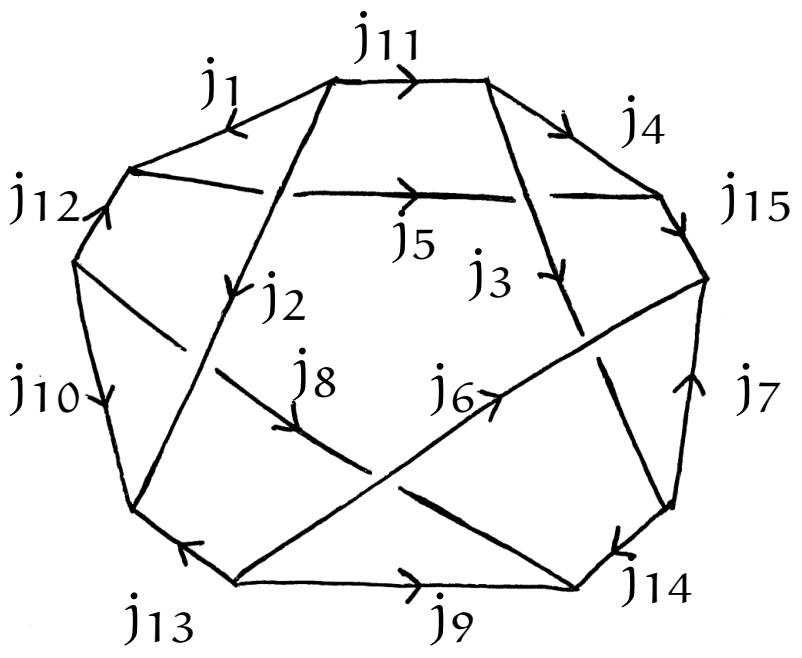 (223)

which is the definition used by [166]. It is different from the convention chosen in [137], which is

$$\begin{Bmatrix} l_1 & l_2 & l_3 & l_4 & l_5 \\ j_1 & j_2 & j_3 & j_4 & j_5 \\ l_{10} & l_9 & l_8 & l_7 & l_6 \end{Bmatrix} = \qquad (224)$$

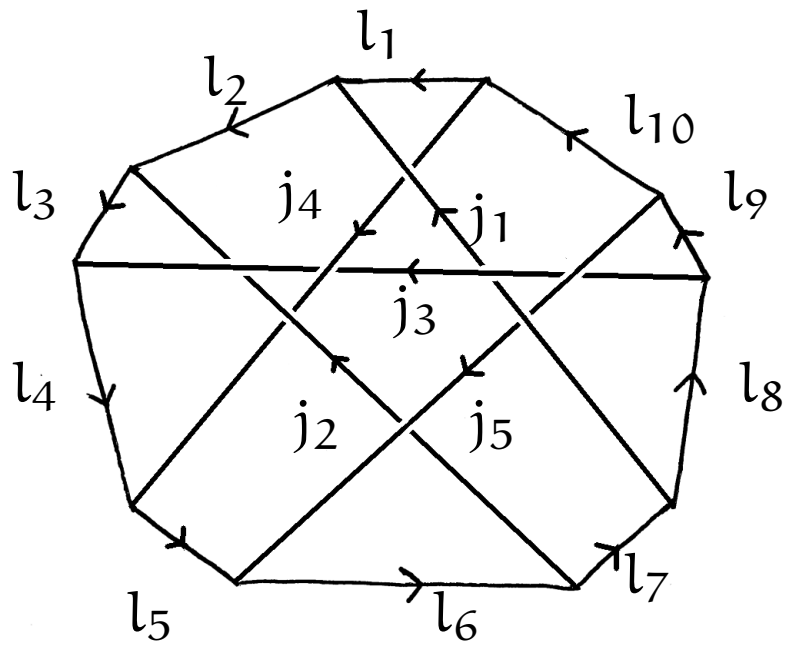

Contrary to the 6j-symbol, there is no consensus about what which convention should be used to define the 15j-symbol, but in all cases it corresponds to an invariant function associated to a 3-valent graph with 15 links. Actually, we can build 5 topologically different 15j-symbols. Here we see the power of graphical calculus: imagine if we had given the analytical formula for it... that is doable, but unreadable.

*For more details on the 15j-symbol, see [197] pp. 65-70.*

In the spirit of the result of Ponzano and Regge (206), Ooguri used the 15j-symbol to provide a model of quantum gravity [137]. It still plays a major role in the EPRL model [174] (see chapter 16).

# 6 HARMONIC ANALYSIS OVER SU(2)

Harmonic analysis is a subfield of mathematics which studies how functions can be decomposed as a sum of *harmonics*. The word 'harmonic' refers initially to beautiful sound waves, described by the theory of *Musica Universalis* of Pythagoras'school, 6th century BCE. In 1619, Kepler published *Harmonices Mundi*, hypothesising that musical intervals describe the motion of the planets. At the beginning of 19th century, Joseph Fourier developed a mathematical theory of heat waves, nowadays known as *Fourier analysis*. It has later been extended in more abstract contexts under the name of *harmonic analysis*. These mathematics involve periodic functions f, which can be decomposed as

$$f(\theta) = \sum_{n \in \mathbb{N}} c_n e^{i\frac{2\pi}{T}n\theta} \tag{225}$$

where T is the period of f, and $c_n \in \mathbb{C}$. Such functions can be seen as functions over the circle U(1). In this short chapter, we first develop the harmonic analysis over U(1) and then generalise it to SU(2). A good and simple reference for this chapter is the thesis [120].

## 6.1 HARMONIC ANALYSIS OVER U(1)

What are the unitary irreps of U(1)? Since U(1) is abelian, the irreps are one-dimensional. So an irrep of U(1) acts as a linear function over $\mathbb{C}$, which is just scalar multiplication. Then, one can show that any irrep of U(1) takes the form

$$\begin{aligned} \chi_m : \quad & U(1) \to \mathbb{C} \\ & z \mapsto z^m \end{aligned} \tag{226}$$

with $m \in \mathbb{Z}$.

Consider a function $f \in L^2(U(1))$. It can be *decomposed into Fourier series* as

$$f(z) = \sum_{m=-\infty}^{\infty} f_m \chi_m(z) \tag{227}$$

with

$$f_m = \int_{U(1)} f(z)\chi_m(z^{-1})d\mu(z) \tag{228}$$

with $d\mu$ the Haar measure over U(1). Concretely this is

$$f_m = \frac{1}{2\pi}\int_{-\pi}^{\pi} f(e^{i\theta})e^{-im\theta}d\theta. \tag{229}$$

This motivates for instance to write the Dirac generalised function as

$$\delta(\theta) = \frac{1}{2\pi} \sum_{m=-\infty}^{\infty} e^{im\theta}. \tag{230}$$

## 6.2 HARMONIC ANALYSIS OVER SU(2)

We can mimic the previous definitions for SU(2). Consider a function $f \in L^2(\mathrm{SU}(2))$. It can be *decomposed into Fourier series* as

$$f(g) = \sum_{j \in \mathbb{N}/2} f_j \chi_j(g) \tag{231}$$

with the *character*

$$\chi_j(g) \overset{\text{def}}{=} \mathrm{Tr}\, D^j(g) \tag{232}$$

and

$$f_j = \int_{\mathrm{SU}(2)} f(g) \chi_j(g^{-1}) d\mu(g). \tag{233}$$

This motivates to write the Dirac generalised function as

$$\delta(g) = \sum_{j \in \mathbb{N}/2} (2j+1)\, \mathrm{Tr}\, D^j(g). \tag{234}$$

# 7 REPRESENTATION THEORY OF $SL_2(\mathbb{C})$

To put it in a nutshell, the kinematics of LQG deal with representations of SU(2), while the dynamics, in its spin-foam formulation, lie in the representation theory of $SL_2(\mathbb{C})$. The current models of spin-foams, like the EPRL one, extensively use the principal series of $SL_2(\mathbb{C})$.

Whether or not all the representations of $SL_2(\mathbb{C})$, including non-reducible ones, have been classified, is unknown to us, but fortunately all the irreps of $SL_2(\mathbb{C})$ are known. In section 7.1, we present the finite-dimensional irreps. In section 7.2, we summarize the infinite-dimensional ones. Finally, in section 7.3, we focus on the principal series, which is of main interest for quantum gravity.

## 7.1 FINITE IRREPS

The finite irreps of $SL_2(\mathbb{C})$ are well-known. They can be obtained from the finite irreps of its 3-dimensional (complex) Lie algebra $\mathfrak{sl}_2(\mathbb{C})$. In section 4.1, we have already seen them: they are indexed by a spin $j \in \mathbb{N}/2$. It is also possible to see $\mathfrak{sl}_2(\mathbb{C})$ as a real Lie algebra of dimension 6, in which case, we will rather denote it $\mathfrak{sl}_2(\mathbb{C})_{\mathbb{R}}$. In this section, we will describe the (real) linear representations of $\mathfrak{sl}_2(\mathbb{C})_{\mathbb{R}}$. We have the following isomorphism between real vector spaces:

$$\mathfrak{sl}_2(\mathbb{C})_{\mathbb{R}} \cong \mathfrak{su}(2) \oplus i\,\mathfrak{su}(2). \tag{235}$$

The algebra $\mathfrak{su}(2) \oplus \mathfrak{su}(2)$ is the Lie algebra of $SU(2) \times SU(2)$. A consequence of Peter-Weyl's theorem is that the irreps of a cartesian product are tensor products of the irreps of the factors. Thus the irreps of $\mathfrak{sl}_2(\mathbb{C})_{\mathbb{R}}$ are given by the usual tensor representation over $\mathcal{Q}_{j_1} \otimes \mathcal{Q}_{j_2}$, abbreviated by $(j_1, j_2)$. The action is given by:

$$\begin{aligned} a \cdot (|j_1, m_2\rangle \otimes |j_2, m_2\rangle) \\ \overset{\text{def}}{=} (a\,|j_1, m_1\rangle) \otimes |j_2, m_2\rangle + |j_1, m_1\rangle \otimes (a\,|j_2, m_2\rangle). \end{aligned} \tag{236}$$

The isomorphism (235) provides naturally a basis of $\mathfrak{sl}_2(\mathbb{C})_{\mathbb{R}}$, given by the three Pauli matrices $\sigma_i \in i\,\mathfrak{su}(2)$ and the three matrices $i\sigma_i \in \mathfrak{su}(2)$. To match the earlier notations introduced in section 4.2, we often denote the *rotation generators* $J_i \overset{\text{def}}{=} \frac{1}{2}\sigma_i$ and the *boost generators* $K_i \overset{\text{def}}{=} \frac{i}{2}\sigma_i$. These generators satisfy the commutation relations:

$$\begin{aligned} [J_i, J_j] &= i\varepsilon_{ijk} J_k \\ [J_i, K_j] &= i\varepsilon_{ijk} K_k \\ [K_i, K_j] &= -i\varepsilon_{ijk} J_k. \end{aligned} \tag{237}$$

Another basis is given by the *complexified generators*. Posing $A_i = \frac{1}{2}(J_i + iK_i)$ and $B_i = \frac{1}{2}(J_i - iK_i)$, the commutation relations become:

$$\begin{aligned} \left[A_i, A_j\right] &= i\varepsilon_{ijk}A_k \\ \left[B_i, B_j\right] &= i\varepsilon_{ijk}B_k \\ \left[A_i, B_j\right] &= 0. \end{aligned} \tag{238}$$

Then the three realisations, which were described in chapter 4 for the action of $\mathfrak{su}(2)$, can be adapted to $\mathfrak{sl}_2(\mathbb{C})_{\mathbb{R}}$:

1. (*Homogeneous*) For $m, n \geqslant 2$, let $\mathbb{C}_{(m,n)}[z_0, z_1; \overline{z_0}, \overline{z_1}]$ be the vector space of homogeneous polynomials of degree $m$ in $(z_0, z_1)$ and homogeneous of degree $n$ in $(\overline{z_0}, \overline{z_1})$. The action of $SL_2(\mathbb{C})$ is given by
$$g \cdot P(\mathbf{z}) = P\left(g^T \mathbf{z}\right). \tag{239}$$
The associated action of the algebra $\mathfrak{sl}_2(\mathbb{C})_{\mathbb{R}}$ is given by
$$\begin{aligned} J_+ &\cong z_0 \frac{\partial}{\partial z_1} + \overline{z_0}\frac{\partial}{\partial \overline{z_1}} \\ J_- &\cong z_1 \frac{\partial}{\partial z_0} + \overline{z_1}\frac{\partial}{\partial \overline{z_0}} \\ J_3 &\cong \frac{1}{2}\left(z_0\frac{\partial}{\partial z_0} - z_1\frac{\partial}{\partial z_1} + \overline{z_0}\frac{\partial}{\partial \overline{z_0}} - \overline{z_1}\frac{\partial}{\partial \overline{z_1}}\right). \end{aligned} \tag{240}$$

2. (*Projective*) Let $\mathbb{C}_{(m,n)}[z;\bar{z}]$ be the space of polynomials of degree at most $m$ in $z$ and at most $n$ in $\bar{z}$. The action is given by
$$g \cdot \phi(\xi) = (g_{12}\xi + g_{22})^m \overline{(g_{12}\xi + g_{22})}^n \phi\left(\frac{g_{11}\xi + g_{21}}{g_{12}\xi + g_{22}}\right). \tag{241}$$

3. (*Spinorial*) Over the space of totally symmetric spinors $\mathfrak{S}^{(A_1 \ldots A_m)(\dot{A}_1 \ldots \dot{A}_n)}$, the action is
$$u \cdot z^{A_1 \ldots A_m \dot{A}_1 \ldots \dot{A}_n} = u^{A_1}{}_{B_1} \ldots u^{A_m}{}_{B_m} \, \overline{u}^{\dot{A}_1}{}_{\dot{B}_1} \ldots \overline{u}^{\dot{A}_n}{}_{\dot{B}_n} \, z^{B_1 \ldots B_m \dot{B}_1 \ldots \dot{B}_n}. \tag{242}$$
See Penrose [143] (p. 142) for details.

The finite representations of $SL_2(\mathbb{C})$ cannot be unitary (except the trivial one), because it is a simply connected non-compact Lie group. If we want unitary representations, we shall turn to infinite ones.

## 7.2 INFINITE IRREPS

In this section, we describe all the infinite-dimensional irreps of $SL_2(\mathbb{C})$.

★ NOTA BENE. All the unitary irreps of the Lorentz group have been found simultaneously in 1946 by Gel'fand and Naimark [77], by Harish-Chandra [95] and by Bargmann [19]. It seems nevertheless that Gel'fand and Naimark were the first to publish (unfortunately their article is only in Russian). The question

remained to find all the irreps, unitary or not, and this was solved also by Naimark in 1954 [131]. In 1963, Gel'fand, Minlos and Shapiro published the first book (with english translation) that reviews all these results [76]. In 1964, Naimark wrote a more detailed and well-written book that wraps up the subject for mathematically-orientated physicists [132].

The infinite irreps of $SL_2(\mathbb{C})$ are parametrised by $(m, \rho) \in \mathbb{Z} \times \mathbb{C}$ with $\operatorname{Im}\rho \geqslant 0$ and $\rho^2 \neq -(|m| + 2n)^2$, with $n \in \mathbb{N}^*$. A realisation is given over the Hilbert space $L^2(\mathbb{C})$ endowed with the scalar product

$$\langle \varphi | \phi \rangle = \frac{i}{2} \int_{\mathbb{C}} \overline{\varphi(\xi)} \phi(\xi) (1 + |\xi|^2)^{-\operatorname{Im}\rho} d\xi d\bar{\xi}, \tag{243}$$

and the action

$$a \cdot f(z) = (a_{12}z + a_{22})^{\frac{m}{2} + \frac{i\rho}{2} - 1} \overline{(a_{12}z + a_{22})}^{-\frac{m}{2} + \frac{i\rho}{2} - 1} f\left(\frac{a_{11}z + a_{21}}{a_{12}z + a_{22}}\right). \tag{244}$$

REMARKS.

1. If $\rho^2 = -(|m| + 2n)^2$ with $n \in \mathbb{N}^*$, the Hilbert space and the action still defines a representation, but a reducible one. Then, if one restricts the action to the subspace of polynomials of degree at most $p = \frac{m}{2} + i\frac{\rho}{2} - 1$ in $z$ and $q = -\frac{m}{2} + i\frac{\rho}{2} - 1$ in $\bar{z}$, the representation is irreducible and equivalent to the finite-dimensional representation $(p, q)$.

2. Not all the representations $(m, \rho)$ are unitary. They are unitary in only two cases: when $\rho \in \mathbb{R}$ (principal series); when $m = 0$ and $i\rho \in ]-2, 0[$ (complementary series), provided another scalar product chosen in the latter case (see below).

3. Among these infinite irreps, only the principal representations $(\rho, k)$ and $(-\rho, -k)$ are equivalent.

4. A proof of the result above can be found in Naimark ([132] pp. 294-295). A sketch of it in the case of the principal series can be found in section 7.3.

PRINCIPAL SERIES. When $\rho \in \mathbb{R}$, the scalar product over $L^2(\mathbb{C})$ becomes the usual one

$$(f_1, f_2) \overset{\text{def}}{=} \frac{i}{2} \int_{\mathbb{C}} \overline{f_1(z)} f_2(z) dz d\bar{z}, \tag{245}$$

and the representation $(\rho, m) \in \mathbb{R} \times \mathbb{Z}$ is unitary. The representations $(\rho, m)$ and $(-\rho, -m)$ are unitarily equivalent. They form the so-called the principal series, parametrised by $(\rho, m) \in \mathbb{R} \times \mathbb{Z}$. This choice of parametrisation is not universal. Here is a table to translate between different authors:

| | |
|---|---|
| Naimark [132], GGV [75] | $(\rho, m)$ |
| Rühl [164] | $(\rho_R, m_R) = (\rho, -m)$ |
| GMS [76] | $(\rho_G, m_G) = (\rho/2, m)$ |
| Rovelli [161], Barrett [26] | $(p, k) = (\rho/2, m/2)$ |

COMPLEMENTARY SERIES. When $m = 0$ and $i\rho \in ]-2, 0[$, the action becomes

$$a \cdot \phi(\xi) = |a_{12}\xi + a_{22}|^{i\rho-2} \phi\left(\frac{a_{11}\xi + a_{21}}{a_{12}\xi + a_{22}}\right). \tag{246}$$

It also defines a unitary representation for the scalar product

$$\langle \varphi | \phi \rangle = \left(\frac{i}{2}\right)^2 \int_{\mathbb{C}^2} \frac{\overline{\varphi(\xi)}\phi(\eta)}{|\xi - \eta|^{2+i\rho}} d\xi d\bar{\xi} d\eta d\bar{\eta}. \tag{247}$$

## 7.3 PRINCIPAL UNITARY SERIES

In this section, we review several ways to build the principal series and we expand on its properties.

### 7.3.1 *Induced representation*

The construction of the principal series by Gel'fand and Naimark is based on the induced representation method, which was introduced in section 2.10 (see [132] for details). The principal series is found as the unitary representations of $SL_2(\mathbb{C})$ induced by the uni-dimensional representations of the upper-triangular subgroup $K_+$.

To prove it, the first step is to notice the following diffeomorphism between differentiable manifolds:

$$SL_2(\mathbb{C})/K_+ \cong \bar{\mathbb{C}}. \tag{248}$$

Then, we can induce the expression of the linear action of $SL_2(\mathbb{C})$ over $\bar{\mathbb{C}}$:

$$a \cdot z = \frac{a_{11}z + a_{21}}{a_{12}z + a_{22}}. \tag{249}$$

It is nothing but the so-called Möbius transformation. Consider the Hilbert space of square integrable complex functions $L^2(\mathbb{C})$ with the scalar product:

$$(f_1, f_2) \stackrel{\text{def}}{=} \frac{i}{2} \int_{\mathbb{C}} \overline{f_1}(z) f_2(z)\, dz \wedge d\bar{z}. \tag{250}$$

Then, we look for a unitary representation over $L^2(\mathbb{C})$ of the form:

$$a \cdot f(z) = \alpha(z, a) f(a \cdot z). \tag{251}$$

After lines of computation, we find that for all $(\rho, m) \in \mathbb{R} \times \mathbb{Z}$, there exists a unitary representation of $SL_2(\mathbb{C})$ over $L_2(\mathbb{C})$ given by

$$a \cdot f(z) = (a_{12}z + a_{22})^{\frac{m}{2}+\frac{i\rho}{2}-1}\overline{(a_{12}z + a_{22})}^{-\frac{m}{2}+\frac{i\rho}{2}-1} \times f\left(\frac{a_{11}z + a_{21}}{a_{12}z + a_{22}}\right). \quad (252)$$

These are called the principal series and we can finally show that they are irreducible!

### 7.3.2 *Homogeneous realisation*

Though rigorous from the mathematical point of view, it is not very intuitive, especially for physicists. In 1962, Gel'fand, Graev and Vilenkin (GGV) published a book where they build the principal series from a space of homogeneous functions, which may seem more natural ([75] pp. 139-201). A beautiful and concise exposure can be found in the article of Dao and Nguyen ([51] pp. 18-21). We present it here.

Consider $\mathcal{F}(\mathbb{C}^2)$, the vector space of the complex functions over $\mathbb{C}^2$. A function $F \in \mathcal{F}(\mathbb{C}^2)$ is said to be *homogeneous of degree* $(\lambda, \mu) \in \mathbb{C}^2$ if it satisfies for all $\alpha \in \mathbb{C}$ :

$$F(\alpha \mathbf{z}) = \alpha^{\lambda}\overline{\alpha}^{\mu}F(\mathbf{z}). \quad (253)$$

To be consistently defined when $\alpha = e^{2i\pi n}$, the degree should satisfy the condition :

$$\mu - \lambda \in \mathbb{Z}. \quad (254)$$

Instead of $(\lambda, \mu)$, we will instead use in the following, the parameters $(p = \frac{\mu+\lambda+2}{2i}, k = \frac{\lambda-\mu}{2})$ (same choice of parameters as Rovelli [161] p. 182). Define $\mathcal{D}^{(p,k)}[z_0, z_1]$ as the subspace of homogeneous functions of degree $(\lambda, \mu)$ infinitely differentiable over $\mathbb{C}^2 \setminus \{0\}$ in the variables $z_0, z_1, \bar{z}_0$ et $\bar{z}_1$ with a certain topology[1]. We define a continuous representation $SL_2(\mathbb{C})$ over $\mathcal{D}^{(p,k)}[z_0, z_1]$ by

$$a \cdot F(\mathbf{z}) \stackrel{\text{def}}{=} F(a^T \mathbf{z}). \quad (255)$$

Now define the following 2-form over $\mathbb{C}^2$:

$$\Omega(z_0, z_1) = \frac{i}{2}(z_0 dz_1 - z_1 dz_0) \wedge (\overline{z_0} d\overline{z_1} - \overline{z_1} d\overline{z_0}).$$

Interestingly, it is invariant for the action of $SL_2(\mathbb{C})$: $\Omega(a\mathbf{z}) = \Omega(\mathbf{z})$. Let $\Gamma$ be a path in $\mathbb{C}^2$ that intersects each projective line exactly once. Then define the scalar product over $\mathcal{D}^{(p,k)}[z_0, z_1]$:

$$(F, G) = \int_{\Gamma} \overline{F(\mathbf{z})} G(\mathbf{z}) \Omega(\mathbf{z}).$$

1 The topology is defined by the following property of convergence: a sequence $F_n(z_0, z_1)$ is said to converge to 0 if it converges to zero uniformly together with all its derivatives on any compact set in the $(z_0, z_1)$-plane which does not contain $(0, 0)$ (see Gel'fand-Graev-Vilenkin (GGV) [75] p. 142).

Thus $\mathcal{D}^{(p,k)}[z_0, z_1]$ is an Hilbert space. Interestingly, the result does not depend on the path $\Gamma$ provided $p \in \mathbb{R}$, which we consider to be the case in the following. This scalar product is invariant for $SL_2(\mathbb{C})$: $(a \cdot F, a \cdot G) = (F, G)$. Thus the representation is unitary. It could be also shown to be irreducible. In the next subsection, we will see that the representation $\mathcal{D}^{(p,k)}[z_0, z_1]$ is equivalent to the representation $(\rho = 2p, m = 2k)$ of the principal series described in section 7.2.

Using the language developed in chapter 3, $\Gamma$ can be understood as a local section of the tautological bundle $\mathcal{O}(-1)$. Then $\bar{f}g\Omega$ is an homogeneous 2-form of degree 0. As explained in section 3.2, it is not surprise that the integral does not depend on the choice of section $\Gamma$. Its computation can be made using what we call the *Gel'fand section*

$$s : d \mapsto (d, (1, \zeta_0(d))). \tag{256}$$

Such a choice directly leads to the projective realisation of subsection 7.3.3.

### 7.3.3 *Projective realisation*

Consider the map

$$\iota : \begin{cases} \mathbb{C} & \to & \mathbb{C}^2 \\ \zeta & \mapsto & (\zeta, 1) \end{cases} \tag{257}$$

$\iota$ is a diffeomorphism from $\mathbb{C}$ to its range. It parametrises a horizontal straight line of $\mathbb{C}^2$. The projective construction consists in restricting the domain of definition of the homogeneous function to this line. If $F \in \mathcal{F}(\mathbb{C}^2)$, define $\iota^* F \in \mathcal{F}(\mathbb{C})$ as

$$\iota^* F(z) \stackrel{\text{def}}{=} F \circ \iota(z) = F(z, 1). \tag{258}$$

The 2-form $\Omega$ becomes similarly

$$\iota^* \Omega(z) = \frac{i}{2} dz \wedge d\bar{z},$$

which is nothing but the usual Lebesgue measure over $\mathbb{C}$. Thus we define the Hilbert space $L^2(\mathbb{C})$ with the the scalar product

$$(f, g) = \frac{i}{2} \int_{\mathbb{C}} \overline{f(z)} g(z) dz \wedge d\bar{z}.$$

Thus we have a map $\iota^* : \mathcal{D}^{(p,k)}[z_0, z_1] \to L^2(\mathbb{C})$. In fact $\iota^*$ is bijective: for all $f \in L^2(\mathbb{C})$, there exists a unique $F \in \mathcal{D}^{(p,k)}[z_0, z_1]$ such that $f = \iota^* F$. F is given explicitly by

$$F(z_0, z_1) = z_1^{-1+ip+k} \bar{z}_1^{-1+ip-k} f\left(\frac{z_0}{z_1}\right). \tag{259}$$

Importantly, $\iota^*$ induces naturally an action of $SL_2(\mathbb{C})$ over $\mathcal{F}(\mathbb{C})$, such that $\iota^*$ becomes an intertwiner between two equivalent representations. After computation, we obtain:

$$a \cdot f(z) \overset{\text{def}}{=} (a_{12}z + a_{22})^{-1+ip+k}\overline{(a_{12}z + a_{22})}^{-1+ip-k} f\left(\frac{a_{11}z + a_{21}}{a_{12}z + a_{22}}\right). \tag{260}$$

This formula is exactly the same formula as (244), with the indices $(p,k) = (\rho/2, m/2)$. Thus we have constructed explicitly the representations of the principal series, and we have shown the equivalence of the realisations $\mathcal{D}^{(p,k)}[z_0, z_1]$ and $L^2(\mathbb{C})$.

### 7.3.4 SU(2)-*realisation*

Following Rühl ([164] p. 57), we are going to build another realisation of the unitary principal representations.

The first step is to observe the following diffeomorphism between manifolds:

$$SL_2(\mathbb{C})/K_+ \cong SU(2)/U(1). \tag{261}$$

Then, instead of constructing a space of functions over $SL_2(\mathbb{C})/K_+$ as was done originally (see subsection 7.3.1), it is equivalent to consider functions $\phi$ over SU(2) satisfying a covariance condition for the group U(1):

$$\phi\left(\begin{pmatrix} e^{i\theta} & 0 \\ 0 & e^{-i\theta} \end{pmatrix} u\right) = e^{in\theta}\phi(u) \tag{262}$$

with $n \in \mathbb{Z}$. The choice of the factor $e^{in\theta}$ corresponds to unidimensional representations of U(1) (see chapter 6).

Concretely, consider the map

$$\kappa : \begin{cases} SU(2) & \to & \mathbb{C}^2 \\ u & \mapsto & (u_{21}, u_{22}) \end{cases} \tag{263}$$

$\kappa$ is a diffeomorphism to its range. In some sense, SU(2) can be seen as the 'unit circle' of $\mathbb{C}^2$, so that $\kappa$ can be seen as the injection of the circle in the plane $\mathbb{C}^2$.

Then, define $\kappa^* : \mathcal{F}(\mathbb{C}^2) \to \mathcal{F}(SU(2))$ such that

$$\kappa^* F(u) \overset{\text{def}}{=} F \circ \kappa(u) = F(u_{21}, u_{22}). \tag{264}$$

If $F \in \mathcal{D}^{(p,k)}[z_0, z_1]$, then we show easily that $\kappa^* F$ satisfies the covariance property

$$\kappa^* F(e^{i\theta\sigma_3}u) = e^{-2i\theta k}\kappa^* F(u). \tag{265}$$

We denote $\mathcal{D}^{(p,k)}[u] \overset{\text{def}}{=} \kappa^* \mathcal{D}^{(p,k)}[z_0, z_1]$. Thus $\kappa^*$ is a bijection from $\mathcal{D}^k[z_0, z_1]$ to $\mathcal{D}^{(p,k)}[u]$. Its inverse is given explicitly by

$$F(z_0, z_1) = (|z_0|^2 + |z_1|^2)^{-1+ip}\phi\left(\frac{1}{\sqrt{|z_0|^2 + |z_1|^2}}\begin{pmatrix} z_1^* & -z_0^* \\ z_0 & z_1 \end{pmatrix}\right). \tag{266}$$

We could also translate the measure $\kappa^*\Omega$, and thus endow $\mathcal{D}^{(p,k)}[u]$ with the structure of a Hilbert space. Interestingly, it is a subspace of $L^2(SU(2))$. As previously, one can translate the action of $SL_2(\mathbb{C})$ over $\mathcal{D}^{(p,k)}[u]$ such that $\kappa^*$ becomes a bijective intertwiner, and we obtain

$$a \cdot \phi(u) = (|\beta_{a,u}|^2 + |\alpha_{a,u}|^2)^{-1+ip} \times \phi\left(\frac{1}{\sqrt{|\beta_{a,u}|^2 + |\alpha_{a,u}|^2}}\begin{pmatrix}\alpha_{a,u} & -\beta^*_{a,u}\\ \beta_{a,u} & \alpha^*_{a,u}\end{pmatrix}\right), \quad (267)$$

with $\alpha_{a,u} \overset{\text{def}}{=} (u_{21}a_{12} + u_{22}a_{22})^*$ and $\beta_{a,u} \overset{\text{def}}{=} u_{21}a_{11} + u_{22}a_{21}$. Thus $\mathcal{D}^{(p,k)}[u]$ is a third equivalent realisation of the unitary principal series. The equivalence with $L^2(\mathbb{C})$ is made through $(\kappa \circ \iota^{-1})^*$ which gives explicitly

$$\phi(u) = u_{22}^{-1+ip+k}\overline{u_{22}}^{-1+ip-k} f\left(\frac{u_{21}}{u_{22}}\right), \quad (268)$$

and conversely

$$f(z) = (1+|z|^2)^{-1+ip}\phi\left(\frac{1}{\sqrt{1+|z|^2}}\begin{pmatrix}1 & -z^*\\ z & 1\end{pmatrix}\right). \quad (269)$$

Notice that this equivalence of representations supervenes on the Hopf bundle $SU(2)/U(1) \cong \mathbb{C}P^1$ (see section 3.3).

### 7.3.5 *Canonical basis*

The advantage of the SU(2)-realisation is that we already know interesting functions over SU(2), namely the coefficients of the Wigner matrix $D^j_{pq}$. Indeed, they are elements of $\mathcal{D}^{(p,k)}[u]$, provided that they satisfy the covariance property (265). We compute easily

$$D^j_{pq}\left(e^{i\theta\sigma_3}u\right) = \sum_{k=-j}^{j} \langle j,p|\, e^{i\theta\sigma_3}\, |j,k\rangle\, D^j_{kq}(u) = \sum_{k=-j}^{j} \delta_{pk} e^{2ik\theta} D^j_{kq}(u) = e^{2ip\theta} D^j_{pq}(u). \quad (270)$$

Thus, the covariance property it is satisfied if $p = -k$, and so

$$\forall j \in \{|k|, |k|+1, ...\}, \quad \forall q \in \{-j, ..., j\}, \quad D^j_{-k,q} \in \mathcal{D}^{(p,k)}[u]. \quad (271)$$

Since the $D^j_{mn}(u)$ form a basis of $L^2(SU(2))$, we show easily that the subset exhibited in (271) form a basis of $\mathcal{D}^{(p,k)}[u]$. Another consequence is the following decomposition of $\mathcal{D}^{(p,k)}[u]$ into irreps of SU(2):

$$\mathcal{D}^{(p,k)}[u] \cong \bigoplus_{j=|k|}^{\infty} \mathcal{Q}_j. \quad (272)$$

We then call *canonical basis* of $\mathcal{D}^{(p,k)}[u]$ the set of functions:

$$\phi^{(p,k)}_{jm}(u) \stackrel{\text{def}}{=} \sqrt{\frac{2j+1}{\pi}} D^{j}_{-k,m}(u),$$
$$\text{with } j = |k|, |k|+1, \dots \text{ and } -j \leqslant m \leqslant j. \quad (273)$$

From (125), we see that they satisfy the orthogonality relations

$$\int_{SU(2)} du \, \overline{\phi^{(p,k)}_{jm}}(u) \phi^{(p,k)}_{ln}(u) = \frac{1}{\pi} \delta_{jl} \delta_{mn}. \quad (274)$$

★ Nota Bene. In Rühl ([164] p. 59), the factor $\frac{1}{\sqrt{\pi}}$ is absent from the definition of the $\phi^{(p,k)}_{jm}$. Thus, the orthogonality relations do not show a factor $\frac{1}{\pi}$ on the RHS. We have chosen this factor so that the canonical basis $f^{(p,k)}_{jm}$ of $L^2(\mathbb{C})$ (see below (276)) is orthonormal for the usual scalar product with the Lebesgue measure $dz$ (for Rühl the measure is $dz/\pi$).

Moreover $\phi^{(p,k)}_{jm}$ could have been defined with a phase factor $e^{i\psi(p,j)}$. This is set to zero in some literature including [26, 164], and we follow that convention here. An alternative phase convention leading to real $SL_2(\mathbb{C})$-Clebsch-Gordan coefficients is obtained for the choice [111, 174] $e^{i\psi(p,j)} = (-1)^{-\frac{j}{2}} \frac{\Gamma(j+i\rho+1)}{|\Gamma(j+i\rho+1)|}$.

An intermediate choice of phase is the one of [51, 152], which has the advantage of simplifying the recursion relations satisfied by the Clebsch-Gordan coefficients [4, 5]. The latter are now either real or purely imaginary.

The intertwiner $\kappa^*$ enables to translate this basis in $\mathcal{D}^{(p,k)}[z_0, z_1]$, and we obtain the canonical basis:

$$F^{(p,k)}_{jm}(z_0, z_1) = \sqrt{\frac{2j+1}{\pi}} (|z_0|^2 + |z_1|^2)^{ip-1} \times D^{j}_{-k,m}\left( \frac{1}{\sqrt{|z_0|^2 + |z_1|^2}} \begin{pmatrix} z_1^* & -z_0^* \\ z_0 & z_1 \end{pmatrix} \right), \quad (275)$$

where an explicit expression for $D^{j}_{-k,m}$ is given by equation (132). The same is done with the intertwiner $\iota^*$ to $L^2(\mathbb{C})$, and we obtain the canonical basis:

$$f^{(p,k)}_{jm}(z) = \sqrt{\frac{2j+1}{\pi}} (1 + |z|^2)^{ip-1-j} D^{j}_{-k,m} \begin{pmatrix} 1 & -z^* \\ z & 1 \end{pmatrix} \quad (276)$$

The constant factors of (273) have been chosen so that

$$\frac{i}{2} \int_{\mathbb{C}} \overline{f^{(p,k)}_{jm}}(z) f^{(p,k)}_{ln}(z) dz d\bar{z} = \delta_{jl} \delta_{mn}.$$

Finally, in ket notations, the canonical basis is denoted $|p, k, jm\rangle$.

### 7.3.6 *Action of the generators*

Similarly to equation (130), the action of the $SL_2(\mathbb{C})$-generators can be computed from the action of the group. The generators of the rotations 'stay inside' the same SU(2)-irreps:

$$\begin{aligned} J_3 \, |p, k, j, m\rangle &= m \, |p, k, j, m\rangle \\ J_+ \, |p, k, j, m\rangle &= \sqrt{(j+m+1)(j-m)} \, |p, k, j, m+1\rangle \\ J_- \, |p, k, j, m\rangle &= \sqrt{(j+m)(j-m+1)} \, |p, k, j, m-1\rangle \, . \end{aligned} \quad (277)$$

The generators of the boost spread over the neighbouring subspaces:

$$K_3\,|p,k,j,m\rangle = \alpha_j\sqrt{j^2-m^2}\,|p,k,j-1,m\rangle + \gamma_j m\,|p,k,j,m\rangle - \alpha_{j+1}\sqrt{(j+1)^2-m^2}\,|p,k,j+1,m\rangle\,, \tag{278}$$

$$\begin{aligned} K_+\,|p,k,j,m\rangle = {}& \alpha_j\sqrt{(j-m)(j-m-1)}\,|p,k,j-1,m+1\rangle \\ &+ \gamma_j\sqrt{(j-m)(j+m+1)}\,|p,k,j,m+1\rangle \\ &+ \alpha_{j+1}\sqrt{(j+m+1)(j+m+2)}\,|p,k,j+1,m+1\rangle \end{aligned} \tag{279}$$

$$\begin{aligned} K_-\,|p,k,j,m\rangle = {}& -\alpha_j\sqrt{(j+m)(j+m-1)}\,|p,k,j-1,m-1\rangle \\ &+ \gamma_j\sqrt{(j+m)(j-m+1)}\,|p,k,j,m-1\rangle \\ &- \alpha_{j+1}\sqrt{(j-m+1)(j-m+2)}\,|p,k,j+1,m-1\rangle \end{aligned} \tag{280}$$

with $\gamma_j \overset{\text{def}}{=} \frac{kp}{j(j+1)}$ and $\alpha_j \overset{\text{def}}{=} i\sqrt{\frac{(j^2-k^2)(j^2+p^2)}{j^2(4j^2-1)}}$. From these expressions, it is possible to compute the action of the two Casimir operators:

$$\begin{aligned} (\vec{K}^2-\vec{J}^2)\,|p,k,j,m\rangle &= (p^2-k^2+1)\,|p,k,j,m\rangle\ , \\ \vec{K}\cdot\vec{J}\,|p,k,j,m\rangle &= pk\,|k,p,j,m\rangle\ . \end{aligned} \tag{281}$$

### 7.3.7 $SL_2(\mathbb{C})$ *Wigner's matrix*

We define the $SL_2(\mathbb{C})$ Wigner's matrix by its coefficients

$$D^{(p,k)}_{j_1q_1j_2q_2}(a) \overset{\text{def}}{=} \langle p,k;j_1q_1|\,a\,|p,k;j_2q_2\rangle\,. \tag{282}$$

These coefficients satisfy the orthogonality relations:

$$\int_{SL_2(\mathbb{C})} dh\, D^{(p_1,k_1)}_{j_1m_1l_1n_1}(h) D^{(p_2,k_2)}_{j_2m_2l_2n_2}(h) = \frac{1}{4(p_1^2+k_1^2)}\delta(p_1-p_2)\delta_{k_1k_2}\delta_{j_1j_2}\delta_{l_1l_2}\delta_{m_1m_2}\delta_{n_1n_2}. \tag{283}$$

To compute it explicitly, it is useful to use Cartan decomposition, $g = ue^{r\sigma_3/2}v^{-1}$, with $u,v\in SU(2)$ and $r\in\mathbb{R}_+$. Then, we have

$$D^{(p,k)}_{jmln}(g) = \sum_{q=-\min(j,l)}^{\min(j,l)} D^j_{mq}(u)\,d^{(p,k)}_{jlq}(r)\,D^l_{qn}(v^{-1}). \tag{284}$$

with the *reduced* $SL_2(\mathbb{C})$ *Wigner's matrix* defined as

$$d^{(p,k)}_{jlm}(r) \overset{\text{def}}{=} D^{(p,k)}_{jmlm}(e^{r\sigma_3/2}). \tag{285}$$

We have the following symmetry properties:

$$d^{(p,k)}_{jlm}(r) = d^{(-p,k)}_{ljm}(-r) = d^{(p,-k)}_{jl,-m}(r) = (-1)^{j-l} d^{(-p,-k)}_{ljm}(r) = \overline{d^{(p,k)}_{ljm}(-r)}. \tag{286}$$

It admits explicit formulae, such as

$$d_{jlm}^{(p,k)}(r) = \sqrt{(2j+1)(2l+1)}\sqrt{\frac{(j-k)!(j+k)!(l-k)!(l+k)!}{(j+m)!(j-m)!(l+m)!(l-m)!}}$$
$$\times \sum_{n_1,n_2} \Bigg[ (-1)^{j+l+2m-n_1-n_2} e^{r(ip-1-2n_2-k+m)}$$
$$\times \binom{j+m}{n_1}\binom{j-m}{j-k-n_1}\binom{l+m}{n_2}\binom{l-m}{l-k-n_2}$$
$$\times \int_0^1 dt \left[1-(1-e^{-2r})t\right]^{ip-1-l} t^{n_1+n_2+k-m}(1-t)^{j+l-n_1-n_2-k+m} \Bigg]. \tag{287}$$

See [122] for other ones.

# 8 RECOUPLING OF $SL_2(\mathbb{C})$

## 8.1 $SL_2(\mathbb{C})$-CLEBSCH-GORDAN COEFFICIENTS

Similarly to the SU(2) case, the tensor product of two irreps of $SL_2(\mathbb{C})$ can be decomposed into a direct sum of irreps:

$$\mathcal{D}^{(p_1,k_1)} \otimes \mathcal{D}^{(p_2,k_2)} \simeq \int_{\mathbb{R}} dp \bigoplus_{\substack{k\in\mathbb{Z}/2\\ k_1+k_2+k\in\mathbb{N}}} \mathcal{D}^{(p,k)}. \tag{288}$$

Kerimov and Verdiev first got interested in the generalisation of the Clebsch-Gordan coefficients to the irreps of $SL_2(\mathbb{C})$ [111]. The $SL_2(\mathbb{C})$-Clebsch-Gordan coefficients are defined by the relation

$$|p,k;j,m\rangle = \int dp_1 dp_2 \sum_{k_1 j_1 m_1} \sum_{k_2 j_2 m_2} C^{pkjm}_{p_1 k_1 j_1 m_1, p_2 k_2 j_2 m_2} |p_1,k_1;j_1 m_1\rangle \otimes |p_2,k_2;j_2,m_2\rangle\,. \tag{289}$$

The coefficients are non-zero only when $k_1 + k_2 + k_3 \in \mathbb{N}$, in addition to the usual triangle inequality $|j_1 - j_2| \leqslant j_3 \leqslant j_1 + j_2$.

We have explicit expression for the $SL_2(\mathbb{C})$-Clebsch-Gordan coefficients but they are a bit tough. First of all, remark that the magnetic part factorises as

$$C^{p_3 k_3 j_3 m_3}_{p_1 k_1 j_1 m_1, p_2 k_2 j_2 m_2} = \chi(p_1,p_2,p_3,k_1,k_2,k_3;j_1,j_2,j_3)\, C^{j_3 m_3}_{j_1 m_1 j_2 m_2}. \tag{290}$$

$\chi$ is a function of 9 variables which can be computed by the following expression (found initially in [111] but corrected slightly in [174]):

$$\begin{aligned}
&\chi(p_1,p_2,p_3,k_1,k_2,k_3;j_1,j_2,j_3) \\
&= \frac{\kappa\, N^{j_1}_{p_1} N^{j_2}_{p_2} \overline{N^{j_3}_{p_3}}}{4\sqrt{2\pi}} (-1)^{(j_1+j_2+j_3+k_1+k_2+k_3)/2} (-1)^{-k_2-k_1} \\
&\times \sqrt{(2j_1+1)(2j_2+1)(2j_3+1)} \left(\frac{(j_1-k_1)!(j_2+k_2)!}{(j_1+k_1)!(j_2-k_2)!}\right)^{1/2} \\
&\times \Gamma(1-\nu_3+\mu_3)\Gamma(1-\nu_3-\mu_3) \\
&\times \sum_{n=-j_1}^{j_1} \left(\frac{(j_1-n)!(j_2+k_3-n)!}{(j_1+n)!(j_2-k_3+n)!}\right)^{1/2} C^{j_3 k_3}_{j_1 n; j_2, k_3-n} \\
&\times \sum_{l_1=\max(k_1,n)}^{\min(j_1,k_3+j_2)} \sum_{l_2=\max(-k_2,n-k_3)}^{j_2} \frac{(j_1+l_1)!(j_2+l_2)!}{(j_1-l_1)!(l_1-k_1)!(l_1-n)!} \\
&\times \frac{(-1)^{l_1-k_1+l_2+k_2}}{(j_2-l_2)!(l_2+k_2)!(l_2-n+k_3)!} \\
&\times \frac{\Gamma(2-\nu_1-\nu_2-\nu_3+\mu_1+l_1+l_2-n)\Gamma(1-\nu_1+\mu_3+l_1)}{\Gamma(2-\nu_1-\nu_2+l_1+l_2)\Gamma(1-\nu_3+\mu_1-n)} \\
&\times \frac{\Gamma(1-\nu_2-\mu_3+l_2)}{\Gamma(2-\nu_1-\nu_3+l_1)\Gamma(2-\nu_3-\nu_2+l_2)}
\end{aligned} \tag{291}$$

with

$$\begin{aligned}
\nu_1 &= \tfrac{1}{2}(1+ip_1-ip_2-ip_3) \\
\nu_2 &= \tfrac{1}{2}(1-ip_1+ip_2-ip_3) \\
\nu_3 &= \tfrac{1}{2}(1+ip_1+ip_2+ip_3) \\
\mu_1 &= \tfrac{1}{2}(-k_1+k_2+k_3) \\
\mu_2 &= \tfrac{1}{2}(k_1-k_2+k_3) \\
\mu_3 &= \tfrac{1}{2}(-k_1-k_2-k_3)
\end{aligned} \tag{292}$$

and a phase

$$\begin{aligned}
\kappa = {} & \frac{\Gamma(\nu_1+\mu_1)\Gamma(\nu_2+\mu_2)\Gamma(\nu_3+\mu_3)}{|\Gamma(\nu_1+\mu_1)\Gamma(\nu_2+\mu_2)\Gamma(\nu_3+\mu_3)|} \\
& \times \frac{\Gamma(-1+\nu_1+\nu_2+\nu_3+\mu_1+\mu_2+\mu_3)}{|\Gamma(-1+\nu_1+\nu_2+\nu_3+\mu_1+\mu_2+\mu_3)|}
\end{aligned} \tag{293}$$

and

$$N^j_p = \frac{\Gamma(1+j+ip)}{|\Gamma(1+j+ip)|}, \tag{294}$$

and the usual gamma function defined over $\mathbb{C}$ by analytic continuation of

$$\Gamma(z) = \int_0^{+\infty} t^{z-1}\, e^{-t}\, dt, \quad \text{with} \quad \mathrm{Re}\, z > 0. \tag{295}$$

★ NOTA BENE. The phase $\kappa$ satisfying $|\kappa| = 1$ was chosen to make the $SL_2(\mathbb{C})$-Clebsch-Gordan coefficients real (equivalent to the Condon-Shortley convention in the SU(2) case). Contrary to the usual SU(2)-Clebsh-Gordan coefficients, there is no consensual convention for this phase. The choice of Kerimov differs from that of Anderson [4] or Speziale [174].

These seemingly intricate expressions have nevertheless been used with much efficiency in [174] to numerically compute spin-foam amplitudes. The formula is indeed interesting because it is expressed with only finite sums.

## 8.2 GRAPHICAL CALCULUS

When willing to define a graphical calculus for $SL_2(\mathbb{C})$ one encounters the difficulty of finding a good $SL_2(\mathbb{C})$-equivalent to the 3jm-symbol of SU(2) recoupling theory, such that it would satisfy the good symmetry relations to be well-represented by a 3-valent vertex. This issue is investigated in [4], but the symmetry relations are intricate and depend on the convention chosen for the phase $\kappa$. As a result there is no consensus about the definition of the rules of graphical calculus for $SL_2(\mathbb{C})$. Following the phase convention of [174], we define then

$$\begin{pmatrix} (p_1, k_1) & (p_2, k_2) & (p_3, k_3) \\ (j_1, m_1) & (j_2, m_2) & (j_3, m_3) \end{pmatrix} \overset{\text{def}}{=} (-1)^{2j_1 - j_2 + j_3 - m_3} C^{p_3 k_3 j_3, -m_3}_{p_1 k_1 j_1 m_1, p_2 k_2 j_2 m_2}. \quad (296)$$

Graphically it corresponds to the 3-valent vertex

$$\begin{pmatrix} (p_1, k_1) & (p_2, k_2) & (p_3, k_3) \\ (j_1, m_1) & (j_2, m_2) & (j_3, m_3) \end{pmatrix} = \quad (297)$$

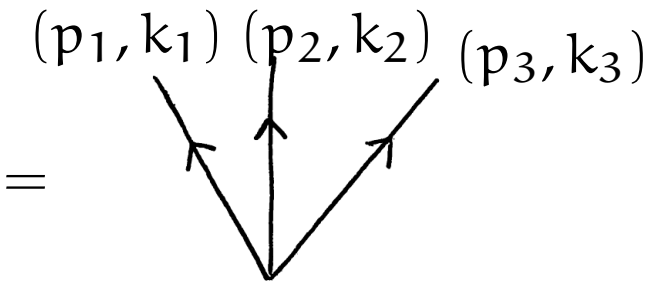

With the same rules of orientation and summation as that of section 5.6, we can then fully develop the graphical calculus of $SL_2(\mathbb{C})$. For instance, we can define $SL_2(\mathbb{C})$-invariant functions, like the $(6p, 6k)$-symbol. The $SL_2(\mathbb{C})$-15j-symbol can be used to define the spin-foam amplitude (see section 16.2).

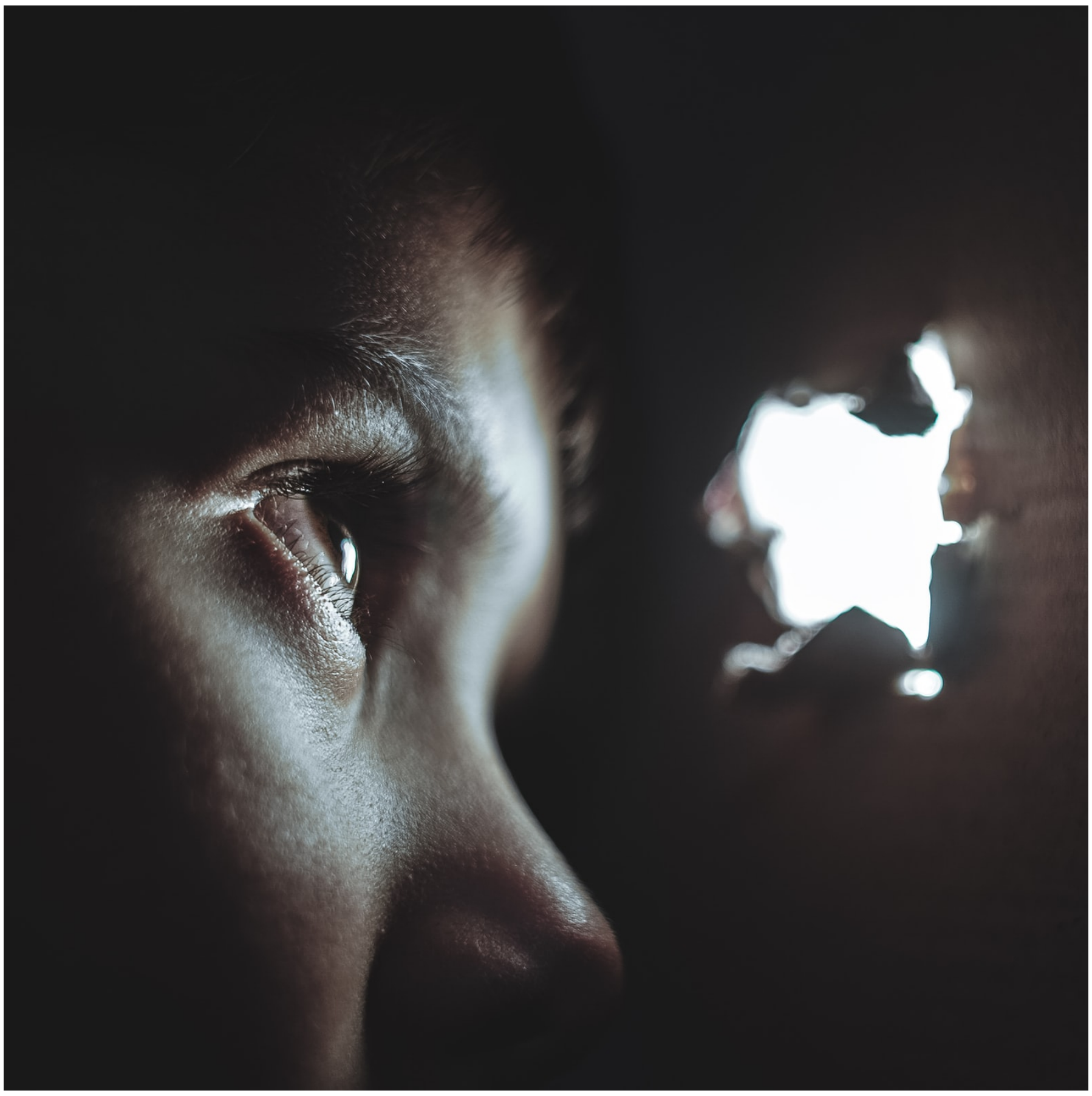

Part II

# OF BLACK AND WHITE HOLES

After a preliminary mathematical part, we now turn to the physical core of this thesis. We describe a scenario where black holes are not eternal, but instead collapsed stars, on the verge of bouncing back into white holes. Their coming explosion would only be a matter of time. Yet, this time may be so long, that they would first evaporate to planckian size before the bounce has a chance to happen. From the outside, the hole now looks like a Planck-mass particle, interacting weakly, but its immense interior volume contains the famous information that was feared to be lost. Before the white hole explodes, a far-away observer would notice outgoing radiation with negative energy, foretelling the cataclysm. Afterwards, the radiation would continue flowing out, this time with positive energy. We will reach this conclusion progressively, laying down its foundation, step-by-step, as follows:

Ch. 9 gathers the essential technical tools of general relativity and black holes, later used in this part.

Ch. 10 is a pedagogical and critical introduction to the pictorial art of conformal diagrams.

Ch. 11 explains why and how black hole evaporates.

Ch. 12 wraps up a long standing debate about black hole information-loss.

Ch. 13 gets white holes out of the astrophysical freak show.

Ch. 14 critically reviews the black-to-white hole transition.

Ch. 15 builds up a mathematical model that supports a new credible scenario.

# 9 GENERAL RELATIVITY LIFEBELT

This chapter is a melting-pot of bells and whistles of general relativity: hypersurface, curvature, action, junction conditions, energy conditions, horizons. We have here gathered technical definitions and results of general interest, that will later be used in this part. Because it is somehow easier for developing the mathematical concepts, we will consider the general case of a space-time of dimension $n$, despite the overwhelming evidence for $n = 4$. For more details, good references are the book of Wald [193], the course of Blau [38] and the toolkit of Poisson [149].

## 9.1 HYPERSURFACE

In general relativity, space-time is described by a $n$-dimensional differentiable manifold $\mathcal{M}$ and a lorentzian metric $g_{ab}$ on top of it. A *hypersurface* $\Sigma$ is a submanifold of $\mathcal{M}$ of codimension 1. Concretely, there are mainly two ways of describing $\Sigma$:

*We choose hereafter the convention $(-, +, ..., +)$ for the signature of $g_{ab}$.*

1. (*Constraint*) If $(x^a)$, with $a \in \{0, ..., n-1\}$, is a coordinate system over $\mathcal{M}$, $\Sigma$ may be defined as the surface of points $(x^a)$ satisfying a constraint equation
$$f(x^a) = 0, \tag{298}$$
with $f$ a real function over $\mathcal{M}$.

2. (*Parametrisation*) The other way is to introduce over $\Sigma$ an internal coordinate system, say $(y^i)$, with $i \in \{1, ..., n-1\}$, so that the embedding of $\Sigma$ in $\mathcal{M}$ is given by a system of parametric equations
$$x^a = x^a(y^i). \tag{299}$$

Both approaches happen to be useful, depending on what we want to compute.

The second description is useful to define the *tangent vector fields*

$$e_i^a \stackrel{\text{def}}{=} \frac{\partial x^a}{\partial y^i}, \quad i \in \{1, ..., n-1\}. \tag{300}$$

Over each point $p \in \Sigma$, the fields $e_i^a$ form a basis of the tangent space $T_p\Sigma$. The index $i$ labels the vectors in the basis, while the index $a$ labels the components of a given vector. A priori, this basis is neither normalised nor orthogonal. $\Sigma$ is said to be space-like if the fields $e_i$ are space-like everywhere. We assume hereafter that such is the case.

The first kind of description is useful to compute the *normal vector field*

$$n^a \stackrel{\text{def}}{=} -\left|g^{cd}\frac{\partial f}{\partial x^c}\frac{\partial f}{\partial x^d}\right|^{-\frac{1}{2}} g^{ab}\frac{\partial f}{\partial x^b}. \tag{301}$$

The seemingly intricate expression is actually simple, when one notices that the first term is only a normalisation factor. The field $n^a$ satisfies the three following properties:

1. $n^a$ is normal to $\Sigma$:
$$\forall i, \quad n_a e_i^a = 0. \tag{302}$$
2. $n^a$ is normalised and time-like:
$$n^a n_a = -1. \tag{303}$$
3. $n^a$ points towards increasing $f$:
$$n^a \frac{\partial f}{\partial x^a} > 0. \tag{304}$$

## 9.2 CURVATURE

The information about the curvature of $\mathcal{M}$ is contained within the metric tensor $g_{ab}$. Its determinant is denoted $g$. The volumes are locally measured by the *volume-form*, given by $\sqrt{|g|}\ d^4x$. One can show that there exists a unique metric-preserving and torsionless connection $\nabla$ over the tangent bundle $T\mathcal{M}$. It is called the *Levi-Civita connection* and it defines a covariant derivative of vector fields as

$$\nabla_a t^b = \partial_a t^b + \Gamma^b_{\ ac} t^c, \tag{305}$$

with $\Gamma^b_{\ ac}$ the *Christoffel symbols*, given by

$$\Gamma^b_{\ ac} = \frac{1}{2} g^{bd}\left(\partial_a g_{cd} + \partial_c g_{ad} - \partial_d g_{ac}\right). \tag{306}$$

The Levi-Civita connection enables to define the *Riemann curvature tensor* $R_{abc}{}^d$ through the equation

$$R_{abc}{}^d\, t_d \stackrel{\text{def}}{=} \nabla_a \nabla_b t_c - \nabla_b \nabla_a t_c, \tag{307}$$

which leads to the formula

$$R_{abc}{}^d = \partial_b \Gamma^d_{\ ac} - \partial_a \Gamma^d_{\ bc} + \Gamma^e_{\ ac}\Gamma^d_{\ eb} - \Gamma^e_{\ bc}\Gamma^d_{\ ea}. \tag{308}$$

By contraction of the indices, one then defines the *Ricci tensor* $R_{ab} \stackrel{\text{def}}{=} R_{acb}{}^c$ and the *scalar curvature* $R \stackrel{\text{def}}{=} R_a{}^a$.

A hypersurface $\Sigma$ is also curved, in two ways: by itself and within $\mathcal{M}$. On the one hand, the *intrinsic curvature* is captured by the *induced metric tensor*

$$h_{ij} \stackrel{\text{def}}{=} g_{ab} e_i^a e_j^b. \tag{309}$$

It is the metric $g_{ab}$ projected down to $\Sigma$. Its determinant is denoted $h$. On the other hand, the *extrinsic curvature tensor* $K_{ab}$ is defined as

$$K_{ab} \overset{\text{def}}{=} \nabla_a n_b. \tag{310}$$

It measures the bending of $\Sigma$ within $\mathcal{M}$, telling how much $n^b$ changes when it is moved along $x^a$. Equivalently, it can be defined as the rate of change of the metric, in the direction $n$,

$$K_{ab} = \frac{1}{2}\mathcal{L}_n h_{ab}, \tag{311}$$

with $\mathcal{L}_n$ the Lie derivative along $n$. In the time-gauge, it reduces to the time derivative of the induced metric. Its trace is denoted $K$.

Both $h_{ab}$ and $K_{ab}$ are:

1. symmetric, i. e.

$$h_{ab} = h_{ba} \qquad\qquad K_{ab} = K_{ba} \tag{312}$$

2. purely spatial, i. e.

$$h_{ab}n^a = 0 \qquad\qquad K_{ab}n^a = 0. \tag{313}$$

## 9.3 ACTION

The *Einstein-Hilbert action* is a real functional of the metric $g_{ab}$, defined by

$$S_{EH}[g_{ab}] \overset{\text{def}}{=} \int_{\mathcal{M}} R\sqrt{|g|}\, d^n x. \tag{314}$$

In general relativity, matter is described by fields on $\mathcal{M}$. Their dynamics is governed by an action $S_M$, which is a functional of both the matter fields and the metric field. Assuming that $\mathcal{M}$ is closed, i. e. compact and without boundary, general relativity asserts that the metric of space-time is a stationnary point of $S_{EH} + S_M$. This demand leads to the Einstein equations:

$$R_{ab} - \frac{1}{2}R\, g_{ab} = 8\pi G T_{ab}, \tag{315}$$

with the *stress-energy tensor*

$$T_{ab} \overset{\text{def}}{=} -\frac{2}{\sqrt{|g|}}\frac{\delta S_M}{\delta g^{ab}}, \tag{316}$$

and $G$ the Newton constant.

Dropping out the assumption that $\mathcal{M}$ is closed, but assuming that its boundary $\partial\mathcal{M}$ is space-like, the same Einstein equations are recovered if a boundary term is added to the action. It is called the *Gibbons-Hawking-York* action and it takes the form

$$S_{GHY}[g_{ab}] \overset{\text{def}}{=} 2\int_{\partial\mathcal{M}} K\sqrt{h}\, d^{n-1}y. \tag{317}$$

## 9.4 JUNCTION CONDITIONS

A hypersurface $\Sigma$ divides $\mathcal{M}$ in two sectors, $\mathcal{M}^+$ and $\mathcal{M}^-$, with $n^a$ pointing towards $\mathcal{M}^+$, as figure 2 shows. Whatever the tensorial quantity $A$, we define the *jump* of $A$ over $\Sigma$ as

$$[A] \stackrel{\text{def}}{=} A|_{\Sigma^+} - A|_{\Sigma^-}, \tag{318}$$

with $\Sigma_+$ and $\Sigma_-$ denoting $\Sigma$ reached by a continuous approach from $\mathcal{M}^+$ and $\mathcal{M}^-$ respectively. If $A$ is continuous across $\Sigma$, then $[A] = 0$. The following *junction conditions* hold:

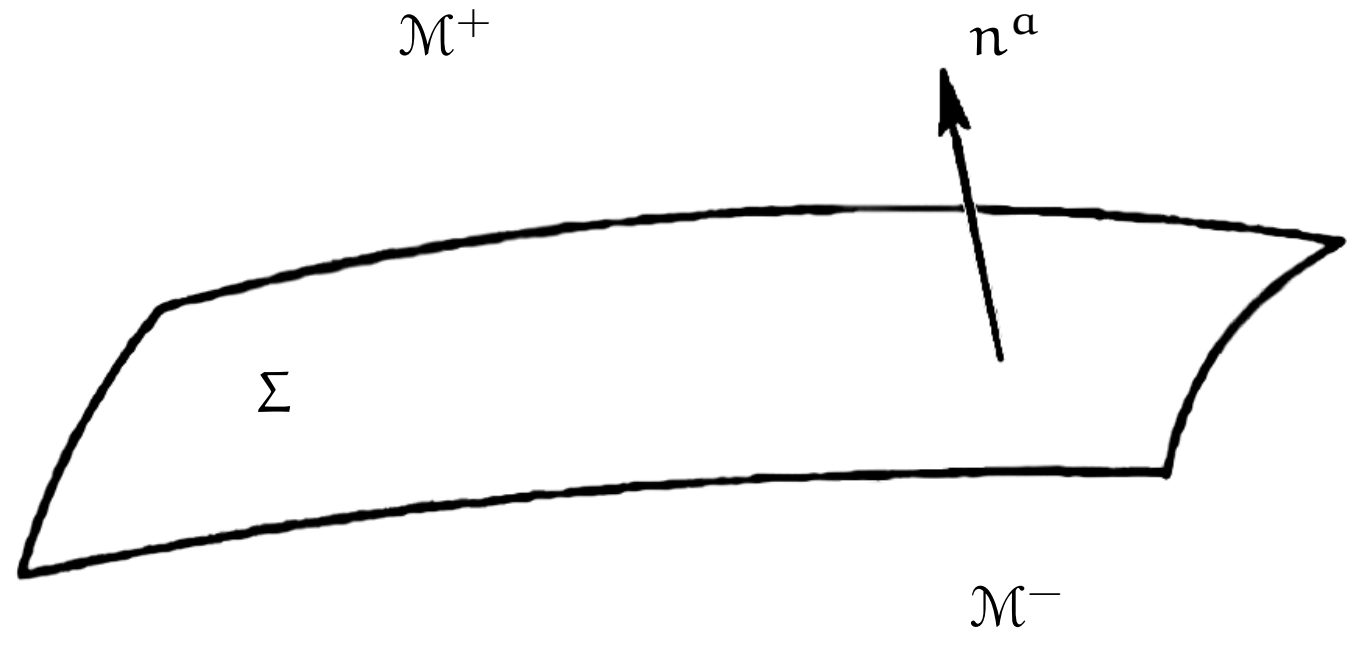

Figure 2: Hypersurface.

1. The jump in the induced metric is

$$[h_{ij}] = 0. \tag{319}$$

2. The jump in the extrinsic curvature is implicitly given by

$$S_{ij} = \frac{1}{8\pi}\left([K_{ij}] - [K]h_{ij}\right) \tag{320}$$

where $S_{ij}$ is the intrinsic stress-energy tensor on $\Sigma$, given by

$$S_{ij} \stackrel{\text{def}}{=} T_{ab} e_i^a e_j^b. \tag{321}$$

If $S_{ij} = 0$, then $[K_{ij}] = 0$.

Conversely, these junction conditions have to be satisfied whenever one constructs a space-time by gluing different patches together.

## 9.5 ENERGY CONDITIONS

In the Einstein equations, the stress-energy tensor, $T_{ab}$, models the matter content of space-time. A priori, general relativity does not prescribe any specific expression for it. Thus, *any* metric $g_{ab}$ is a solution of Einstein equations, with $T_{ab}$ *defined* by them. Of course, life is not so easy. In particular, for $T_{ab}$ to be a *good* stress-energy tensor, it has to satisfy few properties. Importantly, it has to account for *positive energy*. There are several ways to implement this idea. Here is the list of the four main energy conditions found in the literature:

1. (Weak) For any future-directed time-like vector $t^a$,

$$T_{ab}t^a t^b \geqslant 0. \tag{322}$$

   The Left Hand Side (LHS) corresponds to the energy density seen by an observer travelling with the 4-velocity $t^a$.

2. (Strong) For any future-directed time-like vector $t^a$,

$$(T_{ab} - \frac{1}{2}g_{ab}T)t^a t^b \geqslant 0. \tag{323}$$

   It implements the idea that 'gravity is attractive', i. e. time-like geodesics are focusing.

3. (Null) For any future-directed null vector $l^a$,

$$T_{ab}l^a l^b \geqslant 0. \tag{324}$$

   It implements the idea that null geodesics are focusing.

4. (Dominant) In addition to the weak condition, we require

$$T^a_b T_{ac} t^b t^c \leqslant 0. \tag{325}$$

   It implements the idea that energy cannot travel faster than light.

These conditions satisfy the following relation-table.

$$\begin{array}{ccccc} & & & & \text{Strong} \\ & & & & \Downarrow \\ \text{Dominant} & \Longrightarrow & \text{Weak} & \Longrightarrow & \text{Null} \end{array}$$

## 9.6 HORIZON

### 9.6.1 *Event horizon*

Mathematically, a *black hole* is often defined as a region $B$ in a space-time $\mathcal{M}$, that is not contained in the causal past of future-null infinity $\mathcal{I}^+$ (see chapter 10 for a proper definition of $\mathcal{I}^+$). The boundary of $B$ is called the *event horizon*. This definition makes perfect sense and is a good attempt of a formalisation of the idea of 'a region of no escape', invisible to an outside asymptotic observer. However, it is very debatable from an experimental perspective. Indeed, this definition can be considered as *teleological*, meaning the identification of a black hole requires knowledge about future-null infinity. This is disappointing, as it makes the definition untractable for practical purposes, it requires to know the ultimate future of the world to say whether or not we are in a black hole right now. To give an example, an event horizon may well be forming right now across this page, and you would not even notice it before a very long time.

### 9.6.2 *Apparent horizon*

Instead, we prefer a definition of black holes based on some quasi-local geometric properties of space-time. Consider $S$, an orientable two-dimensional space-like surface. At each point $p \in S$, there exists precisely two future-directed null vectors, orthogonal to $S$. They generate two distinct families of future-directed null geodesics orthogonal to $S$. They form the *ingoing* and *outgoing* congruences. Each congruence locally spans a null hypersurface in the neighbourhood of $p$, so that two geodesics of the congruence do not intersect. In our minds, we can have the picture of a shining space-like sphere in flat space-time: some rays converge to the centre (ingoing), while others propagate away from the sphere (outgoing). But this picture is not fully generic as it happens sometimes that both ingoing and outgoing congruences converge to (or diverge from) the centre.

Consider a congruence of null geodesics, and its tangent vector field $l^a$ affinely parametrised, i. e.

$$l^a l_a = 0 \quad \text{and} \quad l^a \nabla_a l^b = 0. \tag{326}$$

We define the *expansion* $\theta$ as

$$\theta \stackrel{\text{def}}{=} \nabla_a l^a. \tag{327}$$

It is the covariant divergence of $l^a$. It measures how infinitesimally nearby null geodesics dilate or contract. In the example of the shining sphere in flat space-time, the ingoing (resp. outgoing) congruence has a negative (resp. positive) expansion.

A *trapped surface* $S$ is a 2-dimensional compact hypersurface, so that for both ingoing and outgoing null geodesics, we have

$$\forall p \in S, \quad \theta(p) < 0. \tag{328}$$

It means that the local area decreases along any future direction. If $\theta(p) < 0$ for one of the two congruences, and $\theta(p) = 0$ for the other, then $S$ is said to be *marginally trapped*.

We can now define a black hole as a *trapped region*, that is a region so that every point lie on a trapped surface. Its boundary is typically a 3-surface foliated by marginally trapped surfaces and is called a *marginally trapped tube*, or an *apparent horizon*.

The general definition is made more concrete in the specific example of spherically symmetric space-time. In such a case, the metric can always be written in double-null coordinates $(u, v)$ as

$$d^2 s = C(u,v) du dv + r^2(u,v) d\Omega^2. \tag{329}$$

with $C(u,v)$ some real function and $d\Omega^2 = d\theta^2 + \sin^2\theta d\phi^2$ the usual metric on the unit sphere. We can consider two radial (only $(u,v)$ components) and spherically symmetric (the components only depend on $(u,v)$) null vector fields $l^a$ and $n^a$ so that

$$l^a l_a = n^a n_a = 0 \quad \text{and} \quad l^a n_a = -1. \tag{330}$$

$l^a$ and $n^a$ are tangent to the ingoing and outgoing null congruences. The expansion can be computed for each:

$$\theta_l = \frac{2}{r}\frac{\partial r}{\partial u} \quad \text{and} \quad \theta_n = \frac{2}{r}\frac{\partial r}{\partial v}. \tag{331}$$

We see in this case, that the radius of trapped surface decreases in both future-null directions.

# 10 CONFORMAL DIAGRAMS

The possibility to visualise space-time models is of great importance for solving problems of general relativity. However, it may be hard to have in mind a clear representation of a 4-dimensional curved space-time. Fortunately, in some simple-enough cases, there exists a convenient way to represent space-time, that goes under the name of *conformal diagrams*, or *Penrose diagrams*, or even *Carter-Penrose diagrams*. We detail a series of examples in this chapter.

## 10.1 CONFORMAL TRANSFORMATION

In general relativity, space-time is described by a pair of a manifold and a metric, $(\mathcal{M}, g_{ab})$. Ideally, a faithful drawing of it would make space-time points correspond to some points on a sheet of paper while preserving the structural relations between them.

But first of all, a sheet of paper is 2-dimensional, whereas space-time has 4 dimensions, so it is clear, from the beginning, that a drawing cannot be entirely faithful. However, if space-time has some symmetries, it carries redundancies that can be used to give a decent picture of it.

Then, a sheet of paper is bounded, so that infinitely remote space-time points will be brought to a finite distance. This will necessarily distort the distances between the points in a non-trivial way, and one may wonder which kind of structural relations are still possibly preserved.

To answer this question, we should first notice that the metric structure can be decomposed into a causal structure and a conformal factor. Given the distance between two points, its sign (positive, negative or null) encodes the causal relation (space-like, time-like or null), while its norm encodes the scaling. It is possible to rescale the metric at each point without changing the causal relations between points. It is called a *conformal transformation*, defined for any strictly positive smooth function $\Omega$ over $\mathcal{M}$, as

$$\tilde{g}_{ab} = \Omega^2 g_{ab}. \tag{332}$$

$\Omega$ is called the *conformal factor*. The space-time $(\mathcal{M}, \tilde{g}_{ab})$ has the same global shape as $(\mathcal{M}, g_{ab})$, but the distances between points are now different. Yet, the kind of distance between two points, whether it is space-like, time-like or null, is unchanged. In other words, the causal structure is preserved by a conformal transformation. Conversely, two space-times with the same global shape and the same causal structure only differ by a conformal factor!

On a space-time diagram, the causal structure is visible in the local shape of light cones. In general, such a shape is not preserved by conformal transformations (although the causal structure is). Ideally, a conformal transformation can straighten up the light cones, so that they are leaned at $\pm 45°$ on the final diagram.

From this preliminary analysis, we can now look for a diagrammatic representation of space-time with the two following requirements

1. Global structure: infinities lie at a finite distance.

2. Causal structure: light rays are at $\pm 45°$.

Both requirements are implemented in a procedure called *conformal compactification*. Let's start with a simple example.

## 10.2 2D MINKOWSKI SPACE-TIME

In two dimensions, Minkowski space-time is $\mathbb{R}^2$ endowed with the metric

$$ds^2 = -dt^2 + dx^2. \tag{333}$$

A standard space-time diagram is given in figure 3a. It uses a cartesian coordinate system. The light cones fulfil the causal structure condition, but the whole space-time does not fit on the page. We need to bring infinities back to a finite distance from the origin. To do so, we can change variables using the tangent function. Indeed, the tangent is a diffeomorphism from $]-\frac{\pi}{2}, \frac{\pi}{2}[$ to $\mathbb{R}$. So we can try

$$\begin{cases} T = \arctan t \\ X = \arctan x \end{cases} \tag{334}$$

Then the graphical representation looks like figure 3b. Now it is the opposite: the global structure condition is fulfilled, but the light cones are completely deformed. This is not surprising after all because this transformation squeezes independently the coordinates along $x$ and $t$ directions.

We need for a conformal transformation that preserves the shape of light cones of figure 3a. It is found by switching to better coordinates, adapted to the causal structure. They are the *light-cone coordinates*, defined by

$$\begin{cases} u = t - x \\ v = t + x \end{cases} \tag{335}$$

In these coordinates, the metric reads

$$ds^2 = -du dv. \tag{336}$$

We see that the null geodesics ($ds^2 = 0$) have a constant $u$ or a constant $v$. This coordinate system is shown in figure 3c. We can now compactify along these directions

$$\begin{cases} U = \arctan u \\ V = \arctan v \end{cases} \tag{337}$$

The diagram finally obtained is in figure 3d. The shape of light cones

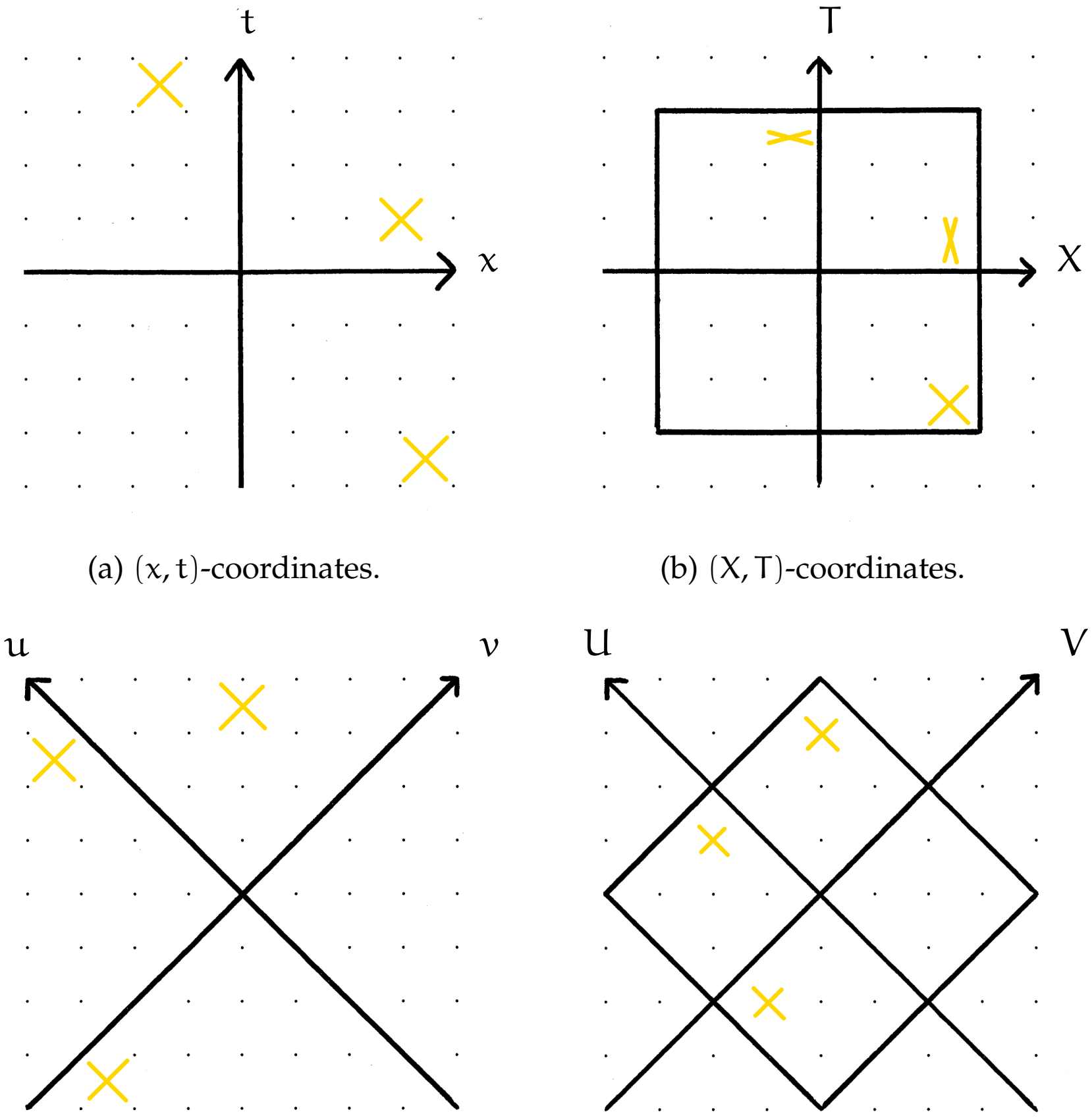

(a) $(x, t)$-coordinates.

(b) $(X, T)$-coordinates.

(c) $(u, v)$-coordinates.

(d) $(U, V)$-coordinates.

Figure 3: Various coordinates on Minkowski space-time. The gold crosses show the local shape of light cones. In figures 3b and 3d, the whole Minkowski space-time fits within the square.

is preserved, as can be seen from the expression of the metric

$$ds^2 = \frac{-dUdV}{\cos^2 U \cos^2 V}. \tag{338}$$

The null geodesics corresponds indeed to constant $U$ or $V$. Both the global and the causal conditions are fulfilled, so that we have obtained a satisfying conformal compactification of 2D Minkowski space-time. The resulting diagram is called a *conformal diagram*, or a *Penrose diagram*. It is drawn with more detail in figure 4. Its (conformal) boundaries correspond to infinitely remote points of space-time. Their classification and denotation is standard:

| | |
|---|---|
| $i^+$ | Future time-like infinity |
| $i^-$ | Past time-like infinity |
| $i^0$ | Space-like infinity |
| $\mathcal{I}^+$ | Future null infinity |
| $\mathcal{I}^-$ | Past null infinity |

*The symbol $\mathcal{I}$ is a script I, named 'scry', which is also a verb meaning 'to predict the future'.*

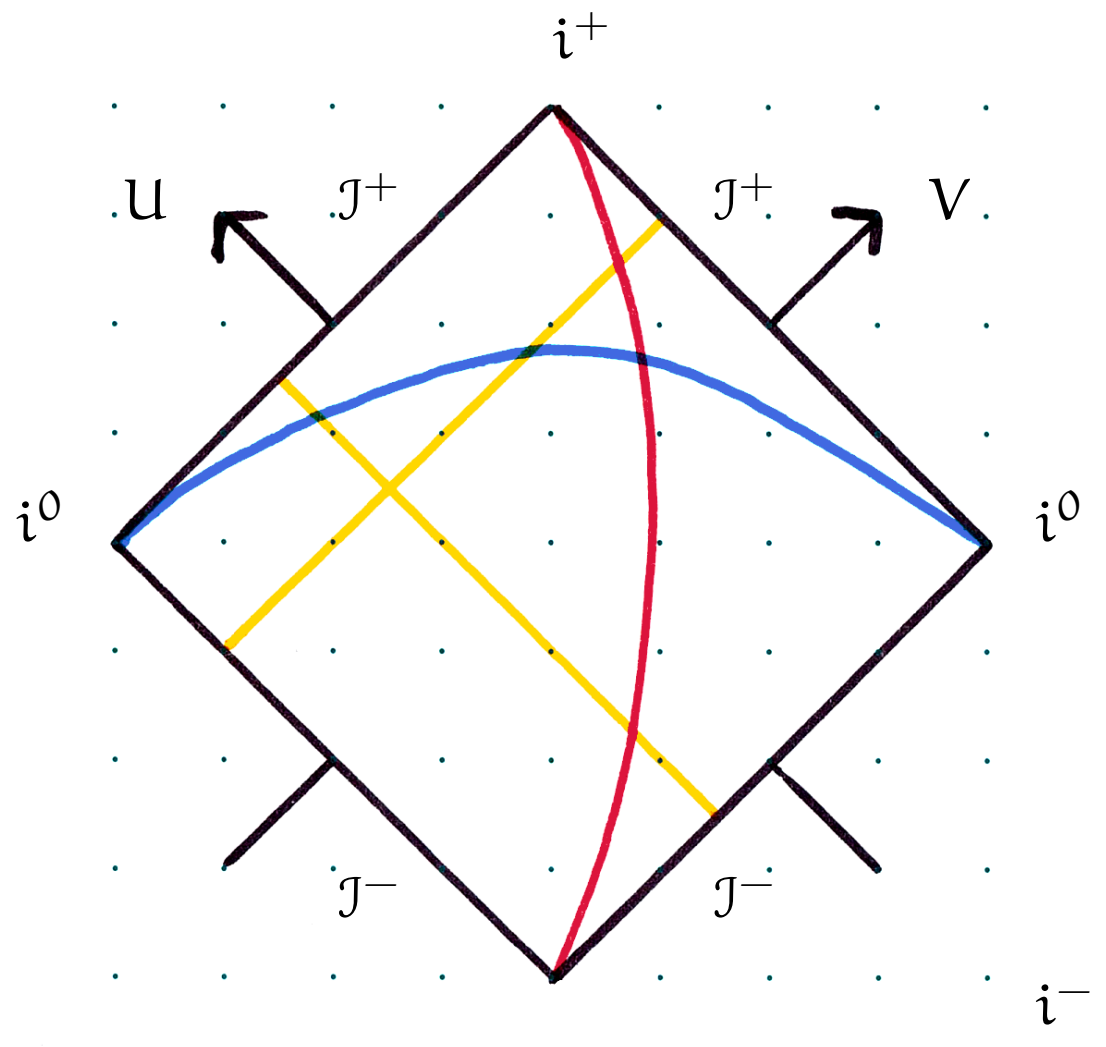

Figure 4: Conformal diagram of 2D Minkowski space-time. The red, blue and gold lines show time-like, space-like and null geodesics respectively.

## 10.3 4D Minkowski space-time

In four dimensions, Minkowski space-time is $\mathbb{R}^4$ together with the metric

$$ds^2 = -dt^2 + dx^2 + dy^2 + dz^2. \tag{339}$$

To draw a conformal diagram, 2 dimensions shall be ignored. A possibility would be to disregard the $y$ and $z$ directions, which would directly bring us back to the previous case, depicted in figure 4. However, it is not a good choice, because, for instance, the null geodesic

$$(t = \lambda, x = 1, y = \lambda, z = 0), \quad \text{with } \lambda \in \mathbb{R}, \tag{340}$$

would be represented by the same curve as the time-like geodesic

$$(t = \lambda, x = 1, y = 0, z = 0), \quad \text{with } \lambda \in \mathbb{R}, \tag{341}$$

like the red curve in figure 4, so that the causal structure is not depicted faithfully. A better choice is found using the spherical symmetry of space-time. In the spherical coordinates, the metric reads

$$ds^2 = -dt^2 + dr^2 + r^2(d\theta^2 + \sin^2\theta \, d\phi^2). \tag{342}$$

Ignoring the angular coordinates $\theta$ and $\phi$, the graphical representation is given in figure 5a. The null geodesics (340) still looks time-like, but at least, it is now clearly distinct from the time-like geodesics (341). After compactification along null directions

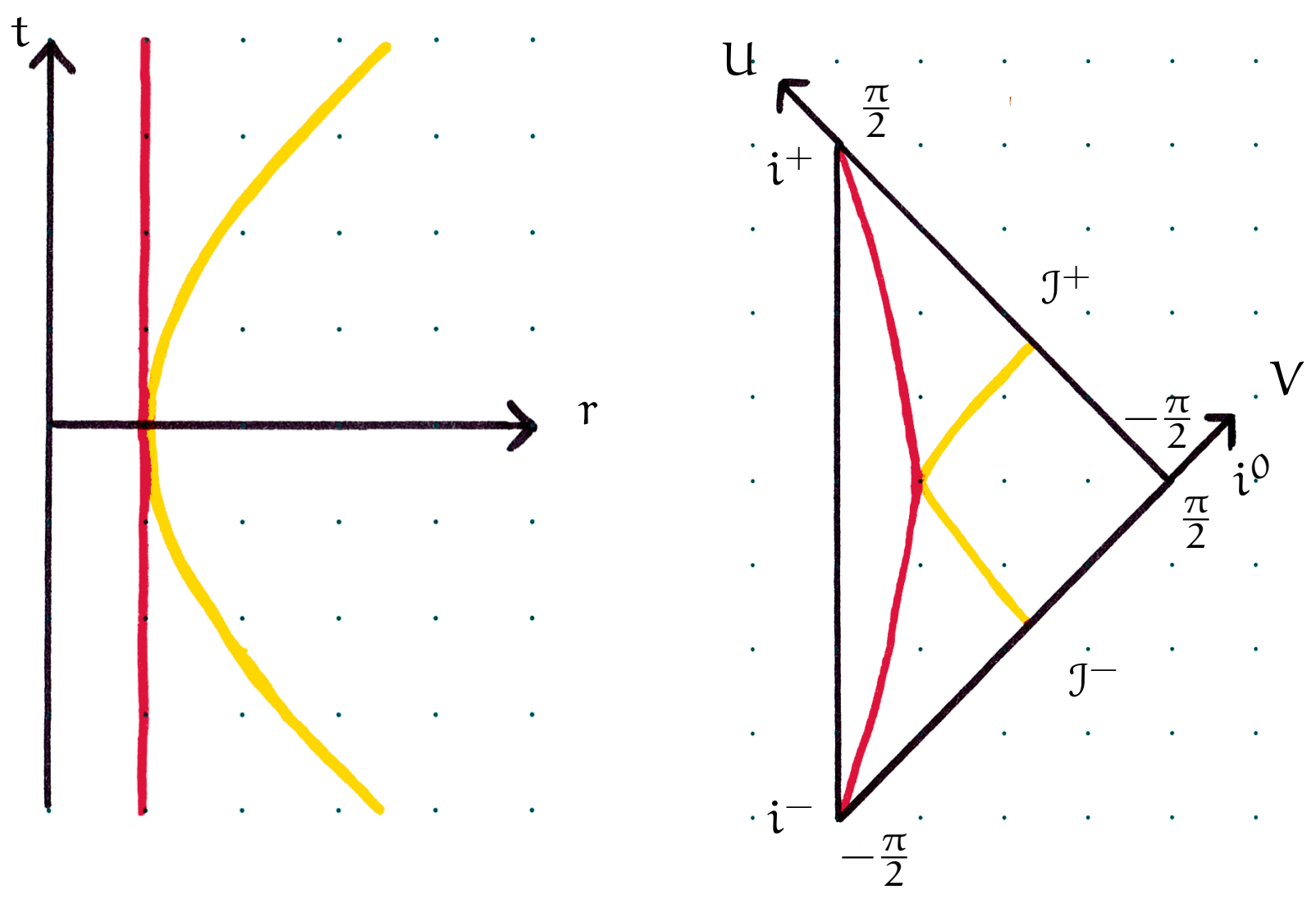

(a) $(r, t)$-coordinates. (b) $(U, V)$-coordinates.

Figure 5: Two diagrams of 4D Minkowski space-time.

$$\begin{cases} U = \arctan(t - r) \\ V = \arctan(t + r). \end{cases} \tag{343}$$

the metric becomes

$$ds^2 = \frac{-dUdV}{\cos^2 U \cos^2 V} + r^2 d\Omega^2 \tag{344}$$

with

$$r = \frac{1}{2}(\tan V - \tan U), \tag{345}$$

and the diagram is in figure 5b. The requirement that light rays should be at $\pm 45°$ is not completely satisfied. Indeed, it is true that $\pm 45°$ straight lines are null, but the converse is false. Some null geodesics, like the one defined by equation (340) and depicted with a gold line in figure 5b, are not $\pm 45°$ straight lines. In fact, the causal condition is only satisfied *asymptotically*, meaning that

1. A geodesic is time-like iff it goes from $i^-$ to $i^+$.

2. A geodesic is space-like iff it ends in $i^0$.

3. A geodesic is null iff it goes from $\mathcal{J}^-$ to $\mathcal{J}^+$.

The lesson is that when we deal with 4-dimensional space-time, there is an irreducible difference between looking-like and being-like. This fact is often overlooked in textbooks dealing with conformal diagrams. It can be summarised in the following relation-table.

| | | | | |
|---|---|---|---|---|
| space-looking | — | space-like | — | null-looking |
| | | \| | | \| |
| time-like | — | time-looking | — | null |

It should be read as 'if a trajectory is looking space-like on the conformal diagram, then it is indeed a space-like trajectory', or 'if a trajectory is space-like, then it may look either space-like, or time-like or null', etc.

## 10.4 SCHWARZSCHILD SPACE-TIME

The Schwarzschild metric is

$$ds^2 = -\left(1 - \frac{2m}{r}\right) dt^2 + \left(1 - \frac{2m}{r}\right)^{-1} dr^2 + r^2 d\Omega^2. \tag{346}$$

As in the previous Minkowski case, we will ignore the angular dimensions on the diagram. Then, we must use null-coordinates, adapted to the conformal compactification, obtained by looking for the equation of null geodesics. A convenient choice are the *retarded* and *advanced* time coordinates, respectively

$$\begin{cases} u = t - r - 2m \log \left| \frac{r}{2m} - 1 \right| \\ v = t + r + 2m \log \left| \frac{r}{2m} - 1 \right|. \end{cases} \tag{347}$$

The metric now reads

$$ds^2 = -\left(1 - \frac{2m}{r}\right) du dv + r^2 d\Omega^2 \tag{348}$$

with the radius

$$r = 2m \left(1 + W\left(e^{\frac{v-u}{4m} - 1}\right)\right). \tag{349}$$

The function $W$ is the upper branch of the *Lambert W-function*. It is an increasing function defined implicitly by the equation

$$x = W(x) e^{W(x)} \tag{350}$$

and its graph is shown in figure 6. Now, we could compatify with the tangent function, as previously, but it is unfortunate that the metric (348) has a fake singularity in $r = 2m$. Space-time is described by two patches separated by the horizon. There is a choice of null-coordinates that resews the two patches in a single one. It is the *Kruskal coordinates*

$$\begin{cases} U = -e^{-\frac{u}{4m}} \\ V = e^{\frac{v}{4m}}. \end{cases} \tag{351}$$

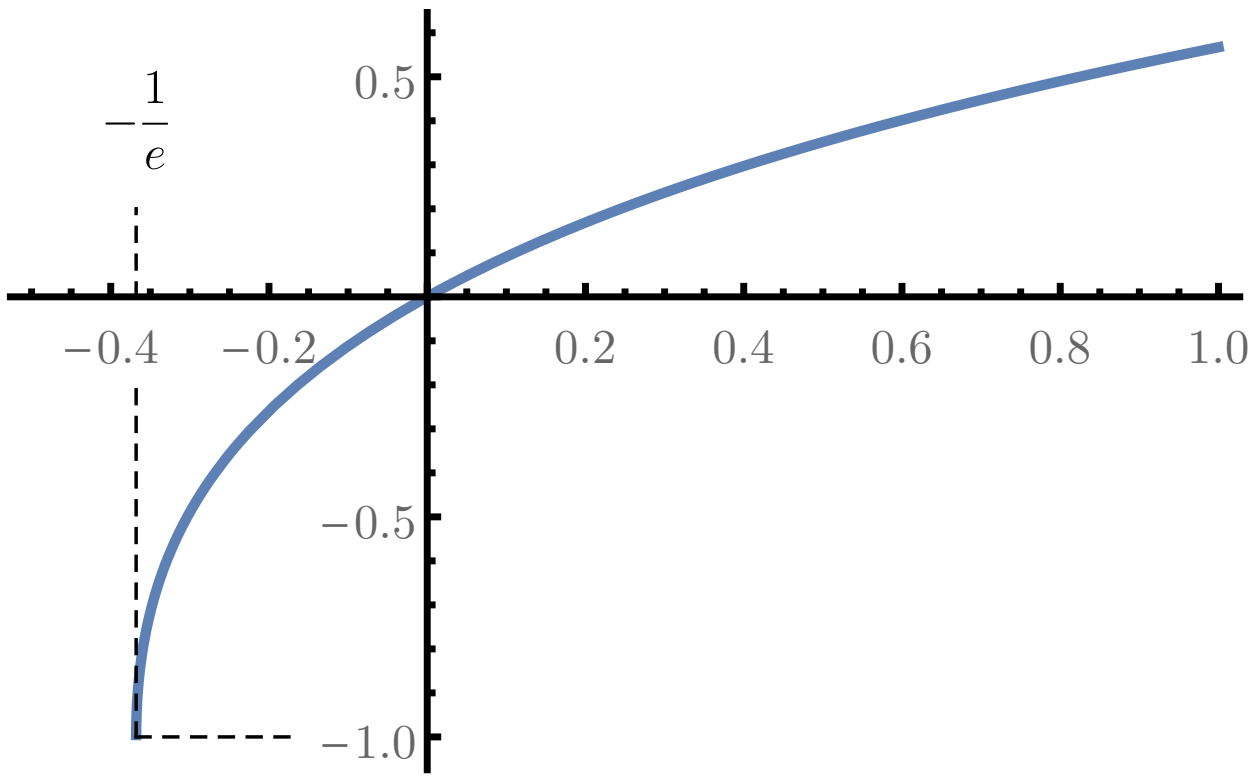

Figure 6: Graph of the upper branch of the Lambert $W$ function.

The metric is given by

$$ds^2 = -32m^3 \frac{e^{-r/2m}}{r} dU\, dV + r^2 d\Omega^2, \tag{352}$$

with

$$r(U, V) = 2m \left[1 + W\left(-\frac{UV}{e}\right)\right]. \tag{353}$$

The event horizon is now described by the equation $U = 0$. Contrary to $v$, $V$ is an affine parameter along the horizon, meaning its tangent vector along the horizon, $t^a$, satisties $t^a \nabla_a t^b = 0$.

Actually, these coordinates are suitable to describe the maximal analytic extension of the Schwarzschild black hole, that is called the *Kruskal-Szekeres extension*. It is maximal in the sense that every affinely parametrised geodesic can either be continued to infinite values of its parameter or it runs into a singularity at a finite value of it. It is compactified with

$$\begin{cases} \tilde{U} = \arctan U, \\ \tilde{V} = \arctan V, \end{cases} \tag{354}$$

and the conformal diagram is shown in figure 7. As we will see later, region II corresponds to the black hole and region IV to the white hole.

## 10.5 FROM STARS TO BLACK HOLES

### 10.5.1 *Spherical star*

The Schwarzschild metric is a good description of the geometry outside a spherical mass, like a star. Inside the star, the metric is different. A toy-model can be given by the metric

$$ds^2 = -F(r)dt^2 + G(r)dr^2 + r^2 d\Omega^2, \tag{355}$$

which is that of an arbitrary static, spherically symmetric spacetime [193]. We take F and G to be smooth functions, that go to 1 when

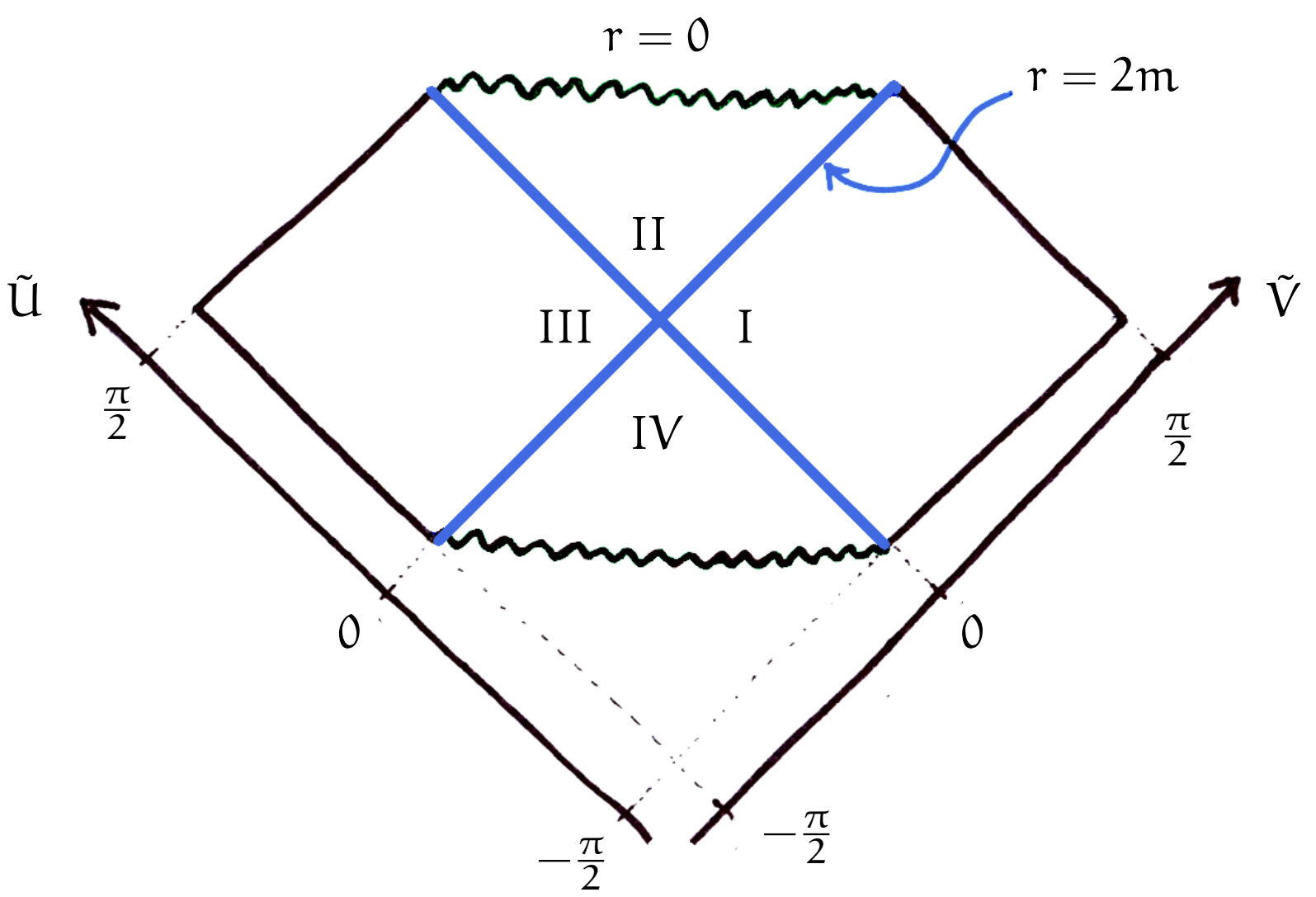

Figure 7: Conformal diagram of the Kruskal spacetime.

$r \to 0$. Then, there exists a change of variable, $r \to r^*$, such that the metric becomes

$$ds^2 = F(r(r_*))(-dt^2 + dr_*^2) + r(r_*)^2 d\Omega^2. \tag{356}$$

Ignoring the angular coordinates, the conformal diagram is given in figure 8a. We are sketchy about its derivation, as it essentially follows the same procedure as before, and we will not use it in the following. In this model, the star is static, i. e. not evolving with time. Of course, this is not realistic, as stars burn, and finally die. For instance, we can consider that the star ultimately collapses into a black hole, as in the conformal diagram of figure 8b.

### 10.5.2 *Collapsing null shell*

An interesting toy-model is provided by the collapse of a thin null shell of matter. Physically, we shall imagine a spherical shell of light collapsing to its center. Inside the metric is flat and it is Schwarzschild outside. The conformal diagram of the model is shown in figure 9. The metric in each patch is

$$(\mathrm{I}) \left[ \begin{array}{l} ds^2 = -du dv + r^2 d\Omega^2 \\ r = \frac{1}{2}(v-u) \end{array} \right. \tag{357}$$

$$(\mathrm{IIa}) \left[ \begin{array}{l} ds^2 = -\left(1 - \frac{2m}{r}\right) du dv + r^2 d\Omega^2 \\ r = 2m\left(1 + W\left(e^{\frac{v-u}{4m}-1}\right)\right) \end{array} \right. \tag{358}$$

$$(\mathrm{IIb}) \left[ \begin{array}{l} ds^2 = \left(1 - \frac{2m}{r}\right) du dv + r^2 d\Omega^2 \\ r = 2m\left(1 + W\left(-e^{\frac{v+u}{4m}-1}\right)\right) \end{array} \right. \tag{359}$$

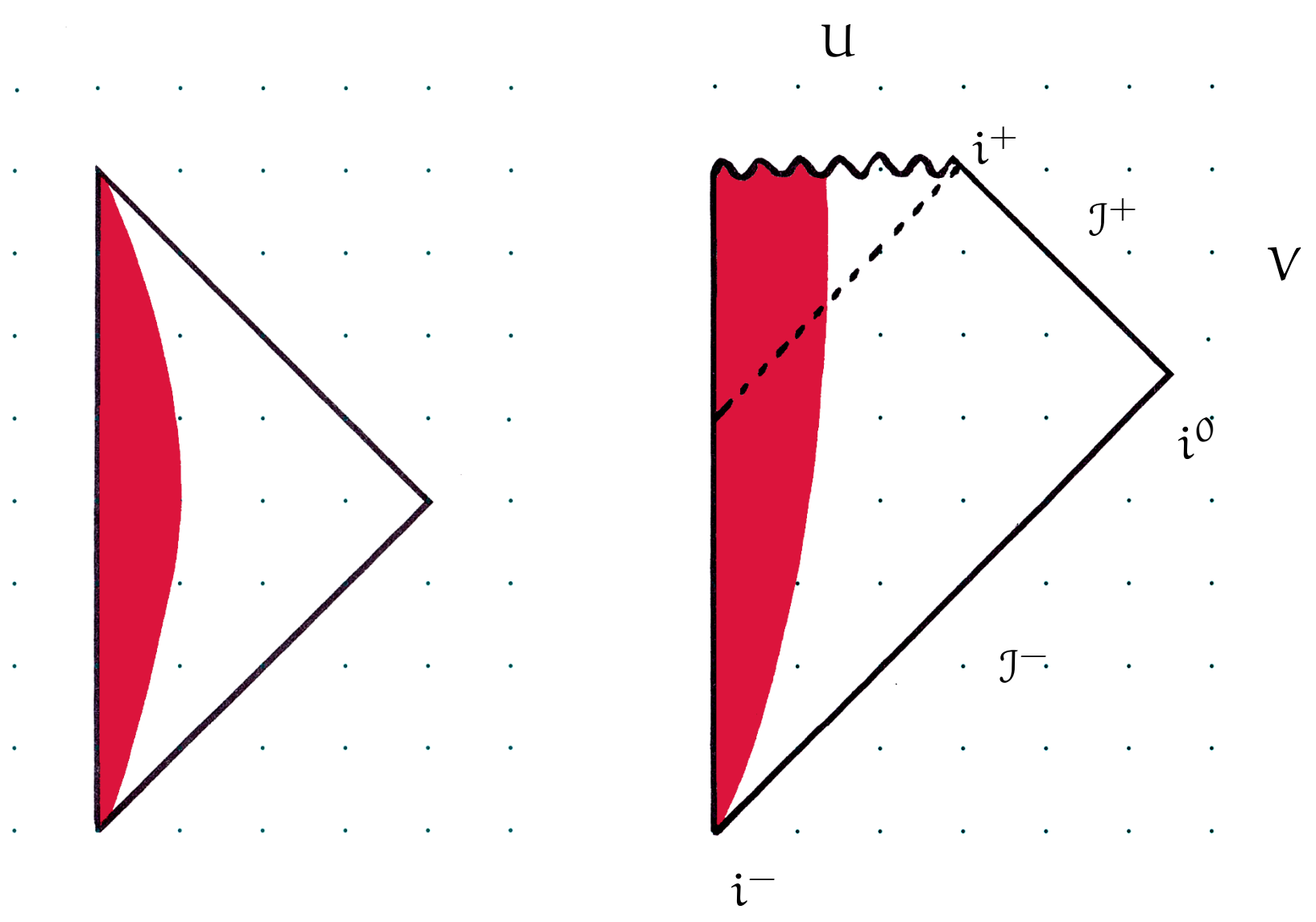

(a) Static spherical star. (b) Collapsing star.

Figure 8: Conformal diagrams of stars.

The map between the metric coordinates $(u, v)$ and the coordinates of the diagram $(U, V)$ is

$$\text{(I)} \left[ \begin{array}{l} u = v_0 - 4m\left(1 + W(-e^{-1}\tan V_0 \tan U)\right) \\ v = v_0 - 4m\left(1 + W(-e^{-1}\tan V_0 \tan(V - 2V_0 + \pi/2))\right) \\ \qquad\qquad\qquad\qquad \text{with } v_0 \overset{\text{def}}{=} 4m \log \tan V_0 \end{array} \right. \tag{360}$$

$$\text{(IIa)} \left[ \begin{array}{l} u = -4m \log(-\tan U) \\ v = 4m \log \tan V \end{array} \right. \tag{361}$$

$$\text{(IIb)} \left[ \begin{array}{l} u = 4m \log \tan U \\ v = 4m \log \tan V \end{array} \right. \tag{362}$$

As the metric is given in three patches, it is important to check the junction conditions, as explained in section 9.4. Both conditions are satisfied between IIa and IIb, but only the first is fulfilled along $V = V_0$. The violation of the second junction condition is the signature of a non-zero stress-energy tensor, which in our case, is the collapsing shell of matter.

### 10.5.3 *Vaidya space-time*

There is an important class of solutions to Einstein equations, called the Vaidya metric [188]. It is expressed in the *outgoing or ingoing*

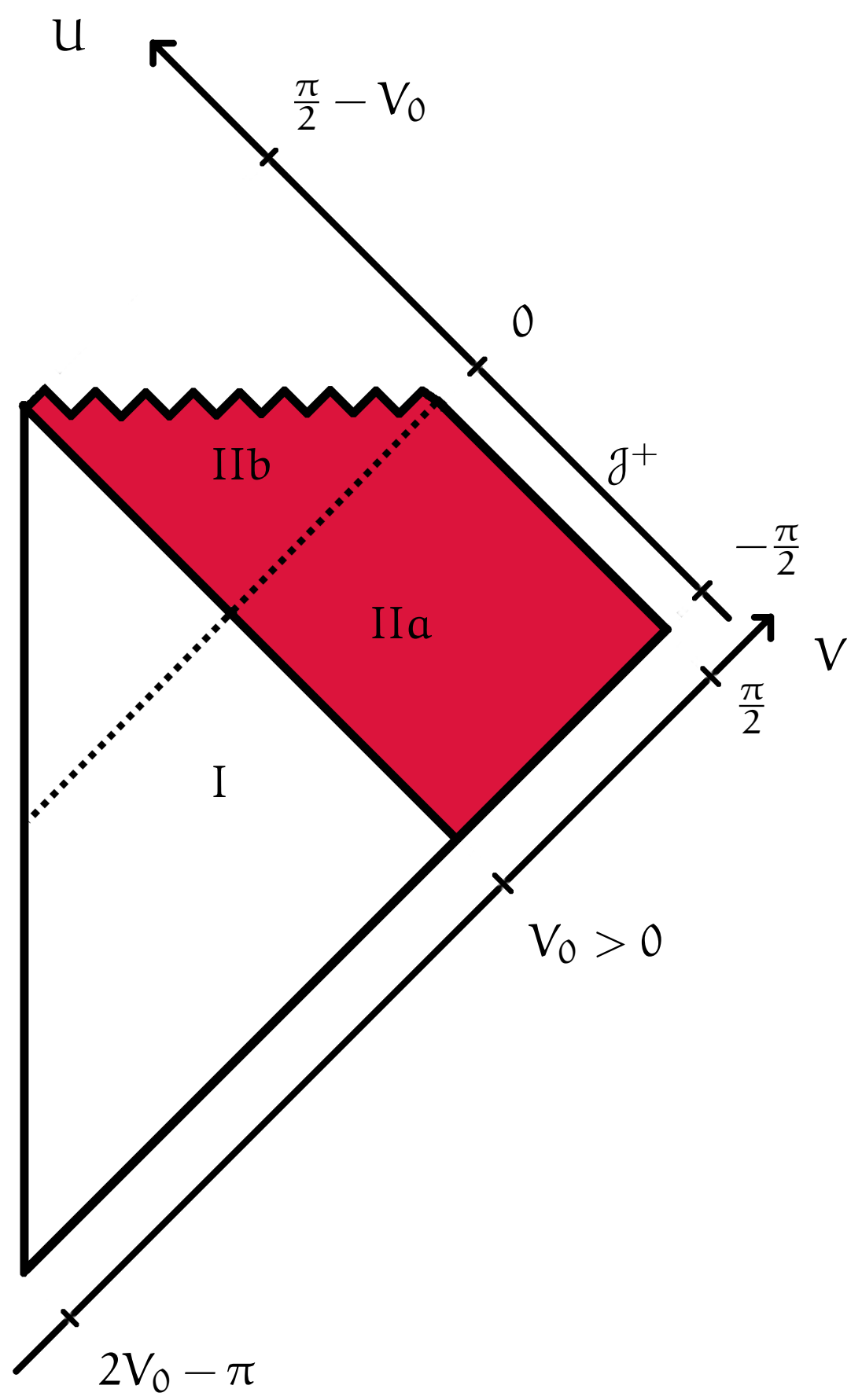

Figure 9: Conformal diagram of a black hole formed by the collapse of a null shell. The event horizon is depicted with a dashed line.

*Eddington-Finkelstein coordinates*, that are respectively $(u, r)$ or $(v, r)$. The *ingoing Vaidya metric* is

$$ds^2 = -\left(1 - \frac{2m(v)}{r}\right) dv^2 + 2dvdr + r^2 d\Omega^2, \tag{363}$$

with $m$ some function of the advanced time coordinate $v$. $m(v)$ is an invariant called the *Misner-Sharp* mass. For a general spherically symmetric space-time, it is defined as

$$M_{MS}(u, v) \overset{\text{def}}{=} \frac{r}{2}\left(1 - g^{ab}\partial_a r \partial_b r\right). \tag{364}$$

It corresponds to the total mass enclosed within a sphere of radius $r$, located by the coordinates $(u, v)$. In the limit $r \to \infty$, it reduces to the *ADM* mass. On $\mathcal{I}^+$ and $\mathcal{I}^-$, it reduces instead to the *Bondi-Sachs* mass. Thus, $m(v)$ is also the Bondi-Sachs mass, that a far-away observer would see along $\mathcal{I}^+$.

The null energy condition requires that $m'(v) \geqslant 0$. It describes the progressive collapse of shells of null matter. The model of the previous subsection 10.5.2 is a particular case of ingoing Vaidya metric with $m(v) = m\,\Theta(v - v_0)$, $\Theta$ being the Heaviside step-function. In this case, the collapsing shell is infinitely thin. Another example is shown in figure 10, with a mass-profile as in figure 11.

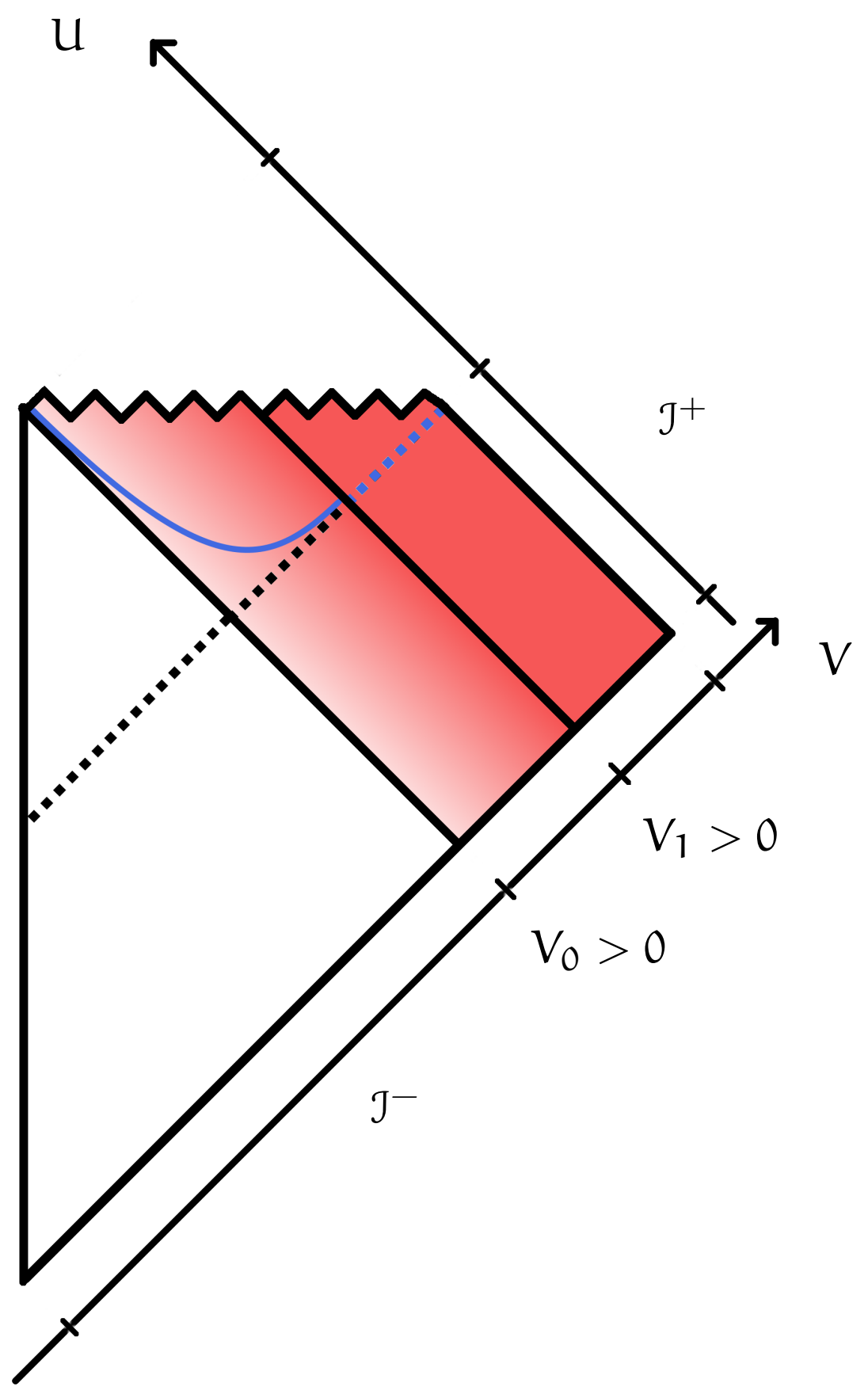

Figure 10: Conformal diagram of ingoing Vaidya space-time. The varying mass is represented by a gradation of reds. The dashed line is still the event horizon, and the blue line is the apparent horizon.

There also exists the *outgoing Vaidya metric*

$$ds^2 = -\left(1 - \frac{2m(u)}{r}\right) du^2 - 2du dr + r^2 d\Omega^2. \qquad (365)$$

In this case, dominant energy condition imposes $m'(u) \leqslant 0$. It describes a radiating star: null shell of matters are progressively expelled out.

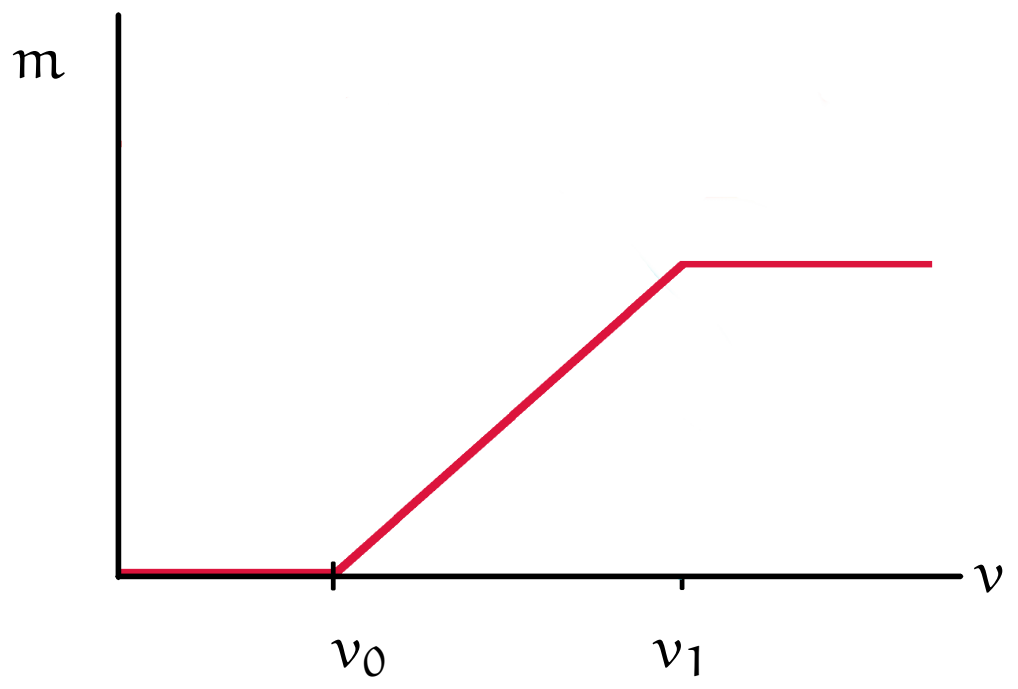

Figure 11: Example of mass-profile $m(v)$.

The ingoing Vaidya metric can be generalised in the form

$$ds^2 = -\left(1 - \frac{2m(v,r)}{r}\right) dv^2 + 2dvdr + r^2 d\Omega^2. \qquad (366)$$

The energy conditions imposes some constraints on the function $m(v,r)$ (see [194]).

## 10.6 ETHICS OF CONFORMAL DIAGRAMS

Conformal diagrams are widely used in general relativity. They offer a visual tool to discuss important issues like the information-loss paradox. However, these diagrams are often drawn hand-wavily, without starting from an explicit metric. For that reason, their soundness is sometimes dubious.

For instance, most of the discussions about the information-loss paradox rely on the conformal diagram depicted in figure 12. However, it is not properly a conformal diagram, because it is not obtained from the conformal compactification of any well-defined metric. When it was first drawn by Hawking, it was more of a guess, than a proper derivation. We can see that it is inspired from the conformal diagram of a collapsing star (figure 8b), to which is vaguely glued a portion of a Minkowski conformal diagram (figure 5b). The details of the construction are absent, although they would matter a lot as they contain the physical content of the evaporation. As pointed out in [168], any concrete construction leading to such a diagram shows that the point B cannot actually be a point of that space-time: it has to be excised for the gluing to be topologically allowed. As a consequence, a Cauchy horizon forms above, so that it is incorrect to assume that any Cauchy surface can exist above it, contrary to many modern discussions (e.g. firewalls, see section 12.5). As a result, the arbitrary drawing of a hand-wavy diagram sways the formulation of the paradox and may put us on the wrong track since the beginning. In chapter 14, we will see that the black-to-white scenario has a different conformal diagram for which the information paradox does not arise.

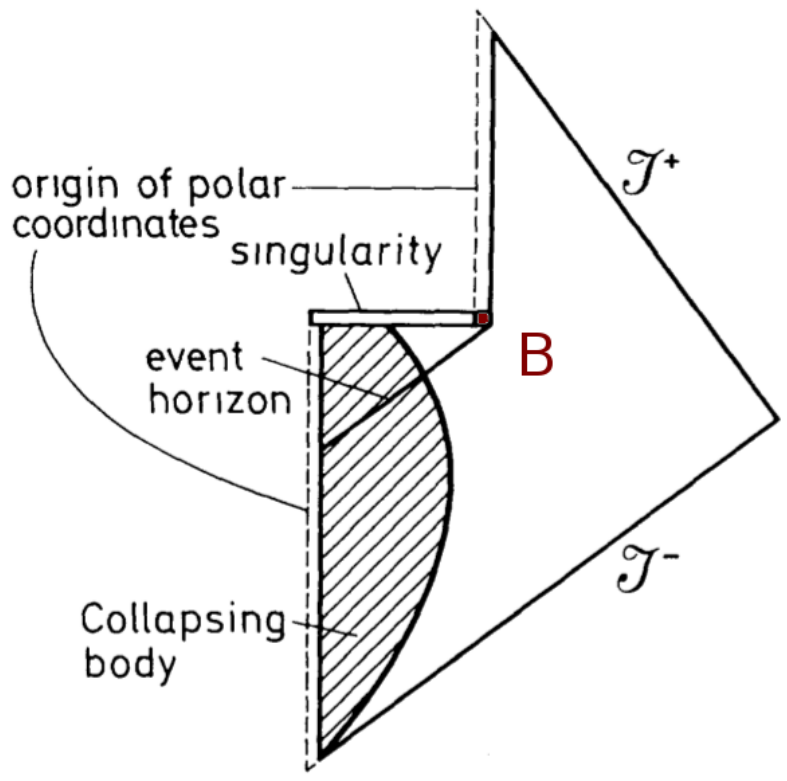

Figure 12: Conformal diagram of an evaporating black hole, as drawn by Hawking in [99], to which has been added the point B.

Against this tendency of drawing diagrams lazily, we advocate a rigorous construction of conformal diagrams. According to our ethics, a conformal diagram should include:

1. A diagram, where points are localised by a global cartesian coordinate system on the plane.

2. A division of the diagram into patches.

3. A metric $ds^2$ for each patch, expressed into some metric coordinates specific to the patch.

4. A coordinate-map for each patch, that translates the metric coordinates to the global cartesian coordinates.

In the literature, the coordinate map is missing most of the time. Our resulting diagram is more than simply 'conformal', as it retains more information than only the causal relations. For instance, a model of rigorous conformal diagram is provided in subsection 10.5.2. The diagram is given in figure 9, together with a global coordinate system $(U, V)$. It is divided into three patches I, II$a$ and II$b$, and the metric is given respectively by equations (357), (358) and (359). Finally, the coordinate-map in each patch is given by equations (360), (361) and (362).

# 11 BLACK HOLE EVAPORATION

In this chapter, we model an evaporating black hole.

## 11.1 HAWKING RADIATION

What is the ultimate fate of a black hole? Classical general relativity disregards all quantum effects and predicts that black holes live forever. By swallowing the rash travellers, they can only grow and never shrink. It is best expressed in the *black hole area theorem*, proved by Hawking in 1971 [97]. It states that, in any classical physical process, the area $A$ of a black hole can only grow with time,

$$\delta A \geqslant 0. \tag{367}$$

It thus appearead as a surprise when Hawking showed, in 1974, that black hole emit radiation [98]. Using QFT in curved space-time, he has proven that this radiation is thermal, like a black body whose temperature is

$$T = \frac{1}{8\pi M}, \tag{368}$$

with $M$ the mass of the black hole, in Planck units $\hbar = G = c = k_B = 1$ (see for instance [192] for a detailed derivation). From the perspective of an observer evolving along $\mathcal{I}^+$ (see figure 9), the evaporation means the detection of particles with positive energy (the Hawking quanta). Two equivalent heuristic pictures can give an idea of the origin of these particles:

1. The collapsed matter, inside the black hole, tunnels out, on the other side of the horizon.

2. The gravitational field 'stretches the vacuum', around the horizon, which creates a pair of particle/anti-particle, one with positive energy emitted outwards, the other with negative energy emitted inwards.

However, there are troubles with this heuristic picture. The typical length-scale $\lambda$ of a quanta, like a photon, is given by the inverse of its frequency, which is proportional to its energy, which is on average given by the temperature $T$, and so

$$\lambda \sim M. \tag{369}$$

We see that the size of a quanta is comparable to the size of the black hole itself, so that it is not really correct to imagine a quantum being created 'near the horizon'.

This question set aside, the total flux of energy observed along $\mathcal{I}^+$ can be estimated from Stefan's law, as

$$F(u) = \frac{\pi^2}{60} A T^4 \propto \frac{1}{M^2}. \tag{370}$$

The back-reaction of this flux on the geometry is hard to compute. However, as a first approximation, energy conservation suggests that the flux must make the black hole slowly to shrink, with a rate of mass loss

$$\frac{dM}{du} \propto -F(u), \tag{371}$$

as was first estimated by Page [138]. A decreasing mass means a decreasing area, which clearly contradicts the area theorem. However, as any theorem, it relies on assumptions, one of which being the null energy condition (see section 9.5). This condition means that the energy of massless matter is always positive and it is violated inside the black hole. Equation (371) is solved for

$$M(u) = M_0 \left(1 - \frac{u}{\tau_H}\right)^{\frac{1}{3}}. \tag{372}$$

where $\tau_H$ is the retarded time at which the black hole evaporates, assuming that it formed at $u = 0$ with an initial mass $M_0$. The mass-profile is depicted in figure 13. If only massless photons and gravitons

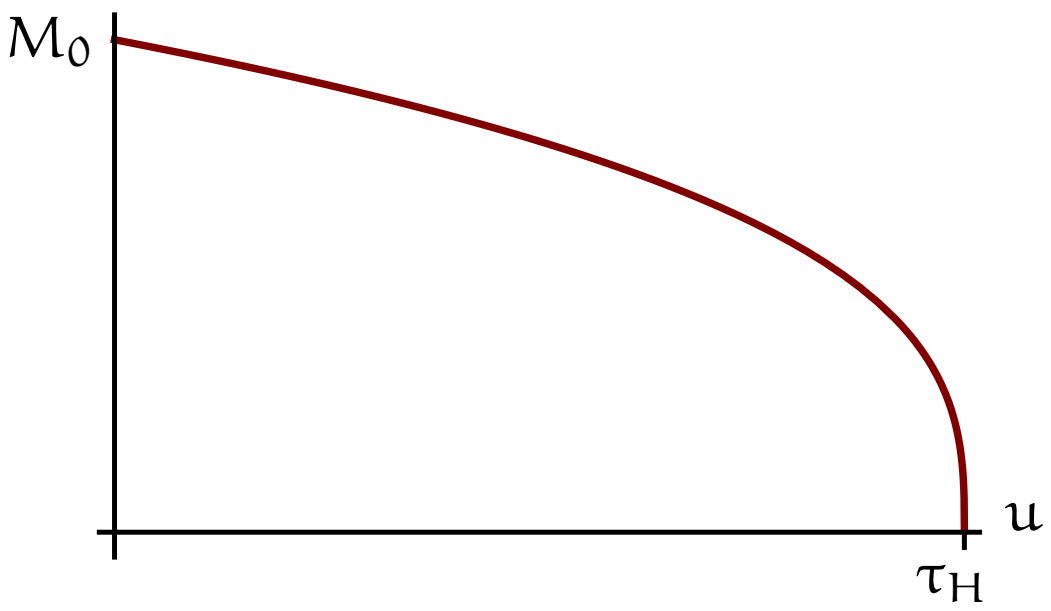

Figure 13: Mass profile along $\mathcal{I}^+$, given by equation (372).

are emitted, we can infer the characteristic time of evaporation [141],

$$\tau_H = 8895\, M_0^3 \approx 10^{74} \left(\frac{M_0}{M_\odot}\right)^3 \text{ s}, \tag{373}$$

where $M_\odot$ is the solar mass. This is long. Very long. The smallest observed black holes, which are also theoretically the smallest that can be formed by gravitational collapse, have an initial mass about $M_0 = 3M_\odot$. So, it would take $10^{58}\tau_U$ for them to evaporate, with $\tau_U$ the age of the universe. Conversely, a black hole evaporating in a time $\tau_U$ should have an initial mass $M = 10^{12}$kg, that is the mass of comet of a few kilometres wide, which could only be a primordial black hole, insofar as they might exist. The existence of primordial black holes would thus be the only hope to observe Hawking radiation. But the troubles are greater from a purely theoretical point of view, when $M$ approaches $0$, at the end of the evaporation process. This leads to the so-called *information-loss paradox*, discussed in chapter 12.

## 11.2 THE BACK-REACTION

The complete description of black hole evaporation requires full quantum gravity, but an approximation can be obtained with QFT on curved space-time, as was done in the original derivation by Hawking [98]. The gravitational degrees of freedom are described by a classical metric, over which evolve the quantum matter fields.

The Hawking quanta, created over a classical space-time, are expected to affect the metric in return: it is the *back-reaction*. If neglected, the Hawking quanta ($T_{ab} \neq 0$), generated over a vacuum solution of the Einstein equations ($G_{ab} = 0$), violate Einstein equations to first-order.

A first expected effect of the back-reaction is the decrease of the black hole mass, as described by equation (371). A refinement of the back-reaction problem is to consider that the classical gravitational field $g_{ab}$ is coupled to quantized matter fields, via the *semi-classical Einstein equations*

$$G_{ab}(g_{ab}) = \langle\psi|\hat{T}_{ab}(g_{ab})|\psi\rangle \,, \tag{374}$$

where $|\psi\rangle$ is a quantum state of matter, and $\hat{T}_{ab}$ is the quantized energy-momentum tensor. Equation (374) was first introduced in 1959 by Møller as a general tool for approaching quantum gravity [129]. An idea to solve it would be the following iterative method:

1. start from a classical background metric $g^0_{ab}$ and a given quantum state $|\psi\rangle$;
2. compute $\langle\psi|\hat{T}_{ab}(g^0_{ab})|\psi\rangle$ using QFT in curved space-time;
3. find $g^1_{ab}$ such that $G_{ab}(g^1_{ab}) = \langle\psi|\hat{T}_{ab}(g^0_{ab})|\psi\rangle$;
4. iterate the procedure to find $g^2_{ab}$;
5. go on until it converges to a self-consistent solution $g^\infty_{ab}$ satisfying equation (374).

## 11.3 VACUUM STATE

In 1976, Davies, Fulling and Unruh, have computed the expectation value $\langle\psi|\hat{T}_{ab}(g_{ab})|\psi\rangle$ for the metric $g_{ab}$ of a two-dimensional space-time $\mathcal{M}$, and a state $|\Psi\rangle$ corresponding to the vacuum of a mass-less scalar field $\phi$ [52].

In two dimensions, the metric can always be written in double-null coordinates as

$$ds^2 = C(\bar{u}, \bar{v}) d\bar{u} d\bar{v}. \tag{375}$$

Then, a classical mass-less scalar field $\phi$ over $\mathcal{M}$ is solution of the wave equation

$$\Box\phi = 0, \tag{376}$$

with $\Box \overset{\text{def}}{=} \nabla_a \nabla^a$ the covariant d'Alembert operator. This fields has to be quantised. To do so, we consider the family $(u_i)$ of solutions of equation (376) which are *stationary*, i. e. eigenvectors of the time derivative operator,

$$\partial_t u_i = -i\omega_i u_i. \tag{377}$$

The $(u_i)$ are called the *modes* and $\omega_i \in \mathbb{R}$ is the frequency of the mode $i$. Note that such a definition requires to choose a privileged time $t$, which unfortunately breaks the general covariance of GR. Depending on the sign of $\omega_i$, the family divides between solutions of positive ($\omega_i > 0$) and negative ($\omega_i < 0$) frequency. Then, any other solution of equation (376), can be decomposed as

$$\phi = \sum_i \left( a_i u_i + a_i^* u_i^* \right). \tag{378}$$

The quantization scheme promotes the coefficients $a_i$ to operators $\hat{a}_i$. The vacuum is then defined as the state $|0\rangle$ annihilated by $\hat{a}_i$,

$$\hat{a}_i |0\rangle = 0. \tag{379}$$

To sum-up, a choice of time-slicing $t$, induces a choice of positive frequency modes $u_i$, which selects a notion of vacuum state $|0\rangle$.

For instance, in flat space-time, the usual Minkowski time $t$, selects the positive frequency modes

$$\frac{1}{\sqrt{4\pi\omega}} e^{-i\omega t}. \tag{380}$$

In this case, it is remarkable that any time-slicing leads to the same notion of vacuum. In other words, the state $|0\rangle$ is invariant under the action of the Poincaré group. Such is not the case in curved space-time, unless there exists a time-like Killing vector field.

Assuming the asymptotic flatness of space-time, Davies, Fulling and Unruh choose the time-slicing of an observer evolving along null infinity. The positive frequency modes are chosen to be

$$\frac{1}{\sqrt{4\pi\omega}} e^{-i\omega\bar{u}} \quad \text{and} \quad \frac{1}{\sqrt{4\pi\omega}} e^{-i\omega\bar{v}}, \tag{381}$$

with $\bar{u}$ and $\bar{v}$ the coordinates used to write the metric as (375). Different choices for $\bar{u}$ and $\bar{v}$ are possible and they will define different notions of vacuum states. For instance, for the Schwarzschild metric, in region I of the Kruskal extension (see figure 7), we could choose alternatively

- $(u, v)$ the retarded and advanced time coordinates (eq. (347)), which are affine parameters respectively along $\mathcal{I}^+$ and $\mathcal{I}^-$, and the resulting vacuum state is called the *Boulware state* $|B\rangle$. Intuitively, it is a vacuum for observers along null infinities.
- $(U, V)$ the Kruskal coordinates (eq. (351)), which are affine parameters respectively along the white hole and the black hole horizons, and the resulting vacuum state is called the *Hartle-Hawking state* $|H\rangle$. Intuitively, it is a vacuum for observers along the horizons.
- $(U, v)$ a mixture of the two previous one, which defines the *Unruh state* $|U\rangle$.

Once a choice of null-coordinates $(\bar{u}, \bar{v})$ is made, it selects a unique vacuum state, generically denoted $|0\rangle$, which is finally used to compute:

$$\langle 0|\hat{T}_{ab}|0\rangle = \theta_{ab} + \frac{R}{48\pi} g_{ab}$$

$$\text{with} \quad \begin{cases} \theta_{\bar{u}\bar{u}} = -\dfrac{1}{12\pi}\dfrac{1}{\sqrt{C}}\dfrac{\partial^2 C^{-1/2}}{\partial \bar{u}^2} \\ \theta_{\bar{v}\bar{v}} = -\dfrac{1}{12\pi}\dfrac{1}{\sqrt{C}}\dfrac{\partial^2 C^{-1/2}}{\partial \bar{v}^2} \\ \theta_{\bar{u}\bar{v}} = \theta_{\bar{v}\bar{u}} = 0 \end{cases} \tag{382}$$

In the next section, we compute it explicitly in the case of a black hole formed by the collapse null shell of matter.

## 11.4 APPLICATION TO BLACK HOLES

Consider the model of black hole formation of subsection 10.5.2. A null thin shell of matter collapses to its center. Inside the shell (region I) the metric $g^0_{ab}$ is Minkowski (eq. (357)) and outside (region II) it is Schwarzschild (eq. (358) and (359)). We choose the following positive frequency modes:

$$\frac{1}{\sqrt{4\pi\omega}} e^{-i\omega v} \quad \text{and} \quad \frac{1}{\sqrt{4\pi\omega}} e^{-i\omega u_{in}} \tag{383}$$

with $v$ the advanced time coordinate along $\mathcal{I}^-$ (eq. (360) and (361)) and $u_{in}$ the retarded time coordinate inside the shell (simply denoted $u$ in eq. (360)). This choice determines the vacuum state $|in\rangle$. It corresponds to the (unique) Minkowski vacuum inside the shell.

Applying the formulas above, $\langle in|\hat{T}_{ab}(g^0_{ab})|in\rangle$, abbreviated $\langle T_{ab}\rangle$, is found to vanish everywhere in region I, so that the semi-classical Einstein equations (374) are satisfied in the Minkowski patch. In the

Schwarzschild patch, region II, the various components are given by [105]

$$\langle T_{uu} \rangle = \frac{\hbar}{24\pi}\left[-\frac{m}{r^3} + \frac{3m^2}{2r^4} + \frac{m}{r(u,v_0)^3} - \frac{3m^2}{2r(u,v_0)^4}\right] \tag{384}$$

$$\langle T_{vv} \rangle = \frac{\hbar}{24\pi}\left[-\frac{m}{r^3} + \frac{3m^2}{2r^4}\right] \tag{385}$$

$$\langle T_{uv} \rangle = -\frac{\hbar}{24\pi}\left(1 - \frac{2m}{r}\right)\frac{m}{r^3}. \tag{386}$$

Notice first that these formulae are valid both outside and inside the hole, although the coordinates $(u, v)$ map the two patches II$a$ and II$b$ in a different way (eq. (361) and (362)).

The interpretation of $\langle T_{ab} \rangle$ is subtle. It divides in two contributions

$$\langle T_{ab} \rangle = \langle B|T_{ab}|B\rangle + \langle in|: T_{ab} :|in\rangle \,. \tag{387}$$

The first term is a contribution of the Boulware state, that matches the vacuum along null infinities. It accounts for the vacuum polarisation. Heuristically, the gravitational field stretches the vacuum which generates particles. The vacuum polarisation is the analogue of the Schwinger effect for the electric field. This contribution exists independently of any gravitational collapse, and so, it does not account properly for Hawking radiation.

The Hawking flux contribution comes only from the second term, the normal ordered stress tensor, whose non-vanishing components are in the outgoing null direction [63]:

$$\begin{aligned} \langle in|: T_{uu} :|in\rangle &= \frac{\hbar}{24\pi}\left[\frac{m}{r(u,v_0)^3} - \frac{3m^2}{2r(u,v_0)^4}\right] \\ \langle in|: T_{vv} :|in\rangle &= 0 \\ \langle in|: T_{uv} :|in\rangle &= 0. \end{aligned} \tag{388}$$

Clearly $g^0_{ab}$ does not solve the semi-classical Einstein equations (374) in region II, since $G_{ab}(g^0_{ab}) = 0$, while $\langle in|\hat{T}_{ab}(g^0_{ab})|in\rangle \neq 0$. The idea of the iterative approach was to propose a corrected metric $g^1_{ab}$, that would ideally solve

$$G_{ab}(g^1_{ab}) = \langle in|\hat{T}_{ab}(g^0_{ab})|in\rangle. \tag{389}$$

Unfortunately, solving this equation seems to be already too hard. Therefore, Hiscock has suggested only to *guess* a metric $g^1_{ab}$, that would violate the semi-classical Einstein equations *less* than the original background $g^0_{ab}$. This has lead him to devise a model for an evaporating black hole that we now recall [106].

## 11.5 HISCOCK MODEL

How to guess a corrected metric? We can take inspiration from the value of $\langle T_{ab} \rangle$ in some regions. In our case two regions are noticeable.

First, along future null infinity, $\mathcal{I}^+$, the only non-vanishing component is

$$\langle T_{uu} \rangle = \frac{\hbar}{24\pi}\left[\frac{m}{r(u,v_0)^3} - \frac{3m^2}{2r(u,v_0)^4}\right]. \tag{390}$$

To understand intuitively what it means, suppose, in 2-dimensional Minkowski space ($ds^2 = -du dv$), that the same kind of stress-energy tensor is due to isolated particles, i.e.

$$T_{ab} = \rho\, u_a u_b \tag{391}$$

with $u^a$ the four-momentum and $\rho$ the energy density. Then, if $T_{uu}$ is the only non-vanishing component, it means that $u_a \propto (1, 0)$, in the $(\partial_u, \partial_v)$ basis, and so $u^a \propto (0, 1)$, which means that particles are going away along the $v$ direction. Since $\langle T_{uu} \rangle > 0$ on $\mathcal{I}^+$, we have the picture of particles of positive energy reaching $\mathcal{I}^+$ along null geodesics directed by $\partial_v$. These are the Hawking quanta: the black hole evaporates. At late times, when $u \to \infty$, the flux becomes

$$\langle T_{uu} \rangle \sim \frac{\hbar}{768\pi m^2}. \tag{392}$$

We thus recover equation (370).

Then, along the horizon, $r = 2m$, the only non-vanishing component is

$$\langle T_{vv} \rangle = -\frac{\hbar}{768\pi m^2}. \tag{393}$$

This time we can have the picture of particles of negative energy leaving from the horizon along null geodesics directed by $\partial_u$.

These two pictures motivate the model of Hiscock. It cleverly uses Vaidya-like metrics to represent the two fluxes of particles. It is made of five patches glued together as shown in figure 14, and the metric (the 'guessed' $g^1_{ab}$) is given by

$$\text{(I)} \left[\begin{array}{l} ds^2 = -du dv + r^2 d\Omega^2 \\ r = \frac{1}{2}(v-u) \end{array}\right. \tag{394}$$

$$\text{(II)} \left[\begin{array}{l} ds^2 = -\left(1-\frac{2m}{r}\right) du dv + r^2 d\Omega^2 \\ r = 2m\left(1 + W\left(e^{\frac{v-u}{4m}-1}\right)\right) \end{array}\right. \tag{395}$$

$$\text{(III)} \left[\ ds^2 = -\left(1-\frac{2N(v)}{r}\right) dv^2 + 2dvdr + r^2 d\Omega^2 \right. \tag{396}$$

$$\text{(IV)} \left[\ ds^2 = -\left(1-\frac{2M(u)}{r}\right) du^2 - 2dudr + r^2 d\Omega^2 \right. \tag{397}$$

$$\text{(V)} \left[\begin{array}{l} ds^2 = -du dv + r^2 d\Omega^2 \\ r = \frac{1}{2}(v-u) \end{array}\right. \tag{398}$$

The metric depends on the intial mass $m$ of the black hole. It also makes use of two functions $M(u)$ and $N(v)$, which represent how the mass decreases with the evaporation. Their value matches along the boundary III/IV, which marks the *apparent horizon*.

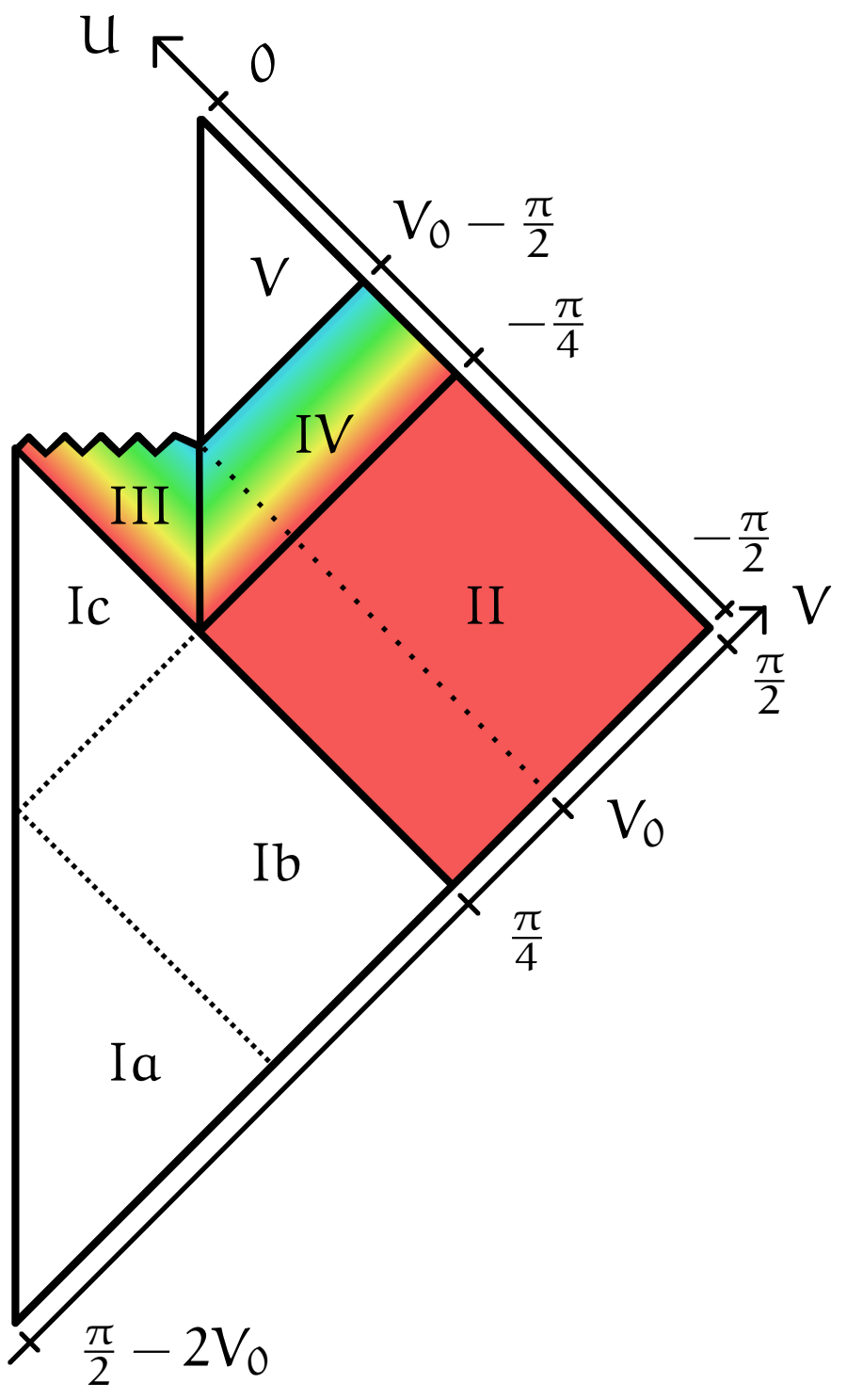

Figure 14: Conformal diagram of the Hiscock model. Everywhere the metric is locally Schwarzschild, characterised by the Misner-Sharp mass. Its value is represented by a color, from white (mass 0, i.e. Minkoswki) to red (initial mass $m$), passing through a gradient ($M(u)$ or $N(v)$). The mass profile along $\mathcal{I}^+$ is shown in figure 15. Region I has been divided in subregions in order to give an explicit expression to the map between the coordinates $(u, v)$ and $(U, V)$.

For completeness of the construction, we shall give the formulae that relates the coordinates $(U, V)$ of the conformal diagram to the coordinates in which the metric of each patch is written. This is not fully done in the original paper of Hiscock [106], but it is a necessary work to show that the conformal diagram of figure 14 correctly represents a consistent space-time model. It is given by the equations:

$$(\mathrm{Ia}) \left[ \begin{array}{l} u = -4m\left[1 + W\left(-\frac{\tan U}{e}\right)\right] \\ v = -4m\left[1 + W\left(-\frac{\tan(V + 2V_0 - \pi)}{e}\right)\right] \end{array} \right. \tag{399}$$

$$(\mathrm{Ib}) \left[ \begin{array}{l} u = -4m\left[1 + W\left(-\frac{\tan U}{e}\right)\right] \\ v = f_1(V) \text{ increasing, such that} \\ \left\{ \begin{array}{l} f_1(-2V_0 + 3\pi/4) = -4m(1 + W(1/e)) \\ f_1(\pi/4) = 0 \end{array} \right. \end{array} \right. \tag{400}$$

$$(\mathrm{Ic}) \left[ \begin{array}{l} u = c_1 + f_1(U - 2V_0 + \pi) \\ v = c_1 + f_1(V) \end{array} \right. \tag{401}$$

$$(\mathrm{II}) \left[ \begin{array}{l} u = -4m \log(-\tan U) \\ v = 4m \log \tan V \end{array} \right. \tag{402}$$

$$(\mathrm{III}) \left[ \begin{array}{l} v = f_2(V) \text{ increasing, such that} \\ \qquad\qquad f_2(\pi/4) = N^{-1}(M(0)) \\ r = g(U,V) \text{ such that} \\ \begin{cases} \frac{\partial g}{\partial V} = \frac{f_2'(V)}{2}\left(1 - \frac{2N(f_2(V))}{g(U,V)}\right) \\ g(U,\pi/4) = -\frac{1}{2} f_1(U - 2V_0 + \pi) \\ g(2V_0 - \pi/2 - V, V) = 0 \end{cases} \end{array} \right. \tag{403}$$

$$(\mathrm{IV}) \left[ \begin{array}{l} u = M^{-1}(N(f_2(U + \pi/2))) \\ r = h(U,V) \text{ such that} \\ \begin{cases} \frac{\partial h}{\partial U} = -\frac{u'(U)}{2}\left(1 - \frac{2M(u(U))}{h(U,V)}\right) \\ h(-\pi/4, V) = 2m\left(1 + W\left(\frac{\tan V}{e}\right)\right) \\ h(U, \pi/2) = \infty \\ h(U, U + \pi/2) = g(U, U + \pi/2) \end{cases} \end{array} \right. \tag{404}$$

$$(\mathrm{V}) \left[ \begin{array}{l} v = M^{-1}(N(f_2(V_0))) + 2h(V_0 - \pi/2, V) \\ u = M^{-1}(N(f_2(V_0))) + 2h(V_0 - \pi/2, U + \pi/2) \end{array} \right. \tag{405}$$

With these expressions, we can check the consistency of the space-time model, and notably the junction conditions, which match the metric along the boundaries of the patches. Moreover, the advanced time $v$ and the retarded time $u$ have been chosen to be both continuous along, respectively, $\mathcal{I}^-$ and $\mathcal{I}^+$.

The metric depends on the parameters $m$ (the initial mass) and $V_0$ (linked to the life-time of the black hole) and an arbitrary constant $c_1$. Besides, the function $f_1, f_2, g, h, M, N$ are not given explicitly:

- $f_1$ and $f_2$ are monotonically increasing functions satisfying the boundary conditions given in eq. (400) and eq. (403).

- $g$ and $h$ are fixed implicitly by the first order differential equations (403) and (404). These equations are obtained from the requirement that the lines of constant $V$ or $U$ are null. No explicit solution is known, except when $M$ and $N$ are constant or linear [195].

- $M$ and $N$ matches along the apparent horizon, between regions III and IV: it is the first equation of (404). Thus, one of them can be freely chosen, depending on the expected phenomenology for the evaporation rate.

A model for the mass profile is given by equation (372). Nevertheless, Hiscock has shown that this behaviour cannot hold until the end of the evaporation, as it would imply an infinite amount of total energy flux. Conversely, a finite total amount of energy flux on $\mathcal{I}^+$ implies that

$$\lim_{M \to 0} \frac{dM}{du} = 0. \tag{406}$$

Therefore, Hiscock proposes a mass profile shown in figure 15.

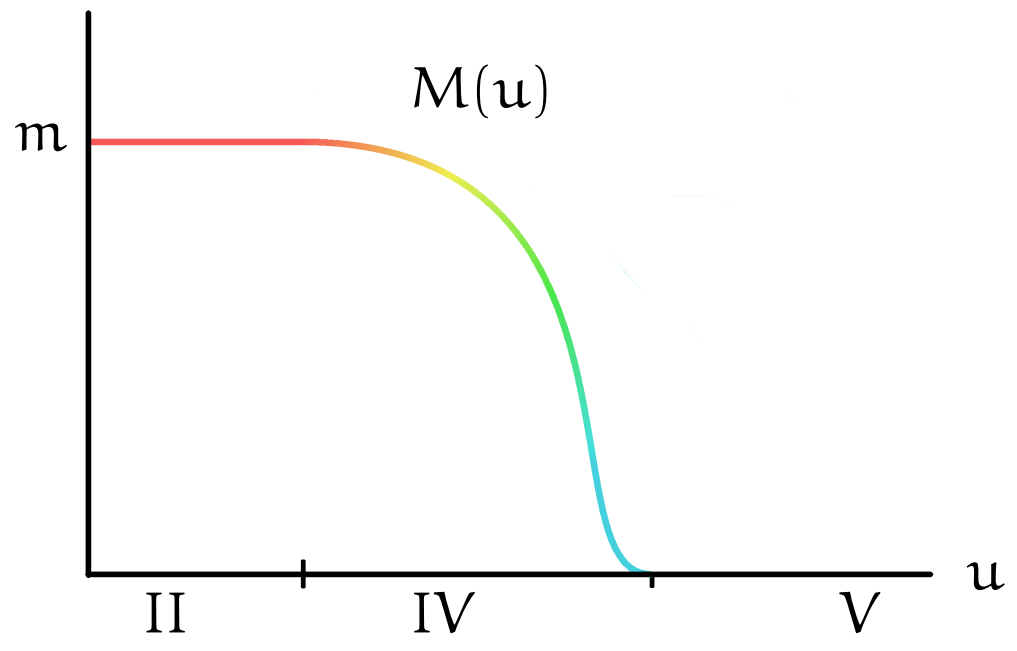

Figure 15: Bondi-Sachs mass function along $\mathcal{I}^+$ for the Hiscock model.

From the perspective of an outside observer, Hiscock model seems to describe correctly the phenomenology expected at the first stages of the evaporation. The decreasing Bondi-Sachs mass $M(u)$ along $\mathcal{I}^+$ corresponds to an outgoing positive energy flux, due to Hawking radiation. According to this model, the black hole evaporates completely and space-time turns to Minkowski back again.

# 12

# INFORMATION PARADOX

In classical general relativity, the mere existence of singularities means a breakdown of predictability: the laws of physics only hold on a smooth space-time background, so that anything could go out of a singularity. However, in quantum gravity, singularities are believed to be resolved, which could save the hope of physicists to predict the future. In quantum mechanics, predictability means unitary evolution of the wave-function. Yet, in 1976, Hawking argued that the unitary evolution is violated for a quantum black hole [100]. This time, the culprit is not the singularity, but the evaporation.

This unitary puzzle, or information(-loss) paradox, has made a big fuss in the theoretical physics community. The number of proposed solutions has been continuously growing, systematically avoiding any possible consensus. However, the situation may not be as dramatic as it seems. Indeed, we are only facing a *paradox*, and not an *antinomy*, meaning that the contradiction within the theory is only superficial, and not fundamental. It appears when the assumption about the semi-classicality of space-time is pushed way beyond its domain of validity. In this chapter, we explore various solutions to the issue, roughly following the historical path.

## 12.1 BLACK HOLE EXPLOSION

Already in his first paper of 1974, Hawking notices that the end of black hole evaporation is not an epiphenomenon. According to the formula (373), a black hole of mass $M = 10^5$ kg (the mass of a blue whale), that is as small as $10^{13}\, l_P$ (a billion times smaller than the nucleus of an atom) would take about $0.01$ s to evaporate. But in this context 'evaporation' is really a euphemism, as it will emit, in such a small amount of time, no less than $10^{23}$ J, that is the energy released by the fall of the asteroid that killed dinosaurs and formed the Chicxulub Crater in the Yucatán Peninsula. It is more of an 'explosion', which was indeed the title of the Hawking article of 1974, *'Black hole explosions?'* [98].

There is no paradox so far, but the suggestion that the evaporation continues until the simple disappearance of the black hole is already a strong assumption. Indeed, the formulae of black hole evaporation are obtained applying QFT in curved space-time, which is only valid as long as the quantum effects of particle creation can be treated as fluctuations over a classical background metric. In particular, when the black hole reaches Planck mass $m_P \approx 2 \times 10^{-8}$ kg, the semi-classical approximation breaks down. In other words, there may well be new physics happening at this scale, that would prevent the complete explosion of black holes.

## 12.2 INFORMATION IS LOST

It would be the matter of quantum gravity to tell what should happen, but in the absence of such a theory, Hawking pointed at the consequences of assuming the complete explosion of black holes. In some respect, it is the simplest assumption to make, naively extending known physics beyond its realm. In this case, the most striking consequence is the violation of the unitarity of time-evolution.

Before the black hole even forms, the state of the fields on a Cauchy slice $\Sigma_0$ is a pure density matrix $\rho_0$, meaning it can written as $\rho_0 = |\psi_0\rangle\langle\psi_0|$. Then, the black hole form and evaporation starts. The purity of the total state is preserved by unitary evolution, but for an observer at infinity, part of the state is hidden by the horizon (particles fall to the singularity), so that the state at infinity is mixed. Wald has shown that the evaporation is independent in each mode of the radiation [192] and the emitted state in each mode is thermal. Finally, an observer at late time along $\mathcal{I}^+$ receives radiation in a state

$$\rho = \bigotimes_i \sum_N e^{-N\frac{\hbar\omega_i}{k_B T}} |\psi_{i,N}\rangle\langle\psi_{i,N}|, \qquad (407)$$

where the tensor product is made over the modes $i$, the sum over the number of particles $N$, and $\psi_{i,N}$ is the $N$ particle state of the $i^{th}$ mode. After the black hole has evaporated, this is the only part of the radiation that remains, so the final state is mixed and the total evolution is non-unitary.

This conclusion is puzzling as it contradicts the principle of unitary evolution in quantum mechanics. It can be understood as a loss of information, as the information, carried in the correlations between the infalling and the outgoing particles, is destroyed in the explosion. To Hawking, this observation suggested that quantum mechanics had to be slightly modified to admit such a possibility. The conclusion has set alight one of the most vivid debate of theoretical physics in the last 45 years. This fact itself is puzzling. Indeed, the idea that information is just lost, in the special case of an ultimate blackout, may not be as radical as it seems. This is at least the opinion supported notably by Unruh and Wald [186].

## 12.3 INFORMATION IS TRAPPED

In 1987, Aharonov, Casher and Nussinov explored another hypothesis [1]. In their scenario, a black hole would certainly explode, but a stable core of planckian mass would remain: the remnant (originally dubbed 'planckon').

Such a remnant could a priori decay into particles whose wave-function would be correlated with the previously emitted quanta, so that in the end, unitarity would be restored. But this possibility is discarded by the authors based on the following argument. It is easy to compute that a spherical black-body of area $A$ and temperature $T$ emits an average number of quanta

$$n \sim AT^3 \tag{408}$$

in Planck units, per unit of time. For a black hole, this is

$$n \sim \frac{1}{M}. \tag{409}$$

Integrating over its whole lifetime, it gives the total number of emitted quanta

$$N \sim M_0^2, \tag{410}$$

with $M_0$ the initial mass of the black hole. To recover the information, a planckian remnant would have to decay in as many quanta. Besides, each of them would have a mass $m \sim \frac{1}{N}$, and so a typical wave-length

$$\lambda \sim N. \tag{411}$$

This is much larger than the Compton wavelength of the remnant, of order 1. The transition amplitude from one remnant to one such quantum should be roughly proportional to the overlap $f$ between the two wave-functions, that is, in 3D,

$$f \sim \frac{1}{N^3}. \tag{412}$$

The rate of emission of $N$ quanta is thus

$$f^N \sim M_0^{-\alpha M_0^2}, \tag{413}$$

with some $\alpha > 0$. This rate goes to zero very quickly when $M_0$ is big enough. So, such a transition is very unlikely to happen, or only after a very long time. As sketchy as the argument may seem (the argument a bit more formal in their paper), it suggests that remnants could be stable. For this reason, they are often called *long-lived remnants*. Information is trapped inside, but the principle of unitarity is preserved. By estimating their very low cross-section with protons and nuclei, the authors even suggested that they may be good candidates for cold dark matter.

The previous argument can be reframed to evaluate the typical lifetime of remnants [151]. There are $M_0^2$ quanta to be released. The total energy is of order 1, and so each quantum carries an energy $1/M_0^2$. This corresponds to a wavelength of $M_0^2$. At each time, the probability to emit a quantum is given by the overlap of wave-functions, which is $1/M_0^2$. Thus the typical time of emission (inverse of the probability) is $M_0^2$. Since all the quanta together are correlated with the early-time radiation, they cannot be correlated between themselves, so that they are emitted independently, one at a time. All in all, it takes a total time

$$\tau_{\text{remnant}} \sim M_0^4. \tag{414}$$

The standard objection against the remnant scenario is the infinite pair production rate [79]. Indeed, remnants can be regarded as particles. For an outside observer, they are characterised only by the mass, the charge, and the spin (no-hair theorem). But they contain a lot of information, that is many internal degrees of freedom, and so many possible quantum states. In fact, considering all the possible scenario ending in a planckian remnant, their state is infinitely degenerated. From a thermodynamical perspective, this means that their entropy is infinite, and so they should be infinitely favoured... in other words, they should pop up everywhere around us! Yet, this argument is not very compelling. It is not because a state is thermodynamically favourable, that it is necessarily realised. For instance, the entropy of a book is much higher in the ashes-state that in the hardcover-state. Yet, most libraries do not burn, until something sets fire. Most of the time, systems are stuck in meta-stable states, and the most thermodynamically favourable states are only realised after a very long time.

In 1992, Callan-Giddings-Harvey-Strominger (CGHS) built a much-discussed toy-model of an evaporating black hole in two dimensions [42]. The classical equations of motion can be solved exactly and the semi-classical one numerically. Based on this model, Banks, Dabholkar, Douglas and O'Loughlin revived the remnant scenario [13]. Their remnants are dubbed 'horned particles' or 'cornucopions'. They are characterised by a very large interior geometry, the 'horn', which can store an infinite amount of quantum states, very difficult to excite with an external probe (see figure 16). The scenario was a nice reply to the infinite pair production rate argument. But it was again counter-attacked in [79] and [177] but we will not enter theses details.

## 12.4 INFORMATION COMES OUT

Let's turn to the third main option, available in the market at the beginning of the 1990s, to solve the information paradox. In a nutshell, the information is carried out within the radiation itself. In other words, the outgoing radiation would not be completely random, but would carry some correlations, and thus carry some information out. At first, this proposition is not possible since Hawking radiation is thermal: no correlations are expected between the emitted quanta.

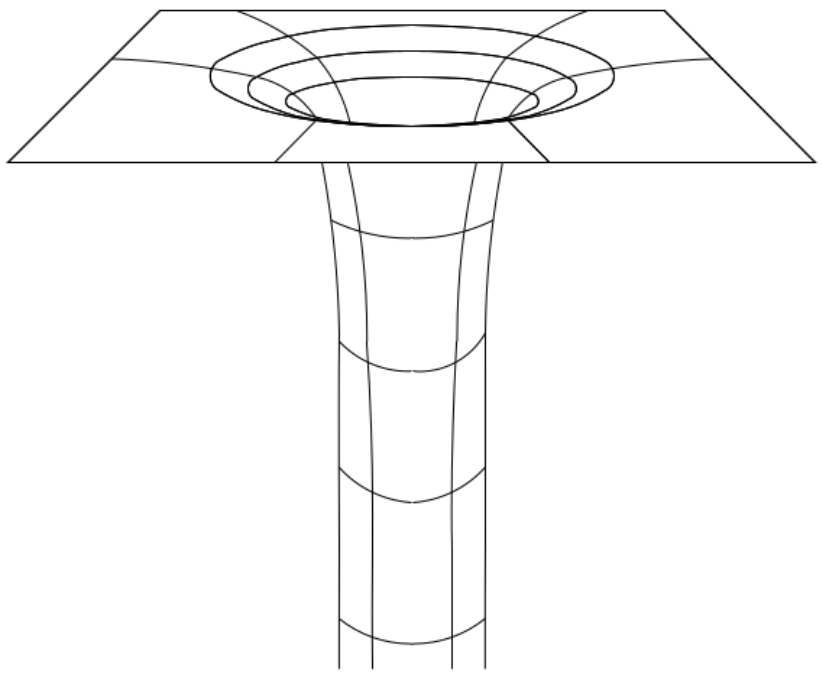

Figure 16: Angular slice of the geometry of a cornucopion. From [14].

However, Hawking computation is only an approximation and one could think that higher-order computation would introduce corrections allowing for such correlations.

This possibility was first proposed by Page in 1980, presented as the most conservative solution [139]. In 1992, a counter-argument was raised by Giddings and Nelson in [81], based on the CGHS model [42]. The toy-model enables to estimate the corrections induced by the back-reaction, and they were seen to be too small to restore the information.

The year after, in 1993, Page raised a strong objection against their argument, which has inadvertently revived the paradox in a more dramatic way [140]. Page intended to show that the computation of Giddings and Nelson was not conclusive. To do so, he estimates the amount of information that the radiation can carry in a typical state.

In a simple idealisation, the whole universe is made only of a black hole and everything else outside, assumed to be only outgoing Hawking quanta. The total Hilbert space is

$$\mathcal{H} = \mathcal{H}_b \otimes \mathcal{H}_r. \tag{415}$$

We denote $b$ (for black hole) and $r$ (for radiation) respectively the dimensions of $\mathcal{H}_b$ and $\mathcal{H}_r$. The overall system is assumed to be isolated and to evolve unitarily, so that the dimension of $\mathcal{H}$ is constant, equal to $b \times r$. However, the dimension of $\mathcal{H}_b$ decreases as the black hole shrinks, while the dimension of $\mathcal{H}_r$ increases as radiation is created. $b$ and $r$ are related to the entropy of the black hole $S_b$ and the entropy of the radiation $S_r$, by

$$b = e^{S_b} \quad \text{and} \quad r = e^{S_r}. \tag{416}$$

The entropy of the black hole is assumed to be given by the Bekenstein-Hawking entropy

$$S_b = \frac{A}{4}. \tag{417}$$

The Hilbert space of the radiation is a tensor product of the Hilbert spaces $\mathcal{H}$ of all the N emitted quanta

$$\mathcal{H}_r = \underbrace{\mathcal{H} \otimes ... \otimes \mathcal{H}}_{N \text{ times}}. \tag{418}$$

Denoting d the dimension of $\mathcal{H}$, the entropy of the radiation is therefore

$$S_r = N \log d. \tag{419}$$

This entropy is a *thermodynamic entropy*, in the sense that is gives a direction to time, as the number of emitted quanta N only increases with time.

The information involved in the paradox is carried in the correlations between the outgoing radiation and the black hole. Such correlations are measured by the entanglement entropy S between the two subsystems. If the total state is $\rho$, S is defined by

$$S \stackrel{\text{def}}{=} -\operatorname{Tr}\left[(\operatorname{Tr}_b \rho) \log (\operatorname{Tr}_b \rho)\right] = -\operatorname{Tr}\left[(\operatorname{Tr}_r \rho) \log (\operatorname{Tr}_r \rho)\right]. \tag{420}$$

The information getting out with the radiation is defined as the deviation of the entanglement entropy from its maximum, which is $S_r$, so

$$I = S_r - S. \tag{421}$$

Initially, when no radiation has yet been emitted, $S = 0$ and $I = 0$. In the investigated scenario, information has entirely come out by the time the black hole has disappeared, so that we also have $S = 0$ at the end, but $I = \log N_{tot}$, with $N_{tot}$ the total number of emitted quanta. In between S shall increase and decrease. Assuming that the overall system is in a typical state, Page shows that, as long as $r \leqslant b$, S increases as

$$S \sim \log r - \frac{r}{2b}. \tag{422}$$

In this first phase, the creation of pairs increases the correlations between the two subsystems. Then, it reaches a maximum for $r = b$, and when $r \geqslant b$, S decreases as

$$S \sim \log b - \frac{b}{2r}. \tag{423}$$

This second phase is the phase of purification: information is restored. During the two phases, the released information I increases, but the rate of release is very different in the two phases. The total behaviour is illustrated in figure 17.

From formulae above, it is not too hard to show that the initial rate of information outflow goes as

$$\frac{dI}{du} \sim e^{-4\pi M_0^2}, \tag{424}$$

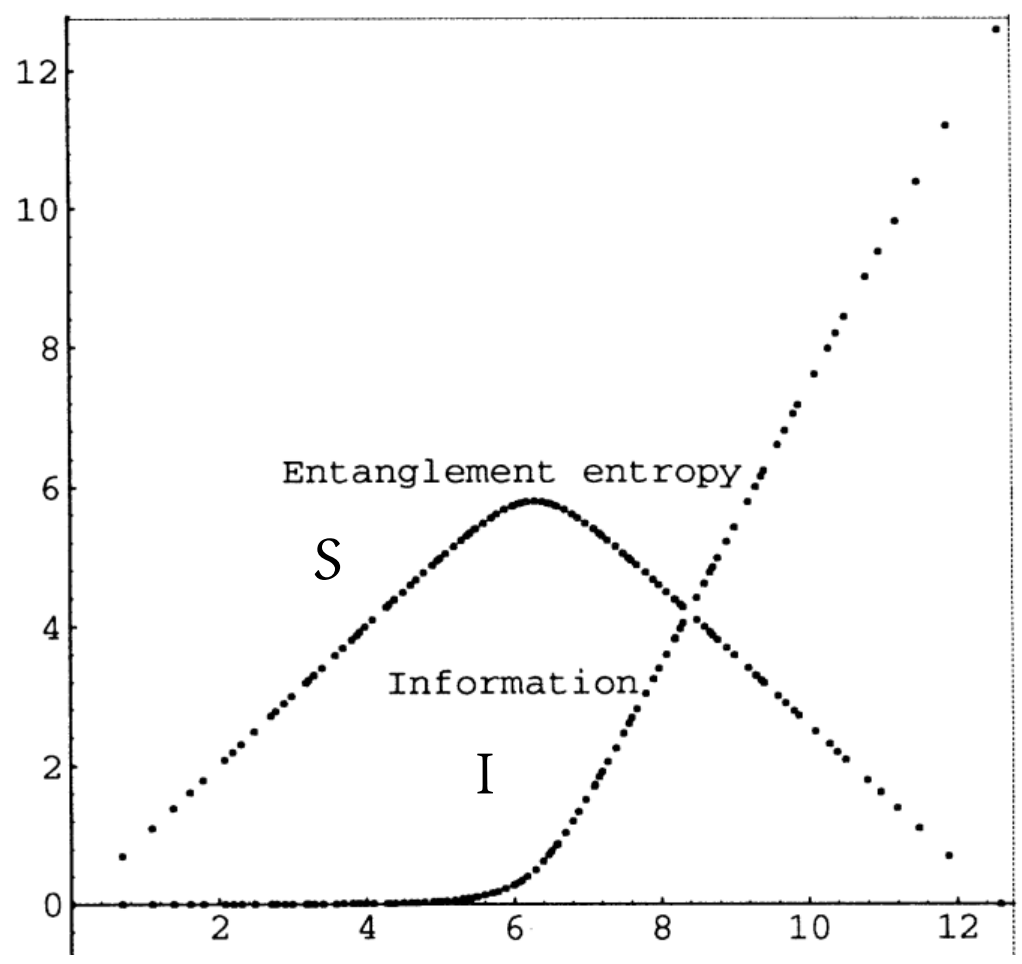

Figure 17: The entanglement entropy S and the information I as functions of $S_r$ (proportional to the number of particles emitted). Taken from [140].

with $u$ the retarded time. The RHS is not an analytic function when $M_0 \to \infty$, so that there is no chance to see it via a perturbative expansion at any order, contrary to what Giddings and Nelson had assumed in their analysis. So their conclusion, that information does not come out, is not proven.

Meanwhile, Page computation has proven an unexpected result, which he does not seem to have noticed in his first paper. What is remarkable is that purification starts much before the end of the evaporation. Indeed, it really starts when S reaches its maximum, that is when $r = b$. This happens at a time $t_*$, called *Page time*, and satisfying

$$S_r(t_*) = S_b(t_*). \qquad (425)$$

On figure 17, we see that it happens when half of the particles have been emitted. In [141], it is estimated about

$$t_* \approx 0.53\tau_H. \qquad (426)$$

At this stage, the rate of information outflow is not negligible anymore, and it is only half-way towards the end. This seems contradictory with the fact that the black hole is still macroscopic and Hawking computation is expected to work very well, and thus to show thermal radiation, carrying no information! Thus, Page has actually traded a paradox for another one: something weird shall happen at Page time. This one is more dramatic has it appears in a regime where nothing special was expected.

## 12.5 FIREWALL

During some time, the dominant view that has prevailed, especially in the string community, is *complementarity* [176, 178]. In a nutshell, it holds together the statements that the evolution is unitary and that the information cannot escape. Then, it claims that the paradox is only apparent, as no inconsistency will ever be observed between an infalling and an external observer, as their results are incomparable. This view is a remake of Bohr's complementarity, developed in the context of the interpretation of quantum mechanics. He had advocated the view that the apparent contradictions of quantum mechanics are themselves an essential feature of reality. They should be accepted as such, noticing that, from the point of view of a definite observer, no contradiction is ever observed. It is a minimalist way to deal with a paradox and one may be disappointed by the fact that it does not solve the paradox, but rather accept it as such.

But the black hole information paradox was shaken up again in 2012, when Almheiri, Marolf, Polchinski and Sully proposed a new argument, in their paper '*Black holes: complementarity or firewalls?*' [2]. The firewall argument is a no-go theorem: if locality, causality, unitarity, low-energy effective field theory and the equivalence principle are assumed, then we run into a contradiction so that one of the previous hypothesis has to be rejected. The authors suggest discarding the equivalence principle so that physics at the horizon would be very different from anything we know. Instead of a smooth and unnoticeable transition from the outside to the interior, the stress-energy tensor would actually diverge along the horizon, so that an infalling observer would literally burn. This divergence is a way to bypass a version of the monogamy theorem. This theorem says that the early outgoing radiation cannot be both entangled with the late outgoing radiation and the infalling radiation unless some high-energy 'firewall' hides one from another. However, the violation of the equivalence principle requires an extreme re-assessment of well-established physical principles. Compared to Hawking and Page paradoxes, the firewall seems to bring it a higher degree in the intensity of the contradiction. Unsurprisingly, the physics community has reacted strongly, and the discussions were revived once again[1].

In my opinion, the progressive scaling in the intensity of the contradiction, from Hawking's first formulation to the firewall, is artificial: if the conclusion seems stronger, the assumptions are stronger too. In other words, it is always possible to make a paradoxical situation look more paradoxical by adding some extra assumptions that will reinforce the unease.

The strong assumption of Page argument is that the entropy of the black hole is given by the Bekenstein-Hawking entropy $S_{BH}$. We have reasons to believe that this entropy only accounts for the degrees of freedom on the horizon and ignores the degrees of freedom

1 The list of alternative proposals is long and we cannot discuss them here. Let us just mention, among others, the fuzzballs (black hole never really forms) [126] or the nonviolent information transfer from black holes [80].

inside. It is at least what is found when the computation is done in LQG [155]. The horizon's area bounds the number of states that are distinguishable from the exterior during a time scale of the order of the black hole lifetime, but it does not bound the number of internal quantum states of the black hole, distinguishable by local quantum field observables inside the hole [159]. Thus it is not a surprise that $S_{BH}$ is proportional to the area. Discarding this assumption repudiates Page paradox and the firewall argument at the same time. In his huge interior, a black hole can store many more bits that those given by $S_{BH}$. The time-evolution of the interior volume of an evaporating black hole has been computed in [46]. For a black hole formed by the collapse of a null shell at the advanced time $v = 0$, the volume evolves as

$$V(v) \approx 3\sqrt{3}M_0^2 v \left(1 - \frac{3M_0}{2\tau_H}\right). \tag{427}$$

Thus, as time goes by, the interior volume increases while the area decreases. This fact is one more argument in favor of the remnant scenario. In chapter 15, we will see that the black-to-white transition, in its evaporating form, revives remnants in an original way.

## 12.6 INFORMATION IS DEGRADED

Ultimately, only a complete theory of quantum gravity could adjudicate the issue, give a definite prediction of the fate of black holes, and tell us how the 'information-loss paradox' is actually solved in nature.

As we will see in chapter 14, the black-to-white hole scenario, motivated by some partial results of quantum gravity, solves it in a very natural way, by resolving the singularity and extending it in the future.

Before going to that, we would like to mention still another seducing possibility: information is not lost, but degraded. It is a quite conservative answer to the unitary puzzle. As was said before, Hawking's argument is not an entirely solid result, as it relies on the strong assumption that the semi-classical computation of QFT in curved space-time is still valid when the black hole is planckian. This is very doubtful, as the strong coupling of matter to the gravitational field may change dramatically the scenario. Motivated by some results of LQG, Perez argues that the correlations between the matter degrees of freedom could be transmitted to the gravitational degrees of freedom at planckian scale [3, 148]. The picture is very similar to what happens when a book is burnt. Noone doubts that the evolution between the book and the ashes is globally unitary, although no experimentalist can read the book in the ashes, because the information is disseminated in the correlations between the many photons emitted during the burning. A macroscopic observer has only a coarse-grained pic-ture of the world so that the information is made inaccessible as it leaks to microscopic degrees of freedom. In the case of black holes, the information would ultimately lie in the correlations between the planckian gravitational degrees of freedom. Information is effectively lost, but unitarity is fundamentally preserved.

# 13 WHITE HOLES

A white hole is the time-reverse of a black hole. Everything that we know about black holes can be translated for white holes under the time transformation $t \to -t$. So similarly to what we have seen in section 9.6 for the case of black holes, there are two ways to formally define a white hole:

1. A region of space-time that does not contain the causal future of past null infinity $\mathcal{I}^-$.
2. An anti-trapped region: everywhere the expansion $\theta$ is positive in both ingoing and outgoing directions.

The discovery of white holes, as theoretically allowed objects, was within easy reach as soon as the Schwarzschild solution was found and Einstein equation were recognised to be time-reversal symmetric. As far as we know, it is not before the discovery of the maximal Schwarzschild extension by Kruskal in 1960 that the white hole patch is explicitly given [115]. The white hole corresponds to region IV in figure 18. Also, the previously constructed conformal diagrams for black hole formation can be inverted. Thus, figure 19 shows the explosion of a white hole. A thin null shell of matter is expelled out. Heuristically, a white hole describes a region of space-time from which you can only be kicked out. This is really the opposite of a black hole, understood as a region where once you are in, your fall is inescapable. Both have long belonged to the theoretical freak show, i. e. perfectly allowed solutions of equations, but considered unphysical for some reason. But while black holes have now won many stripes of respectability, white holes have so far remained undesired guests.

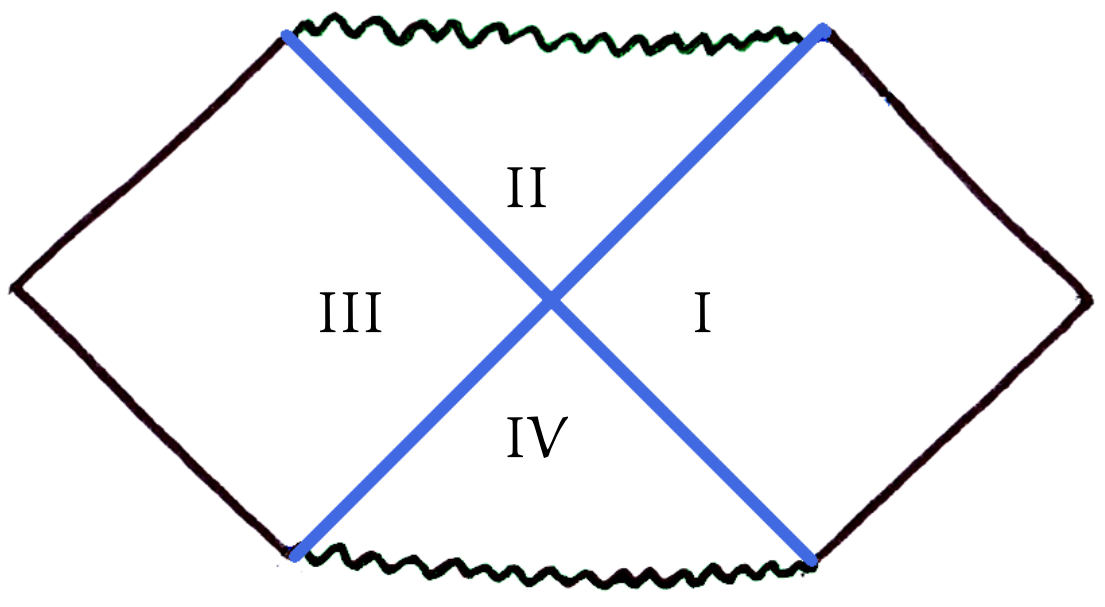

Figure 18: Conformal diagram of the Kruskal extension.

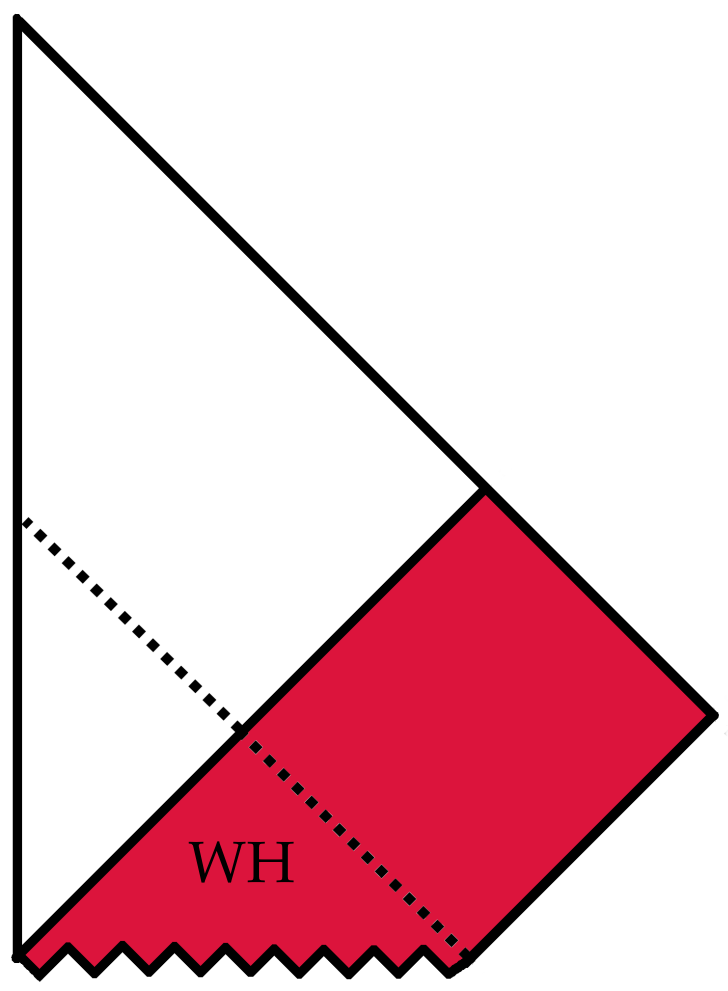

Figure 19: Explosion of a white hole.

## 13.1 WHITE ANALOGUES

Despite the suspicion that they stir up, white holes are closer than it may seem. Whenever you open the tap in your kitchen, you are likely to see the analogue of a white hole down there, at the bottom of the sink.

Analogue gravity is a discipline that models gravitational phenomena using physical systems other than gravity, typically hydrodynamics. It has been demonstrated that a good analogue of white holes is provided by the inner region of a hydraulic jump [110].

A hydraulic jump is that circle that is seen at the bottom of a sink when water is flowing down (see figure 20). It is a typical example of an everyday-life phenomenon, which is very easy to observe, but horrendously hard to explain.

The velocity $v$ of waves propagating on a shallow-water layer of fluid can be estimated, by dimensional analysis, to be

$$v = \sqrt{g\,h} \tag{428}$$

with $h$ the height of the fluid, and $g$ the gravitational acceleration. But in our case, the water itself is moving. In the interior region of the jump, the flow is super-critical, meaning water goes faster than $v$, while it is sub-critical outside. $v$ is the analogue of the speed of light $c$. Inside the white hole, 'space goes faster than light'.

When a wave is created inside the hydraulic jump, with the tip of a finger, it is rapidly expelled out. When it is created outside, even close to the jump, the wave does not enter the circle. The formal analogy has been proven in [190]. However one should keep in mind that the analogy is strict only for the interior region. Indeed in hydrodynamics, there is a real singularity in the model along the horizon, while the metric is actually continuous across the white hole horizon. So, outside the jump, the analogy is not sharp. For instance, white holes are attractive whereas the hydrodynamic pictures may give the impression that they are repulsive since water flows out.

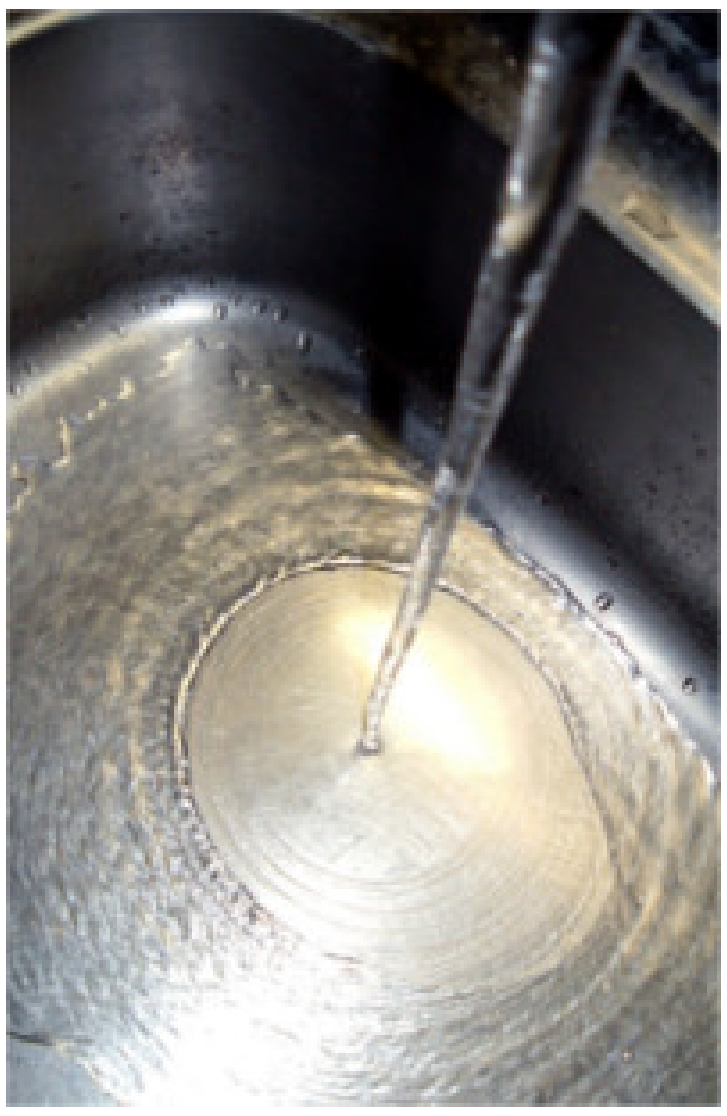

Figure 20: Hydraulic jump at the bottom of a sink.

## 13.2 lagging-cores

In 1964, white holes were first proposed as potentially real observable objects by Novikov [134]. He presented them as a hypothesis to explain quasi-stellar radio-sources, aka *quasars*, which are point-like sources in distant galaxies emitting a large amount of energy at radiofrequency. Such signals were first seen in the 1950s, but their origin remained unknown. It is now admitted that quasars are nothing but supermassive black holes. Their huge luminosity is coming from the extreme heating of their accretion disk. But at the time, Novikov explored the hypothesis of white holes.

In his paper (originally in russian), he does not use the term 'white hole', but rather 'delayed core' or 'lagging-core'. His scenario is that of an expanding Friedmann universe, where the expansion of some local regions would have been delayed. These white holes appear as inhomogeneities of the early expanding universe. Novikov also imagines a previous phase of contraction, very similar in spirit, to the big bounce studied nowadays. He suggests that the delay of the expansion of some regions would originate symmetrically from an anticipated contraction of the same regions, before the big bounce! Similar ideas were suggested in parallel by Ne'eman on the other side of the Iron Curtain [133].

The idea of white holes, as inhomogeneities of the initial big-bang, was killed ten years after it emerged, by the proof of their instabilities, as we see in the next section.

## 13.3 INSTABILITY

It is amusing to notice that they have actually been killed twice the same month of June 1974, in two different ways. The first instability results from the interaction of the white hole with the surrounding matter, while the second deals with the quantum creation of particles in the gravitational field of the hole.

### 13.3.1 *Classical instability*

The first to notice the instability of white holes is Eardley in 1974 [57]. His paper is short and a bit obscure, but it was latter refined in [21, 39, 117]. A modern discussion is found in [16]. We sketch the argument below. Consider a white hole that explodes in a null

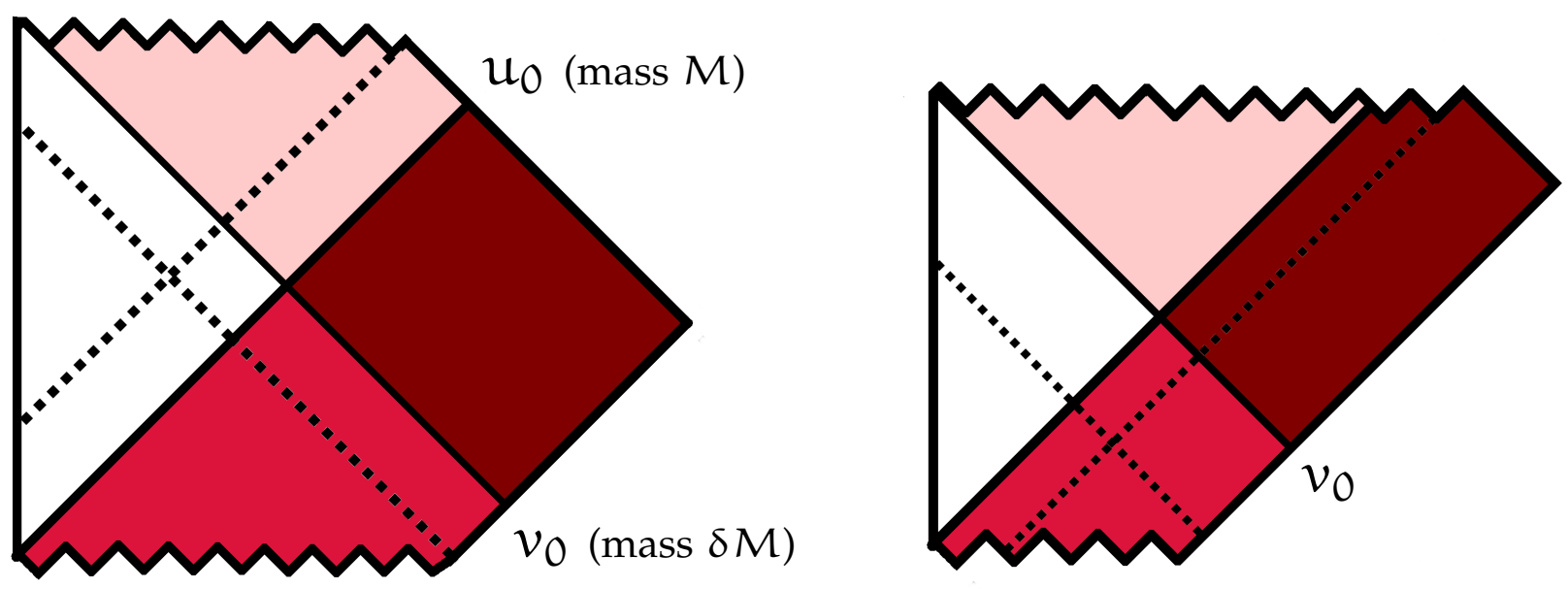

(a) A small mass $\delta M$ falls in the vicinity of a white hole.

(b) If sufficiently close to the white hole, the black hole forms before the white can explode.

Figure 21: Illustration of the classical instability of white holes.

shell of mass $M$ at a retarded time $u = u_0$, as depicted in figure 21a. Imagine an ingoing null shell of mass $\delta M \ll M$ falling towards the center, along a geodesic of equation $v = v_0$. It will necessary cross the exploding shell. The point of intersection is a sphere of radius $r_0$ that contains a total mass $M + \delta M$. If the radius $r_0$ is smaller than the Schwarzschild radius $2(M + \delta M)$, as shown in figure 21b, then matter collapses back into a black-hole and both shells end up on the singularity. The important observation is that however small $\delta M$ may be, there will always be a sufficiently old advanced time $v_0$ such that the infalling shell will intersect the outgoing one at a radius $r_0$ lower than the Schwarzschild radius $2(M + \delta M)$. Formally:

$$\forall \delta M, \quad 0 < \delta M \ll M, \quad \exists v_0, \quad r_0 < 2(M + \delta M). \tag{429}$$

A very small dust forgotten in the very far past provokes the catastrophic re-collapse of the white hole! In short, white holes are unstable.

### 13.3.2 *Quantum instability*

The second kind of instability results from computations of QFT in curved space-time, done by Zel'dovich, Novikov and Starobinskii in [199]. A simplified version of their argument consists in computing the first-order correction to the stress-energy tensor,

$$\left\langle \hat{T}_{ab}(g^0_{ab}) \right\rangle . \qquad (430)$$

When this number is computed along the horizon of the white hole, using the formulae of section 11.3, it is found to be infinite! It means that, even to first-order, the spontaneous creation of particles blows-up along the white hole horizon!

The quantum instability is actually very similar to the classical one. In both cases, it results from the blowing-up of a perturbation. The difference is the nature of the perturbation, whether it is classical or quantum. The blowing-up is related to the blue-shift divergence along the horizon, which was first observed by Szekeres in 1973 [180].

### 13.3.3 *No future?*

After these proofs of instability, the idea to explain quasars as white holes was definitively ruled out. If they ever existed, white holes have already exploded or turned back into black holes. Nevertheless, some white holes still escape the arguments above. They can exist, but with a very short life-time. Or if they are small enough, typically planckian, there may not exist physical perturbations of smaller size to disrupt them. We will explore such a possibility in chapter 15.

For black holes, it is usual to make a distinction between the astrophysical and the primordial ones. The former are formed by the collapse of stars, while the latter were born with the universe, as initial defects of space-time. Similarly, the white holes of Novikov are primordial. But a priori, the existence of astrophysical white holes is very dubious, as a gravitational collapse cannot create anti-trapped regions. However, in chapter 14, we discuss the black-to-white hole scenario, which revives white holes, this time with a potentially astrophysical origin.

# 14 BLACK-TO-WHITE HOLE

The core of this thesis relies on the black-to-white hole transition. The sketch of the scenario is simple: a black hole turns into a white hole. Such a scenario is of course not allowed by the classical theory. Indeed, in General Relativity (GR), black hole singularities have no (predictable) future. But the quantum genie always brings its share of surprises! In chapter 11, we have already seen that black hole actually evaporate. We now explore another possible quantum effect: the quantum tunnelling of black holes!

## 14.1 VARIATIONS ON A THEME

We start with a historical review of several scenarios found in the literature.

### 14.1.1 *Frolov and Vilkovisky*

As early as we can say, the first germs of the black-to-white hole scenario are found in an article by Frolov and Vilkovisky, *Quantum Gravity Removes Classical Singularities and Shortens the Life of Black Holes*, presented at the 2$^{\mathrm{nd}}$ Marcel Grossmann Meeting in Trieste, in 1979 [71].

Their goal is to understand the fate of singularities in quantum gravity. In absence of any complete theory to deal with such an issue, they suggest a rather model-independent approach which consists in using an *effective lagrangian*:

$$\frac{\mathcal{L}}{\sqrt{-g}} \overset{\text{def}}{=} R - \big[A R_{ab} (\log\Box) R^{ab} + B R (\log\Box) R + C R^{*}_{abcd} (\log\Box) R^{*abcd}\big] + \mathcal{O}(\nabla^4) + \mathcal{O}(\hbar^2). \quad (431)$$

with $A, B, C$ constants depending on the number of species of particles, $R^{*}_{abcd}$ the dual Riemann tensor and $\mathcal{O}(\nabla^4)$ contains the weaker terms with four derivatives and more. This lagrangian is the first-order correction to the Einstein-Hilbert one and it includes notably both the vacuum polarisation and the back-reaction of the particle creation. This modified lagrangian provides an effective dynamics to study the fate of gravitational collapse.

In the case of the spherically-symmetric collapse of a null shell of matter, they compute that no singularity ever forms, but instead, when the shell reaches $r = 0$, it is driven back to expand again. The overall scenario is summarised in figure 22. An apparent horizon surrounds a bounded region, which has the metric of a black hole below $t = 0$ and a white hole above.

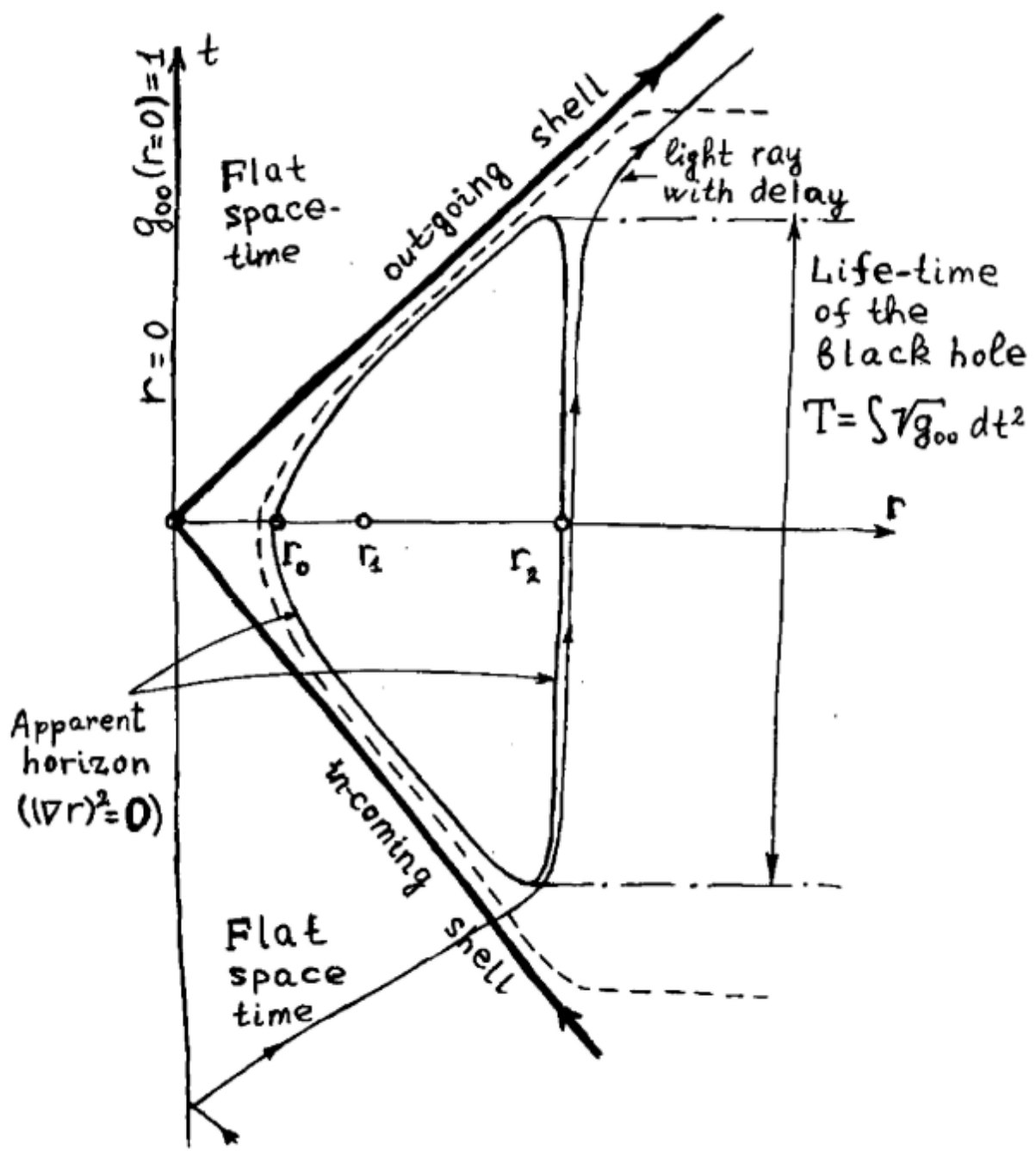

Figure 22: Model of Frolov-Vilkovisky. Taken from [71]. The approximations of their computation are only valid to the left of the dashed line.

Interestingly, they can compute the life-time $\tau$ of the black hole as a function of its initial mass $m$ and they find

$$\tau \sim m\, e^{m}. \tag{432}$$

This number is very large compared to the characteristic time of Hawking evaporation ($\tau_H \sim m^3$). This observation leads them to suggest that their computation may only become relevant in the final stage of the evaporation. It is surprising how close this conclusion is from our scenario (see chapter 15), especially considering that the followed paths are very different!

### 14.1.2 *Hájíček and Kiefer*

Much later, in 2001, another computation, leading to similar conclusions was done by Hájíček and Kiefer [89]. Their computation is not just the repetition of what had been done previously, the crucial difference being that their computation is non-perturbative.

They employ the *reduced quantisation* procedure: instead of starting from a full quantisation of gravity and then applying it to a particular case with few degrees of freedom, they procede the other way around. They start with a model with few degrees of freedom and then quant-ise it exactly. Their model describes the collapse of a null shell. Its dynamics is given by the unitary evolution of a narrow wave-packet. The latter is a function $\Psi$ of time $t$ and radius $r$ and they show that

$$\Psi_{\kappa\lambda}(t,r) = \frac{1}{\sqrt{2\pi}} \frac{\kappa!(2\lambda)^{\kappa+1/2}}{\sqrt{(2\kappa)!}} \left[ \frac{i}{(\lambda + it + ir)^{\kappa+1}} - \frac{i}{(\lambda + it - ir)^{\kappa+1}} \right] \tag{433}$$

with $\kappa$ and $\lambda$ tunable parameters accounting for the energy of the shell. It is noticeable that when $r \to 0$, $\Psi$ vanishes, which is how the singularity is resolved in this case. After its collapse, the shell bounces back. Thus, the quantum unitary evolution naturally resolves the singularity and predicts the evolution of the black hole into a white hole.

### 14.1.3 *Ashtekar, Olmedo and Singh*

An important step in favour of the scenario has been done by Ashtekar, Olmedo and Singh [9]. Contrary to the previous models, they do not consider the gravitational collapse of a null shell, but focus instead on eternal black holes, as described classically by the Kruskal maximal extension.

The interior of the black hole can be foliated by space-like hypersurfaces. The geometry of each slice is described by a point in a phase space, parametrised by the connection $A^i_a$ and the densitised triads $E^a_i$. Then, the full quantum dynamics can be approximated by an effective evolution, in the spirit of Loop Quantum Cosmology (LQC). As a result, it is computed that the expansion $\theta$, initially negative in both null directions, vanishes along a surface $\mathcal{T}$, and then becomes negative in both null directions too. The singularity is thus replaced by a smooth transition surface, between a past trapped and a future anti-trapped region.

The overall scenario is represented in figure 23. Kruskal space-time are piled-up infinitely, and joined smoothly along the transition surfaces $\mathcal{T}$. In this case, the transition from black to white only happens through the interior. An outside observer cannot notice the transition. This version of the black-to-white transition comes nonetheless to strengthen the credibility of the general picture.

## 14.2 A RESEARCH PROGRAMME

The original research presented in this thesis embeds into a wider research programme that started few years ago in the quantum gravity group of Marseilles. Its main goal is to attain a complete description of the black-to-white hole transition, using the tools of covariant LQG. It is not much about predicting the scenario itself, but rather about investigating its observational consequences. The previously discussed versions of the transition have motivated this line of research, by making the scenario a credible one. We now explain the framework of this peculiar model of black-to-white transition, and summarise its main results and failures. It serves as theoretical basis for our later explorations.

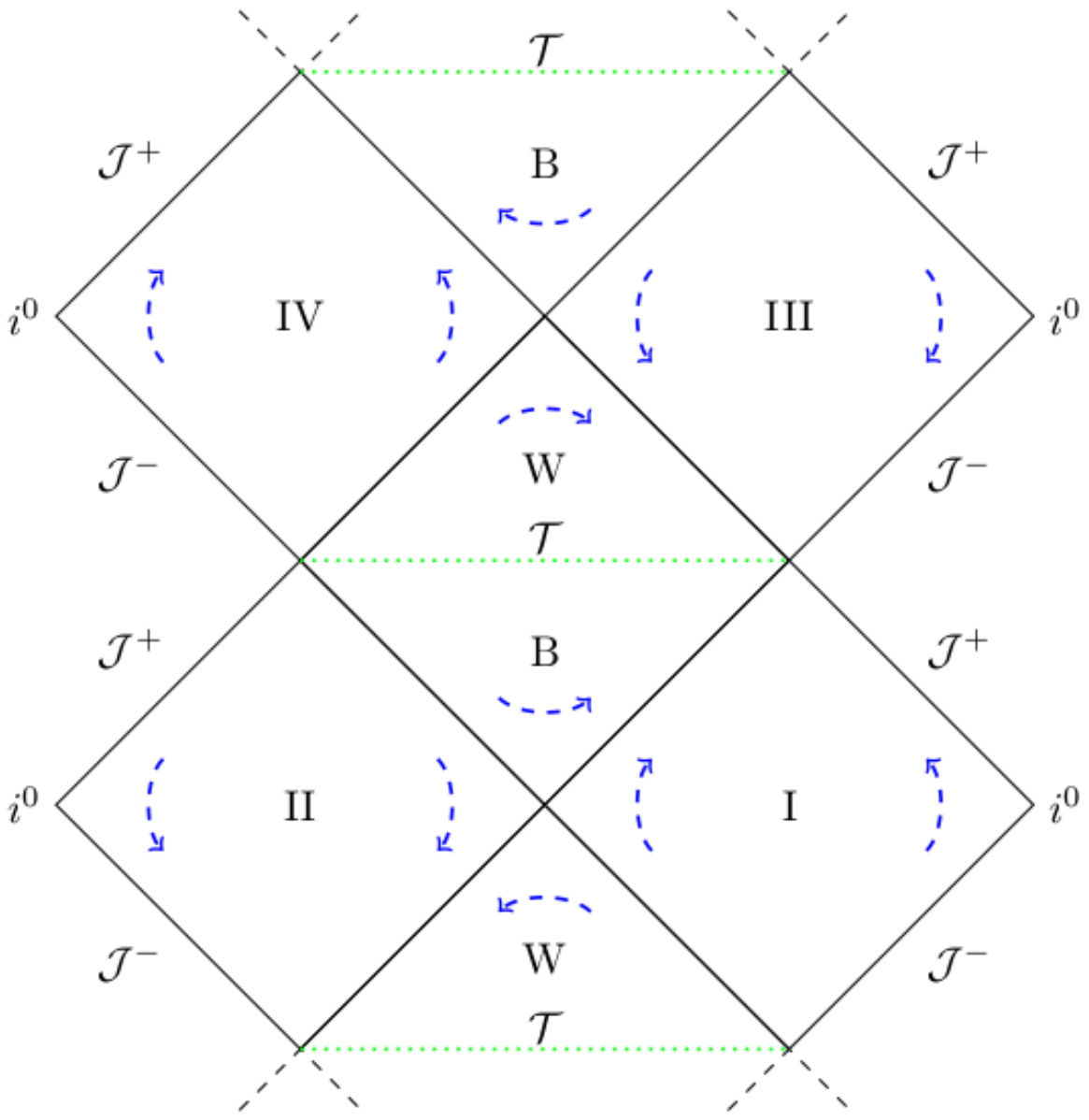

Figure 23: Conformal diagram of the extended Kruskal space-time proposed in [9].

### 14.2.1 *Quantum effects*

If it occurs, the black-to-white hole transition is a quantum phenomenon that comes to modify the classical picture of black holes. It should be emphasised that such quantum effect are actually expected in two separate regions of space-time. Following the naming of [37], the two regions are

A. The planckian-curvature region around the singularity. The latter is expected to be resolved by quantum effects, with an effective 'quantum force' counter-balancing the gravitational collapse, in a similar manner as quantum mechanics prevents the orbiting electron from falling onto the nucleus.

B. The late outside region, where Hawking evaporation, for instance, should be observed. Although, the curvature can be very small in this region, quantum gravity effects are expected provided a sufficient amount of time has elapsed. This piling-up of quantum effects is similar to the quantum radioactive particle: however long the half-life may be, the decay will happen in the long run.

Both regions are depicted in figure 24. We shall not be fooled by the conformal diagram: the two regions A and B do not touch. On the contrary, they are space-like separated very far away from one another. In the two next subsections, we investigate the physics of each of these regions: the Planck stars and the fireworks.

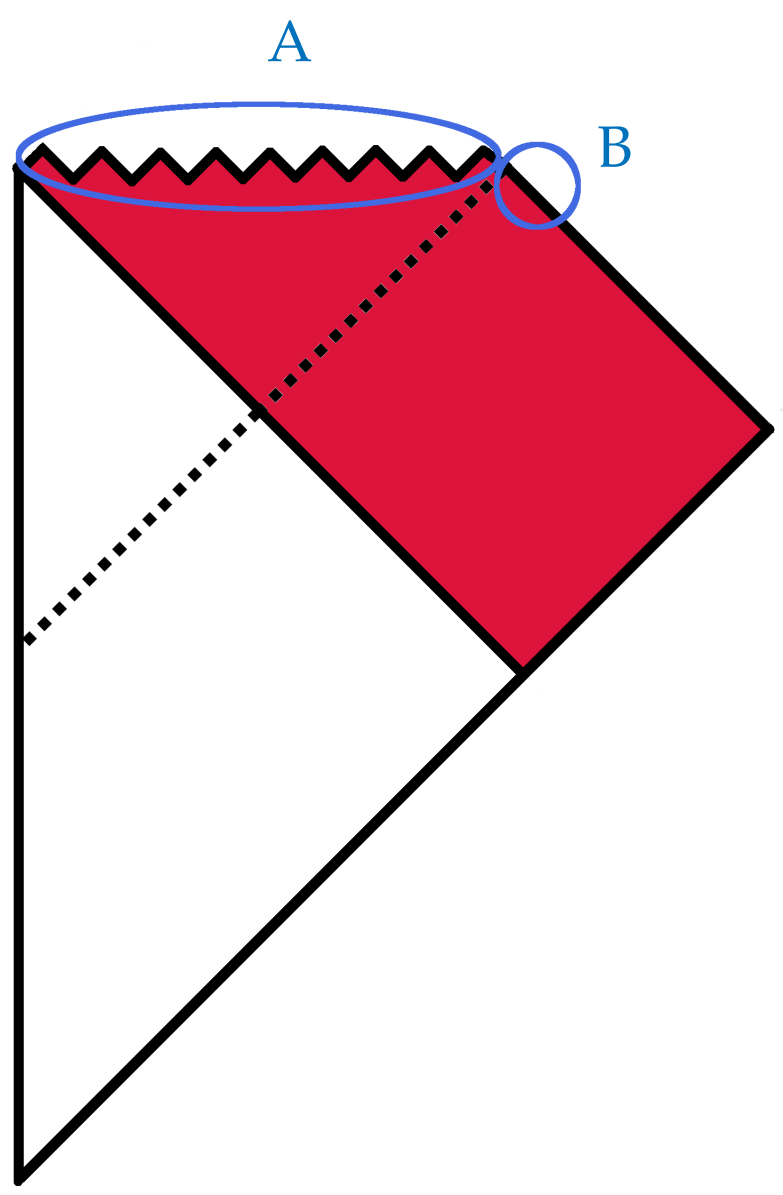

Figure 24: Conformal diagram of a black hole formed by the collapse of a thin null shell. Two different kinds of quantum effects are expected in regions A and B.

### 14.2.2 *Planck stars*

In the classical scenario, whenever a collapsing star has passed below its Schwarzschild radius, the complete collapse is inescapable. The usual laws of physics that guarantee the rigidity of matter cannot resist the ultimate compression of the gravitational force inside a black hole. However, quantum gravity strongly suggests that the classical view is incomplete and that the singularity is actually resolved.

A reasonable analogy is provided by the electron orbiting around its nucleus. Contrary to the prediction of classical electromagnetism, everyday life is made possible by the existence of a minimal non-zero energy level that behaves as a repulsive force against the collapse of atoms. Similarly, we can expect that, at the center of a black hole, the pressure of matter becomes so high that we enter a new regime, yet unknown, where quantum laws counter-balance gravity. This heuristic picture is supported by previous effective descriptions of star collapse in quantum gravity, as the Frolov-Vilkovisky model (subsection 14.1.1) and the Hájíček-Kiefer model (subsection 14.1.2).

As it collapses, the density of matter grows until it reaches Planck density

$$\rho_P \stackrel{\text{def}}{=} \frac{c^5}{\hbar G^2} \approx 5 \cdot 10^{96}\ \text{kg.m}^{-3}. \tag{434}$$

Then it is assumed that the collapse stops, due to quantum gravity effects. Matter is condensed into a highly compressed core. This state of matter at this stage was dubbed 'Planck star' by Rovelli and Vidotto in 2014 [162]. Naming it a 'star' may convey the impression that it is stable, and does not bounce back. In fact, the bounce could well happen, with a short proper time, but it would take a very long time to be observed by an outside observer, due to the huge gravitational time dilation.

The radius of such a Planck star can be estimated from the requirement that the energy density is planckian. Contrary to what could be guessed on first thought, the resulting radius is much bigger than Planck length! The energy density is roughly proportional to the curvature. For a black hole, the curvature, understood as the square-root of the Kretschmann scalar, is

$$\mathcal{R} \sim \frac{m}{r^3}. \tag{435}$$

As a result, the radius of a Planck star, when $\mathcal{R} \sim 1$, is about

$$r \sim m^{\frac{1}{3}}. \tag{436}$$

This radius can be very large! For Sgr A*, the supermassive black hole at the center of the Milky Way, the radius is $r \sim 10^{-10}$ m, that is the size of an atom, which is small, but still 25 orders of magnitude bigger than Planck length.

### 14.2.3 *Fireworks*

Soon after the Planck star paper, Rovelli and Haggard proposed an original analysis of the quantum tunnelling from a black to a white hole [87]. In particular, they provide an explicit exterior metric for the transition, called *fireworks metric*, that is an exact solution of Einstein equations. Of course, the transition cannot satisfy Einstein equations everywhere, because the scenario is prohibited in GR. What Haggard and Rovelli have shown, is that the violation of classical equations can be confined to a compact region of space-time surrounding the singularity (region A) and the late time black hole horizon (region B).

The fireworks metric describes the fall and collapse of a thin null spherical shell of matter, whose thickness is neglected. It bounces at a minimal radius inside its Schwarzschild radius, and then expands forever. The space-time is spherically symmetric, so that it can be well represented in a conformal diagram (see figure 25). Due to Birkhoff's theorem, the shell's interior is a portion of Minkowski space-time (region I), while the exterior is almost everywhere locally isomorphic to a portion of Kruskal space-time (region II). The process is assumed to be invariant under time-reversal. In particular, the dissipative effects such as Hawking radiation are disregarded. Thus, the classicality of space-time is preserved both at large radii and at early and late times. The quantum effects only modify it in a bounded region of space-time (region III).

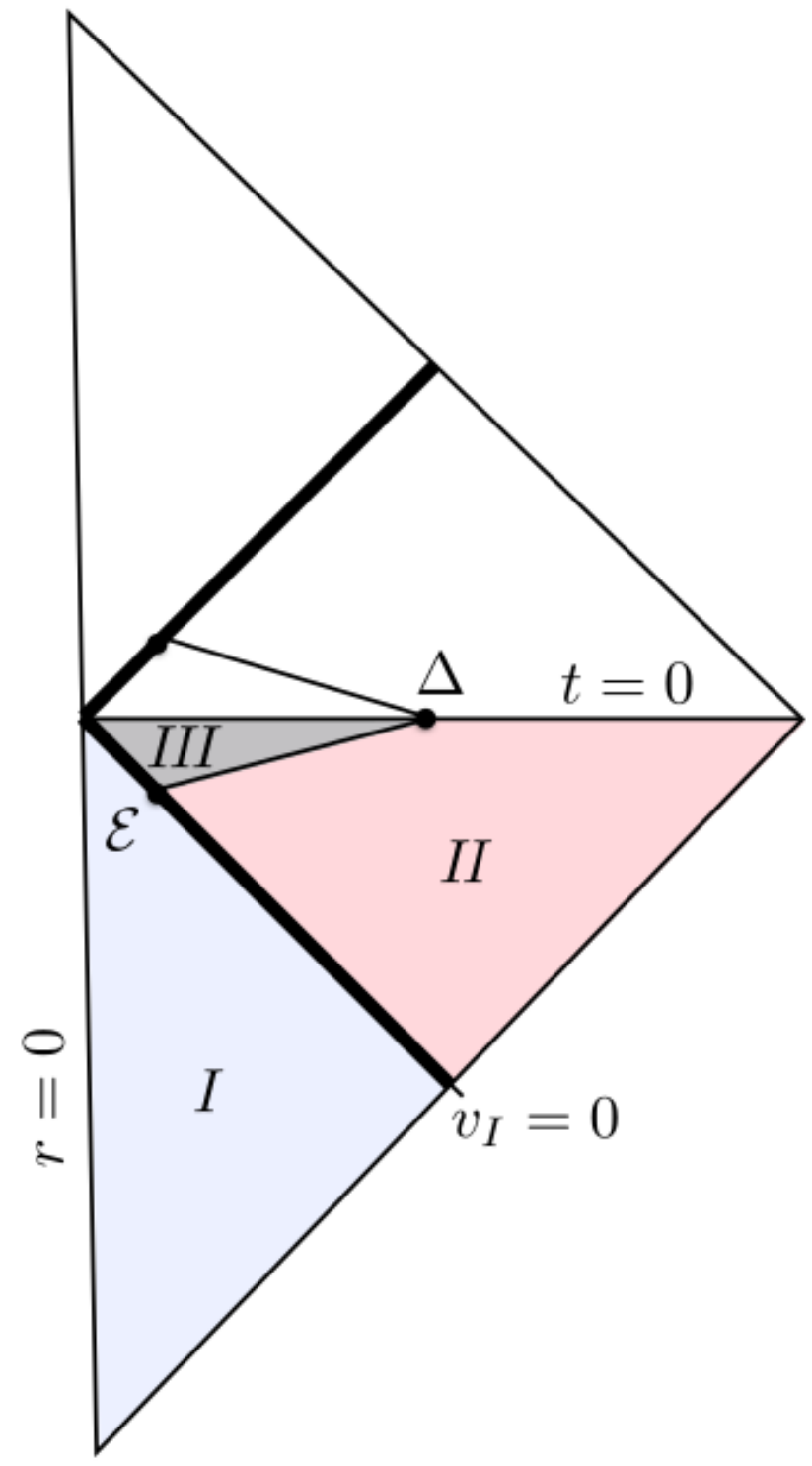

Figure 25: Conformal diagram of the fireworks scenario proposed in [87].

The metric undergoes a quantum tunnelling from black to white hole. Strictly speaking there is no classical metric always in place, like there isn't a physically defined trajectory for a particle tunnelling under a potential barrier. In the case of the particle tunnelling under a potential barrier, it is nevertheless still possible to define an effective trajectory, by connecting the partial semiclassical trajectories of the particle before and after the tunnelling. This effective trajectory of course violates the classical equations of motion during the tunnelling. The tunnelling is therefore modelled by a simple violation of the equations of motion. In a similar fashion, the black to white transition can be modelled by a single classical geometry that violates the classical Einstein equations in compact spatial region during a short time.

Interestingly, they suggest a dimensional analysis to estimate the life-time of the black hole, and find

$$\tau \sim m^2. \tag{437}$$

Interestingly, this time is much smaller compared to the characteristic time of Hawking evaporation $\tau_H \sim m^3$. This means that the evaporation process can be legitimately neglected.

### 14.2.4 *Phenomenological consequences*

Several possible phenomenological consequences of the fireworks scenario have been studied in [22–25, 88]. The main possibly detectable signal in the scenario is coming from the explosion of white holes. So, the observability today really depends on the value of the life-time $\tau$, which should be of the order of the age of the Universe. In the literature, the phenomenology was studied for $\tau$ ranging between $m^2$, the low bound of equation (437), and $m^3$ the high bound of Hawking time.

Such black holes are necessarily primordial black holes, whose mass is thus ranging between $10^{12}$ kg (the total mass of humankind) and $10^{22}$ kg (the mass of the Moon). Then, two kinds of signals are expected:

1. The low energy contribution, with a wave-length comparable to $m$, i. e. with an energy of order $m^{-1}$. Although no precise mechanism explains it, it is natural to expect signals with a wavelength comparable to the size of the astrophysical object, as it is the case, for instance, for Hawking radiation. But unlike Hawking radiation, whose signal extends over a long period, we expect the explosion to occur rapidly.

2. The high energy contribution, coming from the matter itself bouncing out. What comes out should be what came in, with the same energy. This can be estimated from a simple model of formation of a primordial black hole. A black hole of mass $m$ shall have formed when the universe was old as $t \sim m$. At that time, its temperature was typically $T \sim t^{-\frac{1}{2}}$, and so the outgoing signal would have an energy of order $m^{-1/2}$.

Then such signals could appear in two ways:

1. Single event detection: explosions of black holes are detected one at a time.

2. Integrated emission: explosions are many and cause a background signal.

From the analysis of [22], it is observed that

- The low energy channel would be more easily detected in single event detection.
- The integrated emission would manifest itself in a distorted black body radiation law for the high energy contribution.

But no definite conclusion can be drawn. In [25], they study more specifically a signal observed by the Fermi telescope, which is a space telescope launched in 2008 and dedicated to gamma-ray astronomy. Indeed, this telescope has observed an excess of signal with energy in the GeV ($1\,\text{GeV} \approx 10^{-10}\text{J}$), coming from the centre of our galaxy. The physical mechanism beyond this phenomenon is not yet known. In our scenario, the hypothetical high energy contribution cannot explain it, but the low energy could do it in the single detection mode. The signal would originate from the explosion of black holes with a mass of $10^{11}$ kg, and there should be about 100 black holes bouncing per second in the centre of the galaxy to account for the observed flux. Although possible, this hypothesis is one among many others, like the millisecond pulsar origin, maybe more convincing.

## 14.3 CHARACTERISTIC TIME-SCALE

The bounce time $\tau$ is a very important parameter that fixes the scenario. It is an observable that could be observed through its phenomenological consequences. Equation (437) is based on a rough dimensional analysis that should be improved. Theoretically, covariant LQG provides all the necessary tools to compute it. Such a computation was started in [47]. Unfortunately, the calculation is very hard and thus requires many simplifications to be achieved. Two PhD theses, [44] and [49], have been dedicated to such a feat, and the main results can be found in [45].

### 14.3.1 *Tunnelling by path integral*

Let's start by recalling few facts about the tunnelling of a particle of energy $E$ facing a potential barrier $V(x)$, as shown in figure 26. Initially in a state $|x_i\rangle$, the amplitude of probability to tunnel, after a

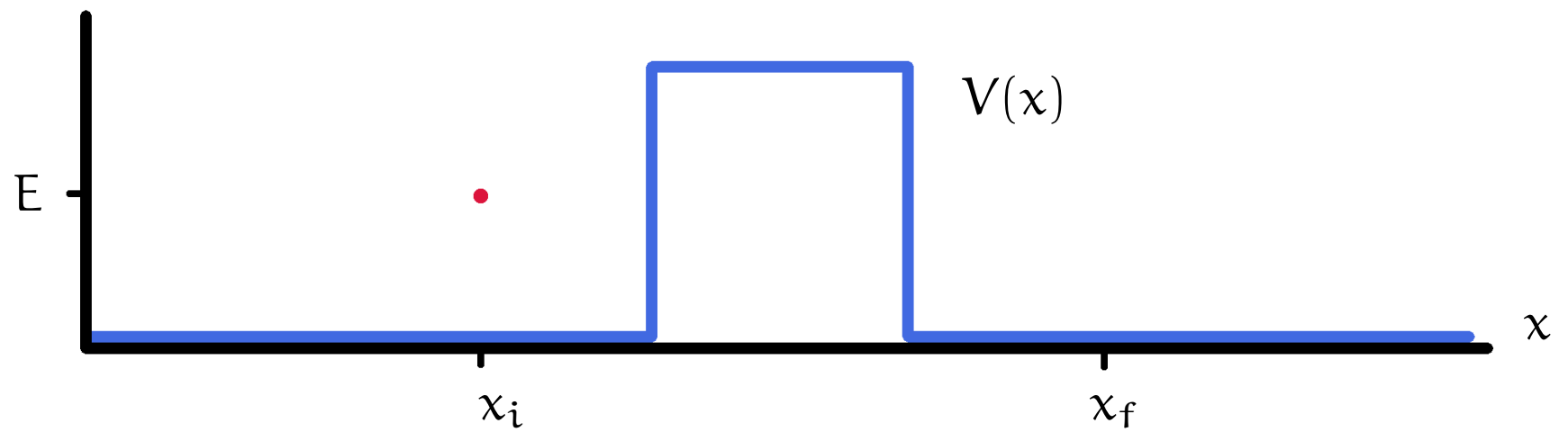

Figure 26: Potential barrier $V(x)$.

time T, in the state $|x_f\rangle$, is computed with the path integral

$$\langle x_f|e^{-i\frac{\hat{H}T}{\hbar}}|x_i\rangle = \int \mathcal{D}x\, e^{iS[x]/\hbar} \tag{438}$$

with the standard action:

$$S = \int_0^T dt \left[\frac{1}{2}\left(\frac{dx}{dt}\right)^2 - V(x)\right]. \tag{439}$$

Normally, in the semi-classical limit, $\hbar \to 0$, the only contribution to the path integral comes from the classical trajectory $x_0$

$$\langle x_f|e^{-i\frac{\hat{H}T}{\hbar}}|x_i\rangle = e^{iS[x_0]/\hbar}. \tag{440}$$

But in this case, there is no classical trajectory through the barrier. Then, the trick is to allow the time T to become complex for a while. It is called a *Wick rotation*, which is a form of analytic extension. After the computation is done, we'll Wick rotate back. So we posit $iT = \tau \in \mathbb{R}$. The probability amplitude becomes

$$\langle x_f|e^{-\frac{\hat{H}\tau}{\hbar}}|x_i\rangle = \int \mathcal{D}x\, e^{-S_E[x]/\hbar} \tag{441}$$

with the *euclidean* action

$$S_E = \int_0^\tau dt \left[\frac{1}{2}\left(\frac{dx}{dt}\right)^2 + V(x)\right]. \tag{442}$$

The stationnary points of $S_E$ are solutions of the equation of motion of a particle in a potential $-V(x)$ (a well), and this time there are such classical trajectories connecting $x_i$ to $x_f$. At the end of the day, we find

$$\langle x_f|e^{-i\frac{\hat{H}T}{\hbar}}|x_i\rangle = e^{-i\frac{ET}{\hbar}} e^{-\frac{1}{\hbar}\int dx\,\sqrt{2(V(x)-E)}}. \tag{443}$$

The modulus square of this amplitude gives the probability $p$ that a particle, coming from $x_i$, crosses the barrier in a first try.

The decay of a radioactive particle can be modelled similarly, with an $\alpha$-particle inside a nucleus, trying to tunnel out with a probability $p$. In the Gamow drop-model [72, 73], the $\alpha$-particle bounces on the inner potential wall with an escape trial frequency $f$

$$f = \frac{v}{R}, \tag{444}$$

with $R$ and $v$ respectively the typical radius of the nucleus and the velocity of the $\alpha$-particle inside the nucleus. The velocity can be estimated roughly as $v \sim \hbar/(Rm)$ with $m$ the mass of an $\alpha$-particle. Finally, the lifetime of the radioactive particle is

$$\tau = \frac{1}{fp}. \tag{445}$$

This model is very useful to understand the meaning of the lifetime of black holes in the black-to-white transition.

### 14.3.2 *Euclidean quantum gravity*

In QFT, the amplitude of transition between different field configurations can also be written in terms of path integrals. Again, the Wick rotation is useful to give meaning to ill-defined path integrals. When applied to GR, it is called *euclidean quantum gravity*. The geometry of space-time is described on a 3D space-like hypersurface, which is time-evolved with the hamiltonian of general relativity. The Einstein-Hilbert and Gibbons-Hawking-York actions $S$ (see section 9.3) are replaced by their euclidean versions $S_E$ defined as

$$S_E \stackrel{\text{def}}{=} -iS[g_E], \tag{446}$$

with $g_E$ a riemannian metric obtained from a lorentzian one via a Wick rotation.

As previously, the black-to-white hole transition amplitude can be estimated as

$$\langle WH|BH\rangle = e^{-S_E[g_{ab}]} \tag{447}$$

with $g_{ab}$ a classical metric connecting a black hole slice $BH$ to a white hole slice $WH$. The Haggard-Rovelli fireworks provide such a metric. In fact, since the scalar curvature vanishes, $R = 0$ inside both the black and the white holes, only the boundary term contributes, that is the Gibbons-Hawking-York action:

$$S_{GHY} = \frac{1}{8\pi}\int_{BH\cup WH} d^3y\,\sqrt{h}\,K. \tag{448}$$

The boundary is made of the black hole past slice $BH$ and the white hole future slice $WH$. By time-reversal symmetry, it is sufficient to compute the action on $BH$. The action on $WH$ will take the same value because both the orientation and the extrinsic curvature have their sign flipped under the time-symmetry. In [47], such slices are defined as constant Lemaître time-coordinate. $BH$ is intrinsically flat ($\sqrt{h} = 1$), and the extrinsic curvature is

$$k_{ab}dx^a dx^b = (2m)^{\frac{3}{2}} r^{-\frac{5}{2}} dr^2 - \sqrt{8mr}\,d\Omega^2. \tag{449}$$

So its trace is

$$K = (2m)^{\frac{3}{2}} r^{-\frac{5}{2}} + \sqrt{8mr} + \sqrt{8mr}\sin^2\theta \tag{450}$$

So

$$S_{GHY} = \frac{1}{4\pi}\int_{r_{min}}^{2m} dr\left[2\pi^2(2m)^{\frac{3}{2}} r^{-\frac{5}{2}} + 2\pi(\pi+2)\sqrt{8mr}\right] \tag{451}$$

The minimal radius $r_{min}$ can be chosen at the surface of the Planck star, i. e. $r_{min} \sim m^{1/3}$. In this case, $S$ is a second-order polynomial in $m$. For the purpose of our rough estimate we keep the first order

$$S_{GHY} \sim m^2, \tag{452}$$

A similar kind of computation can be found in [78]. Thus, the probability of transition is behaving like

$$p \sim e^{-\xi m^2}. \tag{453}$$

with some positive constant $\xi$.

### 14.3.3 *Black hole lifetime*

The computation of the amplitude has been done using covariant LQG in [45]. They find a similar result as equation (453) where $\xi$ admits a geometrical interpretation as a lorentzian angle. To understand how this probability relates to the lifetime of a black hole, it is useful to compare with the Gamow drop-model, seen previously. Heuristically, we can imagine that the collapsed shell is trapped within a potential well, similarly to the $\alpha$-particle in the nucleus. In this case, the escape trial frequency is given by the inverse of the typical size of the black hole, $f \sim 1/m$. Finally, the life-time is estimated as

$$\tau \sim \frac{m}{p} \sim m e^{\xi m^2}. \tag{454}$$

First, we may notice that this is much larger than $m^2$ which was expected by dimensional analysis. Second, it is also much larger than the Hawking evaporation time $m^3$, so that the later should not be neglected in the analysis. Such a conclusion was already obtained in the early work of Frolov and Vilkovisky [71] using a very different path (see subsection 14.1.1).

In [17], the authors pretend to compute the black hole lifetime using euclidean quantum gravity (more precisely than the rough estimate of subsection 14.3.2) and to find $\tau \sim m$. However, as was pointed out in [45], a close look at their work shows that what they compute is improperly dubbed the lifetime and it actually corresponds to some characteristic time, the inverse of the escape trial frequency. So the computations agree, but the interpretations differ. We believe that the interpretation of [45], adopted here, is the right one.

## 14.4 OBJECTIONS

In this section, we present three objections that weaken the Haggard-Rovelli construction.

### 14.4.1 *Conical singularity*

The Haggard-Rovelli metric is built from a portion of the double-cover of Kruslal diagram, as shown in figure 27. It consists of two superposed 'arms', which are diffeomorphically 'squashed', and then glued to Minkowski patches. The local double covering (where the two arms cross) raises no peculiar difficulty, except at point $\Delta$. Here, the mathematical construction of this model is not completely rigorous. In figure 27, it can be seen that the angles are not locally preserved at that point, whereas it is a property expected for any conformal transformation. As a consequence, the metric has a pathological conical singularity in $\Delta$. In other words, the curvature invariants diverge here. It can be cured as we show in the improved construction of section 14.5.

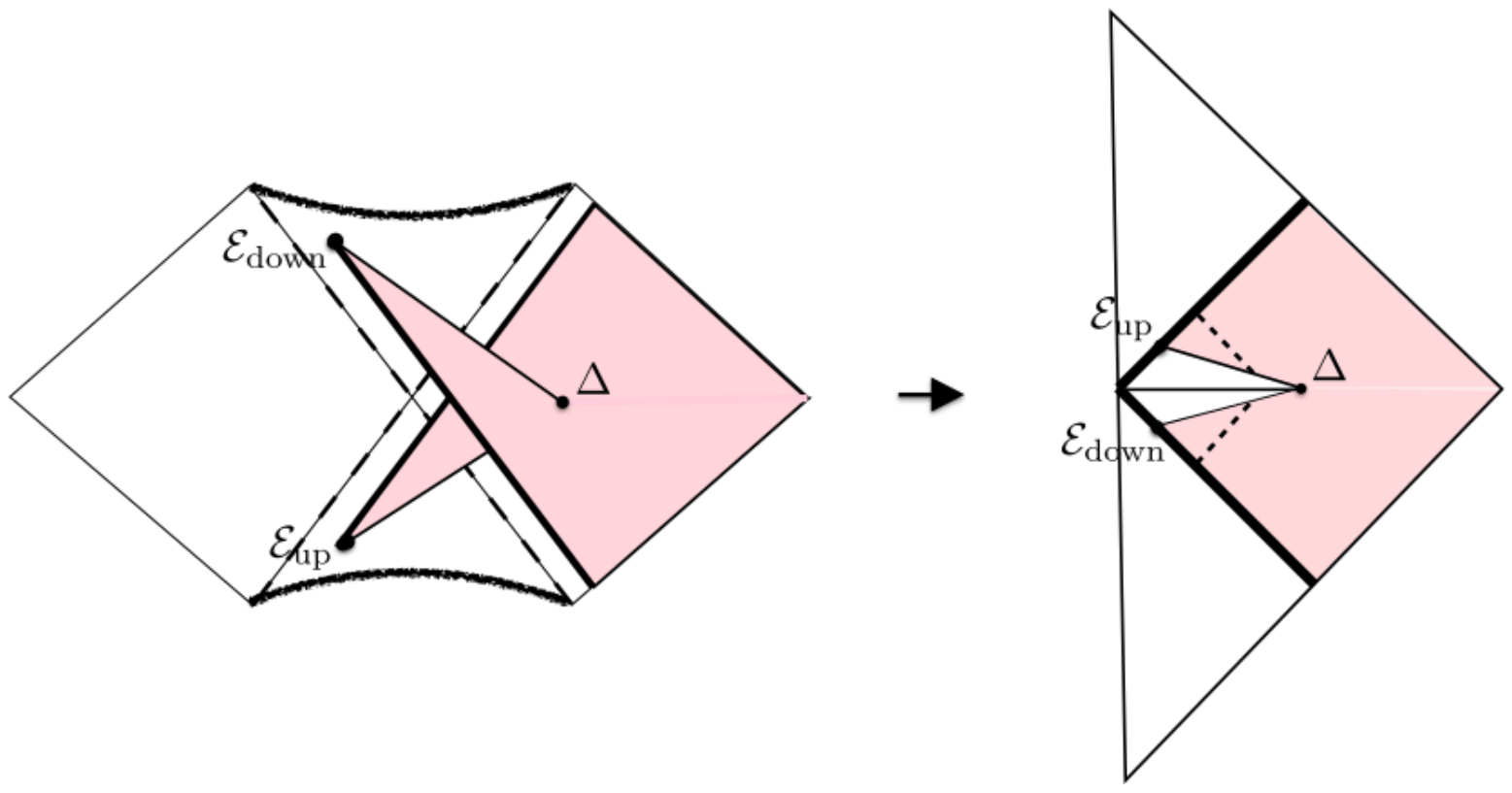

Figure 27: Haggard-Rovelli construction from a portion of the double-cover of Kruskal diagram. From [87].

### 14.4.2 *Instability*

In section 13.3, we have seen that white holes are unstable with respect to small perturbations of the ambient matter (classical instability) and also with respect to particle creation (quantum instability). In [53] it was pointed out that both classical and quantum instabilities may affect the Haggard-Rovelli firework metric. A solution is proposed consisting in making the model time-asymmetric: the black hole phase would last much longer than the white hole phase.

Interestingly, our main work, explained in chapter 15, also consists in building an asymmetric scenario, but the other way around: the white hole phase lasts much longer than the black hole phase. We will see that the instability is also cured in this case!

### 14.4.3 *Dissipative effects*

For completeness, we shall also mention another objection, which regards not only the Haggard-Rovelli construction, but the black-to-white hole scenario in general. As far as we know, this objection has not been carefully formulated in any reference. It was communicated orally through discussions with Andrew Hamilton and Ale-jandro Perez. The objection is that the mathematical description of the bounce in the various models does not take into account all the dissipative effects that could modify the evolution.

A striking picture is provided by the fall of an egg on the ground. Common cooking experience teaches that eggs tend to scratch on the floor. For my purpose, I shall say that it does not bounce. Yet, a simple mathematical model, in one dimension, with the mass of the egg and the height of the fall as the only two parameters, predicts, to a first approximation, an elastic bounce. Such a miracle is never observed in classical kitchens. The fact is that the kinetic energy of the egg, as it reaches the ground, is shared into the many pieces of the egg as it breaks, and finally into thermal energy in all the microscopic degrees of freedom of the floor. The same thing could be true for black holes: the huge gravitational energy of the star, as it collapses, might be lost in the friction of the many 'atoms of space-time'. As for now, it is hard to make the argument more formal and thus to answer it properly. We shall keep it somewhere in mind though.

## 14.5 INTERIOR METRIC

In this section, we present a metric that improves the 'firework' geometry. It is based on our paper [154]. In the original construction of Haggard and Rovelli, the quantum region, where Einstein equations are violated, is bounded, but includes both region $A$ (would-be-singularity) and region $B$ (which surrounds the end of the apparent horizon). In the improved metric, the proper quantum tunnelling is now confined to region $B$ only.

The construction relies on the same set of hypotheses as the original fireworks scenario, together with the additional assumption that the time-like and null geodesics are continuous through the $r = 0$ singularity. The description of the interior metric is obtained by sewing the future singularity of the black hole to the past singularity of the white hole. This smooth joining of two Kruskal space-times is a possibility noted by several authors [50, 142, 179]. As a result, the overall metric satisfies the Einstein equations everywhere except in this small region $B$. Incidentally, we cure the conical singularity at the cusp point of the quantum region.

### 14.5.1 *Kruskal origami*

The pathology of the conical singularity can be removed by excising the point $\Delta$ in figure 27. As any surgical operation, it is safer to remove a bit more than the undesired pathology and we cut a causal diamond along null geodesics around $\Delta$. In addition to that, we extend the arms up to the singularities. The resulting portion of Kruskal space-time is marked out by the red lines in figure 28.

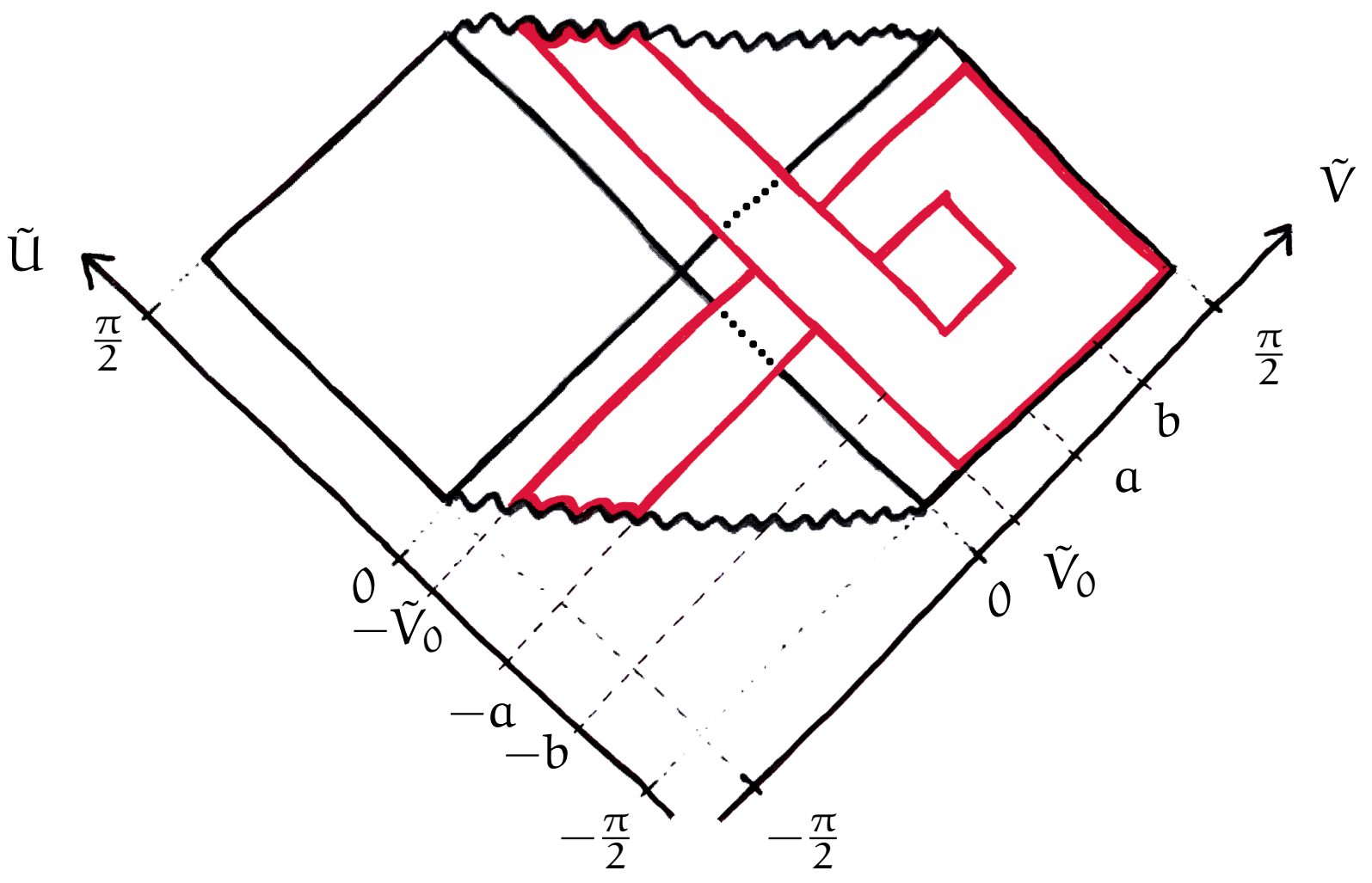

Figure 28: Conformal diagram of the Kruskal space-time. The red straight lines are null, and the two red wavy lines will be identified after 'squashing the arms'. The inside region thus delimited is the space-time of interest for us.

The modelling of the black-to-white hole transition is achieved through the identification between the past and the future singularity. Heuristically, it consists in 'squashing the arms until the hands match'. The conformal diagram of the resulting space-time is represented in figure 29.

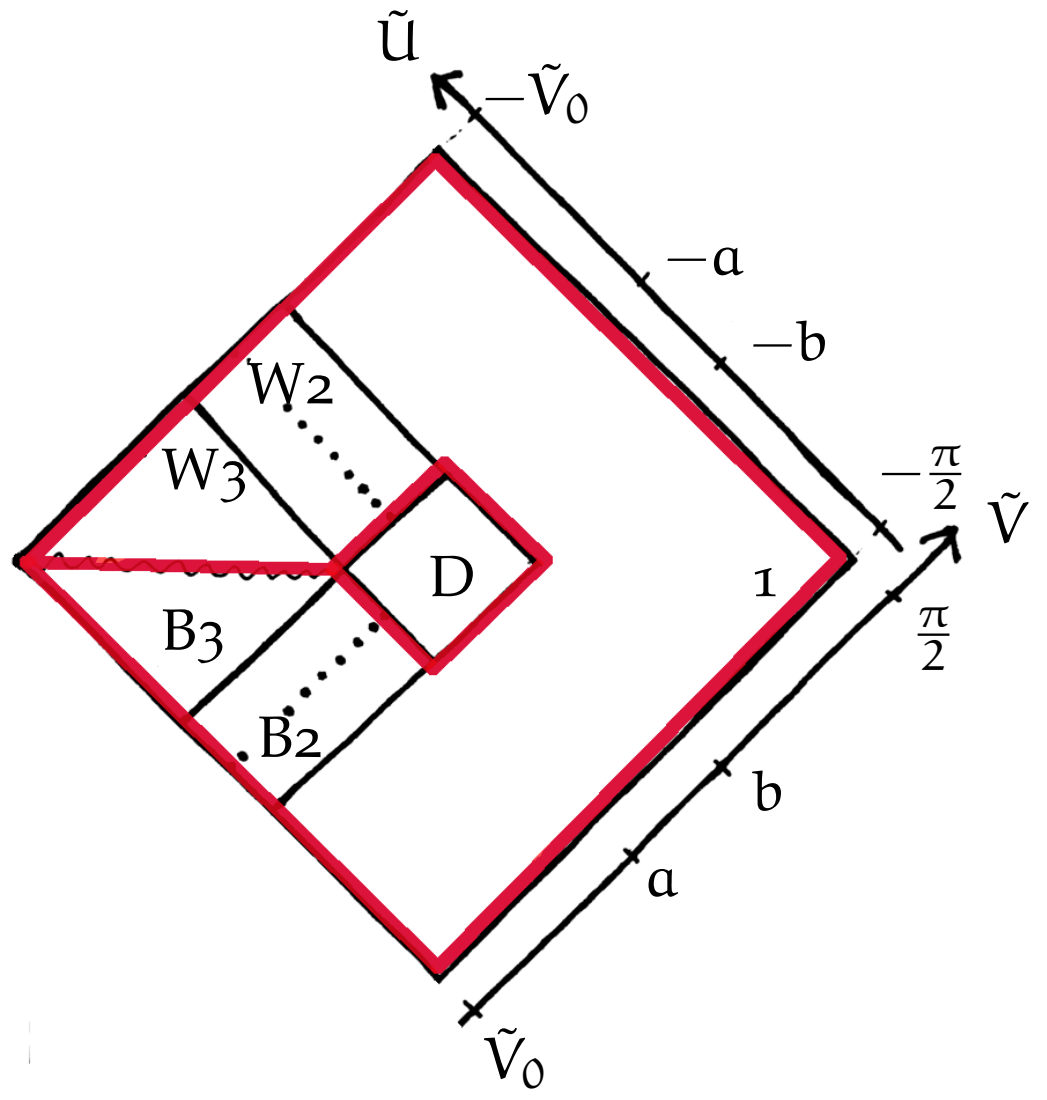

Figure 29: Conformal diagram of the outside of the null shell. The dotted lines are the two horizons at $r = 2m$.

The expression of the metric is given by equations (352) and (353), where the Kruskal coordinates $(U, V)$ are given in terms of the diagram coordinates $(\tilde{U}, \tilde{V})$ by

$$\text{[lower half]} \begin{cases} U = \tan f_B(\tilde{U}) \\ V = \tan \tilde{V} \end{cases} \tag{455}$$

$$\text{[upper half]} \begin{cases} U = \tan \tilde{U} \\ V = \tan f_W(\tilde{V}) \end{cases} \tag{456}$$

where the two functions $f_B$ and $f_W$ are differentiable and defined piecewise such that

$$f_B(\tilde{U}) = \begin{cases} \tilde{U} & \text{for} \quad \tilde{U} \in [-\frac{\pi}{2}, -b] \\ f_B(\tilde{U}) & \text{for} \quad \tilde{U} \in [-b, -a] \\ \tilde{U} + \frac{\pi}{2} & \text{for} \quad \tilde{U} \in [-a, -\tilde{V}_0] \end{cases} \tag{457}$$

and

$$f_W(\tilde{V}) = \begin{cases} \tilde{V} - \frac{\pi}{2} & \text{for} \quad \tilde{V} \in [\tilde{V}_0, a] \\ f_W(\tilde{V}) & \text{for} \quad \tilde{V} \in [a, b] \\ \tilde{V} & \text{for} \quad \tilde{V} \in [b, \frac{\pi}{2}]. \end{cases} \tag{458}$$

For the intermediate intervals ($[-b, -a]$ for $f_B$ and $[a, b]$ for $f_W$) one can choose any continuous and monotonic function which joins 'smoothly enough' with the other pieces.

The minimal smoothness required is $\mathcal{C}^1$. Indeed, the two junction conditions for null hypersurfaces have to be satisfied along the null geodesics $\tilde{V} = a$, $\tilde{V} = b$, $\tilde{U} = -a$ and $\tilde{U} = -b$. The first condition is the continuity of the induced metric on the hypersurface. This requires the continuity of the functions $f_B$ and $f_W$. The second condition is the continuity of the extrinsic curvature, which imposes the continuity of their derivatives. In the following, we will choose, in the intermediate interval, a polynomial of degree 3 (see figure 30).

### 14.5.2 *Across the singularity*

The regions B3 and W3 touch along the singularity. There is no difficulty here. It has been repeatedly noticed [142, 179] that it is possible to match the future singularity of a Kruskal diagram to the past singularity of another (see figure 31). The metric is singular there, but there is a natural prescription for the geodesics to go across the singularity, requiring conservation of linear and angular momentum [142]. As argued in [50], the resulting space-time can be seen as the $\hbar \to 0$ limit of the effective metric of a non-singular space-time where quantum gravity bounds curvature. This effective metric would be given by

$$ds^2 = -\frac{4(\tau^2 + l)^2}{2m - \tau^2} d\tau^2 + \frac{2m - \tau^2}{\tau^2 + l} dx^2 + (\tau^2 + l)^2 d\Omega^2 \tag{459}$$

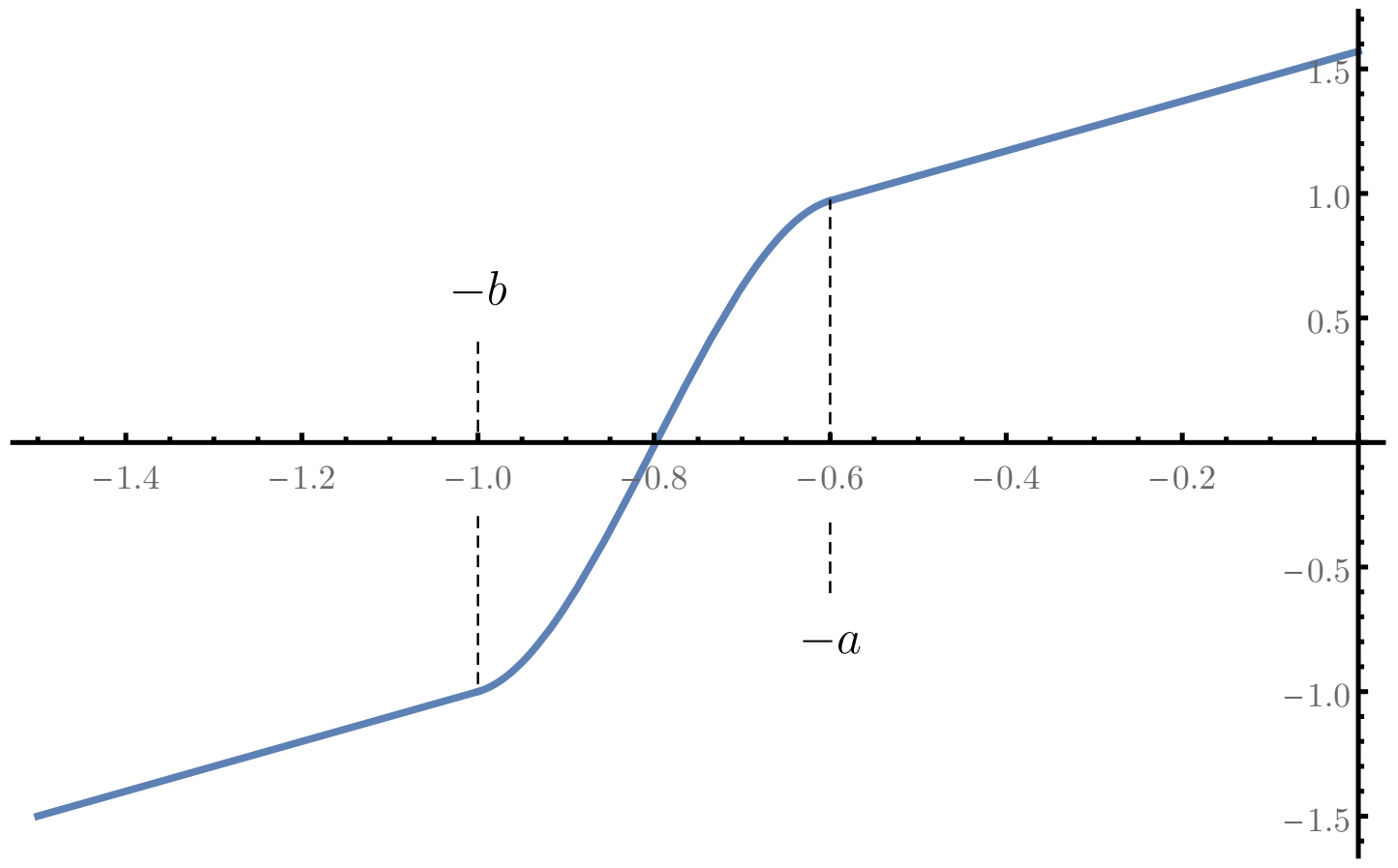

Figure 30: Graph of the function $f_B$. On the interval $[-b, -a]$ it is a polynomial of degree 3. It is linear elsewhere. Here we have chosen $a = 0.6$ and $b = 1$.

with a constant $l \sim (m\hbar)^{1/3}$. This regularisation is not far different from that of Hayward [101]. When $\hbar \to 0$, $l \to 0$, and the the metric boils down to the usual Schwarzschild metric inside the black hole with the usual Schwarzschild time $t$ and radius $r$ given by $t = x$ and $r = \tau^2$. So, there is a sense in which the two glued Kruskal space-times are still a solution of Einstein equations. We take this as a simplified model of the quantum transition across the singularity.

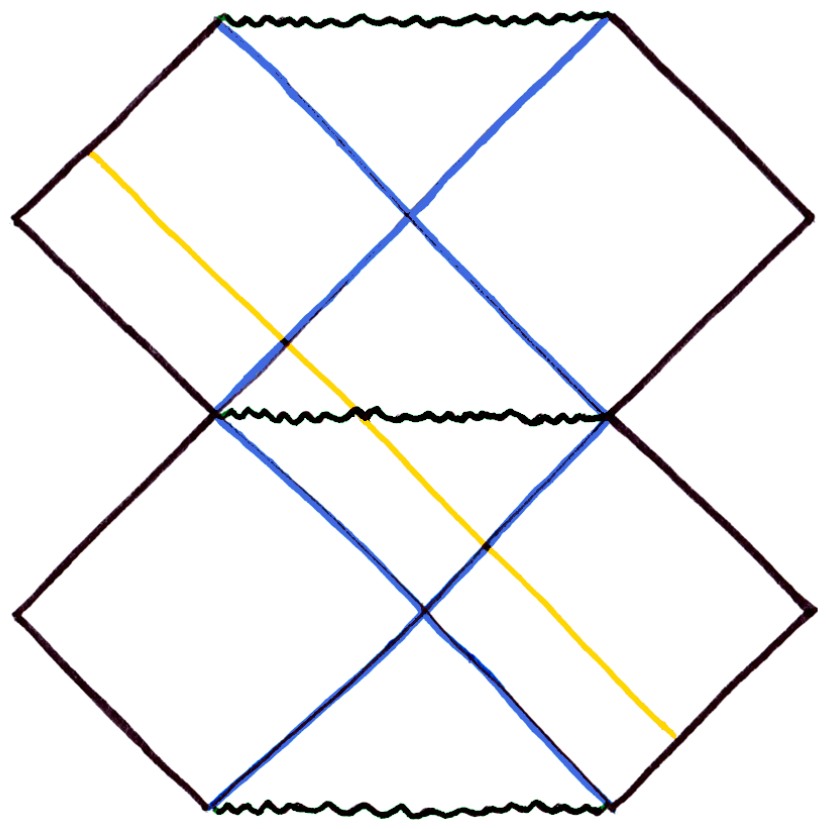

Figure 31: Conformal diagram of the two Kruskal space-times joined at the singularity. The gold line represents an ingoing null geodesic crossing the singularity.

### 14.5.3 *The diamond* B

The metric is well defined all around a central diamond B, the quantum tunnelling region. Compared to the original fireworks metric, its existence removes the conical singularity at the cusp. The simplest possibility to define a metric in this region is to simply extend the metric of B2 and of W2, respectively up to and down to the horizontal line $\tilde{U}+\tilde{V}=0$. Then, the first junction condition along this hypersurface imposes :

$$f_W(x) = -f_B(-x). \tag{460}$$

However, the second junction condition can never be satisfied, because otherwise it would define an exact solution of Einstein's equations with the same past but a different future as a standard collapse metric, which has an event horizon. The discontinuity of the extrinsic curvature therefore encodes the quantum transition in this region, as studied in [45, 47]. The novelty is that now this tunnelling region is confined within the diamond. The discontinuity in the extrinsic curvature could be smoothed out by modifying the metric in a small neighbourhood of the discontinuity. This is possible at the price of violating Einstein's equations in this neighbourhood.

### 14.5.4 *Relighting the fireworks*

The metric constructed in the previous section describes the space-time *outside* the bouncing null shell. Inside the shell, space-time is flat, therefore a portion of Minkowski space-time. What remains to be done is to glue a patch of Minkowski along the collapsing and the emerging null shells. This is done in a similar way to the well-known model of Vaidya presented in subsection 10.5.3.

The Minkowski conformal diagram is described in section 10.3. It is possible to glue a portion of Minkowski spacetime to the Kruskal origami by matching the value of the radius along a null ingoing geodesics ($V=$ constant) for Minkowski with the value of the radius along the line $\tilde{V}=\tilde{V}_0$ of the Kruskal origami. This matching defines a map $U(\tilde{U})$ between the coordinate $U$ of the original Minkowski patch and the coordinate $\tilde{U}$ of the new conformal diagram

$$U(\tilde{U}) = \arctan\left(v_0 - 4m\left[1+W\left(-e^{\frac{v_0}{4m}-1}\tan f_B(\tilde{U})\right)\right]\right), \tag{461}$$

with $v_0 \overset{\text{def}}{=} 4m\log\tan\tilde{V}_0$. Then, the first junction condition is satisfied. The violation of the second is the effect of the stress-energy tensor of the collapsing shell. Finally, the same procedure can be applied for the outgoing null geodesics along the line $\tilde{U}=-\tilde{V}_0$, with the condition

$$V(\tilde{V}) = \arctan\left(-v_0 + 4m\left[1+W\left(-e^{\frac{v_0}{4m}-1}\tan f_B(-\tilde{V})\right)\right]\right). \tag{462}$$

This completes the construction of the new space-time for black-hole fireworks.

### 14.5.5 *Conformal diagrams of the improved fireworks*

A conformal diagram for the new model is drawn in figure 32. To make an easy drawing, we have chosen to impose that the line $r = 0$ should be straight and vertical, which is possible provided the map $V(\tilde{V})$ in BM is given by

$$V(\tilde{V}) = \arctan\left(v_0 - 4m\left[1 + W\left(-e^{\frac{v_0}{4m}-1}\tan f_B(\tilde{V} - 2\tilde{V}_0)\right)\right]\right), \quad (463)$$

and the map $U(\tilde{U})$ in $WM$ is given by

$$U(\tilde{U}) = \left(-v_0 + 4m\left[1 + W\left(-e^{\frac{v_0}{4m}-1}\tan f_B(-\tilde{U} - 2\tilde{V}_0)\right)\right]\right). \quad (464)$$

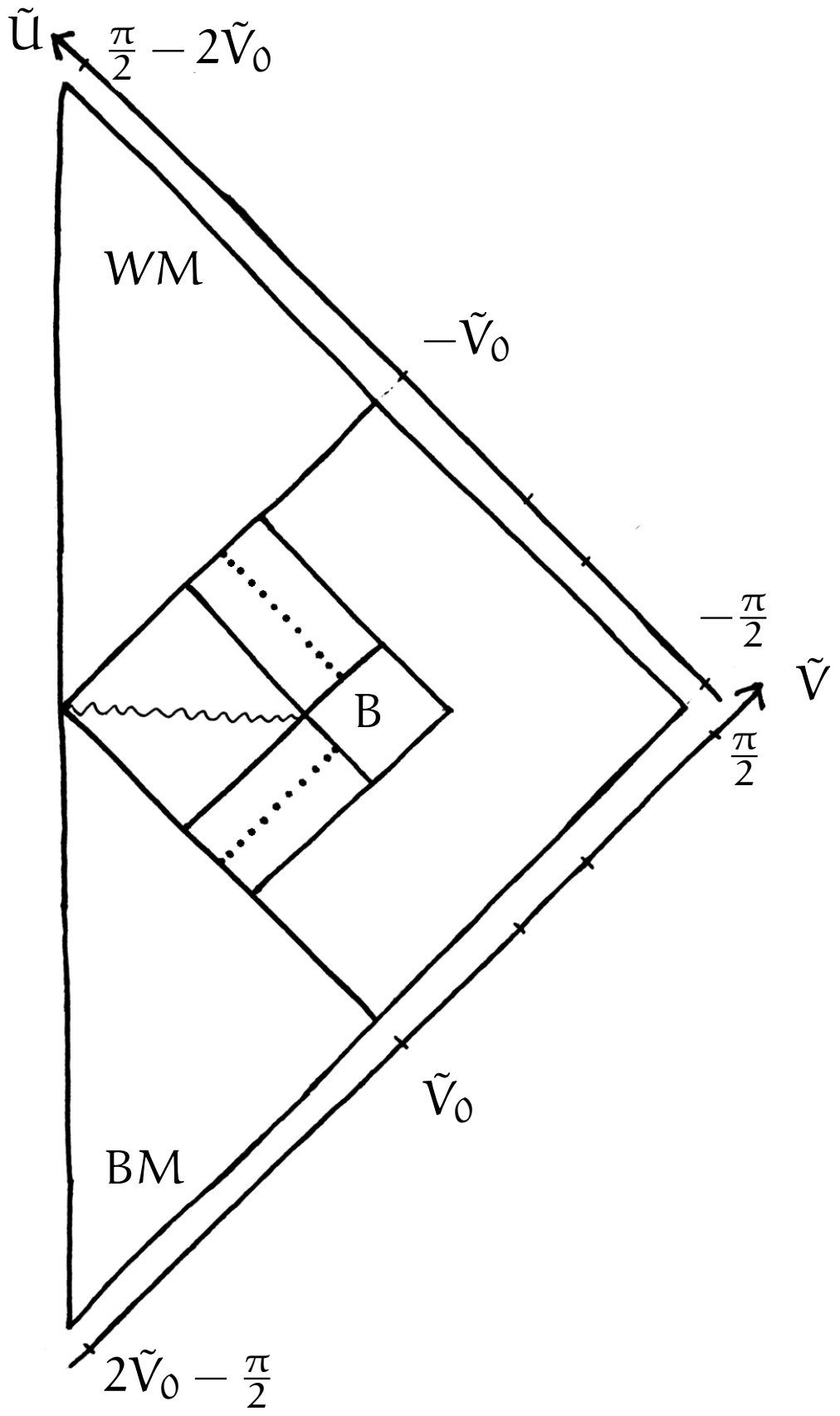

Figure 32: Conformal diagram of the new space-time for fireworks.

The metric outside the shell is Kruskal, described by equations (352), (353), (455), (456), (457) and (458). The metric in the two regions BM and $WM$ is Minkowski, given by equations (344), (345), and respectively, (461) and (463) for BM, and (464) and (462) for $WM$.

Another way to proceed would be to impose

$$\begin{cases} V(\tilde{V}) = \tilde{V} & \text{in BM} \\ U(\tilde{U}) = \tilde{U} & \text{in } WM \end{cases} \quad (465)$$

and then the conformal diagram accordingly looks like figure 33. The only difference is the shape of the line $r = 0$, which is now given by the equation

$$\tan\tilde{V} = v_0 - 4m\left[1 + W\left(-e^{\frac{v_0}{4m}-1}\tan f_B(\tilde{U})\right)\right] \quad (466)$$

in the region BM and the equation

$$\tan \tilde{U} = -v_0 + 4m \left[1 + W\left(-e^{\frac{v_0}{4m}-1} \tan f_B(-\tilde{V})\right)\right] \qquad (467)$$

in the region $WM$.

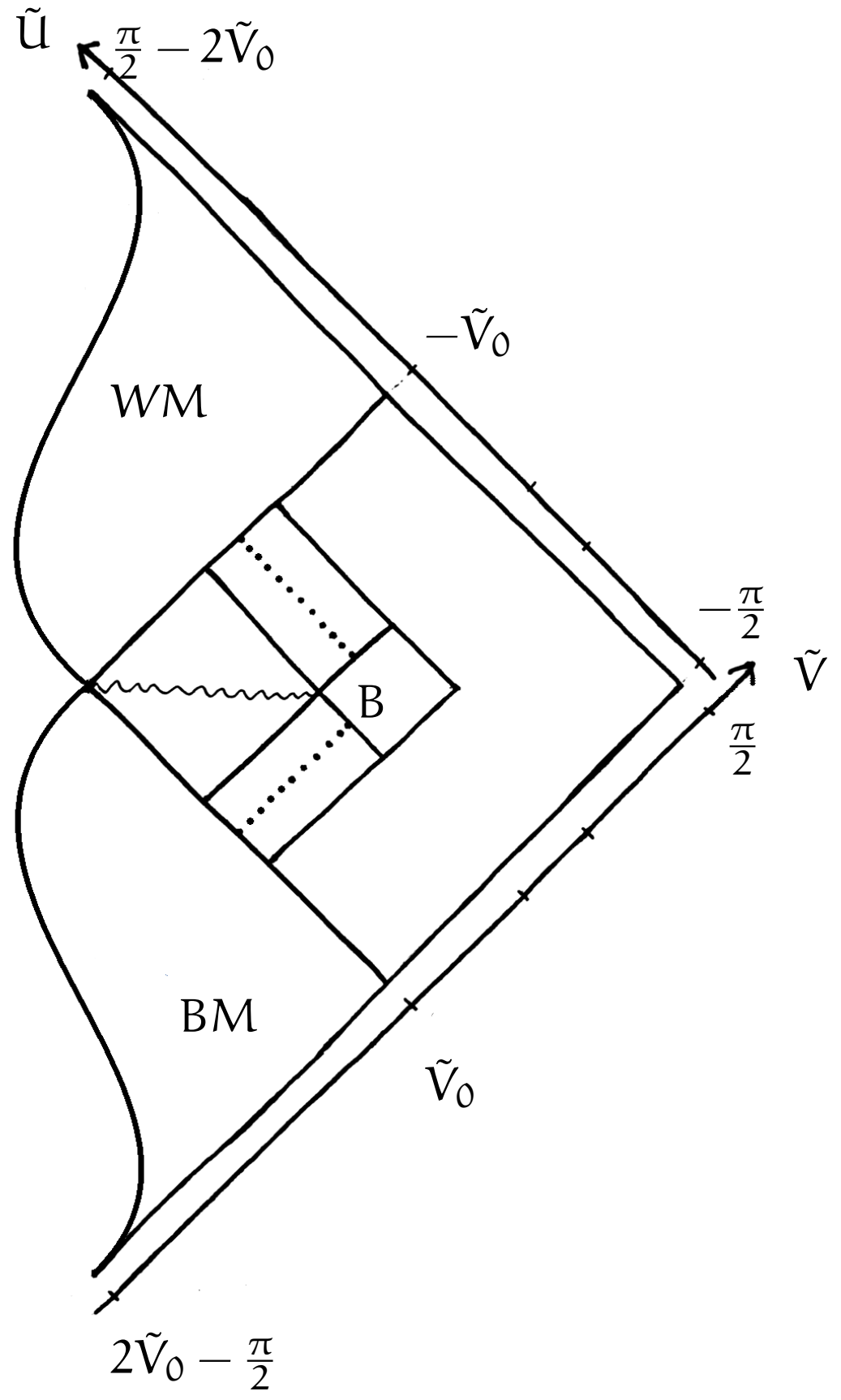

Figure 33: Another possible conformal diagram for fireworks.

We have constructed a complete metric for a space-time describing the tunnelling of a black-hole into a white-hole. It satisfies Einstein equation everywhere except in the small quantum region. Compared to the previous Haggard-Rovelli construction, it introduces two main improvements:

1. the continuity of the transition through the singularity;
2. the resolution of the conical singularity with a diamond B.

However, the instability of Haggard-Rovelli firework metric, pointed out in [53], is not resolved by our metric here. The same analysis could be carried out here, similarly leading to the conclusion that the new metric is still unstable. The resolution of this important flaw will be achieved in the models of chapter 15.

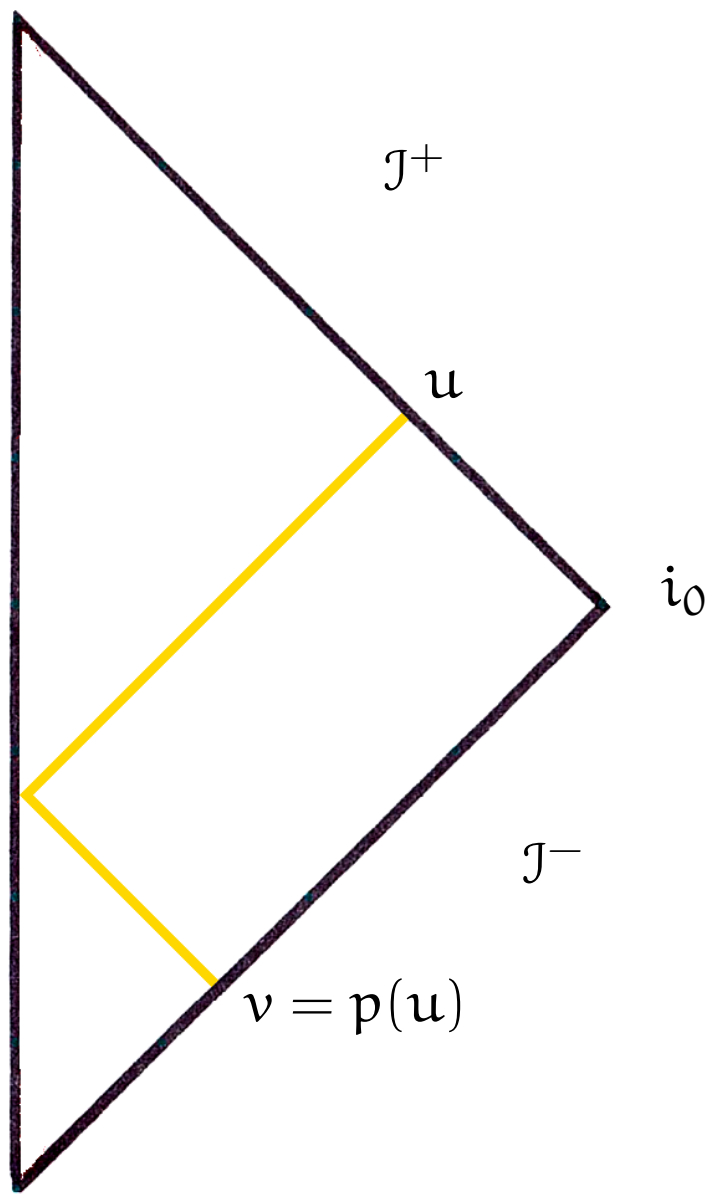

Figure 34: Conformal diagram of an asymptotically flat space-time with null geodesics being reflected at the center.

## 14.6 THE RAY-TRACING MAP

The *ray-tracing map* is the map that sends outgoing light rays of $\mathcal{I}^+$ to incoming ones of $\mathcal{I}^-$, as if they were just reflected backwards by a mirror at the center $r = 0$. Given $u$ and $v$, affine parameters respectively along $\mathcal{I}^+$ and $\mathcal{I}^-$, the ray-tracing map $p$, sends $u$ to $v = p(u)$ so that a null geodesics initiated in $u$ ends up in $v$ (see figure 34).

An advantage of the rigorous construction of conformal diagrams, as advocated in section 10.6, is that it is straightforward to get the ray-tracing map. Indeed, it is directly given by the equation of the line $r = 0$, written in the null-coordinates. Its inverse is

$$p^{-1}(v) = \begin{cases} -4m \log\left[-\tan f_B^{-1}\left(\arctan\left[\left(1 - \frac{v_0 - v}{4m}\right) e^{-v/4m}\right]\right)\right] & \text{if } v \leqslant v_0, \\ \\ -v_0 + 4m\left[1 + W\left(-\tan f_B(-\arctan e^{v/4m}) e^{\frac{v_0}{4m} - 1}\right)\right] & \text{if } v_0 < v. \end{cases} \tag{468}$$

One can check that it is continuous for $v = v_0$ with

$$v_0 = p(-v_0). \tag{469}$$

Usually, the ray-tracing map is defined such that $p(0) = 0$, which is not the case here. It could be easily obtained by a shift of the affine parameters.

The ray-tracing map is plotted in figure 35, for the choice of $f_B$ plotted in figure 30. Unsurprisingly the ray-tracing is symmetric with respect to the line $u = -v$, which was expected from the time-reversal symmetry of the constructed space-time. Moreover, in the limit $v \to \pm\infty$, the ray-tracing map behaves like

$$p^{-1}(v) \sim v - 4m \log \frac{|v|}{4m}, \tag{470}$$

which is also expected from the usual Vaidya space-time. The continuity of the ray-tracing map over all the range of $v$ is an interesting novelty of the metric proposed here compared to the former firework metric, where the ray-tracing map was incomplete around the singularity [31].

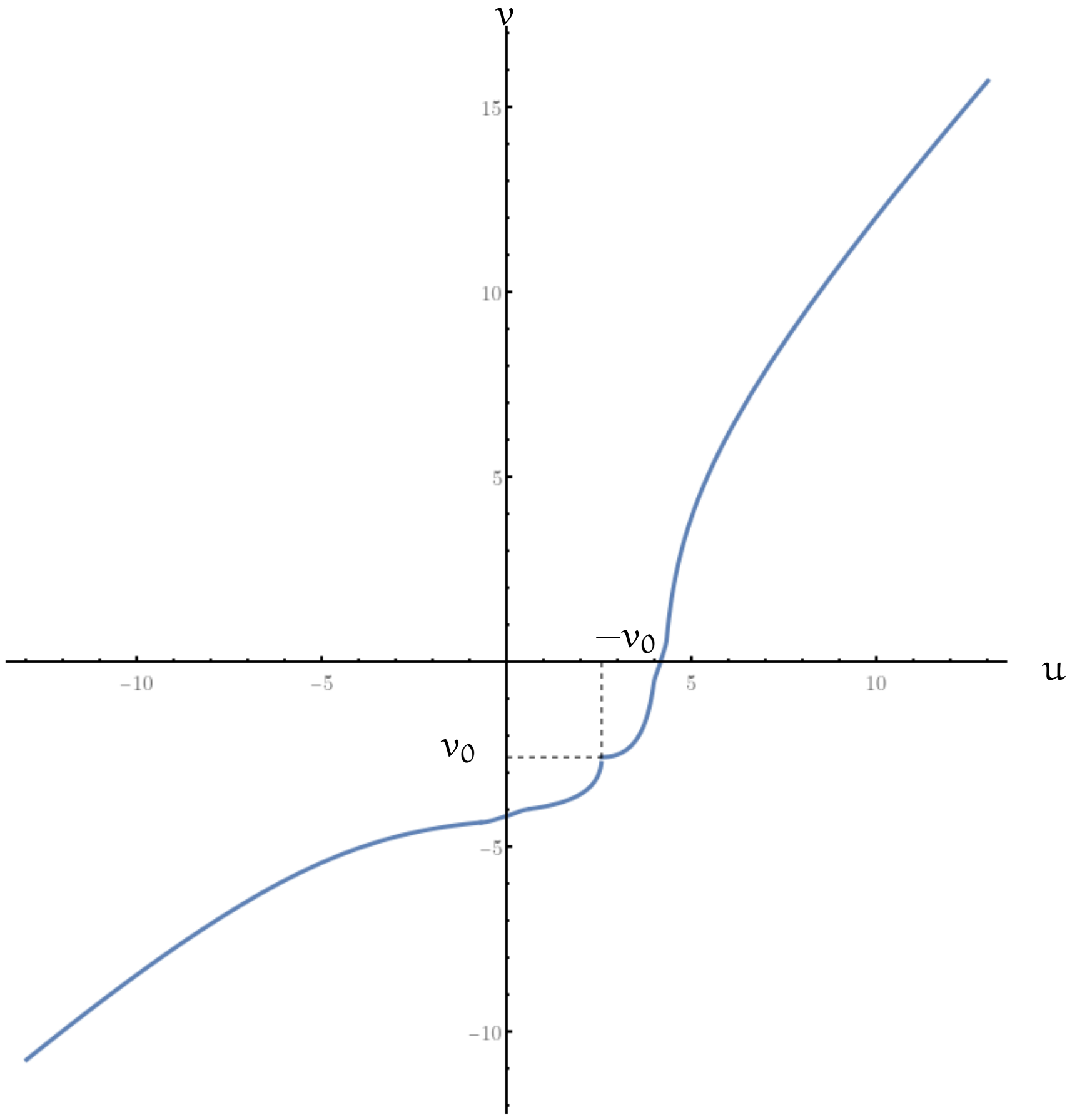

Figure 35: Graph of the ray-tracing map $p(u)$. Here we have chosen the parameters $m = 0.4$, $\tilde{V}_0 = 0.2$, $a = 0.6$ and $b = 1$.

In reference [31], it is shown that the energy flux of the Hawking radiation of a quantum field on a given spherically symmetric metric, can be easily computed, under some approximations, from the ray-tracing map:

$$F(u) = -\frac{1}{24\pi}\left(\frac{\dddot{p}(u)}{\dot{p}(u)} - \frac{3}{2}\frac{\ddot{p}(u)^2}{\dot{p}(u)^2}\right). \tag{471}$$

Similarly, it is easy to compute the *renormalised entanglement entropy* (see [31] for the precise definition) as

$$S(u) = -\frac{1}{12}\log \dot{p}(u). \tag{472}$$

It is a measure of the mixture of the state of the field in the subregion $]-\infty, u]$ of $\mathcal{I}^+$, considering that the state is pure along all $\mathcal{I}^+$.

It is not very interesting to give the analytical expressions, but their graphs are shown in figures 36 and 37. On both graphs a divergence is observed when $u = -v_0$, that is when the null bouncing shell of matter emerges out.

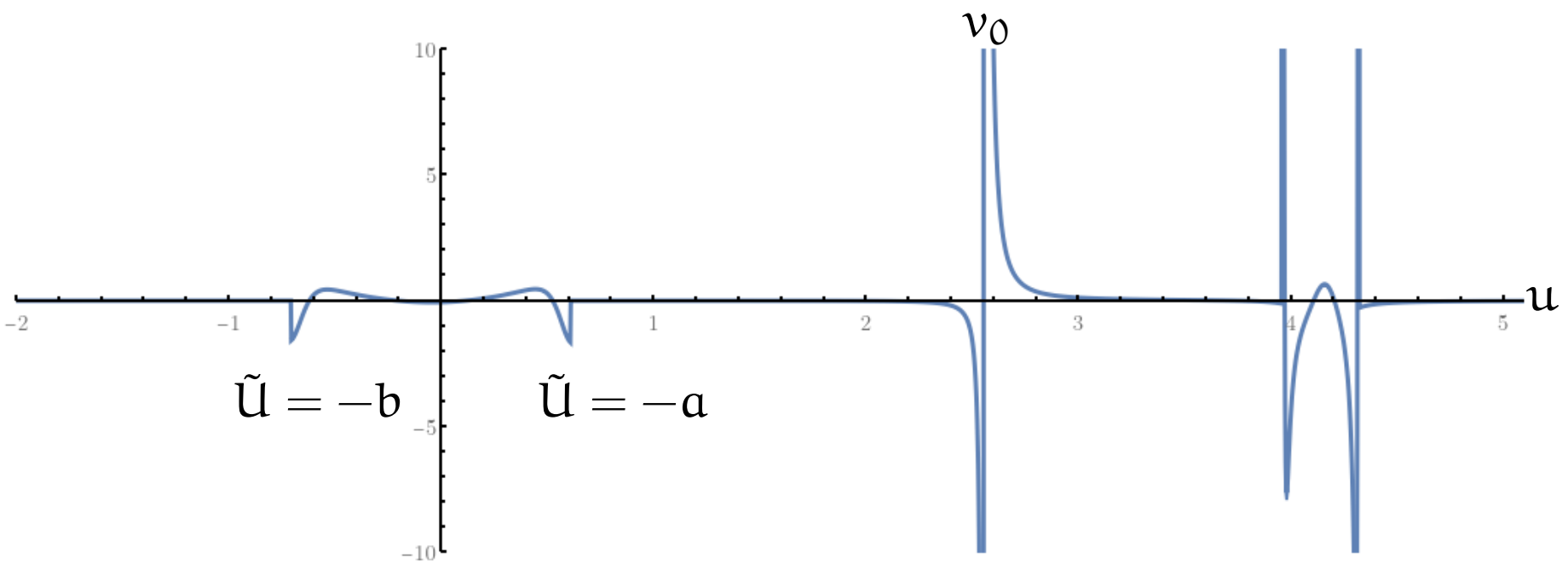

Figure 36: Energy flux $F(u)$.

Before $\tilde{U} = -b$, the flux of energy is positive and slowly growing, as is expected from a usual black hole. After $\tilde{U} = -a$, it is interesting to note that the flux is decreasing and even becomes negative. In other words, negative energy is received from the white hole. At the bounce, $u = -v_0$ a large amount of positive energy is coming out. Then, the situation is symmetric to what happens before the bounce (the high fluctuations in the figure are not divergences, but distortions due to the scale).

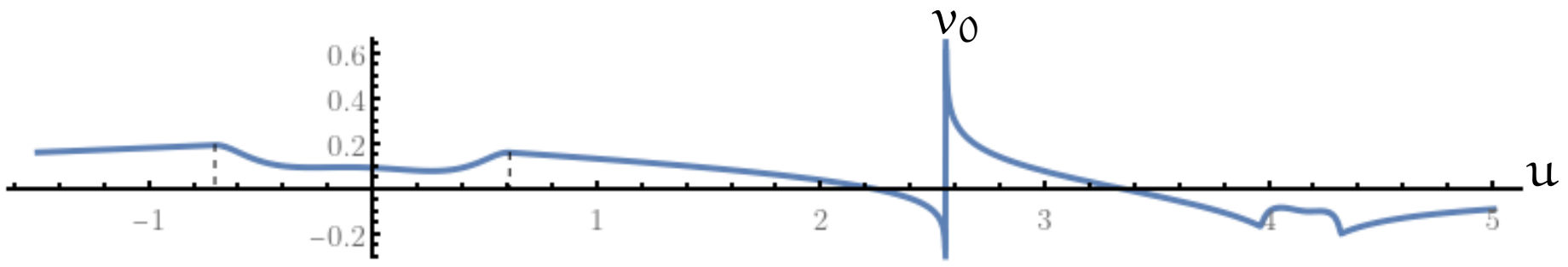

Figure 37: Entanglement entropy $S(u)$.

For the entropy curve, it is noticeable that it comes from $0$ and returns to $0$, which is a sign of a global unitary evolution. In other words, the information is well returned! It is also noticeable that just before the bounce, and later after it, the entropy becomes negative. It means that the quantum fluctuations of the matter field are smaller than the fluctuations of the Minkowski vacuum.

# 15 EVAPORATING BLACK-TO-WHITE HOLE

The black-to-white hole scenario is made possible by the existence of a solution of the classical Einstein equations which is compatible with a black hole undergoing an instantaneous and *local* quantum transition to a white hole [87]. Initially, dimensional arguments suggested that the black hole lifetime $\tau$ behaved, as a function of the initial mass $m$ of the black hole, as

$$\tau \sim m^2. \tag{473}$$

In this case, the tunnelling takes place while the black hole is still macroscopic, and Hawking evaporation can be neglected. However, more rigorous calculations, although debatable for being riddled with approximations, have suggested instead a much longer lifetime, as

$$\tau \sim m e^{m^2}. \tag{474}$$

In such a case, Hawking radiation cannot be neglected anymore and the overall picture changes. In the first stage, Hawking evaporation dominates and the black hole slowly shrinks. The probability of transition to a white hole is about

$$p \sim e^{-m^2}. \tag{475}$$

As the black hole shrinks, the probability of transition increases exponentially, reaching certainty at planckian mass. Thus, the evaporation of a black hole finally results in a planckian white hole! This alternative scenario was first introduced in [37]. Interestingly, it suggests that Planck-mass white holes may be nothing else but long-lived remnants, advocated long before as candidates to the resolution of the information paradox (see section 12.3)!

In this chapter, mainly inspired from [124], we construct and discuss the form that this effective space-time geometry can take. Steps in this direction were taken in [53, 87] and [160], but a crucial element was not taken into account: the Hawking radiation and its back-reaction. Here we improve on the understanding of the physics of the black-to-white hole transition by discussing possible ways of modelling the Hawking radiation and its back-reaction.

In section 15.1, we propose a toy model for an evaporating black-to-white hole, which is then improved, in section 15.2, by a careful study of the evolution of the ingoing Hawking quanta beyond the singularity. In section 15.3, we motivate another possible model describing the evolution of outgoing quanta, and compare it to the previous one.

## 15.1 TOY MODEL

We start with a first naive attempt to build a concrete model for an evaporating black-to-white hole. The initial idea is simple: start with a model of an evaporating black hole, like the one of Hiscock, described in section 11.5, and glue a white hole above the singularity, with an outgoing bouncing null shell, as described in section 14.5.

This can be done easily provided that the Bondi-Sachs mass $M(u)$, observed on $\mathcal{I}^+$, represented in figure 15, does not vanish completely but reaches a small positive value $m_1$. We obtain the conformal diagram of figure 38.

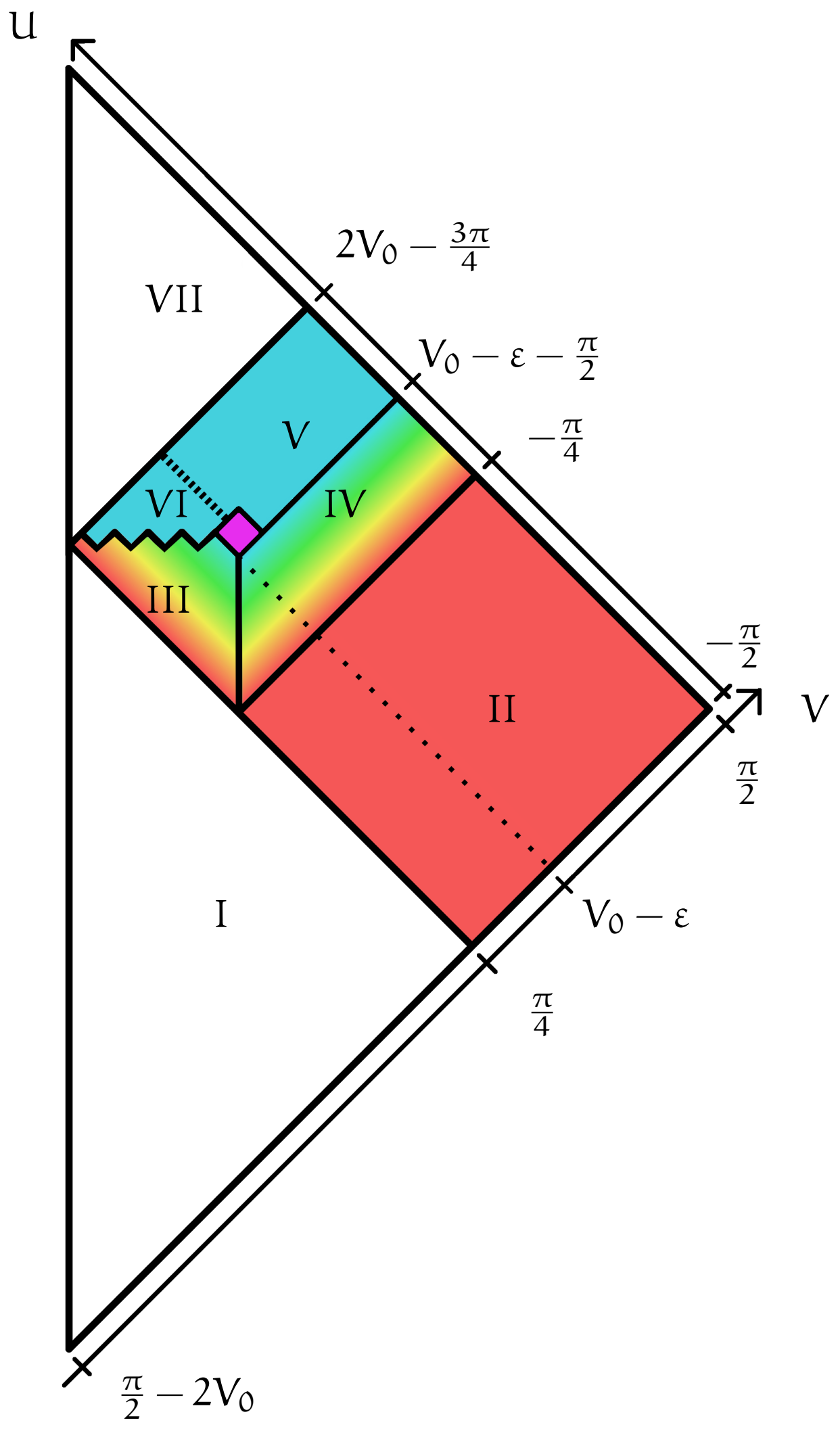

Figure 38: Conformal diagram of a toy model of an evaporating black-to-white hole.

In region I, II, III, IV the metric is the same as the model of Hiscock (see equations (394)-(397) and (399)-(404)). Elsewhere, the metric is given by

$$(\mathrm{V}) \left[ \begin{array}{l} ds^2 = -\left(1 - \frac{2m_1}{r}\right) du dv + r^2 d\Omega^2 \\ r = 2m_1 \left(1 + W\left(e^{\frac{v-u}{4m_1} - 1}\right)\right) \end{array} \right. \tag{476}$$

$$(\mathrm{VI}) \left[ \begin{array}{l} ds^2 = \left(1 - \frac{2m_1}{r}\right) du dv + r^2 d\Omega^2 \\ r = 2m_1 \left(1 + W\left(-e^{-\frac{v+u}{4m_1} - 1}\right)\right) \end{array} \right. \tag{477}$$

$$(\mathrm{VII}) \left[ \begin{array}{l} ds^2 = -du dv + r^2 d\Omega^2 \\ r = \frac{1}{2}(v - u) \end{array} \right. \tag{478}$$

$\varepsilon$ is defined so that the radius at the future endpoint of the apparent horizon is $2m_1$, i.e. $h(V_0 - \varepsilon - \pi/2, V_0 - \varepsilon) = 2m_1$, where $h(U, V)$ is defined in equation (404). For the sake of completeness, we also give explicitly the map between the coordinates $(u, v)$ of the metric and $(U, V)$ of the diagram, which demand the labelling of regions as shown in figure 39 :

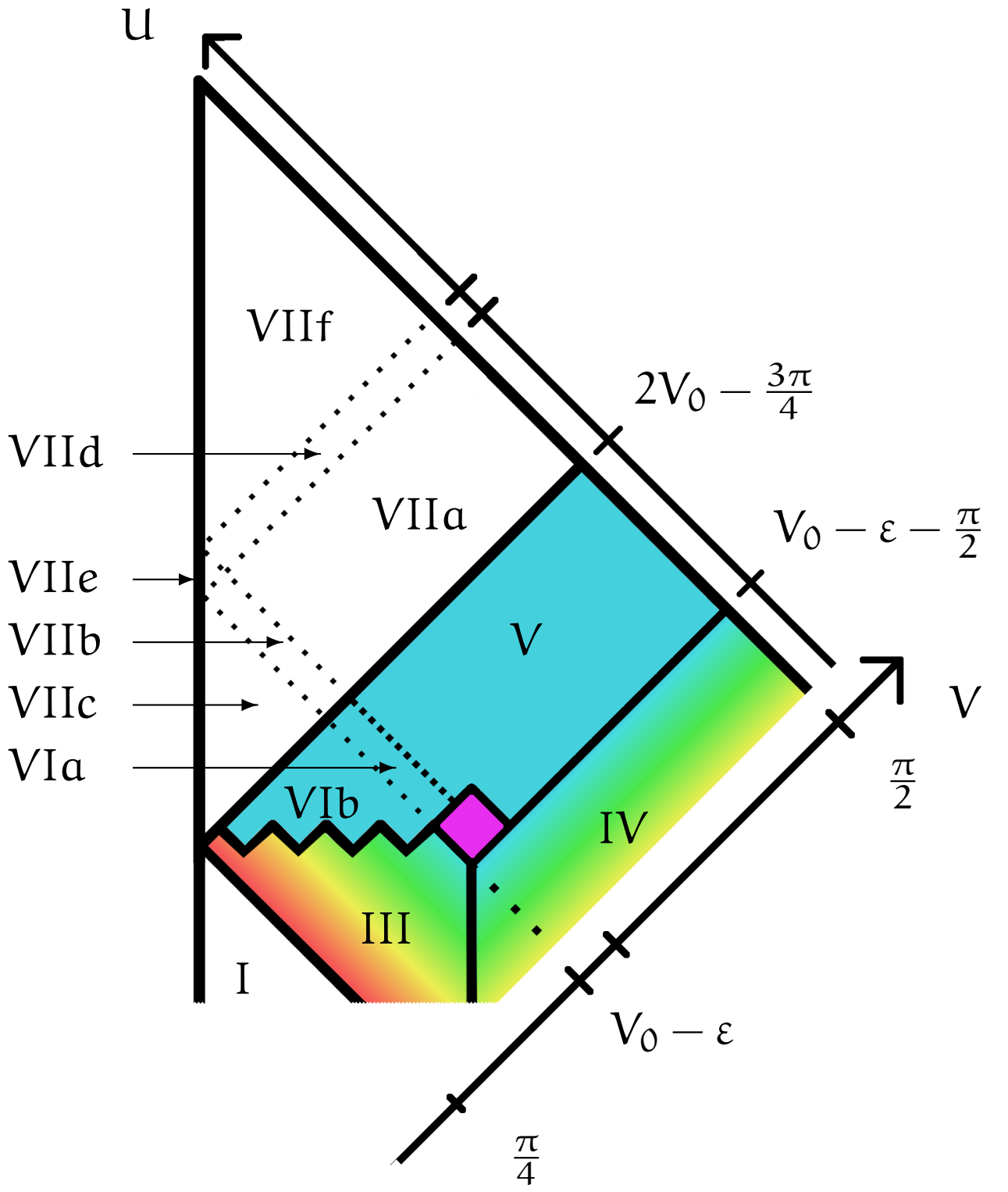

Figure 39: Subdivided close-up of the conformal diagram of a toy model of an evaporating black-to-white hole.

$$(\mathrm{V})\quad\left[\begin{array}{l} u = f_4(U) \text{ increasing, such that} \\ \quad f_4(V_0-\varepsilon-\pi/2) = M^{-1}(N(f_2(V_0-\varepsilon))) \\ v = f_4(V_0-\varepsilon-\pi/2) + 2h(V_0-\varepsilon-\pi/2, V) \\ \qquad\qquad +4m_1 \log\left(\frac{h(V_0-\varepsilon-\pi/2,V)}{2m_1}-1\right) \end{array}\right. \tag{479}$$

$$(\mathrm{VIa})\quad\left[\begin{array}{l} u = f_5(U) \text{ increasing} \\ v = -f_5(V_0-\varepsilon-\pi/2) - 2h(V_0-\varepsilon-\pi/2, V) \\ \qquad\qquad -4m_1 \log\left(1-\frac{h(V_0-\varepsilon-\pi/2,V)}{2m_1}\right) \end{array}\right. \tag{480}$$

$$(\mathrm{VIb})\quad\left[\begin{array}{l} u = c_5 + f_5(U) \\ v = -c_5 - f_5(2V_0-\pi/2-V) \end{array}\right. \tag{481}$$

$$(\mathrm{VIIa})\quad\left[\begin{array}{l} u = f_4(2V_0-3\pi/4) \\ +4m_1\left(1+W(-e^{-\frac{f_5(2V_0-3\pi/4)-f_5(4V_0-3\pi/2-U)}{4m_1}-1})\right) \\ v = f_4(2V_0-3\pi/4) \\ \qquad +4m_1\left(1+W((\frac{h(V_0-\varepsilon-\pi/2,V)}{2m_1}-1)\right. \\ \qquad \left.\times\, e^{\frac{f_4(V_0-\varepsilon-\pi/2)-f_4(2V_0-3\pi/4)}{4m_1}+\frac{h(V_0-\varepsilon-\pi/2,V)}{2m_1}-1})\right) \end{array}\right. \tag{482}$$

$$(\mathrm{VIIb})\quad\left[\begin{array}{l} u = c_6 + 4m_1 W\left(-e^{-\frac{f_5(2V_0-3\pi/4)-f_5(4V_0-3\pi/2-U)}{4m_1}-1}\right) \\ v = c_6 + 4m_1 W\left((\frac{h(V_0-\varepsilon-\pi/2,V)}{2m_1}-1)\right. \\ \left.\times\, e^{\frac{f_5(V_0-\varepsilon-\pi/2)-f_5(2V_0-3\pi/4)}{4m_1}+\frac{h(V_0-\varepsilon-\pi/2,V)}{2m_1}-1}\right) \end{array}\right. \tag{483}$$

$$(\mathrm{VIIc})\quad\left[\begin{array}{l} u = c_7 + 4m_1 W\left(-e^{-\frac{f_5(2V_0-3\pi/4)-f_5(4V_0-3\pi/2-U)}{4m_1}-1}\right) \\ v = c_7 + 4m_1 W\left(-e^{-\frac{f_5(2V_0-3\pi/4)-f_5(2V_0-\pi/2-V)}{4m_1}-1}\right) \end{array}\right. \tag{484}$$

$$(\mathrm{VIId})\quad\left[\begin{array}{l} u = f_4(2V_0-3\pi/4) \\ \qquad +4m_1\left(1+W((\frac{h(V_0-\varepsilon-\pi/2,U-2V_0+\pi)}{2m_1}-1)\right. \\ \left.\times\, e^{\frac{f_5(V_0-\varepsilon-\pi/2)-f_5(2V_0-3\pi/4)}{4m_1}+\frac{h(V_0-\varepsilon-\pi/2,U-2V_0+\pi)}{2m_1}-1}\right) \\ v = f_4(2V_0-3\pi/4) \\ \qquad +4m_1\left(1+W((\frac{h(V_0-\varepsilon-\pi/2,V)}{2m_1}-1)\right. \\ \qquad \left.\times\, e^{\frac{f_4(V_0-\varepsilon-\pi/2)-f_4(2V_0-3\pi/4)}{4m_1}+\frac{h(V_0-\varepsilon-\pi/2,V)}{2m_1}-1})\right) \end{array}\right. \tag{485}$$

$$\text{(VIIe)}\left[\begin{array}{l} u = c_8 + 4m_1 W\left(\left(\frac{h(V_0-\varepsilon-\pi/2,U-2V_0+\pi)}{2m_1}-1\right)\right.\\ \left.\times\, e^{\frac{f_5(V_0-\varepsilon-\pi/2)-f_5(2V_0-3\pi/4)}{4m_1}+\frac{h(V_0-\varepsilon-\pi/2,U-2V_0+\pi)}{2m_1}-1}\right)\\ v = c_8 + 4m_1 W\left(\left(\frac{h(V_0-\varepsilon-\pi/2,V)}{2m_1}-1\right)\right.\\ \qquad \left.\times\, e^{\frac{f_5(V_0-\varepsilon-\pi/2)-f_5(2V_0-3\pi/4)}{4m_1}+\frac{h(V_0-\varepsilon-\pi/2,V)}{2m_1}-1}\right)\end{array}\right. \tag{486}$$

$$\text{(VIIf)}\left[\begin{array}{l} u = f_4(2V_0-3\pi/4)\\ \qquad +4m_1\left(1+W\left(\left(\frac{h(V_0-\varepsilon-\pi/2,U-2V_0+\pi)}{2m_1}-1\right)\right.\right.\\ \left.\left.\times\, e^{\frac{f_4(V_0-\varepsilon-\pi/2)-f_4(2V_0-3\pi/4)}{4m_1}+\frac{h(V_0-\varepsilon-\pi/2,U-2V_0+\pi)}{2m_1}-1}\right)\right)\\ v = f_4(2V_0-3\pi/4)\\ \qquad +4m_1\left(1+W\left(\left(\frac{h(V_0-\varepsilon-\pi/2,V)}{2m_1}-1\right)\right.\right.\\ \qquad \left.\left.\times\, e^{\frac{f_4(V_0-\varepsilon-\pi/2)-f_4(2V_0-3\pi/4)}{4m_1}+\frac{h(V_0-\varepsilon-\pi/2,V)}{2m_1}-1}\right)\right)\end{array}\right. \tag{487}$$

Alright, such a work may seem a bit Stakhanovist, but like house-cleaning, it is necessary and you feel satisfied when it's done.

We have not given an explicit expression for the metric in the central purple diamond, but it would a priori be possible to construct one that matches the boundary conditions around. As in the previous models of chapter 14, it is believed to be a region where quantum effects happen to enable the tunnelling to the white hole. Thus, it would be better described by a quantum geometry, instead of any effective classical metric. The novelty with respect to previous models like [160] is that the region can be very small, typically planckian.

From the perspective of an observer lying on $\mathcal{I}^+$, the Bondi-Sachs mass evolves as depicted on the mass profile of figure 40. It is positive and decreasing all along, going from $m$ to $0$. The white hole manifests itself through a sudden final release of positive energy corresponding to the emergence of the null bouncing shell. In region III the inside Hawking quanta, which carry negative energy, fade over the singularity, and never show up on the other side.

In [35, 36], it was shown that unitary evolution of an evaporating black hole implies a non-monotonic mass loss. To put it differently, a black hole must, at some point, radiate some amount of negative energy (the 'last gasp'), which would be depicted on the mass profile as a momentary increase of the Bondi-Sachs mass. Intuitively, we can understand that the Hawking quanta, that fell inside the black hole, *with negative energy*, are correlated with quanta outside, and should thus come out at some point, to recover unitarity on $\mathcal{I}^+$. The profile of figure 40 does not fulfil the 'last gasp' requirement. Remember from equation (371), the flux of outgoing energy along $\mathcal{I}^+$ is

$$F(u) \propto -\frac{dM(u)}{du}, \tag{488}$$

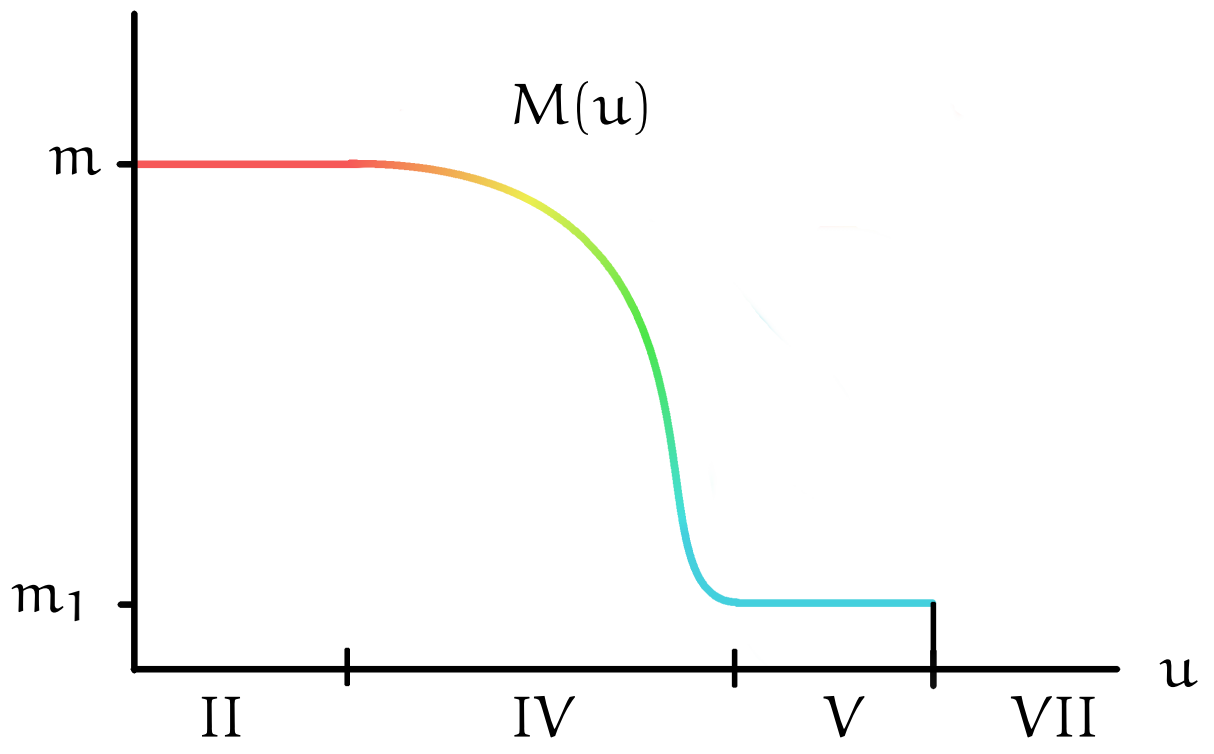

Figure 40: Bondi-Sachs mass function along $\mathcal{I}^+$ for the toy model of evaporating black-to-white hole.

so that a momentary negative energy flux would mean a momentary increase of the Bondi-Sachs mass function. However, preliminary discussions of De Lorenzo and Bianchi (personal communication), suggest that the last gasp theorem may fail in 4D, in which case the mass profile of figure 40 should not be discarded too easily.

Nevertheless, there is another reason why the previous model is not physically satisfying. For simplicity of the construction, we have assumed that the ingoing negative energy fades along the singularity. Quantum gravity results suggest instead that it should cross the singularity. This calls for a refinement of our first toy model.

## 15.2 MODEL I

To do so, we consider that the Hawking quanta cross the singularity. In subsection 14.5.2, we have already noticed that there exists a natural prescription to extend geodesics beyond a singularity. Thus, modelling the crossing of the Hawking quanta through the singularity is the easy part of the refinement. It becomes more intricate afterwards. The negative energy flux is still ingoing, so it will fall upon the emerging bouncing shell. What comes next?

### 15.2.1 *Across the bouncing shell*

The crossing between two null shells has been studied by Dray and t'Hooft in [56]. Their main result is that it is possible to glue four Schwarzschild patches along two null shells (see figure 41), provided that the four masses satisfy the only condition

$$(r_0 - 2m_1)(r_0 - 2m_2) = (r_0 - 2m_3)(r_0 - 2m_4) \tag{489}$$

where $r_0$ is the radius at the intersection.

To a first approximation, the ingoing flux, which was previously modelled continuously by a function $N(v)$, can be approached by a

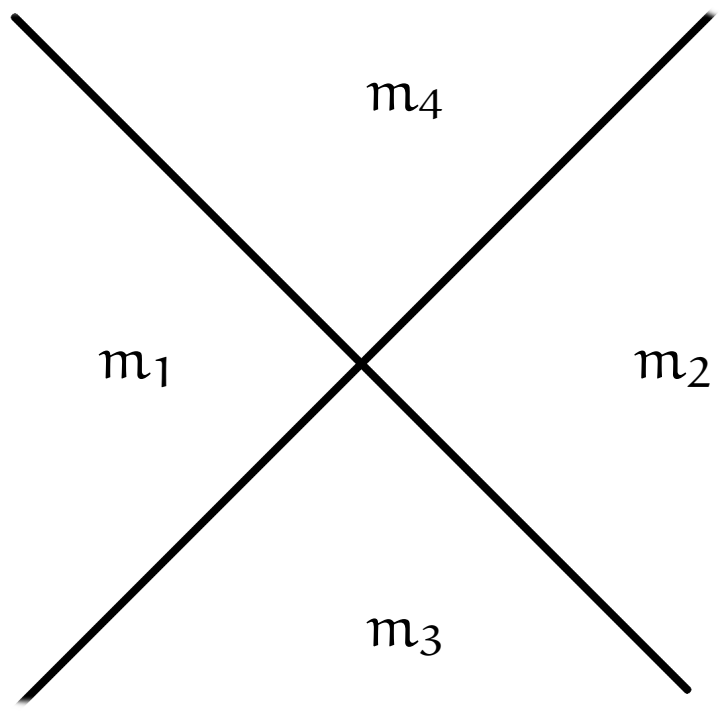

Figure 41: Four Schwarzschild patches can be glued consistently along null geodesics provided the masses satisfy equation (489).

step function made of a number $n$ of slices of constant masses $N_i$. Then, the negative energy is carried by individual Hawking quanta which fall one at a time upon the bouncing shell. The situation is depicted in figure 42 for $n = 5$. In each box, the metric is Schwar-

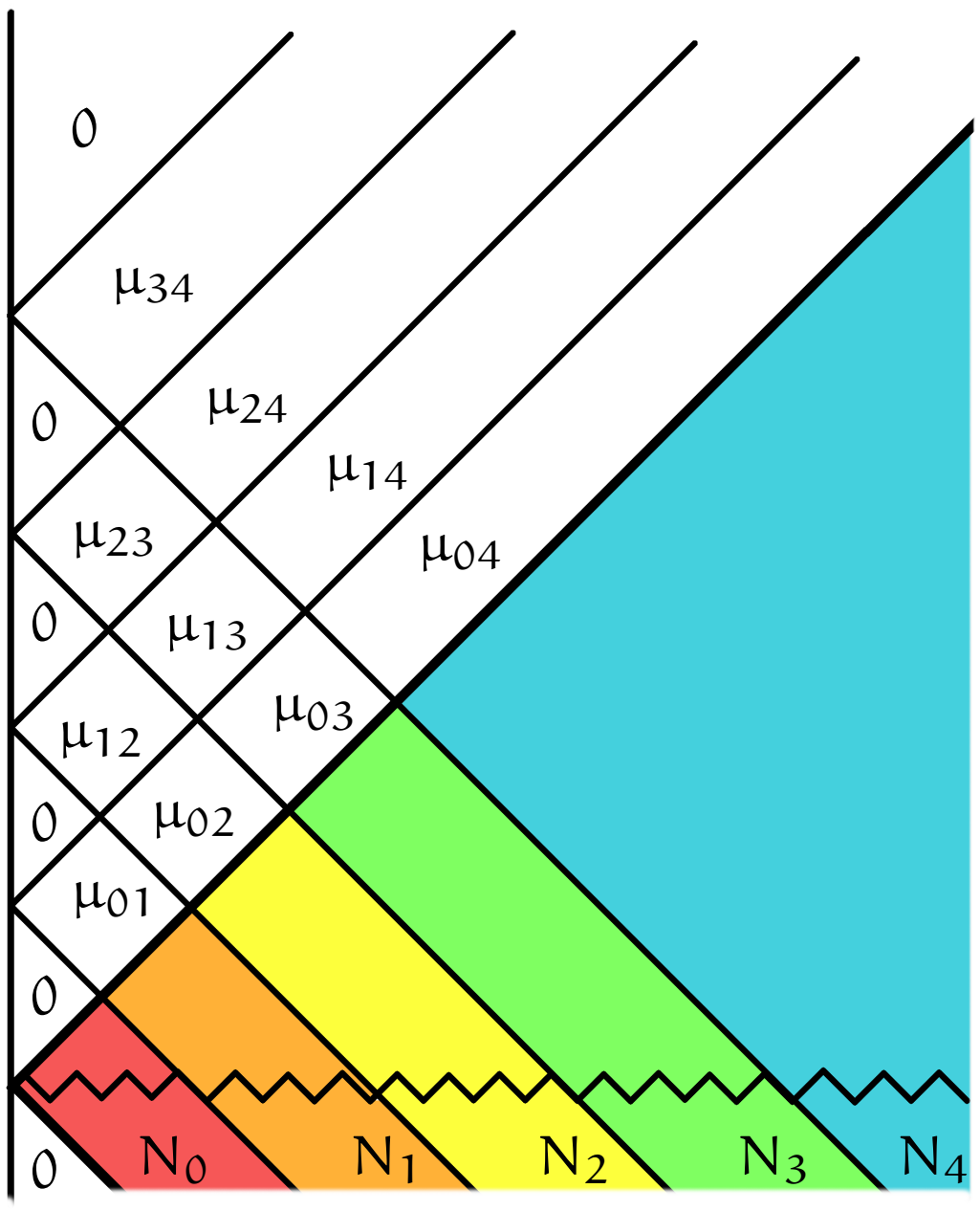

Figure 42: Discrete model for the crossing between the ingoing Hawking quanta and the bouncing shell.

zschild with a constant mass $\mu_{ij}$, which is determined by equation (489) as a function of the three masses in the adjacent boxes below and the value of the radius where the four regions touch. We prove the following theorem:

THEOREM. *Assume that the radius $r$ in the region above the bouncing shell is increasing (resp. decreasing) along outgoing (resp. ingoing) null geodesics. Then $\mu_{ij}$ is a decreasing function of $i$ and an increasing function of $j$. As a result, we have for all $i$ and $j$,*

$$0 < \mu_{ij} < m_1. \tag{490}$$

PROOF. Let us first study $\mu_{0j}$ for varying $j$. Denote $r_j$ the radius at the intersection point at the bottom of the box of mass $\mu_{0j}$. From equation (489), we deduce

$$(\mu_{0,j} - \mu_{0,j-1})(r_j - 2N_j) = (N_j - N_{j-1})(r_j - 2\mu_{0,j}) \tag{491}$$

Since $N_j$ is decreasing, we have, in the RHS, $(N_j - N_{j-1}) < 0$. Then, since the radius $r_j$ is assumed to be increasing, we have $r_j \leqslant 2m_1 \leqslant 2N_j$, so that in the LHS, $(r_j - 2N_j) < 0$. So,

$$\mu_{0,j} > \mu_{0,j-1} \Leftrightarrow r_j > 2\mu_{0,j}. \tag{492}$$

which can be restated saying that for each $j$, one, and only one of the two following must hold:

$$\begin{aligned} \mu_{0,j-1} < \mu_{0,j} < \tfrac{r_j}{2} \\ \mu_{0,j-1} > \mu_{0,j} > \tfrac{r_j}{2}. \end{aligned} \tag{493}$$

Initially, we have $r_0 = 0$. Since $r_1 > 0$ we deduce

$$0 < \mu_{01} < \frac{r_1}{2}. \tag{494}$$

Then, using that $r_{j+1} > r_j$, we show by induction that for any $j$

$$\mu_{0,j-1} < \mu_{0,j} < \frac{r_j}{2}. \tag{495}$$

Thus $\mu_{0j}$ is increasing with $j$ and satisfies

$$0 < \mu_{0j} < m_1. \tag{496}$$

A similar induction shows that $\mu_{1j}$ is also an increasing function of $j$, satisfying.

$$0 < \mu_{1j} < m_1. \tag{497}$$

Then, under the assumption of decreasing $r$ along ingoing null geodesics, an induction over $i$ shows that for any $j$, $\mu_{ij}$ is a decreasing function of $i$. □

The previous discrete model gives a fair description of what can happen when a series of Hawking quanta successively cross the bouncing shell. In the continuum limit, when $n \to \infty$, the resulting metric takes the form

$$ds^2 = -\left(1 - \frac{2\mu(u,v)}{r}\right) du dv + r^2 d\Omega^2 \tag{498}$$

characterised by two functions, namely the radius $r(u,v)$ and the mass $\mu(u,v)$. We cannot give explicitly the change of coordinates from $(u,v)$ to $(U,V)$ but we assume that $v(V)$ and $u(U)$ are increasing. Then, our theorem shows that

$$\frac{\partial \mu}{\partial u} < 0 \qquad \text{and} \qquad \frac{\partial \mu}{\partial v} > 0. \tag{499}$$

As a corollary we have

$$0 < \mu(u,v) < m_1. \tag{500}$$

We have no explicit expression neither for the radius $r(u,v)$ nor for the mass $\mu(u,v)$, for it would require to integrate too difficult equations. However, it is clear for the construction of the discrete setting above that such functions exist.

### 15.2.2 *Conformal diagram I*

To sum-up, the resulting space-time is depicted in figure 43, with the metric given by

$$(\mathrm{V}) \left[ \begin{array}{l} ds^2 = -\left(1 - \frac{2m_1}{r}\right) du dv + r^2 d\Omega^2 \\ r = 2m_1 \left(1 + W\left(e^{\frac{v-u}{4m_1} - 1}\right)\right) \end{array} \right. \tag{501}$$

$$(\mathrm{VIa}) \left[ \begin{array}{l} ds^2 = \left(1 - \frac{2m_1}{r}\right) du dv + r^2 d\Omega^2 \\ r = 2m_1 \left(1 + W\left(-e^{-\frac{v+u}{4m_1} - 1}\right)\right) \end{array} \right. \tag{502}$$

$$(\mathrm{VIb}) \left[ \; ds^2 = -\left(1 - \frac{2\tilde{N}(v)}{r}\right) dv^2 + 2 dv dr + r^2 d\Omega^2 \right. \tag{503}$$

$$(\mathrm{VII}) \left[ \; ds^2 = -\left(1 - \frac{2\mu(u,v)}{r}\right) du dv + r^2 d\Omega^2 \right. \tag{504}$$

$$(\mathrm{VIII}) \left[ \; ds^2 = -\left(1 - \frac{2P(u)}{r}\right) du^2 - 2 du dr + r^2 d\Omega^2 \right. \tag{505}$$

$$(\mathrm{IX}) \left[ \begin{array}{l} ds^2 = -du dv + r^2 d\Omega^2 \\ r = \frac{1}{2}(v - u) \end{array} \right. \tag{506}$$

In regions $\mathrm{I} - \mathrm{IV}$ the metric is the same as the model of Hiscock (see equations (394)-(397) and (399)-404). The mass function $\tilde{N}(v)$ that appears in the metric of region VIb is chosen to match the mass function $N(v)$ along the boundary III/VIb. Similarly, the mass function $P(u)$ of region VIII is chosen to match $\mu(u,v)$ along the boundary VII/VIII. The map between the coordinates $(u,v)$ and $(U,V)$ cannot be given explicitly.

### 15.2.3 *Planckian quantum region*

A word shall be added concerning the size of the central diamond region. The future endpoint of the apparent horizon of the black hole has a radius $r = 2m_1$, which characterises the typical size of the diamond. The mathematical construction of the model requires that $0 < m_1 < m$. However, physically, $m_1$ is believed to be small. How small? Well, remember that in quantum gravity the singularity is expected to be actually a 'thick' singularity, i.e. a Planck star whose radius is given by $r \sim N(v)^{1/3}$ (see subsection 14.2.2). So a Planck star can actually be quite big. Now, along the apparent horizon, the radius is given by $r \sim 2N(v)$. Then, evaporation can last at most until the 'thick singularity' and the apparent horizon meet, i.e. when $m_1^{1/3} \sim 2m_1$. This condition means the mass $m_1$ should be planckian. Without surprise, it is a planckian $m_1$ that marks a lower bound for our model. In this case, the size of the diamond itself is planckian, so it is really just 'one quantum of space'.

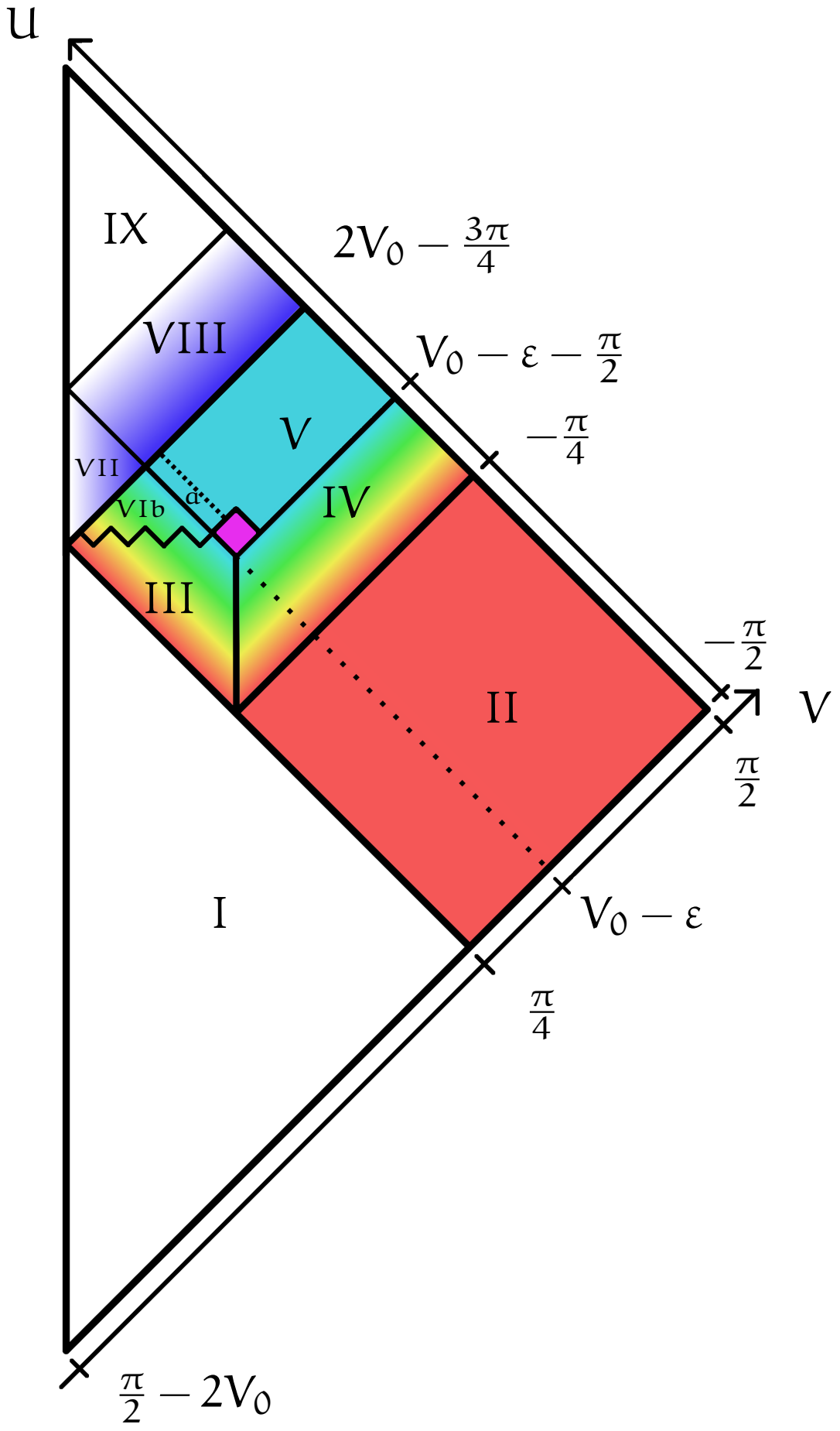

Figure 43: Conformal diagram of an evaporating black-to-white hole with ingoing energy flux that crosses first the singularity and then the bouncing shell. The dashed boundary V/VIa represents the apparent horizon of the white hole, characterised by $r = 2m_1$.

### 15.2.4 *Positive energy flux along $\mathcal{I}^+$*

The resulting mass profile along $\mathcal{I}^+$ is shown in figure 44. Instead of a sharp release of energy when the shell bounces out, as in the previous toy model (see figure 40), the Bondi-Sachs mass slowly decreases to zero. It can be interpreted as the emergence of the Hawking quanta that finally reach $\mathcal{I}^+$. It should be noticed however that they carry positive energy (since the Bondi-Sachs mass is decreasing all along), whereas they were known to carry negative energy after they formed at the apparent horizon. This change of sign is due to the exchange of energy that occurs when the quanta cross the bouncing shell: positive energy from the shell is transferred to the quanta. The final long-dying tail on the mass profile enables energy (and information) to be slowly released.

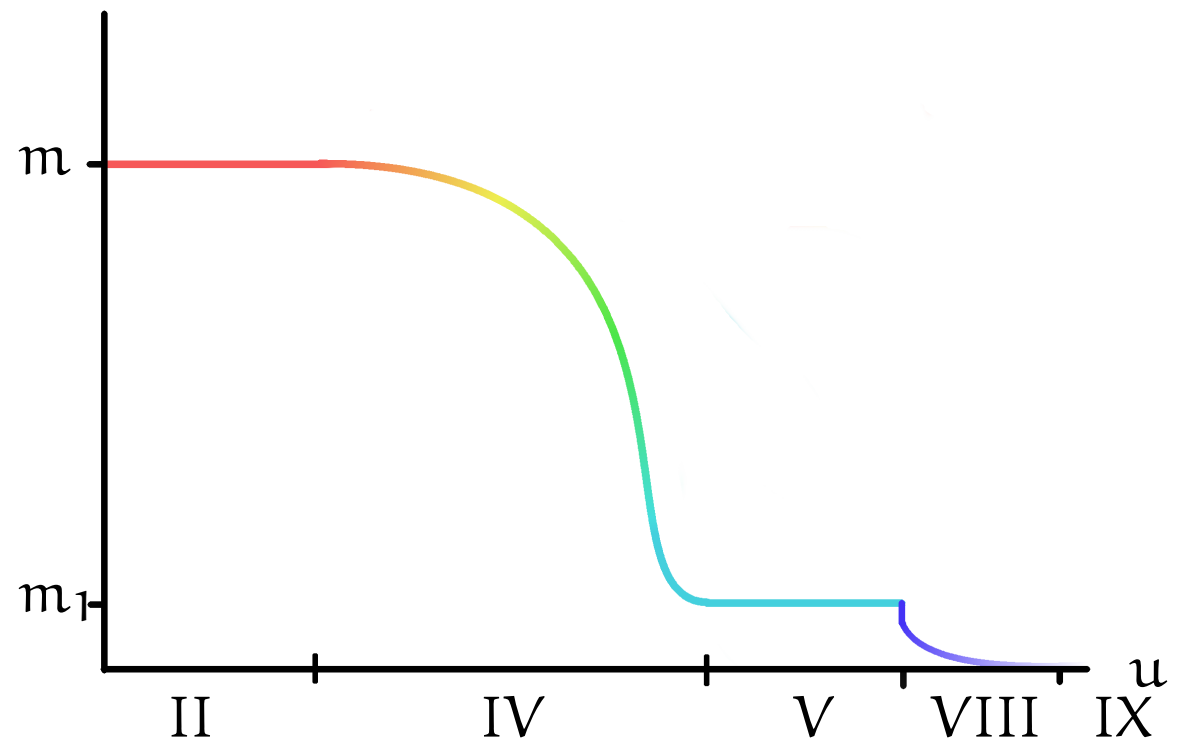

Figure 44: Bondi-Sachs mass function along $\mathcal{I}^+$ for an evaporating black-to-white hole (model I).

### 15.2.5 *Stability*

Starting from a macroscopic black hole, the tunnelling happens when the black hole has reached planckian size. The evaporation takes a very long time of order $m^3$. As we have already mentioned in chapter 12, an old Planck-mass black hole is not the same as a young Planck-mass black hole. From the outside, it is indeed the same. But as it grows older, the interior volume of a black hole grows too. After a long time of evaporation, our Planck-mass black hole thus has a huge volume. In that respect, they are very much like the cornocupions or horned particles of [13]. After the transition, the resulting white hole can thus be seen as a long-lived remnant. The stability of long-lived remnants makes no doubt, but the stability of white holes does a lot, so how is it now?

The lifetime of long-lived remnants is about $m^4$, which is the time necessary for the information to be released. This is much bigger than the lifetime of black holes $m^3$. The newly built model is thus time-asymmetric. In [53], it was pointed out that the instability of the Haggard-Rovelli metric could be cured in an asymmetric scenario. However, the asymmetry proposed in [53] is the inverse to what we find here: they suggest the lifetime of white holes to be much shorter to that of black holes...

Interestingly, the instability problem is still cured in this case. The stability of Planck-mass white holes is discussed in [163]. The main idea is that the proof of white hole instability does not hold anymore in the case of Planck-mass white holes. Indeed, in the macroscopic case, the instability arises due to the blueshift amplification of tiny quantum perturbations (see chapter 13). In our case, the hole itself is planckian, and it seems fair to contest the existence of transplanckian perturbations. Thus is saved the life of white holes.

## 15.3 MODEL II

### 15.3.1 *Inside outgoing flux*

As can be seen from equations (388), Hawking flux is outgoing even inside the black hole. In other words, Hawking quanta are well falling towards the singularity, but they are *out*-falling, i.e. falling along outgoing null geodesics. This has led some people to doubt the credibility of the previous Hiscock model, where the correction inside the hole only corresponds to an *in*-falling negative energy flux. Nevertheless, this objection is not correct because the iterative approach to the semi-classical Einstein equations requires to consider the full $\langle in|T_{ab}|in\rangle$, including both the Hawking flux contribution $\langle in|: T_{ab} :|in\rangle$ and the vacuum polarization part $\langle B|T_{ab}|B\rangle$. The formulae are given by equations (384)-(386) and we see that all of the components play a role.

In section 11.5, the Hiscock model was motivated by looking at the direction of the flux along the horizon and along $\mathcal{I}^+$. We noticed that along the horizon, the only non-vanishing component is $\langle T_{vv}\rangle$, which corresponds to an ingoing flux. However, it is true that, as we move away from the horizon, towards the singularity, the components $\langle T_{uu}\rangle$ and $\langle T_{uv}\rangle$ come into play. In fact, on the singularity itself, when $r \to 0$, all the components of $\langle T_{ab}\rangle$ diverge, but all with the same behaviour:

$$\begin{aligned} \langle T_{uu}\rangle &\sim -\frac{\hbar}{24\pi}\frac{m}{r^3} \\ \langle T_{vv}\rangle &\sim -\frac{\hbar}{24\pi}\frac{m}{r^3} \\ \langle T_{uv}\rangle &\sim -\frac{\hbar}{24\pi}\frac{m}{r^3}. \end{aligned} \tag{507}$$

One of our goals is to investigate the fate of the negative energy after it has reached the singularity. To that aim, the direction of the energy when it reaches the singularity matters. We have thus considered equally important to study the case of an outgoing energy flux inside the hole. This has motivated the design of another model of evaporating black hole. It is a slight modification of the Hiscock model inside the hole. The idea is simple: after the ingoing flux of particles has been created along the apparent horizon, as in the Hiscock model, they are scattered by the gravitational field and change direction. This scattering is sketched by introducing a space-like surface inside the hole (boundary III/VI) along which particles are separated. The model is represented as a conformal diagram in figure 45.

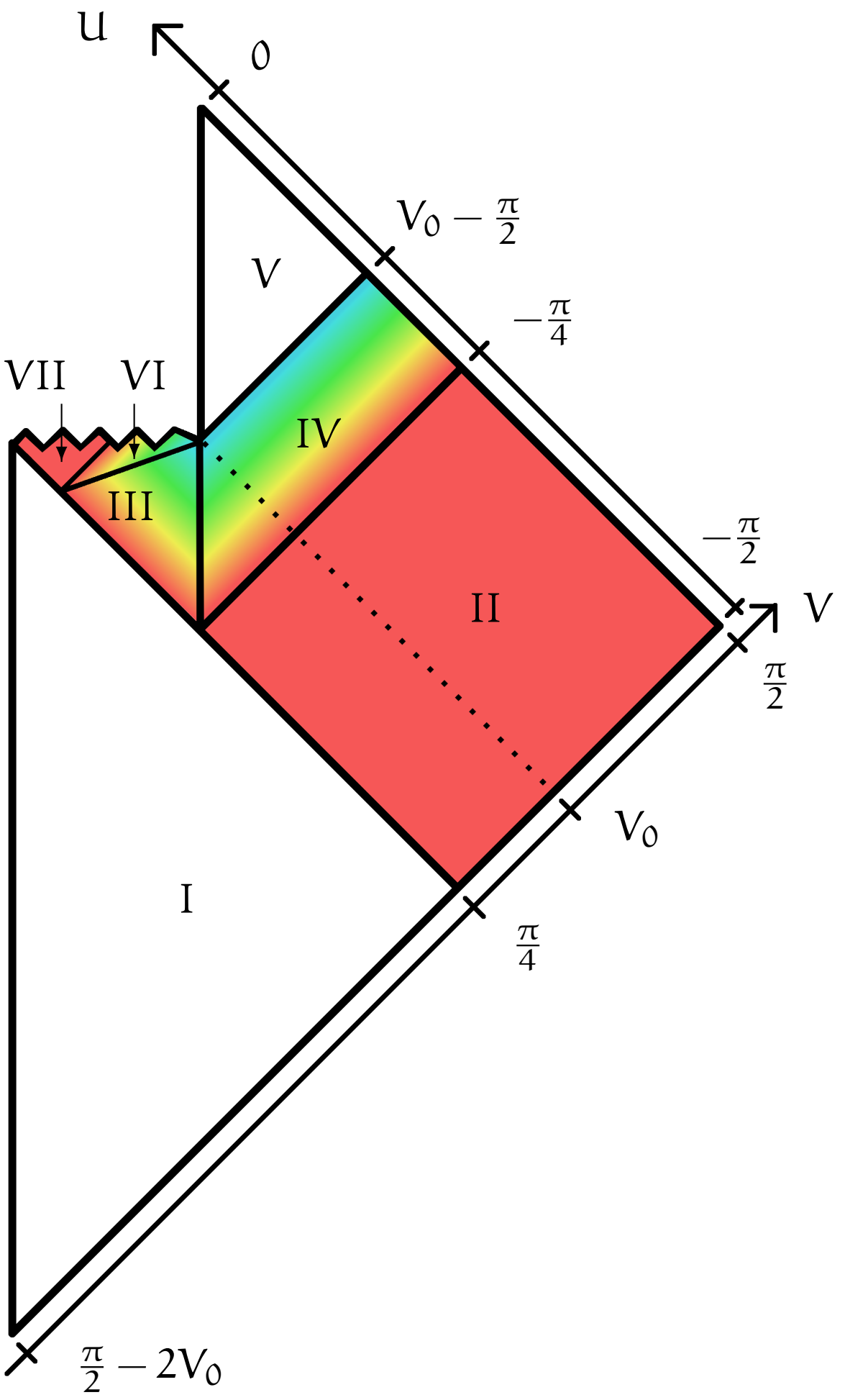

Figure 45: Conformal diagram of an evaporating black hole with inside *outgoing* flux.

The metric is given in 7 patches. In regions I – V, the metric is the same as Hiscock model, given by equations (394)-(398). The space-like boundary III/VI is chosen arbitrarily. In regions VII and VIII, the metric is given by

$$\text{(VI)} \left[ \; ds^2 = -\left(1 - \frac{2Q(u)}{r}\right) du^2 - 2dudr + r^2 d\Omega^2 \right. \tag{508}$$

$$\text{(VII)} \left[ \begin{array}{l} ds^2 = \left(1 - \frac{2m}{r}\right) dudv + r^2 d\Omega^2 \\ r = 2m\left(1 + W\left(-e^{\frac{v+u}{4m}-1}\right)\right) \end{array} \right. \tag{509}$$

The mass function $Q(u)$ is chosen so that it matches the mass function $N(v)$ along the boundary III/VI. The metric is written in terms of coordinates $(u, v)$ or $(u, r)$, which are related to the coordinates $(U, V)$ of the conformal diagram by the following formulae, based on the subdivision of figure 46.

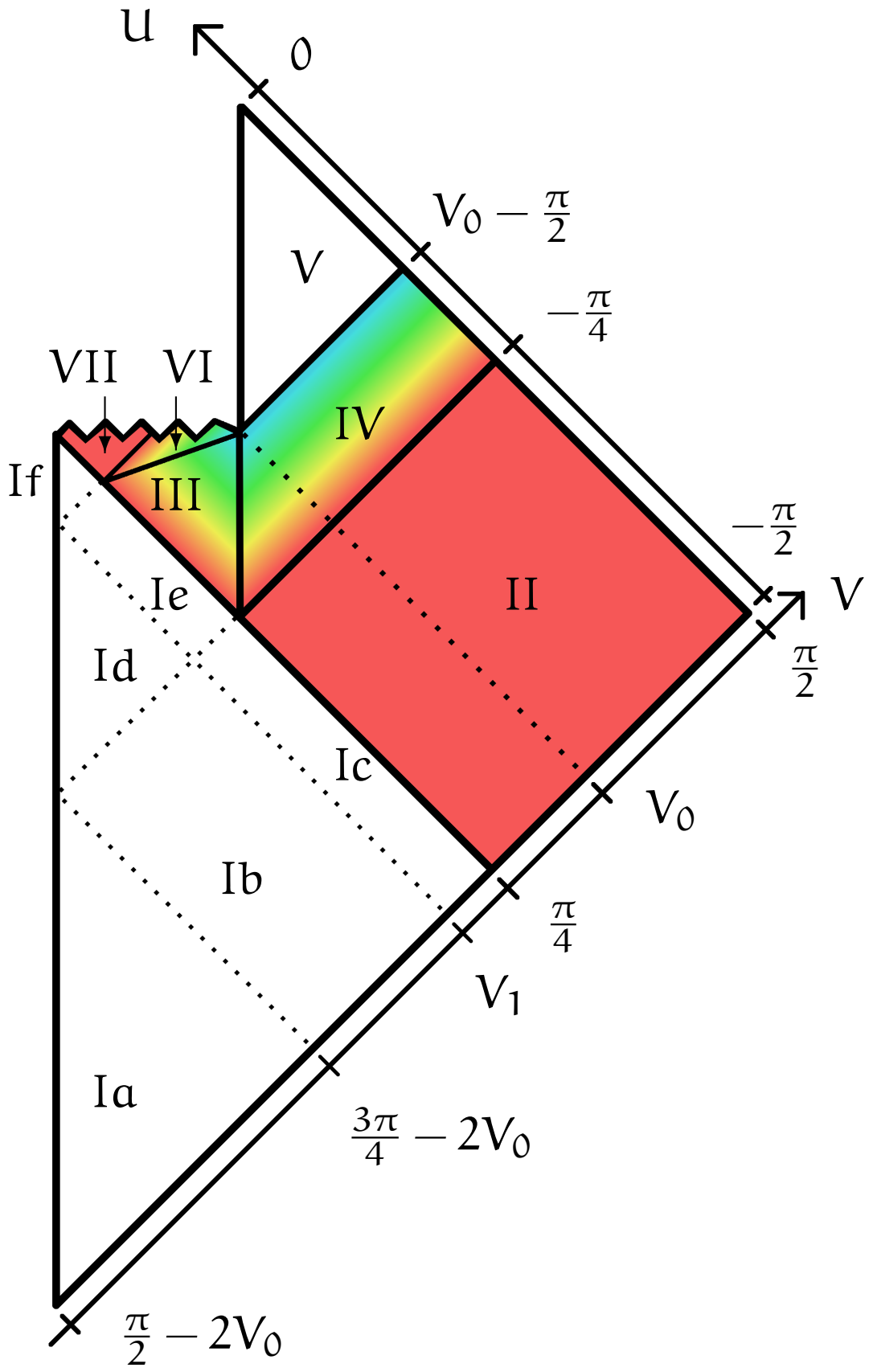

Figure 46: Subdivided conformal diagram of an evaporating black hole with inside *outgoing* flux.

$$\text{(Ia)} \quad \left[\begin{array}{l} u = -4m\left[1 + W\left(-\frac{\tan U}{e}\right)\right] \\ v = -4m\left[1 + W\left(-\frac{\tan(V + 2V_0 - \pi)}{e}\right)\right] \end{array}\right. \tag{510}$$

$$\text{(Ib)} \quad \left[\begin{array}{l} u = -4m\left[1 + W\left(-\frac{\tan U}{e}\right)\right] \\ v = f_1(V) \text{ increasing, such that} \\ f_1\left(\frac{3\pi}{4} - 2V_0\right) = -4m(1 + W(1/e)) \end{array}\right. \tag{511}$$

$$\text{(Ic)} \quad \left[\begin{array}{l} u = -4m\left[1 + W\left(-\frac{\tan U}{e}\right)\right] \\ v = f_2(V) \text{ increasing, such that} \\ \begin{cases} f_2(V_1) = f_1(V_1) \\ f_2\left(\frac{\pi}{4}\right) = 0 \end{cases} \end{array}\right. \tag{512}$$

$$(\mathrm{Id})\left[\begin{array}{l} u = c_1 + f_1(U - 2V_0 + \pi) \\ v = c_1 + f_1(V) \end{array}\right. \tag{513}$$

$$(\mathrm{Ie})\left[\begin{array}{l} u = c_2 + f_1(U - 2V_0 + \pi) \\ v = c_2 + f_2(V) \end{array}\right. \tag{514}$$

$$(\mathrm{If})\left[\begin{array}{l} u = c_3 + f_2(U - 2V_0 + \pi) \\ v = c_3 + f_2(V) \end{array}\right. \tag{515}$$

$$(\mathrm{II})\left[\begin{array}{l} u = -4m \log(-\tan U) \\ v = 4m \log \tan V \end{array}\right. \tag{516}$$

$$(\mathrm{III})\left[\begin{array}{l} v = f_3(V) \text{ monotonous, such that} \\ \qquad\qquad f_3(\pi/4) = N^{-1}(M(0)) \\ r = g(U, V) \text{ such that} \\ \begin{cases} \frac{\partial g}{\partial V} = \frac{f_3'(V)}{2}\left(1 - \frac{2N(f_3(V))}{g(U,V)}\right) \\ g(U, \pi/4) = -\frac{1}{2} f_1(U - 2V_0 + \pi) \end{cases} \end{array}\right. \tag{517}$$

$$(\mathrm{IV})\left[\begin{array}{l} u = M^{-1}(N(f_3(U + \pi/2))) \\ r = h(U, V) \text{ such that} \\ \begin{cases} \frac{\partial h}{\partial U} = -\frac{u'(U)}{2}\left(1 - \frac{2M(u(U))}{h(U,V)}\right) \\ h(-\pi/4, V) = 2m\left(1 + W\left(\frac{\tan V}{e}\right)\right) \\ h(U, \pi/2) = \infty \\ h(U, U + \pi/2) = g(U, U + \pi/2) \end{cases} \end{array}\right. \tag{518}$$

$$(\mathrm{V})\left[\begin{array}{l} v = M^{-1}(N(f_3(V_0))) + 2h(V_0 - \pi/2, V) \\ u = M^{-1}(N(f_3(V_0))) + 2h(V_0 - \pi/2, U + \pi/2) \end{array}\right. \tag{519}$$

$$(\mathrm{VI})\left[\begin{array}{l} u = P^{-1}(N(f_3(\mathcal{C}^{-1}(U)))) \\ r = j(U, V) \text{ such that} \\ \begin{cases} \frac{\partial j}{\partial U} = -\frac{u'(U)}{2}\left(1 - \frac{2P(u(U))}{j(U,V)}\right) \\ j(2V_0 - \pi + V_1, V) = 2m\,(1 \\ \quad + W\left(-\frac{4m + f_2(V_1)}{4m + f_2(\pi/2 - V)} e^{-\frac{f_2(V_1)}{4m} + \frac{f_2(\pi/2 - V)}{4m} - 1}\right)\Big) \\ j(2V_0 - \pi/2 - V, V) = 0 \\ j(\mathcal{C}(V), V) = g(\mathcal{C}(V), V) \end{cases} \end{array}\right. \tag{520}$$

$$(\mathrm{VII})\left[\begin{array}{l} u = -c_4 - f_2(U - 2V_0 + \pi) \\ +4m \log\left(1 + \frac{f_2(U - 2V_0 + \pi)}{4m}\right) \\ v = c_4 + f_2(\pi/2 - V) \\ \quad -4m \log\left(1 + \frac{f_2(\pi/2 - V)}{4m}\right) \end{array}\right. \tag{521}$$

There are three parameters $m$, $V_0$ and $V_1$. The constants $c_1, c_2, c_3$ and $c_4$ are arbitrary. The functions $g, h$ and $j$ are fixed implicitly by the differential equations. the functions $f_1$, $f_2$ and $f_3$ are arbitrary. The functions $P, M, N$ encode the phenomenology of the evaporation (two fix the position of the two pseudo-horizons, while the third fixes the rate of evaporation). The curve $\mathcal{C}$ parametrise the boundary III/VI in the diagram coordinates (it is space-like).

Above, we have introduced the outgoing flux as a consequence of the scattering of the ingoing flux. Another, maybe simpler, physical intuition can be given for the outgoing flux provided a better system of coordinates is used to represent the region surrounding the horizon. Indeed, due to the distortion of distances, the conformal diagram does not properly depict the fact that the space-like boundary III/VI and the time-like apparent horizon III/IV may actually be very close. Using Eddington time coordinate,

$$\tilde{t} \stackrel{\text{def}}{=} t + 2m \log\left|\frac{r}{2m} - 1\right| \quad \text{with } t = \frac{u+v}{2}, \tag{522}$$

a small region around the horizon, thin in $V$, but large enough in $U$ to include the two boundaries, looks like figure 47. Regions IV

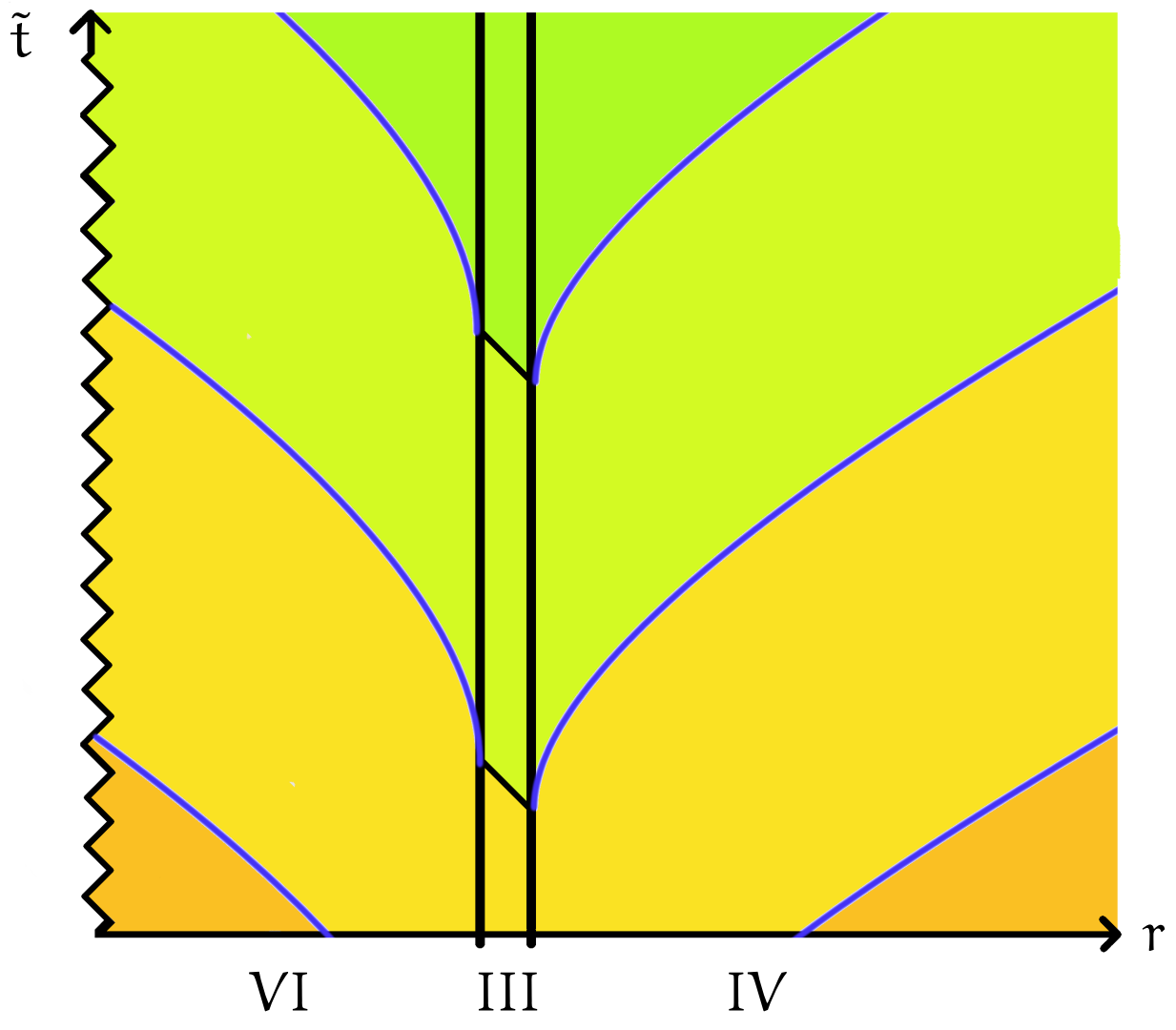

Figure 47: The region surrounding the horizon in Eddington time coordinate. Three pairs of Hawking quanta are represented by blue lines.

and VI surround a very small region III. Pairs of Hawking quanta are created alongside the null event horizon. Both quanta, inside and outside the hole, are outgoing, i.e. following the same side of the light cone (remember that the light cones are tilted in the Eddington time representation). However sketchy this description may be, we see that the modified model proposed in this section, with outgoing inside radiation, can be related to the usual intuitive idea of pairs of particles created along the event horizon.

### 15.3.2 *Conformal diagram II*

The new model proposed for an evaporating black hole extends naturally to the black-to-white hole scenario. The inside energy flux cross the singularity, and goes ahead towards $\mathcal{I}^+$. The corresponding conformal diagram is drawn in figure 48. The metric in regions I – IV is given by the equations (394)-(397). In regions VI and VII it is given by equations (508) and (509). Elsewhere, the metric is given by

$$\text{(V)} \left[ \begin{array}{l} ds^2 = -\left(1 - \frac{2m_1}{r}\right) du dv + r^2 d\Omega^2 \\ r = 2m_1 \left(1 + W\left(e^{\frac{v-u}{4m_1} - 1}\right)\right) \end{array} \right. \tag{523}$$

$$\text{(VIII)} \left[ \; ds^2 = -\left(1 - \frac{2R(u)}{r}\right) du^2 - 2du dr + r^2 d\Omega^2 \right. \tag{524}$$

$$(\text{IX}) \left[ \begin{array}{l} ds^2 = -\left(1 - \frac{2m}{r}\right) du dv + r^2 d\Omega^2 \\ r = 2m\left(1 + W\left(e^{\frac{v-u}{4m}-1}\right)\right) \end{array} \right. \tag{525}$$

$$(\text{X}) \left[ \begin{array}{l} ds^2 = -du dv + r^2 d\Omega^2 \\ r = \frac{1}{2}(v-u) \end{array} \right. \tag{526}$$

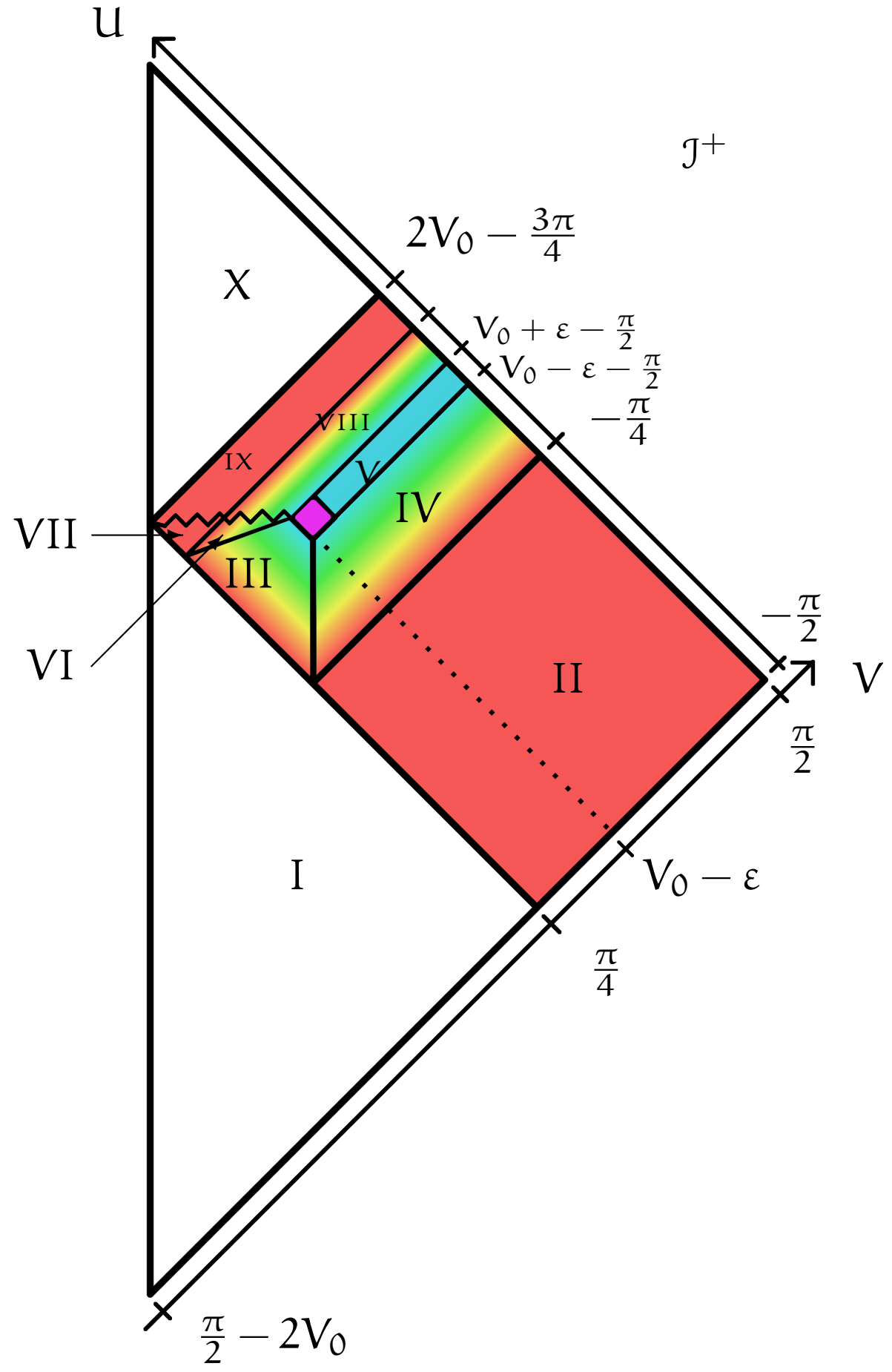

Figure 48: Conformal diagram of the second model of evaporating black-to-white hole. Energy flux now crosses the singularity along outgoing directions.

The mass function $R(u)$ is such that it matches with that of region VII along the singularity. The metric inside the purple central diamond has been already discussed in subsection 15.2.3. We do not give explicitly the map between the coordinates $(u, v)$ and $(U, V)$, but there is no doubt that the construction is consistent, and that the gluing can be performed along all boundaries.

In this model, the information-loss paradox is obviously solved since the inside quanta, which are correlated to those emitted outside the hole, finally reach $\mathcal{I}^+$. In [18], James M. Bardeen had independently proposed a scenario quite similar to this second model presented here.

### 15.3.3 *Negative energy flux along $\mathcal{I}^+$*

In this scenario, the Bondi-Sachs mass is shown in figure 49. Contrary to the previous model, the mass function is not monotonic. After the black hole has shrunk from $m$ to $m_1$, the transition to a white hole occurs, but then the mass increases again from $m_1$ to $m$. All these outgoing quanta carry a negative energy, so that the energy conditions are strongly violated in this case. This feature makes the scenario consistent with the expectation of a 'last gasp' [35, 36], but the violation is clearly too strong to be physically acceptable.

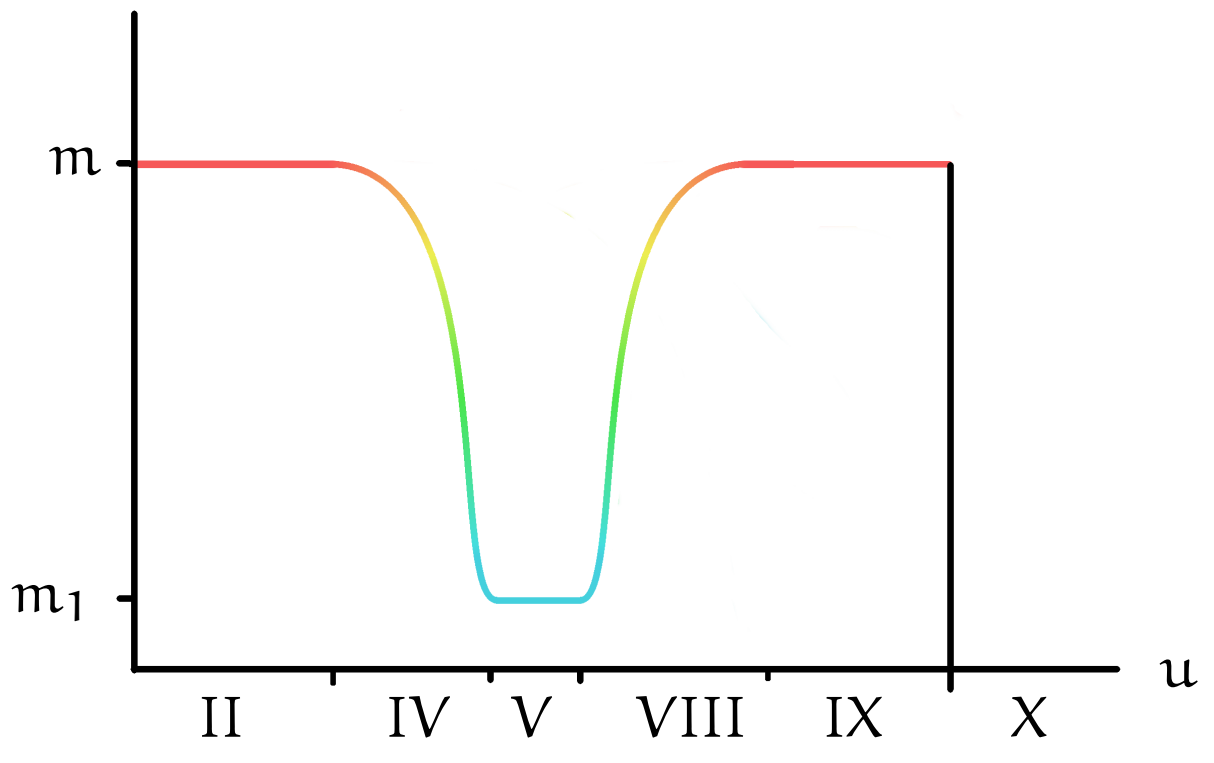

Figure 49: Bondi-Sachs mass function along $\mathcal{I}^+$ for an evaporating black-to-white hole (model II).

Although energy is always positive in classical physics, the energy conditions can be violated by quantum effects. However, the violation is not arbitrary and is constrained by some quantum inequalities. In short, energy cannot be too negative for too long. Such an inequality has been shown by Flanagan in 2D curved space-time [66]. We will apply it to our model.

Define a smearing function $\rho$ as any smooth, non-negative function, such that

$$\int_{-\infty}^{\infty} \rho(\xi)d\xi < \infty. \tag{527}$$

It shall be regarded as a window of observation of some observer: the measuring apparatus is turned on and off. Given a null geodesics $\gamma$ with affine parameter $\lambda$, and a quantum state $|\psi\rangle$, the average energy observed along $\gamma$ is

$$\mathcal{E}[\gamma,\rho,\psi] \stackrel{\text{def}}{=} \langle\psi|\int_{\gamma} d\lambda\, \rho(\lambda)\hat{T}_{ab}k^a k^b|\psi\rangle\,, \tag{528}$$

with $k^a$ the tangent vector field of $\gamma$. Flanagan has shown that for any null geodesic $\gamma$, smearing function $\rho$, and state $|\psi\rangle$,

$$\mathcal{E}[\gamma,\rho,\psi] \geqslant -\frac{1}{48\pi}\int_{\gamma} d\lambda\, \frac{\rho'(\lambda)^2}{\rho(\lambda)} \tag{529}$$

and the lower bound is reached for some $\rho$, $\gamma$, $|\psi\rangle$. We can apply it in our case for a the null geodesic along $\mathcal{I}^+$, and a smearing function $\rho = e^{-\frac{u^2}{2\sigma}}$, with $\sigma \in \mathbb{R}$, the typical time during which the particle detector is turned on. We center this time-window in region VIII to capture the maximum of the negative energy. So $\sigma$ is the duration of region VIII in the retarded affine time along $\mathcal{I}^+$. Then, we will assume that there is a state $|\psi\rangle$ such that

$$\mathcal{E}[\gamma,\rho,\psi] = \int du\, \rho(u)T_{ab}k^a k^b \tag{530}$$

where $T_{ab}$ is the classical stress-energy tensor, given by our model. In Blau [38] (eq. 40.36), it is shown that, for an outgoing Vaidya metric

$$T_{ab}k^a k^b = -\frac{1}{G}m'(u). \tag{531}$$

So Flanagan's inequality (529) becomes

$$-\frac{1}{G}\int du\, e^{-\frac{u^2}{2\sigma}} m'(u) \geqslant -\frac{1}{48\pi}\frac{\sqrt{2\pi}}{\sigma}. \tag{532}$$

Approximating roughly $\rho$ with a door function in the LHS, the inequality boils down to

$$\sigma\Delta m \leqslant \frac{\sqrt{2\pi}}{48\pi}G, \tag{533}$$

with $\Delta m = m - m_1$. By restoring the dimensional constants and getting rid of the inessential constants, we get finally

$$\sigma\Delta m \leqslant \frac{\hbar}{c^3}. \tag{534}$$

This equation means that during a time $\sigma$, the energy conditions cannot be violated by a too big amount $\Delta m$. Moreover the longer the time $\sigma$, the smaller the violation $\Delta m$ is allowed. This is because over a long period of time, the local violations are expected to compensate on average. It shall be compared to the more familiar Heisenberg inequality for time and energy:

$$\Delta t \Delta E \geqslant \frac{\hbar}{2}. \tag{535}$$

This one can be interpreted saying that during a time $\Delta t$, the fluctuations of the energy $\Delta E$ cannot be too small. On the contrary, Flanagan's inequality says that the fluctuation cannot be too big!

In this over-simplified model, all inside quanta are outgoing while it is known from equations (507) that only part of them reaches the singularity with this direction. A fully satisfying model would then lie in-between the ingoing model of section 15.2 and the outgoing one of section 15.3. As a result, the mass profile itself should lie somewhere between figure 44 and figure 49. Some negative energy radiation shall be observed before the emergence of the bouncing shell, and some positive energy radiation afterwards.

## 15.4 CONCLUSION

In this chapter, we have constructed and discussed two main effective models that describe an evaporating black-to-white hole. Based on a first construction by Hiscock, we have emphasised the double contribution from vacuum polarisation and Hawking quanta to the expectation value of the energy-momentum tensor that enters the semi-classical Einstein equations. This justifies that we should consider both models where the inside radiation is ingoing and outgoing. Then, we have shown how an evaporating black hole can be naturally extended to a white hole future, as quantum gravity suggests. The consistent mathematical models finally obtained display two main different profiles for the Bondi-Sachs mass along $\mathcal{I}^+$, but it is believed that the actual phenomenology should lie in-between the two.

Few main results shall be underlined:

1. The models are rigorous conformal diagrams in the sense of section 10.6. They describe a well-defined metric field satisfying Einstein equations.

2. The black hole interior evolves into a white hole interior, as expected from computation in canonical quantum gravity [8].

3. The transition results from quantum effects confined in a region $\mathcal{B}$ of planckian size.

4. The information paradox is solved with a long-lived remnant. Such objects were not found convincing as long as their physical nature remained obscure. The black-to-white scenario gives credibility to the idea.

5. The resulting Planck-mass white hole is stable. The phenomenology of such objects is yet to be explored. Preliminary works suggest that they could account for dark matter [163].

6. Along $\mathcal{I}^+$, an observer shall observe some negative energy radiation before the bounce (model II), and some positive energy radiation after it (model I).

*Mais je ne m'arrête point à expliquer ceci plus en détail,*
*à cause que je vous ôterais le plaisir de l'apprendre de vous-même,*
*et l'utilité de cultiver votre esprit en vous y exerçant,*
*qui est à mon avis la principale qu'on puisse tirer.*[1]

— René Descartes, La Géométrie.

1 Translation: *But I don't stop to explain this in more detail, because I would take you the pleasure of learning it yourself, and the usefulness of cultivating your mind by practicing it, which is in my opinion the main one we can draw.*

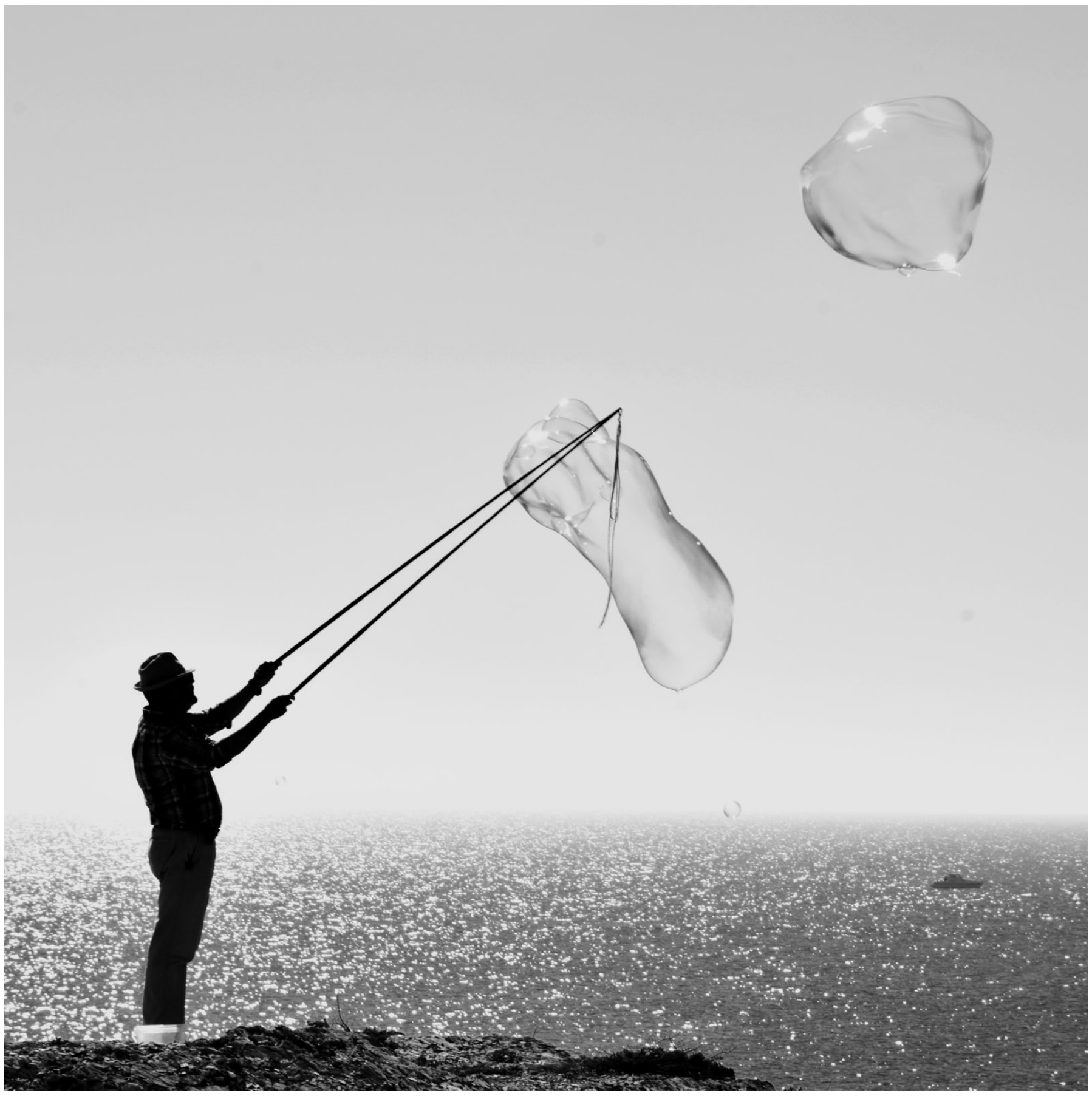

# Part III

## LOOPS, FOAMS AND COHERENT STATES

If it exists, the black-to-white hole transition is thought to be a quantum tunnelling phenomenon. It is thus expected of any theory of quantum gravity to provide tools to compute the quantum amplitude of the transition. Loop Quantum Gravity offers such tools, relying over the definition of a boundary surrounding the region where quantum effects are expected to be dominant. In our models, this region is a central diamond, and we have shown how its size could be reduced to planckian scale. The effective computation of the transition amplitude would ultimately confirm or not previous estimations of the probability of transition and the lifetime of black holes. Unfortunately, we have fallen short of time to complete this task, but work is ongoing. Great symphonies are unfinished, aren't they? Nevertheless, we raise here two first scaffoldings for the poliorcetics of the computation.

Ch. 16 summarises the formal framework of spin-foam computations.

Ch. 17 is an historical, conceptual and technical overview of the essential notion of coherent state.

# 16

# LOOPS AND FOAMS IN A NUTSHELL

Loop Quantum Gravity (LQG) is a good candidate theory for quantum gravity. It is obtained by the canonical quantisation of general relativity and describes the quantum states of space with the so-called spin-networks. Spin-foam theory is a later spinoff of both LQG and the sum-over-histories approach to quantum gravity. It describes quantum space-time, seen as the time evolution of spin-networks.

Most of the main textbooks provide a derivation of the theory, following more or less its historical developments through the process of quantisation [12, 54, 161]. Here we will only introduce the general mathematical framework of the theory, trying to be as concise as possible, since we believe that a full-fledged fundamental theory should come to a point where it stands on its own, with its mathematical framework and physical principles, without any reference to older approximate theories like general relativity or non-relativistic quantum mechanics.

## 16.1 SPIN-NETWORK

As any good quantum theory, LQG comes with an Hilbert space. It is the mathematical space of the various possible states of physical space. A very convenient basis is parametrised by the so-called spin-networks that we first define.

SPIN-NETWORK. An *abstract*[1] *directed graph* $\Gamma$ is an ordered pair $\Gamma = (\mathcal{N}, \mathcal{L})$, where $\mathcal{N} = \{n_1, ..., n_N\}$ is a finite set of $N$ nodes, and $\mathcal{L} = \{l_1, ..., l_L\}$ a finite set of $L$ links[2], endowed with a target map $t : \mathcal{L} \to \mathcal{N}$ and a source map $s : \mathcal{L} \to \mathcal{N}$, assigning each link to its endpoints (respectively the head or the tail, defined by the orientation). We denote $\mathcal{S}(n)$ (resp. $\mathcal{T}(n)$) the set of links for which the node $n$ is the source (resp. the target). The valency of a node $n$ is the number of links which have $n$ as a endpoint. A graph is said to be $p$-valent if the valency of each node is $p$. Given a directed graph $\Gamma$, we denote

1 Strictly speaking 'LQG' refers to the canonical approach for which the spin-networks are embedded in a space-like hypersurface. Here, we adopt a more abstract point of view, sometimes called 'covariant LQG', which is motivated by spin-foams. This alternative construction raises difficulties for defining the hamiltonian, but they are circumvented by the spin-foam formalism.

2 Mathematicians usually say *edge* or *arrow*, but not 'link', which has another meaning in knot theory. The terminology of LQG keeps 'edge' for spin-foams (see below), and uses 'link' for spin-networks.

$\Lambda_\Gamma$ the set of labellings $j$ that assign to any link $l \in \mathcal{L}$, an SU(2)-irrep $j_l \in \mathbb{N}/2$. Given a labelling $j \in \Lambda_\Gamma$, we denote

$$\mathrm{Inv}(n, j) \stackrel{\text{def}}{=} \mathrm{Inv}_{SU(2)} \left( \bigotimes_{l \in \mathcal{S}(n)} \mathcal{Q}_{j_l} \otimes \bigotimes_{l \in \mathcal{T}(n)} \mathcal{Q}^*_{j_l} \right). \tag{536}$$

The tensor product above assumes the prescription of an ordering of the links around a node, i.e. a sense of rotation and a starting link. A *spin-network* is a triple $\Sigma = (\Gamma, j, \iota)$, with $\Gamma$ a directed graph, a labelling $j \in \Lambda_\Gamma$, and $\iota$ is a map that assigns to any $n \in \mathcal{N}$ an intertwiner $|\iota_n\rangle \in \mathrm{Inv}(n, j)$. Figure 50 shows a pictorial representation of a 4-valent spin-network.

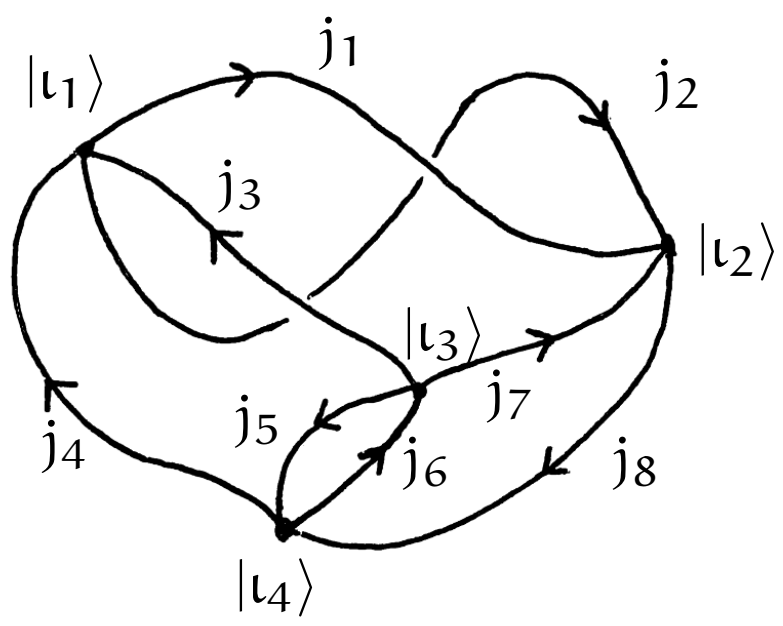

Figure 50: A 4-valent spin-network.

HILBERT SPACE. The Hilbert space of LQG is given by

$$\mathcal{H}_{LQG} = \bigoplus_{\Gamma} \mathcal{H}_\Gamma \tag{537}$$

where the direct sum is made over all possible directed 4-valent graphs $\Gamma$, and $\mathcal{H}_\Gamma$ is

$$\mathcal{H}_\Gamma \stackrel{\text{def}}{=} \bigoplus_{j \in \Lambda_\Gamma} \bigotimes_{n \in \mathcal{N}} \mathrm{Inv}(n, j) \tag{538}$$

It is spanned by the set of *spin-networks states*

$$|\Gamma, j, \iota\rangle = \bigotimes_{n \in \mathcal{N}} |\iota_n\rangle \tag{539}$$

where $\Gamma$ range over all possible 4-valent graphs, $j$ over $\Lambda_\Gamma$, and $|\iota_n\rangle$ over an orthonormal basis of $\mathrm{Inv}(n, j)$. By definition of the invariant space $\mathrm{Inv}(n, j)$, it is straightforward to see that 'the action of any $g_n \in$ SU(2) over a node $n$', i.e. over $\mathrm{Inv}(n, j)$, leaves the spin-network states invariant:

*The designation of 'Gauss constraint' comes from an analogy with Maxwell theory of electromagnetism.*

$$g_n \cdot |\Gamma, j, \iota\rangle = |\Gamma, j, \iota\rangle . \tag{540}$$

With this property, the spin-network states are said to satisfy the *Gauss constraint* at each node.

Since we only consider 4-valent graphs, an orthonormal basis of $\mathrm{Inv}(n, j)$ is given by the states of equation (196). Thus, instead of writing the abstract states $|\iota\rangle$, it is equivalent to split each 4-valent node (according to the prescribed ordering of the links around the nodes), like

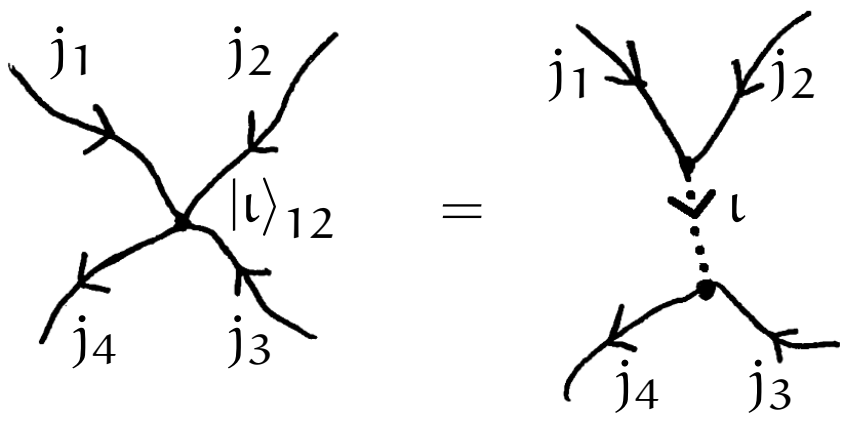

(541)

and then associate to the virtual link the spin $\iota \in \{\max(|j_1 - j_2|, |j_3 - j_4|), ..., \min(j_1 + j_2, j_3 + j_4)\}$, which parametrises the basis $|\iota\rangle_{12}$ of equation (196). By metonymy the spin $\iota$ is also called an intertwiner. Thus the spin-network of figure 50, becomes

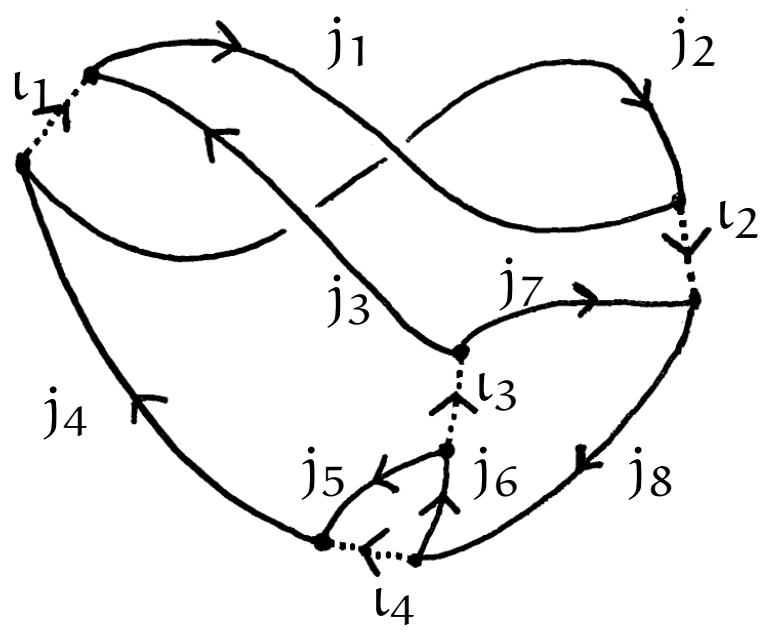

(542)

SPIN-NETWORK WAVE FUNCTION. The isomorphism (127), deduced from Peter-Weyl's theorem, offers another possible realisation of $\mathcal{H}_\Gamma$, as a subspace of $L^2(SU(2)^L)$, denoted[3] $L^2_\Gamma(SU(2)^L/SU(2)^N)$. A spin-network state $|\Gamma, j, \iota\rangle$ becomes a *spin-network wave function*

$$\Psi_{(\Gamma,j,\iota)}(g_{l_1}, ..., g_{l_L}) \in L^2_\Gamma(SU(2)^L/SU(2)^N), \tag{543}$$

obtained with the following procedure:

1. Associate to each link $l$

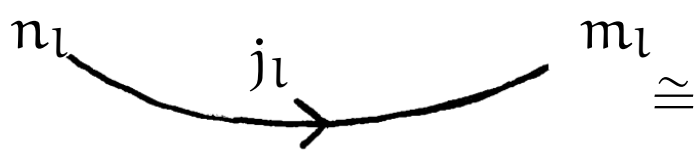

$$\cong D^{j_l}_{m_l n_l}(g_l) \tag{544}$$

with the magnetic indices $m_l$ or $n_l$, depending of the orientation, and the variable $g_l \in SU(2)$.

3 This subspace is sometimes denoted $L^2\left(SU(2)^L/SU(2)^N\right)$, but this is not mathematically rigorous.

2. Associate to each (splitted) node a 4jm symbol, like

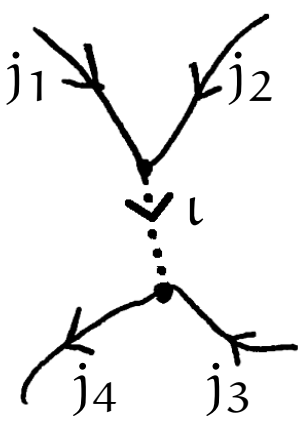

$$\cong (-1)^{j_4-n_4}\begin{pmatrix} j_1 & j_2 & j_3 & j_4 \\ m_1 & m_2 & m_3 & -n_4 \end{pmatrix}^{(\iota)} \quad (545)$$

with an index $-n$ and a phase $(-1)^{j-n}$ for outgoing links.

3. Finally multiply all together, and sum over all the magnetic indices.

For instance, the spin-network

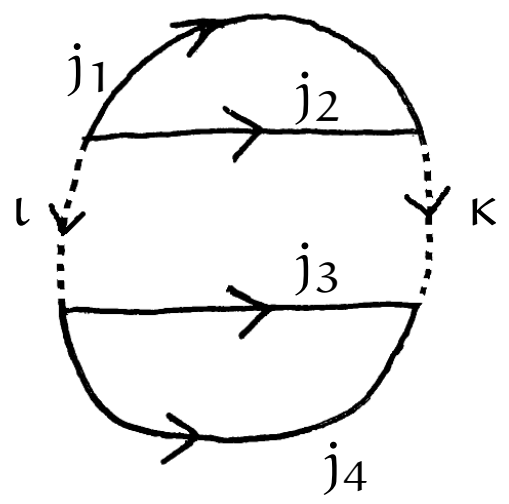

(546)

encodes the function

$$\begin{aligned}\Psi(u_1,u_2,u_3,u_4) \\ &= \sum_{m_i,n_i}(-1)^{\sum_i(j_i-n_i)}\begin{pmatrix} j_1 & j_2 & j_3 & j_4 \\ -n_1 & -n_2 & -n_3 & -n_4 \end{pmatrix}^{(\iota)} \\ &\quad\times\begin{pmatrix} j_1 & j_2 & j_3 & j_4 \\ m_1 & m_2 & m_3 & m_4 \end{pmatrix}^{(\kappa)}\prod_{i=1}^{4} D^{j_i}_{m_i n_i}(u_i). \quad (547)\end{aligned}$$

From the isomorphism (127), we can express the Gauss constraint (540) as an invariance of the functions $\Psi_{(\Gamma,j,\iota)}(g_{l_1},\dots,g_{l_L})$: for all sets $(u_n)\in SU(2)^L$, parametrised by the nodes $n\in\mathcal{N}$, we have

$$\Psi_{(\Gamma,j,\iota)}(g_{l_1},\dots,g_{l_L}) = \Psi_{(\Gamma,j,\iota)}(u_{s(l_1)}g_{l_1}u^{-1}_{t(l_1)},\dots,u_{s(l_L)}g_{l_L}u^{-1}_{t(l_1)}) \quad (548)$$

with $s$ and $t$ the source and target map of the graph. In fact, the space $L^2_\Gamma(SU(2)^L/SU(2)^N)$ can be characterized as the subspace of functions of $L^2(SU(2)^L)$ that satisfy this property.

Notice finally that evaluating the function at the identity on all links result in the graphical calculus previously defined in section 5.6.

ALGEBRA OF OBSERVABLES. In fact, there is not much information in the Hilbert space itself. What really matters physically is the algebra of observables $\mathcal{A}$ acting upon it. The observables of LQG are obtained by the principle of correspondence. Thus, they come with a geometrical interpretation: they correspond notably to measurements of area or measurement of volume. The Hilbert space $\mathcal{H}_{LQG}$ is built from the building block spaces $\mathcal{Q}_{j_l}$, where $j_l$ labels a link $l$. Similarly, the algebra of observables is built from the action of $\mathfrak{su}(2)$ (the flux) and SU(2) (the holonomy) over $\mathcal{Q}_{j_l}$. Notice that an observable should not 'go out' of $\mathcal{H}_{LQG}$: in other words, an observable should commute with the Gauss constraint.

Given a graph $\Gamma$, the observable of area associated to a link $l$ is

$$\hat{A}_l \overset{\text{def}}{=} 8\pi \frac{\hbar G}{c^3} \gamma \sqrt{\vec{J}_l^2}, \tag{549}$$

where $\gamma$ is a real parameter called the *Immirzi parameter*, and $\vec{J}_l$ are the generators of SU(2) acting over $\mathcal{Q}_{j_l}$. The spin-network basis diagonalises $\hat{A}_l$:

$$\hat{A}_l \,|\Gamma, j, \iota\rangle \overset{\text{def}}{=} 8\pi \frac{\hbar G}{c^3} \gamma \sqrt{j_l(j_l+1)}\,|\Gamma, j, \iota\rangle\,. \tag{550}$$

It also diagonalises the observable $(\vec{J}_1 + \vec{J}_2)^2$, acting over a node $n$,

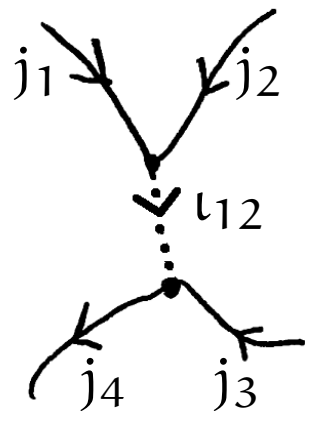

, (551)

so that

$$(\vec{J}_1 + \vec{J}_2)^2 \,|\Gamma, j, \iota\rangle \overset{\text{def}}{=} \iota_{12}(\iota_{12}+1)\,|\Gamma, j, \iota\rangle\,. \tag{552}$$

The latter observable encodes a notion of 'angle' between the links $j_1$ and $j_2$. Given a graph $\Gamma$, the set of area observables associated to each link and the set of 'angle operators' like $(\vec{J}_1 + \vec{J}_2)^2$ (one per each node), define a Complete Set of Commuting Observables (CSCO) over $\mathcal{H}_\Gamma$, diagonalised by the spin-network basis.

On each node like (551), we can also define the volume operator

$$\hat{V}_n = \frac{\sqrt{2}}{3}(8\pi G\hbar\gamma)^{3/2}\sqrt{|\vec{J}_1 \cdot (\vec{J}_2 \times \vec{J}_3)|}. \tag{553}$$

It is not diagonalised by the spin-network basis, but its eigenvalues can be computed numerically. It does not commute with $(\vec{J}_1 + \vec{J}_2)^2$ but it does with the areas, so that the areas $\hat{A}_l$ and the volumes $\hat{V}_n$ form another CSCO (diagonalised by another basis than that of spin-networks).

These geometric operators of area, volume or angle, built from the principle of correspondence, suggest a vision of the 'quantum geometry'. It is obtained as the dual picture of a graph $\Gamma$: a tetrahedron is associated to each node, and they glue together along faces (whose area is given by the eigenvalue of $\hat{A}_l$) dual to links.

## 16.2 SPIN-FOAM

DYNAMICS. The latter mathematical framework of LQG is obtained through the canonical quantisation of general relativity: the spin-network states represent quantum states of space. The time evolution of these states should be found by looking for the subspace formed by the solution to the hamiltonian constraint $\hat{H}|\Psi\rangle = 0$, where $|\Psi\rangle$ is a superposition of spin-network states, $\hat{H}$ the quantized hamiltonian. This hard path of finding the dynamics was followed notably by Thiemann [185]. Below we present a way to short-circuit the issue, called spin-foams, which takes inspiration from former sum-over-histories approaches to quantum gravity. Spin-foams can be seen as the time evolution of spin-networks, or also as quantum states of space-time.

SPIN-FOAMS. Spin-foams can be seen both as a higher dimensional version of Feynman diagrams propagating the gravitational field, and as the time evolution of spin-networks. Spin-foams are built out of combinatorial objects, which generalise graphs to higher dimensions, called *piecewise linear cell complexes*, often abbreviated as *complexes*. An *oriented 2-complex* is an ordered triple $\kappa = (\mathcal{E}, \mathcal{V}, \mathcal{F})$, with a finite set $\mathcal{E} = \{e_1, ...e_E\}$ of edges, a finite set $\mathcal{V} = \{v_1, ...v_V\}$ of vertices, and a finite set $\mathcal{F} = \{f_1, ..., f_F\}$ of faces, such that they all 'glue consistently'[4]. The orientation is given on the edges by a target map $t : \mathcal{E} \to \mathcal{V}$ and a source map $s : \mathcal{E} \to \mathcal{V}$, and the orientation of each faces gives a cyclic ordering of its bounding vertices.

Given an oriented 2-complex $\kappa$, we denote $\Lambda_\kappa$ the set of labellings $j$ that assign an SU(2)-irrep $j_f \in \mathbb{N}/2$ to any face $f \in \mathcal{F}$. Similarly we denote $I_\kappa$ the set of labellings $\iota$ that assign to each edge $e$ an intertwiner $|\iota_e\rangle$,

$$|\iota_e\rangle \in \mathrm{Inv}(e, j) \overset{\text{def}}{=} \mathrm{Inv}_{SU(2)}\left(\bigotimes_{f \in \mathcal{F}(e)} \mathcal{Q}_{j_f} \otimes \bigotimes_{f \in \mathcal{F}^*(e)} \mathcal{Q}^*_{j_f}\right), \tag{554}$$

where $\mathcal{F}(e)$ and $\mathcal{F}^*(e)$ are the set of faces adjacent to the edge $e$, whose orientation respectively matches and does not match that of $e$. A *spin-foam* is a triple $F = (\kappa, j, \iota)$. where $\kappa$ is an oriented 2-complex, $j \in \Lambda_\kappa$, and $\iota \in I_\kappa$. We can stick to a purely 'abstract' combinatorial definition of 2-complexes, but we can also adopt a geometrical 'hypostasis' that represents 'faces' as polygons. For instance, figure 51 shows a spin-foam embedded into 3-dimensional euclidean space. Notice that such a graphical representation is not always possible in 3 dimensions, and sometimes a fourth dimension can be required. In-

4 There is a way to give a precise meaning to this gluing, but it will be sufficient to keep it intuitive below, and to avoid these technicalities.

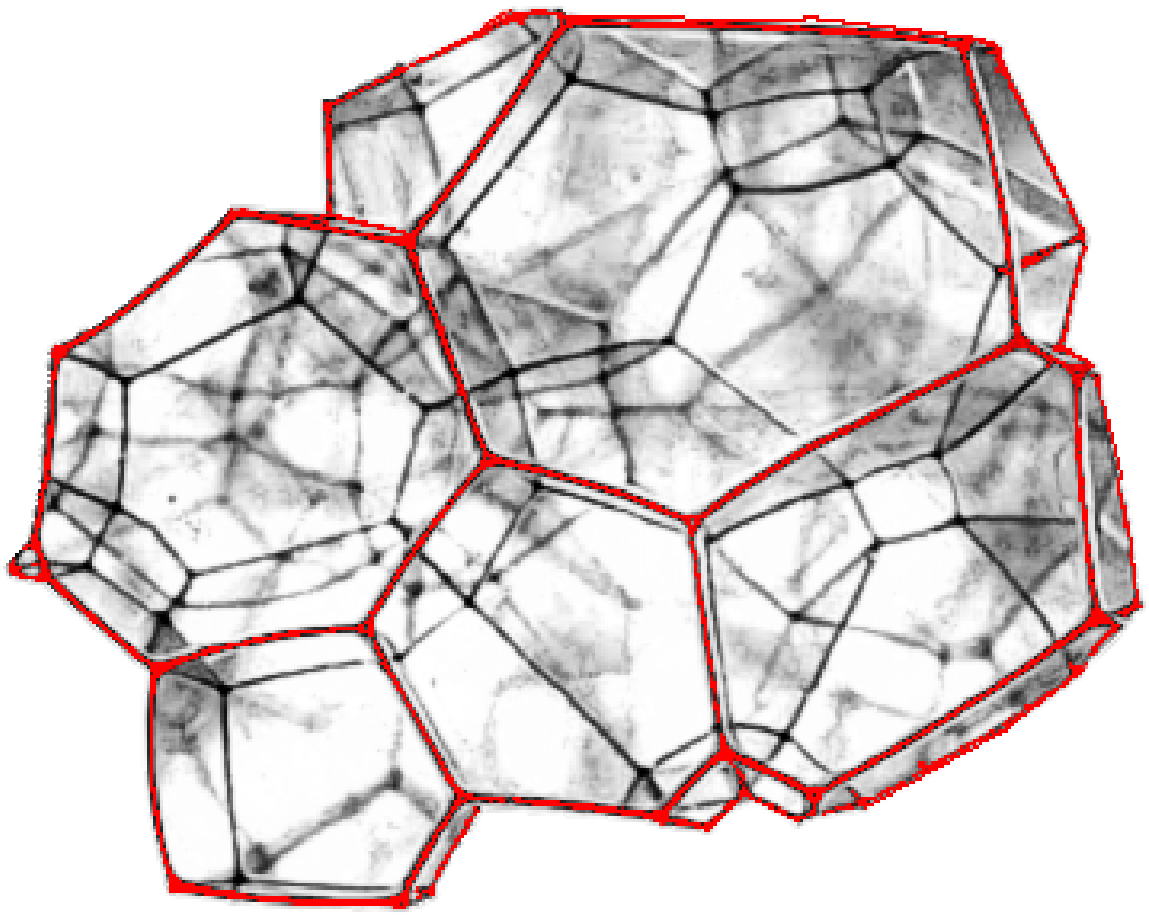

Figure 51: A 2-complex embedded in 3-dimensional euclidean space. Its boundary is a graph (in red).

terestingly, the boundary of a 2-complex[5] is a graph, as can be seen in figure 51. Thus, the boundary of a spin-foam is a spin-network. The vertices and the edges of the boundary are called respectively nodes and links. Each link bounds an inside face, so that the spin of the link is also the spin of the face. Similarly, each node is an endpoint of an inside edge, so that the associated intertwiners match.

SPIN-FOAM AMPLITUDE. To each spin-foam we associate an amplitude, which is like the propagator associated to a Feynman diagram. Its interpretation is made precise below. Given a spin-foam $(\kappa, j, \iota)$, we define the spin-foam amplitude as

$$\mathcal{A}(\kappa, j, \iota) = \left(\prod_{f \in \mathcal{F}} (2j_f + 1)\right) \left(\prod_{e \in \mathcal{E}} (2\iota_e + 1)\right) \left(\prod_{v \in \mathcal{V}} A_v(j, \iota)\right). \quad (555)$$

$A_v$ is called the vertex amplitude. In the short history of spin-foam amplitudes there has already been many various formulae proposed for the vertex amplitude. First, let us say that for quantum gravity, it is sufficient to consider spin-foams whose vertices are 5-valent (5 edges attached to it) and whose edges are 4-valent (4 faces attached to it). This restriction comes from the fact that the 2-complexes of quantum gravity are built by dualising the triangulation of a 4-dimensional manifold. Unfortunately there is no possible nice picture as figure 51 to visualise such a 2-complex since it cannot be embedded into 3-dimensional euclidean space. However it is sufficient to get an idea of the combinatorial structure of each vertex by representing the ad-

5 The notion of boundary of an abstract 2-complex requires a formal definition, but we keep it intuitive below for simplicity. We can admit that any 2-complex comes with a boundary.

jacent edges with dots and the faces with lines, so that we draw the *vertex graph*

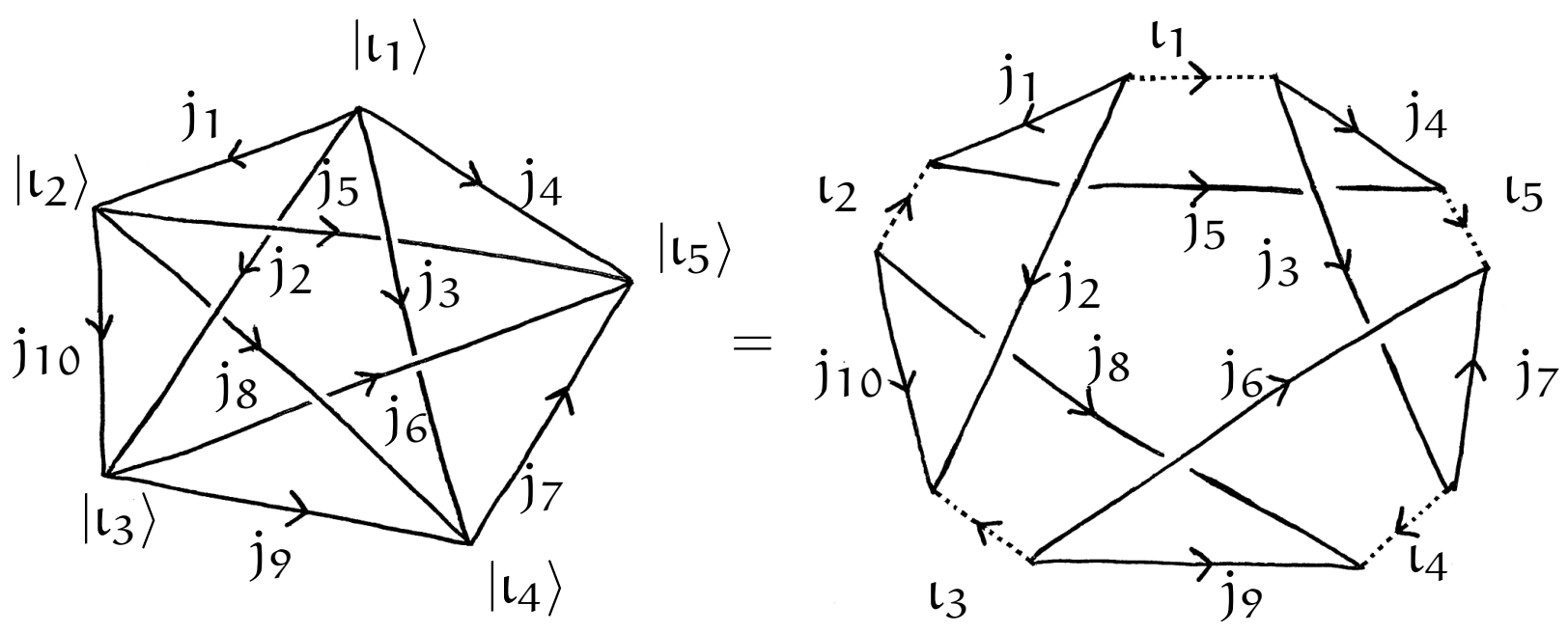

(556)

The orientation and the spin of the links, and the intertwiners of the nodes are naturally inherited from the underlying spin-foam, so that the vertex graph is a spin-network.

★ NOTA BENE. To avoid confusion, let us recap. Each spin-foam comes with a *boundary spin-network*, and also with a *vertex graph* for each of its vertex. If the spin-foam is made of only one vertex, then the boundary spin-network and the vertex graph coincides. Contrary to the the boundary spin-network, there is in general no interpretation of the vertex graphs in terms of quantum states of space.

The combinatorial shape of each vertex suggests to define the amplitude $A_v$ as the value obtained with the rules of graphical calculus of SU(2) recoupling theory, defined in section 5.6. This is precisely what Ooguri did in [137] by defining the vertex amplitude as the $\{15j\}$-symbol, but it later appeared not to be a good candidate for quantum gravity. Since then many other models were suggested [147]. They all consist in finding other rules than that of SU(2) recoupling theory to assign a value to the vertex graph (556).

The EPRL model, introduced in [61], is a model that is still considered a good candidate for quantum gravity. The vertex amplitude is computed from the vertex graph (556) with the following rules:

1. Compute the spin-network wave function as shown in the previous section. We obtain a function of $L^2(SU(2)^{10})$ which satisfies the Gauss constraint (548):

$$\Psi_{(\Gamma,j,\iota)}(g_{l_1},\dots,g_{l_{10}}) \tag{557}$$

2. Apply the so-called $Y_\gamma$-*map*, which is the linear map

$$Y_\gamma : L^2(SU(2)^{10}) \to \mathcal{F}(SL_2(\mathbb{C})^{10}) \tag{558}$$

defined over the canonical basis of Wigner matrix coefficients by

$$Y_\gamma\left(\prod_i D^{j_i}_{m_i n_i}\right) = \prod_i D^{(\gamma j_i, j_i)}_{j_i m_i j_i n_i}, \tag{559}$$

where $\gamma$ is the Immirzi parameter. We thus obtain a function of $\mathcal{F}(SL_2(\mathbb{C})^{10})$

$$Y_\gamma \Psi_{(\Gamma,j,\iota)}(h_{l_1}, ..., h_{l_{10}}). \tag{560}$$

It still satisfies the invariance of the Gauss constraint (548) for SU(2) action, be not for $SL_2(\mathbb{C})$.

3. Project down to the $SL_2(\mathbb{C})$-invariant subspace on each node with the projector $P_{SL_2(\mathbb{C})}$ acting as

$$P_{SL_2(\mathbb{C})} Y_\gamma \Psi_{(\Gamma,j,\iota)}(h_{l_1}, ..., h_{l_{10}}) = \int_{SL_2(\mathbb{C})} \delta(a_{n_5}) \prod_{n \in \mathcal{N}} da_n$$
$$\times \Psi_{(\Gamma,j,\iota)} \left( a_{s(l_1)} h_{l_1} a^{-1}_{t(l_1)}, ..., a_{s(l_{10})} h_{l_{10}} a^{-1}_{t(l_1)} \right). \tag{561}$$

with $n_5$ any of the 5 nodes. The delta function $\delta(a)$ (only non-vanishing when $a = \mathbb{1}$) is required to avoid the divergence of the integration, but the final result does not depend on the choice of node $n_5$. To put it differently the integration is only effective over (any) four nodes, while the fifth $a_{n_5}$ is fixed to the identity $\mathbb{1}$.

4. Evaluate all the variables $h_l$ to $\mathbb{1}$. So if $(\Gamma, j, \iota)$ is the vertex graph of a vertex $v$ in a spin-foam $(\kappa, j, \iota)$, we can finally write in a nutshell

$$A_v(j, \iota) = \left( P_{SL_2(\mathbb{C})} Y_\gamma \Psi_{(\Gamma,j,\iota)} \right) (\mathbb{1}). \tag{562}$$

Thus we have fully defined the spin-foam amplitude $\mathcal{A}(\kappa, j, \iota)$ of the EPRL model. The specificity of this model is the $Y_\gamma$-map which selects only the irreps $(p = \gamma j, k = j)$ among the principal series of $SL_2(\mathbb{C})$. It implements the so-called *simplicity constraints*, which enable the formulation of general relativity as a BF theory [12]. Besides, the apparently sophisticated procedure should not hide the fact that the value of $A_v(j, \iota)$ is the same as that obtained from the $SL_2(\mathbb{C})$ graphical calculus, defined in chapter 8, when the simplicity constraint is applied.

★ NOTA BENE. For those only interested in the actual computation of the amplitude of a given vertex graph, we can summarise the previous procedure with the following algorithm:

1. Associate a variable $h_p \in SL_2(\mathbb{C})$ to each intertwiner $\iota_p$.
2. Associate to each link

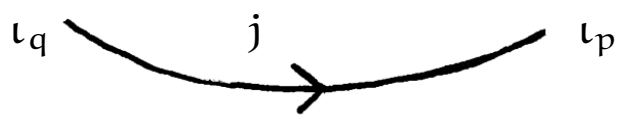

$$\cong D^{(\gamma j, j)}_{jmjn}(h_p^{-1} h_q). \tag{563}$$

3. Associate a $3jm$-symbol to each node as in usual graphical calculus (eq. (207)).
4. Multiply everything together and sum over all the magnetic indices $m$ and $n$.
5. Integrate over (any) four of the five $SL_2(\mathbb{C})$ variables $h_p$, and fix the fourth to the identity $\mathbb{1}$.

INTERPRETATION. The interpretation of spin-foams relies on the general boundary formulation of quantum mechanics which was introduced by Oeckl [135, 136]. Consider a finite region of spacetime. Its boundary $\Sigma$ is a 3-dimensional hypersurface which constitutes the quantum system under consideration. Its space of states is the Hilbert space of LQG, $\mathcal{H}_{LQG}$, spanned by the spin-network states. An observer $\mathcal{O}$ may know some partial information about the state $\psi$ of $\Sigma$, which can be expressed by the fact that $\psi \in \mathcal{S}$, where $\mathcal{S}$ is a linear subspace of $\mathcal{H}_{LQG}$. Then $\mathcal{O}$ can carry out measurements with the operators of the algebra, to determine information about $\psi$. If $\mathcal{A}$ is a linear subspace of $\mathcal{S}$, then the probability to find $\psi \in \mathcal{A}$ is

$$P(\mathcal{A}|\mathcal{S}) = \frac{\sum_{i\in I} |\rho(\xi_i)|^2}{\sum_{j\in J} |\rho(\zeta_j)|^2}, \tag{564}$$

where $\xi_i$ (resp. $\zeta_j$) is an orthonormal basis of $\mathcal{A}$ (resp. $\mathcal{S}$). $\rho : \mathcal{H}_{LQG} \to \mathbb{C}$ is a linear map, called the *transition amplitude* defined for a spin-network state $\Psi$ by

$$\rho(\Psi) \stackrel{\text{def}}{=} \sum_{\sigma} W_\sigma(\Psi) \tag{565}$$

where the sum is done over all possible spin-foams $\sigma$ which have $\Psi$ as a boundary, and $W_\sigma(\Psi)$ is the 2-complex amplitude defined as

$$W_\sigma(\Psi) = \sum_j \sum_\iota \mathcal{A}(\sigma, j, \iota) \tag{566}$$

where the sum is done over all the possible spin labellings $j \in \Lambda_\kappa$, and intertwiner labellings $\iota \in I_\kappa$, that are compatible with the spin-network $\Psi$ at the boundary.

This completes the mathematical formulation of the theory and its probabilistic interpretation. Of course, much remains to be discovered. In particular, the theory has yet to meet the benchmark of experimental evidence!

# 17 COHERENT STATES

The black-to-white hole amplitude can be computed with the spin-foam formalism described in chapter 16. This requires defining a spin-foam whose boundary spin-network matches the geometry of the black-to-white hole. In chapter 15, we have proposed a model of classical metric for the black-to-white hole transition. Thus, we are looking for spin-network states that approximate this geometry. The task of finding quantum states which approximate classical physics has a long history. Such states are called *coherent states*.

In this chapter, almost entirely taken from [123], we review their history, from their origins to quantum gravity. We discuss the notion of coherent states from three different perspectives: the seminal approach of Schrödinger, the experimental take of quantum optics, and the theoretical developments in quantum gravity. This comparative study tries to emphasise the connections between the approaches, and to offer a coherent short story of the field, so to speak.

## 17.1 INTRODUCTION

Coherent states are essential tools in theoretical physics. Since their early introduction by Schrödinger in 1926, they have served practical purposes in quantum optics, while several mathematical generalisations of the notion have been proposed, and some of them applied to quantum gravity. The present chapter was initially motivated by the following observations:

- The existing reviews on coherent states, like [74] or [146], do not deal with quantum gravity. So, we would like to summarise the various coherent states introduced in quantum gravity.
- The quantum gravity literature is very technical and does not insist much on the conceptual motivations behind the definitions. We would like to show that the semi-classical properties of coherent states are expected rather than magical.
- The many approaches to coherent states convey the impression of a disparate field made of arbitrary definitions. On the contrary, we would like to insist on the unity of the landscape and expose the big picture.

Thus we offer a journey among coherent states, from Schrödinger to quantum gravity, passing by quantum optics.

Along the way, we will notably answer the following puzzling questions:

- Coherent states are sometimes introduced as the states $|\psi\rangle$ such that $\langle\psi|\hat{x}(t)|\psi\rangle$ and $\langle\psi|\hat{p}(t)|\psi\rangle$ satisfy the classical equations of motion. It is, for instance, the impression conveyed in the seminal paper of Schrödinger [169], but also in the recent reference book [74]. However such a property cannot be a characterisation of coherent states whatsoever, since it is clear, from Ehrenfest theorem, that *any* time-evolved state $|\psi(t)\rangle$, coherent or not, satisfies it. Is there a way to make this first intuition of classicality rigorous?

- Coherent states are also often introduced in courses as eigenstates of the annihilation operator, but this does not seem to be the best pedagogical way as the physical motivation of this approach may seem rather obscure at first sight. Indeed, doing so, the classical properties that can be checked afterwards appear as magical, rather than expected. What could be a better pedagogical introduction to the topic? Also, can we define coherent states for other simple quantum systems like the free particle?

- Coherent states can also be generated by the action of the Heisenberg group over the vacuum state. This group is sometimes called the dynamical symmetry group of the harmonic oscillator (see [64, 82]), although it is very unclear a priori in which sense the group is 'dynamical', a 'symmetry group', or even specific to the harmonic oscillator. Can we make the statement precise?

- Coherent states are wanted to be quasi-classical states, but in quantum optics, for instance, the coherent states of light are those that maximise the interference pattern, which is paradoxically regarded as a very quantum feature, produced by lasers far from being classical sources of light as an incandescent bulb may be. Is the paradox of designation only superficial?

- The definition of coherent states in quantum gravity is covered by a jungle of technicalities, far from the experimental point of view of quantum optics. Can we nevertheless summarise the story to keep the key physical idea and make our way through the jungle?

To start with, we go back to the initial ideas of Schrödinger in section 17.2, and propose a modern follow-up in section 17.3. Then, in section 17.4, we enlarge the discussion with a kinematical characterisation of coherent states in terms of annihilation operators. We explain the physical meaning of these operators in quantum optics in section 17.5, which motivates an algebraic generalisation of coherent states presented in section 17.6. In section 17.7, we present an independent geometric generalisation, which was later applied in quantum gravity, as we show in section 17.8.

## 17.2 SCHRÖDINGER COHERENT STATES

Historically, the initial motivation for introducing coherent states is to demonstrate how classical mechanics can be recovered from quantum mechanics. It is done in 1926 in a short seminal paper by Erwin Schrödinger [169], translated in English in [170], entitled *The Continuous Transition from Micro- to Macro-Mechanics*. The title is rather explicit about its goal, although one may discuss whether it has been achieved or not.

Interestingly, Schrödinger does not use the word 'coherent' anywhere, but he aims at constructing mathematically

> *a group of proper vibrations [that] may represent a 'particle', which is executing the 'motion', expected from the usual mechanics.*

Neither does he use the words 'quasi-classical' or 'semi-classical', but the latter would convey his intuition probably better than 'coherent'. The paper does not shine by its clarity, but one can understand the overall logic, that we present below in modernised terms and notations.

### 17.2.1 *Quantum harmonic oscillator.*

Let's consider the quantum harmonic oscillator in one dimension, with mass $m$ and frequency $\omega$. Its Hilbert space is $L^2(\mathbb{R})$ over which are acting the position operator $\hat{x}\psi(x) = x\psi(x)$ and the momentum $\hat{p}\psi = -i\hbar\,\partial_x\psi$ . The dynamics is provided by the hamiltonian which reads

$$\hat{H} = \frac{\hat{p}^2}{2m} + \frac{1}{2}m\omega^2\hat{x}^2. \tag{567}$$

The eigenstates of $\hat{H}$ form an orthonormal basis, indexed by $n \in \mathbb{N}$,

$$\psi_n(x) = \sqrt[4]{\frac{m\omega}{\pi\hbar}}\frac{1}{\sqrt{2^n n!}}e^{-\frac{m\omega}{2\hbar}x^2}H_n\left(\sqrt{\frac{m\omega}{\hbar}}x\right) \tag{568}$$

where $H_n(x)$ are Hermite's polynomials[1] and the associated eigenvalues are

$$E_n = \hbar\omega\left(n + \frac{1}{2}\right). \tag{570}$$

1 Wikipedia mentions two conventions for Hermite's polynomial. We use the physicist one, i.e.

$$H_n(x) \overset{\text{def}}{=} (-1)^n e^{x^2}\frac{d^n}{dx^n}e^{-x^2}. \tag{569}$$

### 17.2.2 *Schrödinger coherent states.*

Then, Schrödinger defines[2], out of the blue, the following family of states, indexed by time $t \in \mathbb{R}$ and another parameter $\alpha \in \mathbb{R}$:

$$\psi_\alpha(x,t) \stackrel{\text{def}}{=} e^{-\frac{\alpha^2}{2}} \sum_{n=0}^{\infty} \frac{\alpha^n}{\sqrt{n!}} e^{\frac{i}{\hbar}E_n t} \psi_n(x). \tag{571}$$

In Dirac notation, it is the $x$-representation of

$$|\beta\rangle = e^{-\frac{|\beta|^2}{2}} \sum_{n=0}^{\infty} \frac{\beta^n}{\sqrt{n!}} |n\rangle, \quad \text{with } \beta \in \mathbb{C}, \tag{572}$$

such that $\psi_\alpha(x,t) = e^{i\omega t/2} \langle x | \alpha e^{i\omega t} \rangle$. It is immediate to see that $\psi_\alpha(x,t)$ is the temporal evolution of $\psi_\alpha(x,0)$ by the unitary operator $e^{\frac{i}{\hbar}\hat{H}t}$, as

$$e^{\frac{i}{\hbar}\hat{H}t} |\alpha_0\rangle = e^{i\omega t/2} \left|\alpha_0 e^{i\omega t}\right\rangle. \tag{573}$$

Then Schrödinger argues that these states approximate the 'macro-mechanics', what we would call in modern language, being semi- or quasi-classical. More precisely, he highlights three properties:

1. First, the average position satisfies the law of classical motion:
$$\langle \hat{x} \rangle = \alpha \cos \omega t. \tag{574}$$
2. Secondly, the average energy is almost the classical one:
$$\langle \hat{H} \rangle \approx \alpha^2 m \omega^2. \tag{575}$$
3. Third, he argues (without any explicit computation) that the wave packet does not 'spread out', but 'remains compact', like a particle.

The two first properties provide physical meaning to the parameter $\alpha$ as the amplitude of some corresponding classical wave. Thus the Schrödinger coherent states are parametrised by an amplitude $\alpha$ and an instant $t$.

### 17.2.3 *Wrong characterisation of quasi-classicality.*

The three arguments above appear as a first attempt to formalise the property of 'quasi-classicality', and have been the basis of the later developments of coherent states. Unfortunately, it has been hardly never noticed that the first property cannot characterise quasi-classicality in any way. Indeed *all* the quantum states of the harmonic oscillator satisfy this property! More precisely, given any initial state $|\psi_0\rangle$, its time

2 Compared to the strictly original definition of Schrödinger, we have here chosen to normalise the states, with a factor $e^{-\alpha^2/2}$ in front of the sum.

evolution will be so that it satisfies equation (574). It is a consequence of Ehrenfest theorem, that drives the evolution of the expected value of an observable $\hat{A}$ in any state $|\psi(t)\rangle$, according to the equation

$$\frac{d\langle\hat{A}\rangle}{dt} = \frac{1}{i\hbar}\left\langle[\hat{A},\hat{H}]\right\rangle. \tag{576}$$

In the case of the harmonic oscillator, the equations for $\hat{x}$ and $\hat{p}$ are

$$\frac{d\langle\hat{x}\rangle}{dt} = \frac{1}{m}\langle\hat{p}\rangle \qquad \text{and} \qquad \frac{d\langle\hat{p}\rangle}{dt} = -m\omega^2\langle\hat{x}\rangle. \tag{577}$$

These are actually the classical equations of motion for $\langle\hat{x}\rangle$ and $\langle\hat{p}\rangle$, and so all solutions $\langle\hat{x}\rangle$ take the form of equation (574). It is completely generic and so cannot be used as a characterisation of quasi-classicality. Thus, there is no such constraining property in Schrödinger first statement, except maybe the implicit demand that the time evolution of a 'quasi-classical state' should still be 'quasi-classical'. It is surprising that this fact has not been much recognised. Of course, with a more complicated hamiltonian, the property is not a trivial statement, but for the harmonic oscillator, it is.

Let us now analyse the two other properties, first 3 then 2, which may at first be disappointing for their vague formulation. When it is made precise, we show that each of them, alone, is a sufficient condition that fully characterises the family of coherent states.

## 17.3 DYNAMICAL CHARACTERISATION

A characteristic feature of quantum mechanics is the fact that the position of a particle is not given by a classical trajectory, but rather by a probability density that evolves with time. Thus, a 'quasi-classical' state could be one for which a quantum particle is *well localised in space*, and *remains localised as time goes by*. Let us try to formalise it, and see how this programme fails in the case of the free particle and succeeds for the harmonic oscillator.

### 17.3.1 *Free particle*

Consider the free particle in one dimension. Its Hilbert space is $L^2(\mathbb{R})$. The Dirac delta function $\delta(x)$ describes the state of a particle perfectly well localised at $x = 0$. The uncertainty about its position is zero: $\Delta\hat{x} = 0$. For that reason, it may seem a good candidate for being a quasi-classical state.

However, this first attempt fails because the particle does not remain localised as time goes by. Indeed, the hamiltonian of the free particle

$$\hat{H} = \frac{\hat{p}^2}{2m}, \tag{578}$$

drives the time-evolution of $\delta(x)$ to

$$\psi(x,t) = \sqrt{\frac{m}{2\pi\hbar|t|}} e^{-i\,\mathrm{sgn}(t)\,\frac{\pi}{4}} e^{i\frac{mx^2}{2\hbar t}}. \tag{579}$$

The probability distribution $|\psi(x,t)|^2$ is now completely spread, $\Delta\hat{x} = \infty$, and not even normalised! Thus, a wave function which is infinitely well localised at initial time, turns instantaneously into an infinitely spread state[3].

So consider instead a more reasonable initial state, like a gaussian curve

$$\psi_0(x) = \frac{1}{\sqrt{\sigma\sqrt{2\pi}}} e^{-\frac{x^2}{4\sigma^2}}, \tag{580}$$

which is spread as $\Delta_0\hat{x} = \sigma$. It evolves as a free particle to

$$\psi(x,t) = \frac{1}{\left(2\pi(\sigma^2 + i\hbar t/m)\right)^{1/4}} e^{-\frac{x^2}{4(\sigma^2+i\hbar t/m)}}. \tag{581}$$

It is also a gaussian which is spread like

$$\Delta\hat{x} = \sqrt{\sigma^2 + \frac{t^2\hbar^2}{4m^2\sigma^2}}, \tag{582}$$

so that it irremediably spreads with time and looses its initially compact shape.

In fact, whatever the initial state $\psi_0$ at time $t = 0$, it evolves as a free particle to a state $\psi(x,t)$ which satisfies[4]:

$$(\Delta\hat{x})^2 = (\Delta_0\hat{x})^2 + \frac{(\Delta_0\hat{p})^2}{m^2}t^2 + \frac{t}{m}\left(\langle\hat{x}\hat{p} + \hat{p}\hat{x}\rangle_0 - 2\langle\hat{x}\rangle_0\langle\hat{p}\rangle_0\right). \tag{583}$$

It is a second order polynomial in t. A necessary condition to prevent the time spreading would be to have $(\Delta_0\hat{p})^2 = 0$, but this implies, through Heisenberg inequality, that $\Delta_0\hat{x} = \infty$, i.e. a maximally spread state in space... So, for the free particle, the spreading is unavoidable. From this perspective, there is no 'quasi-classical state' for the free particle.

### 17.3.2 *Harmonic oscillator*

Let us now consider the more sophisticated hamiltonian of the harmonic oscillator:

$$\hat{H} = \frac{\hat{p}^2}{2m} + \frac{1}{2}m\omega^2\hat{x}^2. \tag{584}$$

A priori, there is more chance to find coherent states now, because we have added a potential well in the hamiltonian that can help to

3 This matter of fact seems even to contradict the postulate according to which two successive measurements should give the same result. But these pathologies can be imputed to the already suspicious Dirac delta function.

4 A proof can be found in [93] p. 104.

confine the wave-function and prevent it from spreading. In this case, the spreading of a general solution $\psi(x,t)$ is given by[5]

$$\begin{aligned}(\Delta\hat{x})^2 = \frac{1}{2}\left((\Delta_0\hat{x})^2 + \frac{(\Delta_0\hat{p})^2}{m^2\omega^2}\right) \\ -\frac{1}{2}\left(\frac{(\Delta_0\hat{p})^2}{m^2\omega^2} - (\Delta_0\hat{x})^2\right)\cos(2\omega t) \\ +\frac{1}{2m\omega}\left(\langle\hat{x}\hat{p}+\hat{p}\hat{x}\rangle_0 - 2\langle\hat{x}\rangle_0\langle\hat{p}\rangle_0\right)\sin(2\omega t)\end{aligned} \quad (587)$$

It is noticeable that the spreading oscillates. There is no irresistible increasing of the spreading. Instead, whatever the state we start with, the wave packet will stay confined within a finite range, and even come back periodically to its initial spreading $\Delta_0\hat{x}$.

So a first lesson to draw from this computation is that one should not be surprised by the fact that Schrödinger coherent states do not spread out, as a free particle would do, because no state of the harmonic oscillator does it!

### 17.3.3 *Constant and minimal*

Then, one can try to express the third property of Schrödinger in more precise terms. For instance, one can look for states for which $\Delta\hat{x}$ is constant in time. This requires the two conditions

$$\begin{cases} (\Delta_0\hat{p})^2 = m^2\omega^2(\Delta_0\hat{x})^2 \\ \langle\hat{x}\hat{p}+\hat{p}\hat{x}\rangle_0 = 2\langle\hat{x}\rangle_0\langle\hat{p}\rangle_0 \end{cases} \quad (588)$$

One can check that the Schrödinger coherent states do satisfy these conditions. However, these two conditions are not sufficient to characterise them. For instance all the eigenstates of the hamiltonian, $\psi_n(x)$, also satisfy these conditions. Another condition is still required: the minimisation of $\Delta_0\hat{x}$.

---

5 PROOF. Using Ehrenfest theorem we have

$$\begin{aligned} &\tfrac{d}{dt}\langle\hat{x}\rangle = \langle\hat{p}\rangle/m \\ &\tfrac{d}{dt}\langle\hat{p}\rangle = -m\omega^2\langle\hat{x}\rangle \\ &\tfrac{d}{dt}\left\langle\hat{x}^2\right\rangle = \langle\hat{x}\hat{p}+\hat{p}\hat{x}\rangle/m \\ &\tfrac{d}{dt}\left\langle\hat{p}^2\right\rangle = -m\omega^2\langle\hat{x}\hat{p}+\hat{p}\hat{x}\rangle \\ &\tfrac{d}{dt}\langle\hat{x}\hat{p}+\hat{p}\hat{x}\rangle = 2\left\langle\hat{p}^2\right\rangle/m - 2m\omega^2\left\langle\hat{x}^2\right\rangle \end{aligned} \quad (585)$$

from which we show the differential equation

$$\frac{d^4}{dt^4}\left\langle\hat{x}^2\right\rangle = -4\omega^2\frac{d^2}{dt^2}\left\langle\hat{x}^2\right\rangle \quad (586)$$

which is finally solved easily, and leads to our expression for $(\Delta\hat{x})^2 \stackrel{\text{def}}{=} \left\langle\hat{x}^2\right\rangle - \langle\hat{x}\rangle^2$.

The family of states for which $\Delta\hat{x}$ is constant in time is foliated by the value of $\Delta_0\hat{x}$, with a minimal value being strictly positive. Indeed, from the Heisenberg inequality

$$\Delta_0\hat{x}\Delta_0\hat{p} \geqslant \frac{\hbar}{2}. \tag{589}$$

and from the first condition in (588), we have

$$\Delta_0\hat{x} \geqslant \sqrt{\frac{\hbar}{2m\omega}}. \tag{590}$$

Now one can show that the only states minimising this inequality are the coherent states of Schrödinger! We have thus found a characterisation of them: they are these states whose spreading in position $\Delta\hat{x}$ is *constant and minimal*. Both conditions are important. Otherwise, there are states whose spreading is momentarily smaller but will grow later to a larger value. There are also states whose spreading is constant, but not minimal (like the $\psi_n$). Geometrically, the two conditions select a 2-dimensional submanifold out of the infinite-dimensional space $L^2(\mathbb{R})$.

### 17.3.4 *Minimal time average*

There is another way to make the third property of Schrödinger more precise. Consider the time average of $\Delta\hat{x}$:

$$T[\Delta\hat{x}] = \sqrt{\frac{(\Delta_0\hat{x})^2}{2} + \frac{(\Delta_0\hat{p})^2}{2m^2\omega^2}}. \tag{591}$$

Now, from Heisenberg inequality, this time average is bounded by

$$T[\Delta\hat{x}] \geqslant \sqrt{\frac{\hbar}{2m\omega}}. \tag{592}$$

This inequality is saturated for the coherent states and only for them. So we have a second characterisation of coherent states as the states which minimise $\Delta\hat{x}$ on average (as time goes by).

### 17.3.5 *'Almost classical energy'*

Let us now turn to the second property underlined by Schrödinger: 'the average energy is almost classical'. As we said before, the classical behaviour of $\langle\hat{x}\rangle$ cannot reasonably be taken as evidence for the classicality of coherent states. So one may wonder whether the same holds for $\langle\hat{H}\rangle$. It is easy to see that

$$\left\langle\hat{H}(\hat{x},\hat{p})\right\rangle = H(\langle\hat{x}\rangle, \langle\hat{p}\rangle) + \frac{1}{2m}(\Delta\hat{p})^2 + \frac{1}{2}m\omega^2(\Delta\hat{x})^2, \tag{593}$$

where

$$H(x,p) \stackrel{\text{def}}{=} \frac{1}{2m}p^2 + \frac{m\omega^2}{2}x^2 \tag{594}$$

is the classical hamiltonian, function over the phase space $\mathbb{R}^2$ with coordinates $(x, p)$. From Heisenberg inequality one deduces then that

$$\langle \hat{H}(\hat{x}, \hat{p}) \rangle - H(\langle \hat{x} \rangle, \langle \hat{p} \rangle) \geqslant \frac{\hbar\omega}{2}. \tag{595}$$

One can say in precise terms that a state is quasi-classical with respect to the energy if it saturates this inequality. And happily, this condition alone is sufficient to define the coherent states!

### 17.3.6 *Conclusion*

We have cleaned up the properties of coherent states underlined by Schrödinger in his seminal paper. First, we have realised that the first property was very generic and not specific to coherent states. Second a careful analysis of the two other properties has led us to formulate three equivalent definitions of coherent states of the harmonic oscillator:

1. Constant and minimal $\Delta\hat{x}$
2. Minimal temporal average of $\Delta\hat{x}$
3. Minimal $\big(\langle \hat{H}(\hat{x}, \hat{p}) \rangle - H(\langle \hat{x} \rangle, \langle \hat{p} \rangle)\big)$.

We regard these definitions as better suited for a pedagogical introduction to coherent states, compared to the abstract definition as eigenstates of the annihilation operator that one finds in most textbooks.

These three definitions are *dynamical* in the sense that they make use of the temporal evolution of states or the hamiltonian. A priori, if another hamiltonian is used, like for the free particle, another family of states will be found. In this sense, one can talk indeed of the coherent states *of the harmonic oscillator*, and not *of the free particle*. The stability of the family of coherent states under temporal evolution is made obvious with the two first definitions, but not so much with the third one.

In an attempt of generalisation of the notion of coherent states, we are now going to relax this dynamical aspect of the definitions and propose a purely kinematical characterisation.

## 17.4 KINEMATICAL CHARACTERISATION

Let's start all over again, from a general quantum system. Its states form a Hilbert space $\mathcal{H}$, and we consider the problem of finding the states which are 'quasi-classical' in a sense to be determined.

### 17.4.1 *Geometrical formulation*

As regards its kinematical features, the departure of quantum mechanics from classical mechanics can be understood geometrically, by the so-called *geometrical formulation of quantum mechanics* [167, 198].

In quantum mechanics, we are used to systems whose states are taken to be vectors of a Hilbert space $\mathcal{H}$ endowed with a scalar product $\langle .|. \rangle$, and the algebra of observables $\mathcal{B}_{\mathbb{R}}(\mathcal{H})$ consists of the (bounded) self-adjoint linear operators over $\mathcal{H}$. In fact, we only consider the normalised vectors of $\mathcal{H}$, up to a global phase, so that the space of physical states really is the projective Hilbert space $P\mathcal{H}$. $P\mathcal{H}$ is a Khäler manifold which means that it is naturally endowed with two geometric structures: a symplectic 2-form $\omega$ (coming from the imaginary part of $\langle .|. \rangle$) and a riemannian metric $g$ (from the real part of $\langle .|. \rangle$). Then, the algebra of observables $\mathcal{B}_{\mathbb{R}}(\mathcal{H})$ can be recast as the space of functions of $\mathcal{C}^\infty(P\mathcal{H}, \mathbb{R})$ which preserve both geometric structures, i.e. whose hamiltonian vector fields are also Killing vector fields.

Although it may look a bit abstract, this formulation frames quantum mechanics in very similar terms to classical mechanics, where the space of states is a symplectic manifold $(\mathcal{P}, \omega)$, and the observables are functions of $\mathcal{C}^\infty(\mathcal{P}, \mathbb{R})$. In this framework, both classical and quantum spaces of states are symplectic manifolds, but the quantum case bears the additional structure of a riemannian manifold. One does not need to know the details of the geometrical formulation to understand the point we want to make, that is, classical mechanics can be seen as the particular case of quantum mechanics when the riemannian structure is trivial, i.e. $g = 0$!

This fact is of importance because the riemannian metric gives precisely a measure of the uncertainty of observables. An observable $\hat{A} \in \mathcal{B}_{\mathbb{R}}(\mathcal{H})$, defines a function $A$ over $P\mathcal{H}$ by $A : |\psi\rangle \mapsto \langle A \rangle$, and thus a hamiltonian vector field $X_A$. One can prove that

$$\Delta \hat{A} = g(X_A, X_A). \tag{596}$$

In the classical case ($g = 0$), we have $\Delta \hat{A} = 0$ for any observable $\hat{A}$ and state $|\psi\rangle$. Classical mechanics is quantum mechanics without uncertainty.

One may wonder whether it was necessary to appeal to the abstract geometrical formulation to reach this conclusion. Indeed the result goes along very well with the intuitive idea that the quantum is fuzzy, while the classical is peaked. However, the geometrical formulation brings clarity and precision to the debate, and points towards a definite mathematical direction where to look for classicality inside the quantum realm.

The quest for 'quasi-classical' states can now be reformulated in the following terms. Are there states $|\psi\rangle$ for which $\Delta \hat{A} = 0$ for any observable $\hat{A}$?

### 17.4.2 *Eigenstates*

Start considering a single observable $\hat{A}$. What are the states that satisfy $\Delta\hat{A} = 0$? It is easily shown that they are all, and only, the eigenstates $|a\rangle$ of $\hat{A}$. Thus, an eigenstate shows some classical features, which is not a surprise after all: the eigenstate $|a\rangle$ of $\hat{A}$ is very peaked *with respect to* $\hat{A}$. Similarly, $|x\rangle$ is classical in the sense that $\Delta\hat{x} = 0$, i.e. it is very peaked with respect to $\hat{x}$, which was indeed our first attempt to define 'quasi-classical state' in section 17.3. So the last question of the previous paragraph, admits a direct answer: no. Because if $\Delta\hat{A} = 0$ for all $\hat{A}$, then $|\psi\rangle$ is an eigenstate of all $\hat{A}$ which is not possible.

Our expectations have to be qualified, and one can look instead for states which satisfy $\Delta\hat{A} = 0$ for *some* observables $\hat{A}$, i.e. a common eigenstate of a subset $\mathcal{A} \subset \mathcal{B}_{\mathbb{R}}(\mathcal{H})$. Such an eigenstate can be said 'quasi-classical' *with respect to* $\mathcal{A}$. If $\mathcal{A}$ is commutative, then its common eigenstates form a basis of $\mathcal{H}$. Such are the eigenstates of a CSCO which may be regarded in this respect as the most classical states of a given quantum system. However, if $\mathcal{A}$ is non-commutative, there will be generically no common eigenstates. The question is now shifted to the definition of 'quasi-classical' states with respect to a non-commutative set of observables.

### 17.4.3 *Squeezed coherent states*

Let's consider two non-commutative observables $\hat{A}$ and $\hat{B}$. Generically, they do not share any common eigenstates, so that we cannot have both $\Delta\hat{A} = 0$ and $\Delta\hat{B} = 0$. One has to find instead a fair trade-off between $\Delta\hat{A}$ and $\Delta\hat{B}$, so that they are both *small*, although non zero. The trade-off is ruled by Heisenberg inequality which reads

$$\Delta\hat{A}\Delta\hat{B} \geqslant \frac{1}{2}\left|\langle[\hat{A},\hat{B}]\rangle\right|. \qquad (597)$$

One can show[6] that this inequality is saturated precisely when $|\psi\rangle$ is an eigenstate either of $\hat{A}$, or of $\hat{B}$, or of

$$\hat{A} + i\gamma\hat{B} \qquad (598)$$

with $\gamma \in \mathbb{R}$. The meaning of this $\gamma$ is understood with the following corollary:

$$\gamma = \frac{\Delta\hat{A}}{\Delta\hat{B}} \qquad (599)$$

It ponders the respective weights of $\hat{A}$ and $\hat{B}$. The eigenstates of $\hat{A} + i\gamma\hat{B}$ are called the *$\gamma$-squeezed coherent states with respect to $\hat{A}$ and $\hat{B}$.*

6 See for instance [93] p. 244.

### 17.4.4 *Application to* $\hat{x}$ *and* $\hat{p}$

Let's apply the result to $\hat{A} = \hat{x}$ and $\hat{B} = \hat{p}$. Heisenberg inequality reads

$$\Delta\hat{x}\Delta\hat{p} \geqslant \frac{\hbar}{2}. \tag{600}$$

The normalised eigenstate of $\hat{x} + i\gamma\hat{p}$, with eigenvalue $z \in \mathbb{C}$, is

$$\psi_{z,\gamma}(x) = \frac{e^{-\frac{(\operatorname{Im} z)^2}{2\gamma\hbar}}}{\sqrt[4]{\pi\gamma\hbar}} e^{-\frac{(x-z)^2}{2\gamma\hbar}} \tag{601}$$

The normalisation is only possible for $\gamma > 0$. Thus, the squeezed coherent states (with respect to $\hat{x}$ and $\hat{p}$) form a 3-dimensional submanifold of $PL^2(\mathbb{R})$, parametrised by $\gamma$ and $z$.

The Schrödinger coherent states of equation (571) are recovered by fixing $\gamma = \frac{1}{m\omega}$. More precisely, we have

$$\psi_\alpha(x,t) = e^{i\frac{\alpha^2}{2}\sin(2\omega t)} e^{i\frac{\omega t}{2}} \psi_{z,\gamma}(x) \tag{602}$$

with

$$\gamma = \frac{1}{m\omega} \qquad \text{and} \qquad z = \sqrt{\frac{2\hbar}{m\omega}}\alpha e^{i\omega t}. \tag{603}$$

We have found an equivalent definition of Schrödinger coherent states: they are the states which minimise $\Delta\hat{x}\Delta\hat{p}$, with equal weight[7] for $\hat{x}$ and $\hat{p}$.

### 17.4.5 *Kinematics vs dynamics*

This new characterisation of coherent states differs from the two previous one of Schrödinger in a central aspect: it only refers to the *kinematics*, and not to the *dynamics*. Indeed, the previous definitions were involving the specific form of the hamiltonian $\hat{H}$ of the harmonic oscillator, while now the definition only uses two observables $\hat{x}$ and $\hat{p}$ acting on the Hilbert space $L^2(\mathbb{R})$.

The move is noticeable because, for instance, the free particle and the harmonic oscillator have the same kinematics, and only differ by their dynamics. From this new kinematical characterisation, the previous coherent states of the harmonic oscillator could be equally called coherent states of the free particle, while the original dynamical characterisation was dismissing such a possibility.

To be clear, the family of coherent states, as a whole, can be fully characterised by kinematical considerations, but it will only exhibit nice dynamical properties in a particular case. Indeed, the family is

7 I.e. $\Delta\hat{p} = m\omega\Delta\hat{x}$. The constant $m\omega$ guarantees the homogeneity of the physical dimension.

stable under the harmonic oscillator evolution, whereas it is not with the free particle one.

The transitional role from the dynamical to the kinematical perspective is played by the operator

$$\hat{a} \overset{\text{def}}{=} \sqrt{\frac{m\omega}{2\hbar}} \left( \hat{x} + \frac{i}{m\omega} \hat{p} \right). \tag{604}$$

The task of minimising Heisenberg inequalities has been shown to reduce to that of finding the eigenstates of this operator, which are precisely the Schrödinger coherent states. This way, the operator $\hat{a}$ has arisen through purely kinematical considerations. However, the same operator plays an important role in the dynamics of the harmonic oscillator, where it is known as the *annihilation operator*, for it acts destructively over the eigenstates of $\hat{H}$:

$$\hat{a}\, \psi_n = \sqrt{n}\, \psi_{n-1}. \tag{605}$$

It is important to keep in mind this double role of $\hat{a}$ to understand better later generalisations of coherent states.

## 17.5 OPTICAL COHERENCE

In the previous section, we have been looking for quasi-classical states and ended up with an abstract definition of coherent states as eigenstates of the so-called annihilation operator $\hat{a}$. This definition is the one used in many textbooks to define coherent states at first. It is disappointing to see that it is often not motivated by physical considerations.

We have shown how it could be motivated in a rather abstract way, from the geometrical formulation and the minimisation of Heisenberg inequalities. We are now going to show a more experimentally grounded way to introduce it, which is also the path that was followed historically. It is the way of Glauber when he revived coherent states, thirty years after Schrödinger, in the concrete context of quantum optics. Besides, we will understand why the adjective 'coherent' can be preferred to 'quasi-classical'. The formulas of this section are taken from the collection book [84].

### 17.5.1 *Quantum optics*

Quantum optics describes light using the theory of quantum electrodynamics (QED). The electromagnetic field is described by a state in a Hilbert space, and the observable quantities are described by the electric and magnetic hermitian operators, $\hat{\vec{E}}(\vec{r}, t)$ and $\hat{\vec{B}}(\vec{r}, t)$. In absence of any source, the time evolution of states is driven by the hamiltonian

$$\hat{H} = \frac{1}{2} \int d\vec{r}\, (\hat{\vec{E}}^2 + \hat{\vec{B}}^2). \tag{606}$$

In the time gauge, the observables can in fact be derived from a vector potential $\hat{\vec{A}}$ such that

$$\hat{\vec{E}} = -\frac{1}{c}\frac{\partial \hat{\vec{A}}}{\partial t} \qquad \text{and} \qquad \hat{\vec{B}} = \nabla \times \hat{\vec{A}}. \tag{607}$$

Assuming the field is confined within a cubic box of side L, the vector potential $\hat{\vec{A}}$ can be decomposed into a superposition of modes k such that

$$\hat{\vec{A}}(\vec{r},t) = c\sum_k \left(\frac{\hbar}{2L^3\omega_k}\right)^{1/2} \Big(\hat{a}_k\,\vec{e}_\lambda e^{i(\vec{k}\cdot\vec{r}-\omega_k t)} + \hat{a}_k^\dagger\,\vec{e}_\lambda e^{-i(\vec{k}\cdot\vec{r}-\omega_k t)}\Big), \tag{608}$$

where the sum is made over an index k, used as a shorthand for $(\lambda,\vec{k})$, where $\vec{e}_\lambda$ ($\lambda \in \{1,2\}$) is the polarisation vector, perpendicular to $\vec{k}$, and $\vec{k}$ ranges over a discrete set of values permitted by the boundary conditions. Then $\hat{a}_k$ and $\hat{a}_k^\dagger$ are operators associated respectively to the positive and negative frequency part of $\hat{\vec{A}}$. They satisfy

$$[\hat{a}_k,\hat{a}_l] = [\hat{a}_k^\dagger,\hat{a}_l^\dagger] = 0 \qquad \text{and} \qquad [\hat{a}_k,\hat{a}_l^\dagger] = \delta_{kl}. \tag{609}$$

The hamiltonian can be rewritten as

$$\hat{H} = \sum_k \hbar\omega_k \left(\hat{a}_k^\dagger\hat{a}_k + \frac{1}{2}\right). \tag{610}$$

Thus, the electromagnetic field is mathematically equivalent to an assembly of one-dimensional harmonic oscillators (one per mode k), so that $\hat{a}_k$ and $\hat{a}_k^\dagger$ are properly annihilation and creation operators. The basis of eigenstates of $\hat{H}$ is immediately deduced:

$$\bigotimes_k |n_k\rangle \tag{611}$$

where $n_k$ is the number of photons in the mode k.

### 17.5.2 *Interacting theory*

The basis of stationary states of the free theory is not best suited to describe the states of light coming out of photon beams. Instead, the family of coherent states is much more convenient, and it appears naturally once one considers interactions.

So far, we have described the free theory of the electromagnetic field. But light is created by sources, like lamps or antennae, which consist of charges exciting the electromagnetic field. It is thus crucial to describe the interaction of light with charged matter. A simple model consists in the description of the photon field radiated by a classical electric current. A classical current $\vec{j}(\vec{r}, t)$ is assumed to interact with the vector potential through the following hamiltonian of interaction:

$$\hat{H}_I = \frac{1}{c}\int \vec{j} \cdot \hat{\vec{A}}\, d\vec{r}. \tag{612}$$

Starting at initial time in the vaccum state $|0\rangle$, the field gets excited, and ends up at time t in a state

$$|t\rangle = e^{i\phi(t)} e^{\frac{i}{\hbar}\int_0^t \hat{H}_I(t')dt'}|0\rangle \tag{613}$$

where the phase $\phi(t)$ admits a definite expression, but irrelevant for our purposes. It can be rewritten

$$|t\rangle = e^{i\phi(t)}\prod_k |\alpha_k(t)\rangle, \tag{614}$$

where $|\alpha_k(t)\rangle$ is the coherent state with

$$\alpha_k(t) = \frac{i}{\sqrt{2L^3\hbar\omega_k}}\int_0^t \vec{j} \cdot \vec{e}_\lambda e^{-i(\vec{k}\cdot\vec{r}+\omega_k t')} d\vec{r} dt'. \tag{615}$$

This hamiltonian of interaction is a good model for most of the macroscopic sources where radiation is generated by a charge current $\vec{j}(\vec{r}, t)$ whose expression is known. In practice, lasers indeed produce coherent states of light, but incandescent bulbs do not, for they consist of many independent and chaotic sources which break the overall coherence.

Thus coherent states have appeared as the most natural states of the electromagnetic field when it is minimally coupled to a classical source. In addition to this special role in the production of light, coherent states exhibit major features from the point of view of its detection, that we explain now.

### 17.5.3 *Maximising interference*

To detect light, one uses a photon counter. Typically, a photon counter is sensible to the intensity of the electric field E, that we now assume to be only a scalar field E, for simplicity. If you assume that the electromagnetic field is in a state $|\psi\rangle$, then the intensity in $x = (r, t)$ (the spacetime point where the detector is) is on average

$$\langle\psi|\hat{E}^{(-)}(x)\hat{E}^{(+)}(x)|\psi\rangle\,. \tag{616}$$

where $\hat{E}^{(-)}$ is the negative frequency part of the electric field, conjugate to the positive frequency part $E^{(+)}$, which is

$$\hat{E}^{+}(\vec{r}, t) = i\sum_k \left(\frac{\hbar\omega_k}{2L^3}\right)^{1/2} \hat{a}_k\, e^{i(\vec{k}\cdot\vec{r}-\omega_k t)}. \tag{617}$$

It is a fancy way of writing that the energy of the electric field is $|E|^2$.

Let's define the first-order (or two-point) correlation function as

$$G^{(1)}(x_1,x_2) \stackrel{\text{def}}{=} \langle\psi|\hat{E}^{(-)}(x_1)\hat{E}^{(+)}(x_2)|\psi\rangle\,. \tag{618}$$

In a double-slit experiment, the interference pattern observed on the screen is a measure of $G^{(1)}(x,x)$ along the screen. In fact the electric field $E(x)$ on the screen is the linear superposition of the electric field $E(x_1)$ and $E(x_2)$ that was emitted by each of the two slits at spacetime points $x_1$ and $x_2$:

$$\hat{E}(x) \propto \hat{E}(x_1) + \hat{E}(x_2). \tag{619}$$

As a consequence,

$$G^{(1)}(x,x) = G^{(1)}(x_1,x_1) + G^{(1)}(x_2,x_2) + 2\,\mathrm{Re}\,G^{(1)}(x_1,x_2). \tag{620}$$

The two first terms are the independent contributions from each slit. The last term is responsible for the interference. When $G^{(1)}(x_1,x_2) = 0$, no fringes are observed. In fact, the visibility of the fringes is given by

$$\nu = \frac{I_{max} - I_{min}}{I_{max} + I_{min}} = \frac{2|G^{(1)}(x_1,x_2)|}{G^{(1)}(x_1,x_1) + G^{(1)}(x_2,x_2)} \tag{621}$$

Now one can show the inequality:

$$|G^{(1)}(x_1,x_2)|^2 \leqslant G^{(1)}(x_1,x_1)G^{(1)}(x_2,x_2) \tag{622}$$

so that, keeping $G^{(1)}(x_1,x_1)$ and $G^{(1)}(x_2,x_2)$ fixed, the maximum of interference is obtained for

$$|G^{(1)}(x_1,x_2)| = \sqrt{G^{(1)}(x_1,x_1)G^{(1)}(x_2,x_2)}. \tag{623}$$

When this condition is assumed to be valid for all $x_1$ and $x_2$, one can show that there exists a function $\mathcal{E}(x)$ so that

$$G^{(1)}(x_1,x_2) = \mathcal{E}^*(x_1)\mathcal{E}(x_2). \tag{624}$$

$G^{(1)}(x_1,x_2)$ factorises, and the state $|\psi\rangle$ is said to be *optically coherent*.

It is easy to see from the definition (618) that a sufficient condition for the factorisability of $G^{(1)}(x_1,x_2)$ is that $|\psi\rangle$ is an eigenstate of $\hat{E}^{(+)}(x)$ for all $x$. From equation (617), we see that it is equivalent to say that $|\psi\rangle$ is an eigenstate of $a_k$ for all $k$. And here we land on our feet! This is indeed the definition of coherent states that was given previously but here applied to an assembly of independent harmonic oscillators. Here the definition is motivated on strong physical ground: coherent states are those which maximise the inference pattern or said differently, that factorise the 2-point correlation function[8]!

8 Technically, being a coherent state is only a sufficient, and not a necessary condition to be optically coherent, i.e. to factorise the 2-point correlation function, and so maximise the interference pattern, but we ignore this subtlety here.

### 17.5.4 *Coherence, Classicality, and Purity*

At this stage, the origin of the word 'coherent' has been brought to light: the state $|\psi\rangle$ is such that the values of the field at different points of space-time 'conspire' together to maximise the interference pattern. Meanwhile, we have lost sight of the sense in which they can be seen as 'quasi-classical'. Even worse, coherence and classicality may seem contradictory. Indeed, an example of coherent light is that produced by a laser, which is usually presented as a very quantum device, far from anything classical. On the contrary, ordinary light produced by incandescent bulbs, close to black-body radiation, is optically very incoherent, while it seems to be much more 'classical' than lasers. Where is the catch?

The paradox arises from the confusion of two layers of 'classicality'. The first layer is classicality as the minimisation of the uncertainties of some non-commuting observables. Compared to the previous example of the harmonic oscillator, the vector potential $\hat{A}$ and the electric field $\hat{E}$ play now respectively the role of the position $\hat{x}$ and the momentum $\hat{p}$. Coherent states of light are quasi-classical in the sense that they minimise $\Delta\hat{A}$ and $\Delta\hat{E}$ together.

The second layer of classicality is the difference between pure and mixed states. Classical physics is usually very noisy, that is very mixed, due to the difficulty to control interactions with the environment. For instance, the light of an incandescent bulb is a very mixed state (black-body radiation is maximally mixed), so that it is tempting to say that it is more classical with respect to a pure state, which is much more difficult to create in a lab.

The two layers, coherent/incoherent and pure/mixed, shall not be confused and are actually independent. Many pure states are incoherent, while some mixed states can be coherent. For instance, the state of an ideal laser is actually a mixed state, which reads

$$\rho = \frac{1}{2\pi}\int_0^{2\pi} \left| |\alpha|e^{i\theta} \middle\rangle\middle\langle |\alpha|e^{i\theta} \right| \, d\theta, \tag{625}$$

although it is optically coherent (it factorises the 2-point correlation function). This distinction is sometimes overlooked, especially in the context of quantum gravity, where one usually exclusively considers pure coherent states. This state of research may surprise as it is reasonable to believe that quantum states of space are horrendously hard to isolate. Considering mixed states instead may have some relevance in solving some of the hard problems in the field [3].

## 17.6 ALGEBRAIC APPROACH

### 17.6.1 *Displacement operator*

The vacuum state is the only coherent state which is also an eigenstate of the hamiltonian of the harmonic oscillator:

$$|\alpha = 0\rangle = |n = 0\rangle \,. \tag{626}$$

For this reason, there is no ambiguity[9] and one can write $|0\rangle$. In the previous section, we have seen how the other coherent states are generated from the vacuum $|0\rangle$ by the unitary evolution of a simple hamiltonian of interaction $\hat{H}_I$. Equation (613) can be rewritten

$$|t\rangle = e^{i\phi(t)} \prod_k \hat{D}_k(\alpha_k(t))\,|0\rangle \tag{627}$$

with $\alpha_k(t)$ given by equation (615), and $\hat{D}$ a unitary operator-valued function over $\mathbb{C}$, called the *displacement operator*, and defined by

$$\hat{D}(\alpha) \stackrel{\text{def}}{=} e^{\alpha\hat{a}^\dagger - \alpha^*\hat{a}}. \tag{628}$$

It is easy to check indeed that[10]

$$|\alpha\rangle = \hat{D}(\alpha)\,|0\rangle \,. \tag{629}$$

### 17.6.2 *Heisenberg group*

The set of displacement operators almost form a group:

$$\hat{D}(\alpha+\beta) = e^{-i\,\mathrm{Im}(\alpha\beta^*)}\hat{D}(\alpha)\hat{D}(\beta) \tag{630}$$

In other words they form a group up to a phase. More precisely, they form a subset of a group, called the Heisenberg group. Let's see what it is.

The position and momentum operators $\hat{x}$ and $\hat{p}$ generate a Lie algebra called the *Heisenberg algebra* (or also the *Weyl algebra*). It is the smallest algebra generated by $\hat{x}$ and $\hat{p}$, by linear combination and

$$\begin{aligned}\hat{D}(\alpha)\,|0\rangle &= e^{-|\alpha|^2/2} e^{\alpha a^\dagger} e^{-\alpha^* a}\,|0\rangle \\ &= e^{-|\alpha|^2/2} e^{\alpha a^\dagger}\,|0\rangle \\ &= e^{-|\alpha|^2/2}\sum_n \frac{\alpha^n}{\sqrt{n!}}\,|n\rangle \\ &= |\alpha\rangle\end{aligned}$$

---

9 Whereas it would be ambiguous to write for instance $|1\rangle$, since $|\alpha = 1\rangle = e^{-1/2}\sum_{n=0}^{\infty} \frac{1}{\sqrt{n!}}|n\rangle \neq |n = 1\rangle$.

10 PROOF.Use the formula $e^{A+B} = e^A e^B e^{-[A,B]/2}$.

Lie bracketing with $i[.,.]$. It is a 3-dimensional non-commutative real algebra, denoted $\mathfrak{h}$ and consisting of elements of the form

$$a\hat{x} + b\hat{p} + c\hat{I} \quad a, b, c \in \mathbb{R}, \tag{631}$$

where $\hat{I}$ is the identity operator.

Exponentiating the Heisenberg algebra gives a 3-dimensional real Lie group, which is called, wisely, the *Heisenberg group* (or also the *Weyl group*), denoted $H_3$. It consists of elements of the form

$$e^{i(a\hat{x}+b\hat{p}+c\hat{I})} \quad a, b, c \in \mathbb{R}. \tag{632}$$

The displacement operators are just a subset of this group, such that

$$\hat{D}(\alpha) = e^{i\sqrt{2}\,\mathrm{Im}\,\alpha\,\hat{x} - i\sqrt{2}\,\mathrm{Re}\,\alpha\,\hat{p}}. \tag{633}$$

But since the global phase of states is irrelevant, one also gets the family of coherent states!

In fact, from equation (630), it is easy to see that one can generate the whole family of coherent states from any $|\alpha\rangle$, and not only $|0\rangle$. One says that the action of the Heisenberg group is transitive: any two coherent states are related by a transformation of the Heisenberg group.

### 17.6.3 *Generalisation*

The previous analysis motivates a generalisation of coherent states for any Lie group $G$ acting over a Hilbert space $\mathcal{H}$. It was first proposed in parallel by Perelomov [145, 146] and Gilmore [82].

Let $G$ be a Lie group, and $T$ a unitary irrep of $G$ over a Hilbert space $\mathcal{H}$. Choose $|\psi_0\rangle \in \mathcal{H}$, and denote $H$ the subgroup which stabilises $|\psi_0\rangle$ up to a phase, i.e.

$$H \stackrel{\text{def}}{=} \left\{ g \in G \mid \exists \phi \in \mathbb{R}, \quad T(g)\,|\psi_0\rangle = e^{i\phi}\,|\psi_0\rangle \right\}. \tag{634}$$

The family of generalised coherent states is defined as the orbit of $|\psi_0\rangle$ under the action of the (left) quotient space $G/H$. More precisely, for each class $x \in G/H$, choose a representative $g(x) \in G$, and define the generalised coherent states as

$$|x\rangle = T(g(x))\,|\psi_0\rangle\,. \tag{635}$$

Thus, the generalisation of coherent states depends a priori on many choices: a group $G$, a unitary irrep $T$, a vector $|\psi_0\rangle$ and a set of representatives $g(x)$. Of course, when it is projected down to the projective Hilbert space $P\mathcal{H}$, the set of coherent states does not depend on the choice of the representatives $g(x)$. The choice of initial state $|\psi_0\rangle$ is a priori arbitrary, but can be motivated by another criterion, like being a ground state or minimising some uncertainty relations.

In the case of Schrödinger coherent states $|\alpha\rangle$, the group is $H_3$, with stabiliser U(1), the initial state is the ground state $|0\rangle$ of the harmonic oscillator, and the representatives are the displacement operators $D(\alpha)$.

### 17.6.4 *Bloch states*

When we apply the method to the Lie group SU(2), we obtain what quantum opticians call *Bloch states*. The unitary irreps of SU(2) are the Hilbert spaces $\mathcal{Q}_j$ (see chapter 4. As an initial state we choose[11] $|j, -j\rangle$. Then, we can show that the stabiliser is U(1), and we have the following diffeomorphism, reminiscent of Hopf fibration (see chapter 3),

$$SU(2)/U(1) \cong S^2. \tag{636}$$

The unit sphere $S^2$ can be parametrised by a complex number $\zeta \in \mathbb{C}$ (except for one point), by (the inverse of) the stereographical projection $\vec{n}(\zeta)$ (eq. (77)). The representative $u \in SU(2)$ for each class $\vec{n}(\zeta) \in S^2$ is given by what is sometimes called the *Hopf section*

$$u(\zeta) \stackrel{\text{def}}{=} \frac{1}{\sqrt{1+|\zeta|^2}} \begin{pmatrix} 1 & \zeta \\ -\zeta^* & 1 \end{pmatrix}. \tag{637}$$

Finally we define the SU(2) coherent states as

$$|j, \vec{n}(\zeta)\rangle \stackrel{\text{def}}{=} u(\zeta)\,|j, -j\rangle\,. \tag{638}$$

In terms of the magnetic basis, one can show that

$$|j, \vec{n}(\zeta)\rangle = \frac{1}{(1+|\zeta|^2)^j} \sum_{m=-j}^{j} \begin{pmatrix} 2j \\ j+m \end{pmatrix}^{\frac{1}{2}} \zeta^{j+m}\,|j, m\rangle\,. \tag{639}$$

Among the important properties that these states satisfy, we should note that the SU(2) coherent states are eigenstates of $\vec{n} \cdot \vec{J}$

$$\vec{n} \cdot \vec{J}\,|j, \vec{n}\rangle = -j\,|j, \vec{n}\rangle\,. \tag{640}$$

Also they saturate the following Heisenberg inequality

$$\Delta J_1 \Delta J_2 \geqslant \frac{1}{2} |\,\langle J_3\rangle\,|. \tag{641}$$

Finally, they also satisfy the following resolution of the identity

$$\frac{2j+1}{4\pi} \int_{S^2} d\vec{n}\,|j, \vec{n}\rangle\,\langle j, \vec{n}| = \mathbb{1}, \tag{642}$$

with $d\vec{n}$ being the usual measure on the unit sphere $S^2$. The Bloch states have latter been used in quantum gravity as we shall see in section (17.8).

11 The choice $|j, j\rangle$ is often made too.

### 17.6.5 *Dynamical group*

From the perspective of experimentalists, the Lie group G is not something abstract but something very concrete, for its action drives the unitary time evolution of states. In quantum optics, one deals typically with some effective model of perturbed hamiltonian:

$$\hat{H} = \hat{H}_0 + \hat{H}_{pert}. \tag{643}$$

The initial state is chosen to be the ground state of $\hat{H}_0$, and the coherent states are generated through the time evolution induced by the perturbation $\hat{H}_{pert}$, which can be due to the coupling to some classical current as in equation (612).

In this context, the group G is sometimes called a *dynamical symmetry group*, so that for instance $H_3$ is said to be *the* dynamical (symmetry) group of the harmonic oscillator [64, 82]. This naming is confusing because it conflicts both with the notion of 'dynamical group', as defined by Souriau in [173], and with the usual notion of 'symmetry group' of a hamiltonian.

Usually, a classical system is given by a phase space $(\mathcal{P}, \omega)$ and a hamiltonian H. Souriau defines a dynamical group as any Lie group G acting as a symplectomorphism (canonical transformation) over $\mathcal{P}$. Then one can consider the symmetries of the hamiltonian, i.e. the functions $C_i \in \mathcal{C}^\infty(\mathcal{P}, \mathbb{R})$ such that

$$\{C_i, H\} = 0, \tag{644}$$

where $\{.,.\}$ is the Poisson bracket associated to $\omega$. The $C_i$ generate a Lie algebra of conserved quantities, which can be exponentiated into a Lie group, which is called the *symmetry group* of the hamiltonian H. This group is acting over the phase space as symplectomorphism and for that reason, it is sometimes emphasised as the *dynamical symmetry group* of H. In this sense, $H_3$ is not the dynamical symmetry group of the harmonic oscillator!

This should not be too much of a surprise because, as we said earlier, the Schrödinger coherent states can be defined independently of the *dynamics* of the harmonic oscillator. What might be regarded as dynamical in them is the initial choice of the ground state $|0\rangle$, but the Heisenberg group that further generates them is built from a choice of coordinates $(x, p)$ over the phase space, that is, independently of the specific form of the hamiltonian of the harmonic oscillator.

## 17.7 GEOMETRIC APPROACH

In the two previous sections, we have seen how some initial quantum optical work by Glauber in the 1960s [84], has lead in the 1970s, to a generalisation of coherent states, by Perelomov [145] and Gilmore [82], using Lie groups. In this section, we explore a second and independent path of generalisation in more geometrical terms, which was proposed in the 1990s by Hall [90, 91], based on some earlier works by Segal [171] and Bargmann [20] in the 1960s. Both approaches have their relevance for quantum gravity, as we shall see in section 17.8.

### 17.7.1 *Phase space vs Hilbert space*

The classical phase space of the harmonic oscillator is $T^*\mathbb{R}$, endowed with the usual symplectic structure given by the determinant. The quantum analogue is the Hilbert space $L^2(\mathbb{R})$, with the usual scalar product. The family of coherent states constitute a 2-dimensional submanifold of $L^2(\mathbb{R})$, parametrised by amplitude and time $(\alpha, t)$, or the complex number $z = \alpha e^{i\omega t}$. It could be as well parametrised by position $x = \alpha \cos \omega t$ and momentum $p = -\alpha\omega \sin \omega t$, so that the diffeomorphism between the classical phase space and the family of coherent states is made explicit. Any point in phase space determines uniquely a coherent state in the Hilbert space, and conversely. This fact is not a coincidence, but rather a crucial aspect of coherent states that sheds further light on the classicality of these states.

It is a general feature of coherent states that they define a natural injection of the classical phase space $\mathcal{P}$ into the Hilbert space $\mathcal{H}$. A priori, there are many possible such injections, but coherent states provide a natural one. To work this out, we shall first see how one can build a Hilbert space $\mathcal{H}$ from a phase space $\mathcal{P}$. Well, this is the whole point of quantisation, and so one may know that the subject is not easy. Nevertheless *geometric quantisation*[12] is a prototypical such method, rather technical, but we can skip the details and keep the general idea.

### 17.7.2 *Geometric quantisation*

Start with a configuration space $\mathcal{M}$ and build the phase space $T^*\mathcal{M}$. There are actually many ways in which $\mathcal{M}$ can be seen as a subspace of $T^*\mathcal{M}$. Each way consists in choosing what is called a *polarisation*. Then one can build a complex line bundle over $T^*\mathcal{M}$, denoted L. It has $\mathbb{C}$ as fibre. The *prequantum Hilbert space*, $\mathcal{G}$, is the space of equivalence classes of square-integrable sections of L, where two sections are said equivalent when they are equal almost everywhere. Roughly, it is the square-integrable functions over $T^*\mathcal{M}$, i. e. $L^2(T^*\mathcal{M})$. It is much too big to be a good quantum Hilbert space. Now a choice of polarisation enables to select a subspace of $\mathcal{G}$ and to build the good quantum Hilbert space $\mathcal{H} \cong L^2(\mathcal{M})$. This construction really enables to see the Hilbert space $\mathcal{H}$ as a subspace of $L^2(T^*\mathcal{M})$. The important point is

12 Not to be confused with the previously discussed geometrical formulation of [167].

that a state of the Hilbert space $|\psi\rangle \in \mathcal{H}$ can be seen as a complex function $\psi$ over the phase space $T^*\mathcal{M}$!

Then, for each point in phase space, $z \in T^*\mathcal{M}$, we define the coherent state $|z\rangle$ as the unique state in $\mathcal{H}$ such that

$$\forall\, |\psi\rangle \in \mathcal{H}, \quad \langle z|\psi\rangle = \psi(z). \tag{645}$$

This definition is both elegant and confusing. Elegant, because the equation is very simple, confusing, because it is too simple. On the LHS, we have a scalar product between two states in $\mathcal{H}$, while on the RHS we have the evaluation of a function $\psi$ in one point $z$ of the phase space $T^*\mathcal{M}$. A practical example should clarify the matter.

### 17.7.3 *Segal-Bargmann transform*

The classical phase space of the harmonic oscillator is $T^*\mathbb{R} \cong \mathbb{C}$, so that the prequantum Hilbert space is roughly $L^2(\mathbb{C})$. By using the Kähler polarisation, the quantum Hilbert space finally obtained is the Segal-Bargmann space[13] [90], denoted $\mathcal{SB}$. It is made of functions over $\mathbb{C}$ which are both holomorphic and square-integrable, with the (Gaussian) scalar product

$$(f_1, f_2) = \int_{\mathbb{C}} \bar{f_1}(z) f_2(z) \frac{i}{2\pi\hbar} e^{-|z|^2/\hbar} dz \wedge d\bar{z} \tag{646}$$

$\mathcal{SB}$ is isomorphic to $L^2(\mathbb{R})$, which is not a big surprise after all, since all Hilbert spaces of the same dimension are isomorphic. What is more interesting is that there is actually an isometry between them two, given by

$$\hat{\phi}(z) = \int_{\mathbb{R}} K(z, x)\phi(x)\, dx \tag{647}$$

with the kernel

$$K(z, x) = \sqrt[4]{\frac{m\omega}{\pi\hbar}} e^{\frac{1}{\hbar}\left(\frac{z^2}{2} - (\sqrt{\frac{m\omega}{2}}x - z)^2\right)} \tag{648}$$

This isomorphism is called the *Segal-Bargmann transform*. Its inverse is given by

$$\phi(x) = \int_{\mathbb{C}} \overline{K(z, x)} \hat{\phi}(z) \frac{i}{2\pi\hbar} e^{-|z|^2/\hbar} dz \wedge d\bar{z} \tag{649}$$

Equation (647) is in fact a concrete instantiation of equation (645), so that the kernel K is actually a coherent state! More precisely, it matches the expression of equation (601), provided a rescaling of $z$, and up to a phase and a normalisation factor:

$$K(z, x) = e^{\frac{i}{\hbar} \operatorname{Re} z \operatorname{Im} z} e^{\frac{1}{\hbar}|z|^2} \psi_{\sqrt{\frac{2}{m\omega}} z, \frac{1}{m\omega}}(x). \tag{650}$$

13 Also called the Fock-Bargmann space in [74].

For this reason the Segal-Bargmann transform is also called the *coherent-state transform*.

In the case of Schrödinger coherent states, the Lie group approach, applied to the Heisenberg group $H_3$, generates the same coherent states as the phase space approach, applied to $T^*\mathbb{R}$. Both approaches are secretly linked by the fact that the Heisenberg group $H_3$ is naturally obtained by exponentiating the quantised coordinates $\hat{x}$ and $\hat{p}$ of the phase space $T^*\mathbb{R}$. However, the two methods do not always match. In the case of the phase space $S^2$, the exponentiation of the quantised coordinates $\sigma_1, \sigma_2, \sigma_3$, generates SU(2). However the coherent states obtained from geometric quantisation of $S^2$ are different from the SU(2) coherent states of equation (638) (see [94]).

### 17.7.4 *Resolution of the identity*

The fact that the coherent-state transform is an isometry is equivalent to the following *resolution of identity* for the coherent states

$$\mathbb{1} = \frac{1}{\pi}\int_{\mathbb{C}} |\alpha\rangle\langle\alpha| \, \mathrm{d}\,\mathrm{Re}\,\alpha \, \mathrm{d}\,\mathrm{Im}\,\alpha. \tag{651}$$

This equation should be understood in the sense of weak convergence, that is, for any two given states $|\phi_1\rangle$ and $|\phi_2\rangle$,

$$\langle\phi_1|\phi_2\rangle = \frac{1}{\pi}\int_{\mathbb{C}} \langle\phi_1|\alpha\rangle \langle\alpha|\phi_2\rangle \, \mathrm{d}\,\mathrm{Re}\,\alpha \, \mathrm{d}\,\mathrm{Im}\,\alpha. \tag{652}$$

This resolution of the identity is similar to the more familiar one of any orthonormal basis of $\mathcal{H}$, like

$$\mathbb{1} = \sum_n |n\rangle\langle n| , \tag{653}$$

up to the crucial difference that the coherent states are parametrised by a *continuous* parameter $\alpha \mapsto |\alpha\rangle$.

The (strong) continuity of the map $\alpha \mapsto |\alpha\rangle$ together with the (weak) resolution of identity are so important that they are often regarded as *the* two properties that coherent states should have to deserve such a designation. It is a bit surprising because it seems too generic, that is not restrictive enough[14], but this is the point of view defended for instance by Klauder in his collection of papers [112].

### 17.7.5 *Heat kernel*

When the geometric quantisation is performed using the Khäler polarisation, the coherent states finally obtained by equation (645) can be expressed in terms of the more romantic notion of the heat kernel

14 In the same manner as the condition of being an orthonormal basis is far from being a sufficient property to define the $|n\rangle$ basis.

[91]. We explain it below for it has played a role in quantum gravity as we will see in the next section.

The heat kernel $\rho_t(x)$ is the solution of the heat equation

$$\frac{d\rho}{dt} = \partial_x^2 \rho \tag{654}$$

that satisfies $\rho_0(x) = \delta(x)$. Its explicit expression is[15]

$$\rho_t(x) = \frac{1}{\sqrt{4\pi t}} e^{-\frac{x^2}{4t}} \tag{655}$$

The kernel K can be rewritten in term of the *heat kernel* $\rho_t(x)$ such that

$$K\left(z\sqrt{\hbar}, x\sqrt{\frac{\hbar}{2m\omega}}\right) = \sqrt[4]{\frac{2m\omega}{\hbar}} \frac{\rho_{\frac{1}{2}}(x-z)}{\sqrt{\rho_{\frac{1}{2}}(x)}}. \tag{656}$$

This observation has suggested a construction of coherent states when the configuration space is a connected compact Lie group G, instead of $\mathbb{R}$ [90]. The heat equation over G reads

$$\frac{d\rho}{dt} = \Delta\rho, \tag{657}$$

where $\Delta$ is the Casimir operator, and $\rho$ a function of $t \in \mathbb{R}$ and $g \in G$. One can show the existence of a smooth and strictly positive solution, called the *heat kernel*, which is a delta over the identity at $t = 0$. It admits the following expansion

$$\rho_t(g) = \sum_\pi \dim \pi \, e^{-\lambda_\pi t} \chi^\pi(g), \tag{658}$$

where the sum is taken over all classes of equivalence of irreps $\pi$, and $\lambda_\pi$ is the characteristic Casimir (non-negative) number of the representation, and $\chi^\pi$ is the character (see chapter 6). For instance, in the case of SU(2) one gets

$$\tilde{\rho}_t(g) = \sum_{j \in \mathbb{N}/2} (2j+1)\, e^{-j(j+1)t} \operatorname{Tr} D^j(g), \tag{659}$$

with $D^j(g)$ the Wigner matrix of g in the spin-j irrep. Then Hall defines the *complexification* $G_\mathbb{C}$ of the Lie group G. For instance, for SU(2) it is $SL_2(\mathbb{C})$. From this he shows that there is a unique analytic continuation of the heat kernel $\rho_t$ from G to $G_\mathbb{C}$. In analogy[16] with

15 The heat equation is just Schrödinger equation with complex time, so that the solution can be easily recovered from (579). However, the solutions in (579) were discarded as coherent states. It seems to be a pure coincidence that the same equation with complex time now gives proper coherent states.

16 The square-root in the denominator in equation (656) is only a normalisation factor, of which one can get rid of, provided a good choice of measure in the definition of the scalar product.

equation (656), the coherent states are defined as the functions over $G$, indexed by $x \in G_{\mathbb{C}}$, as

$$K_{x,t}(g) \stackrel{\text{def}}{=} \rho_t(x^{-1}g), \quad \text{with } g \in G. \tag{660}$$

In [91], it is shown how the heat kernel construction with a group $G$ matches the geometric quantisation approach over the phase space $T^*G$. The latter approach shows how the coherent states provide a natural embedding of the phase space within the corresponding Hilbert space, and in this sense, it points towards their quasi-classical properties. The heat kernel approach presents the advantage of offering more analytical formulas like (660), compared to (645), but it is then harder to see how the quasi-classical properties can arise.

## 17.8 QUANTUM GRAVITY

Quantum gravity is still at a very speculative stage compared to quantum optics, but the two fields of research can speak to one another. Indeed, the experimental control of the latter has motivated many theoretical developments, like the coherent states, which can now be reinvested into the former. At least, having in mind the concrete set-up of quantum optics may help theoreticians of quantum gravity to keep their feet on the ground, so to speak.

Both the algebraic (Lie group) and the geometric (heat kernel) approaches to coherent states have been applied in LQG. Historically, first has come the heat kernel method, starting with an extension of Hall's construction to diffeomorphism invariant gauge theories [10]. Bloch states were later used in [121] and helped in providing a semi-classical picture of the geometry of space.

As explained in chapter 16, the Hilbert space of LQG is

$$\mathcal{H}_{LQG} = \bigoplus_{\Gamma} L^2_{\Gamma}(SU(2)^L/SU(2)^N), \tag{661}$$

where the direct sum is made over all the different abstract finite directed graphs $\Gamma$, with different number of links $L$ and nodes $N$. Over a single link $l$, the space $L^2(SU(2))$ is the quantisation of the phase space $T^*SU(2)$. Actually, since SU(2) is a Lie group, its cotangent bundle can be trivialised as $T^*SU(2) \cong SU(2) \times \mathfrak{su}(2)$, and the coordinates $(h, E)$ are respectively called *holonomy* and *flux*. Subtlety left aside, this description of phase space suggests that it is possible to apply directly the heat kernel method for the Lie group SU(2)! Thus we expect the coherent states of LQG to be indexed by elements of the complexification of SU(2), that is $SL_2(\mathbb{C})$.

### 17.8.1 *Coherent spin-networks*

The rigorous construction of coherent states for LQG was done in [181–184], where they are called *complexifier coherent states*, but we prefer to call them coherent spin-network states as in [33, 34]. They are not properly coherent states for the full theory, but only for its truncation on a fixed graph $\Gamma$. They are denoted $\Psi_{(\Gamma, H_l, t_l)}$, parametrised by an element $H_l \in SL_2(\mathbb{C})$ and a positive number $t_l \in \mathbb{R}^+$, per each link $l$ of $\Gamma$. They are defined by

$$\Psi_{(\Gamma, H_l, t_l)}(g_l) \overset{\text{def}}{=} \int_{SU(2)^N} \left( \prod_l \tilde{\rho}_{t_l}(h_{s(l)}\, g_l\, h^{-1}_{t(l)}\, H_l^{-1}) \right) dh_n \tag{662}$$

with $\tilde{\rho}_t(g)$ the analytic continuation of (659) to $SL_2(\mathbb{C})$. $s(l)$ and $t(l)$ denote respectively the source and the target of the link $l$. The logic of this formula is the following: to each link $l$ is associated the coherent state $\tilde{\rho}_{t_l}(g_l\, H_l^{-1})$, which are then multiplied together, and finally projected it down to $\mathcal{H}_{LQG}$ by a SU(2)-integration over each node to impose the Gauss constraint.

The formal jungle of techniques that helps to define these coherent states in [181–184] should not make us forget about the underlying physical intuition that originally motivates them: they constitute a natural embedding of the classical phase space into the quantum Hilbert space. The embedding is parametrised by $H_l \in SL_2(\mathbb{C})$ which admits two possible semi-classical interpretations as we now explain. Each interpretation depends on a different decomposition of $SL_2(\mathbb{C})$ (see section 2.4).

### 17.8.2 *Holonomy and flux*

The first and original interpretation is based on the polar decomposition of $H_l$. Dropping the $l$ index for readability, it is

$$H = h\, e^{i \frac{E}{8\pi G\hbar\gamma} t}, \tag{663}$$

with $h \in SU(2)$ and $E \in \mathfrak{su}(2)$ being respectively the holonomy and the flux over a link, while $\gamma$ is the Immirzi parameter. They are peaked on the classical holonomy-flux configuration $(h_l, E_l)$ on each link $l$ [165, 182].

### 17.8.3 *Twisted geometry*

The second interpretation is based on the Cartan decomposition

$$H = u(\zeta_s)\, e^{-iz\sigma_3}\, u(\zeta_t) \tag{664}$$

with $\zeta_s, \zeta_t \in \mathbb{C}$, and $u(\zeta) \in SU(2)$ given by equation (637), and

$$z = \xi + i\, a\, t \tag{665}$$

with $\xi, a \in \mathbb{R}$.

With this parametrisation, the state is peaked on a discrete kind of geometry, called *twisted geometry* [70]. The phase space $T^*SU(2)$ on each link is parametrised by $(\vec{n}(\zeta_s), \vec{n}(\zeta_t), \xi, a)$, with the map $\vec{n}(\zeta)$ given by equation (77). Thus each node is surrounded by a set of unit vectors $\pm\vec{n}_i$, one per attached link, with the sign $\sigma = \pm$ chosen depending on whether the link is ingoing or outgoing. At the semi-classical level, the Gauss constraint imposes the closure of the vectors $\vec{n}_i$ surrounding a node, such as

$$\sum_{l \in n} \sigma_l \, a_l \, \vec{n}_l = 0 \tag{666}$$

where the sum is done on all the links $l$ surrounding a node $n$. It is a theorem by Minkowski that there is a unique convex polyhedron such that its faces have unit exterior-pointing normals $\sigma_l \, \vec{n}_l$ and areas $a_l$ [32, 128]. So that each node of the coherent state can really be seen, in the semi-classical picture of twisted geometry, as a polyhedron. Two neighbouring polyhedra are glued along a link, that is a face of same area, although the shape of the faces may not match (reason for which the geometry is said to be twisted). Finally, the number $\xi$ can be used to encode the extrinsic curvature on the common face, when the polyhedra are embedded into 4-dimensional spacetime (but the issue is delicate, see [6]).

### 17.8.4 *Coherent intertwiners*

The previous parametrisation is also interesting for it connects with the Bloch states $|j, \vec{n}\rangle$, obtained by the algebraic approach. The *coherent intertwiners* $|\{j, \vec{n}\}\rangle_n$ were defined in [121] as a tensor product of Bloch states $|j_l, \vec{n}_l\rangle$ (one per link $l$ attached to the node $n$), projected down to the invariant subspace:

$$|\{j, \vec{n}\}\rangle_n \overset{\text{def}}{=} P \bigotimes_{l \in n} |j_l, \vec{n}_l\rangle \tag{667}$$

with $P$ the projector of equation (180). Over a graph $\Gamma$, the tensor product of coherent intertwiners (one per node $n$ of the graph),

$$\bigotimes_{n \in \Gamma} |\{j, \vec{n}\}\rangle_n \,, \tag{668}$$

can be seen as a state

$$\Psi_{\Gamma, j_l, \vec{n}_l} \in L^2_\Gamma(SU(2)^L / SU(2)^N). \tag{669}$$

The semi-classical interpretation of $\Psi_{\Gamma, j_l, \vec{n}_l}$ is the juxtaposition of all polyhedra, i.e. without the extrinsic angle that glues them together. For this reason they are called intrinsic coherent states by Rovelli [161], as opposed to the extrinsic coherent states $\Psi_{(\Gamma,H_l,t_l)}$. For large values of $a$, the two are related by

$$\Psi_{(\Gamma,H_l,t_l)}(g_l) \sim \sum_{j_l}\left(\prod_l (2j_l+1)e^{-t_l(j_l-\frac{a_l}{2}+\frac{1}{2})^2}e^{-i\gamma\theta_l j_l}\right)\Psi_{\Gamma,j_l,\vec{n}_l}(g_l), \quad (670)$$

(see [33] for details).

We have given a quick overview of the two main definitions of coherent states in the context of quantum gravity. There are also other proposals under investigation like the $U(N)$ coherent states [68], the $SO^*(2N)$ coherent states [83], or the coherent intertwiners of [67].

## 17.9 Conclusion

In this chapter, we have tried to gather in few pages, the core ideas of coherent states, as they have emerged in the head of Schrödinger, later fuelled by quantum optics, and applied to quantum gravity. These three steps of developments are, to some extent, independent, but we have tried on the contrary to weave them together, to show the beautiful and consistent landscape that they create. Of course, we have only been scratching the surface, as the general subject of coherent states is now branching in many directions. We refer to [74] or [146] for a more exhaustive treatment, although quantum gravity is absent in them. Let's finally conclude by remembering the anecdote found in [153], that the term 'quantum gravity' has first been used in 1969 in the title of a talk by John R. Klauder, who is now especially remembered for his contribution to the theory of coherent states, which may, after all, not be a surprise.

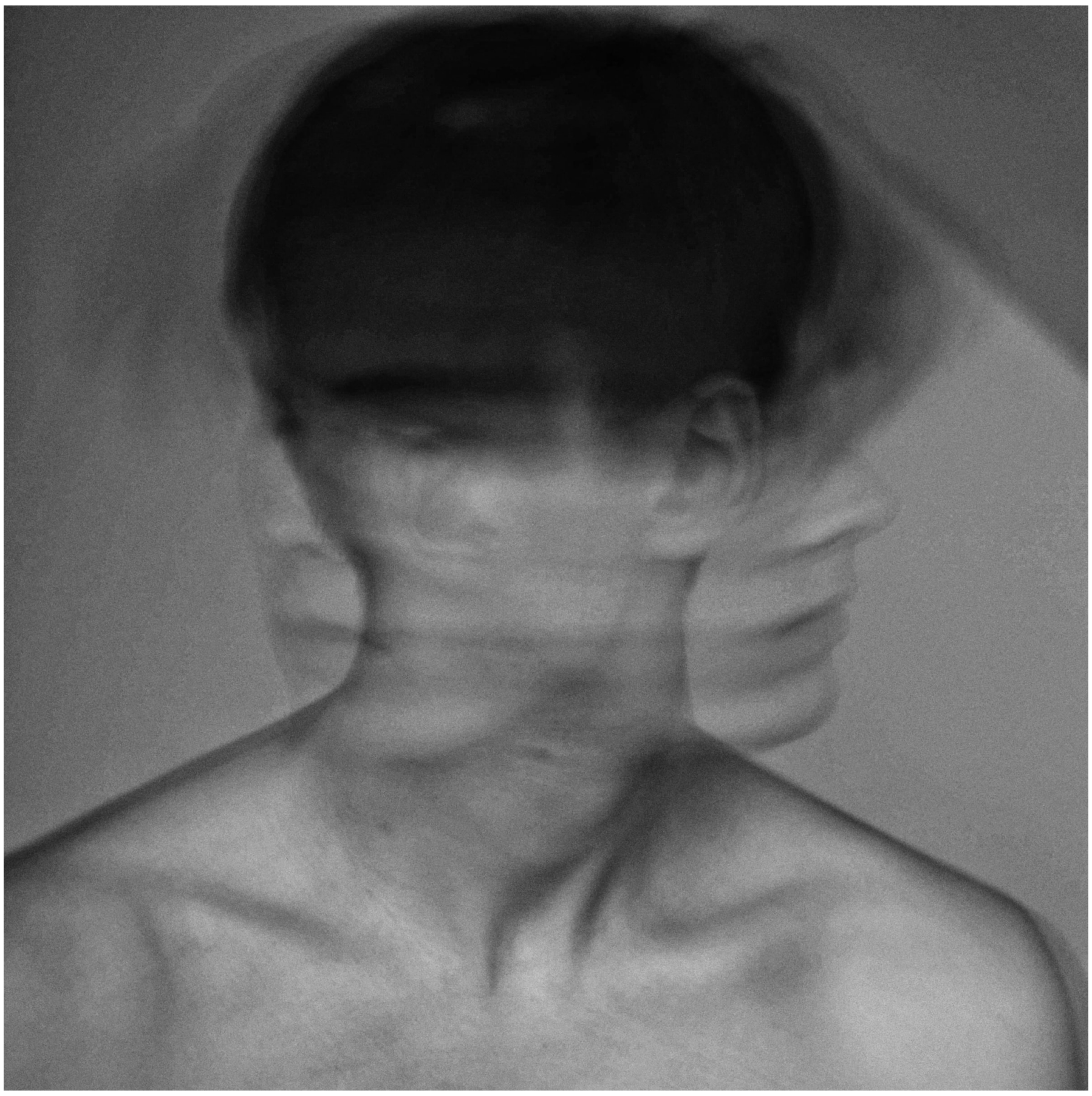

Part IV

# RELATIONAL PHYSICS

Quantum Gravity is a challenge for theoreticians and experimentalists, but also for philosophers. In the folkloric fantasy of physicists, philosophers often play the role of the backward show-off. The disdain for philosophy is best illustrated in this Feynman's quote

> *The philosophy of science is as useful to scientists as ornithology is to birds.*[17]

Unfortunately for him, he has been a major philosopher of physics in spite of himself. Whenever the goal of physics is recognised as 'understanding Nature', any technical result is vain as long as it is not articulated to our naive preconceptions about reality. In quantum gravity, it means relating the mathematical formalism to the familiar notions of space and time of every-day life. Observing that some operator has a discrete spectrum is point-less, unless it is said precisely how this operator relates to the actual process of area measurement. Otherwise, speaking of an 'area-operator' is just keyword snake oil. Then, the first mark of honesty of a theoretical physicist is to acknowledge that his technical results are grounded in prejudices about the nature of reality. It is the task of physicists themselves to demystify their own formalism. Such a stance has lead the mathematician Jacques Harthong to write one of the best introductory course to probability and statistics [96]. Although officially a math book, it is full of physical and philosophical insights, which break any attempt of categorisation. He writes in the *Avertissement*

> *Il est devenu courant de dévaloriser comme "philosophique" toute question non strictement technique, mais les questions auxquelles je tente de répondre dans ce livre ne sont pas de la philosophie, elles sont partie intégrante de la science au sens le plus strict.*[18]

Our personal journey through quantum gravity has lead us to face with puzzlement and excitement the relational point of view about physics. To sum it up, the physics until 1925 is the physics of *things*. The unquestioned philosophical framework is *cartesian* –there is a world out there– and *realist* –mathematical entities correspond to physical objects. It has made the glory of galilean and newtonian physics; but in 1925, Heisenberg causes a breach writing

*In der Arbeit soll versucht werden, Grundlagen zu gewinnen für eine quantentheoretische Mechanik, die ausschließlich auf Beziehungen zwischen prinzipiell beobachtbaren Größen basiert ist.*[19] [104]

The physics after 1925 is a physics of *relations* (*Beziehungen*). These two ways of viewing Nature, as things or relations, constitute the core of the debate on the nature of reality between Bohr and Einstein [116]. The new perspective challenges our preconceptions so much that it is tempting to reject growing confusion as triviality or non-sense. At this point, philosophers come to rescue lost physicists, and avoid them to succumb to the *shut-up-and-calculate* temptation.

Ch. 18 deals with relationality, and is based on a talk which was given at the Quantum Information Structure of Space-time (QISS) conference in Hong Kong in January 2020.

Ch. 19 tackles the issue of (non)-locality, which was much discussed at the Rethinking Workshop 2018 in Dorfgastein (Austria), and resulted in a paper [122].

17 Attributed to Feynman by the historian of science Brian Cox (according to Wikipedia).

18 Translation: *It has become common to discard as 'philosophical' any non strictly technical question, but the questions I am addressing in this book are not philosophy, they are integral part of science in its strictest acceptation.*

19 Translation (from [103]): *The present paper seeks to establish a basis for theoretical quantum mechanics founded exclusively upon relationships between quantities which in principle are observable.*

# 18

# RELATIONAL ASPECTS BETWEEN GRAVITY AND THE QUANTUM

It is a common place to point at the many differences between General Relativity (GR) and Quantum Mechanics (QM). This gap motivates the construction of a bridge that one would call *quantum gravity*. To achieve such a quest it is valuable to look, not only at the differences, but also at the points where the two theories can meet. And there are a few. For instance, people study black holes or the early cosmology, where both realms of physics come into play. Here we would like to ask the question at a more conceptual level, and focus on a notion which appears in both theories, although it takes different aspects. It is the notion of *relationality*.

Relationality appears crucially in GR, where it takes the form of diffeomorphism invariance. It also appears in an essential manner in QM, at least if one is ready to buy a relational interpretation of it... One may wonder whether the two notions of relationality are actually the same. Or to put it maybe more precisely, suppose we have a theory of quantum gravity, which kind of relationality should we expect at this level? To my knowledge, the question was first raised in [157].

In this chapter, we will not reach many definite answers, but we will try to be conceptually clear and to ask relevant questions. To start with, we need to be precise about what we mean by relationality, because it is a very general notion, and it is easy to get confused by a misuse of terms, especially when people from various communities use the same words with different meanings. The many notions of relationality are subtle, and unlike poetry, science should not let its words go faster than its thoughts.

## 18.1 CLASSICAL MECHANICS

Let's start we a familiar example: the principle of relativity. In his book *Questions on the Four Books on the Heavens and the World of Aristotle*, the 14th century philosopher Buridan writes

> *If anyone is moved in a ship and he imagines that he is at rest, then, should he see another ship which is truly at rest, it will appear to him that the other ship is moved. This is so because his eye would be completely in the same relationship to the other ship regardless of whether his own ship is at rest and the other moved, or the contrary situation prevailed.* ([15], p. 203)

It is surprising how closely Buridan's ideas are from the formulation of galilean relativity, as it is taught at school. Did people know about galilean relativity before Galileo himself? Not really. What Buridan is describing is *kinematic relativity*. It is the observation that whenever one tries to describe concretely a motion, it has to be done with respect to some reference, which is postulated at rest. And this is true whatever the kind of motion, not only uniform straight line motion. Kinematic relativity is almost a linguistic fact about what motion means, and it was observed much before Galileo.

This initial observation raises an issue, that Julian Barbour in [15] calls the *fundamental problem of motion*: if all motion is relative and everything in the universe is in motion, how can one ever set up a determinate theory of motion?

The answer is provided by Newton's bucket thought experiment. Imagine a bucket of water, and nothing else in the universe. Does the bucket turn? If it does, it must turn with respect to something else, according to the principle of kinematical relativity. For instance the bucket could turn with respect to the water. What if there is no relative motion of the water with respect to the bucket? Newton notices two possibilities, as observed experimentally on earth (see figure 52):

1. The bucket is hanging to a gallows, still, and the surface of the water is flat.

2. The bucket is turning with respect the gallows, and the surface of the water is concave.

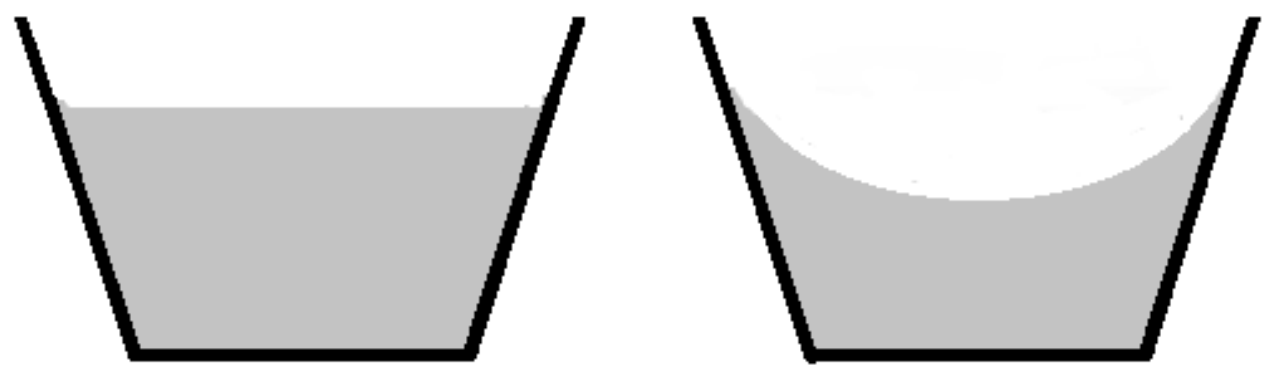

Figure 52: Newton's bucket.

Now forget about the gallows and forget about the earth. In both cases, the water is at rest with respect to the bucket. Nevertheless, by looking at the shape of the surface only, it is possible to say in which case the bucket is turning or not. It seems that there is no need of a reference system to say that the bucket is turning. For Newton it must be turning with respect to something, and he concludes by calling this something *absolute space*. Motion with respect to absolute space is called *absolute motion*. Such a statement, which could be discarded as metaphysical, is actually essential, as it gives meaning to Newton's laws. Indeed, the equation of dynamics are written with coordinates which carry the physical meaning of a cartesian mapping of absolute space. This solves the fundamental problem of motion.

Now comes galilean relativity. I deliberately reverse chronological order to restore conceptual clarity. Galilean relativity states a fundamental experimental restriction on the description of absolute motion: although one can indeed detect absolute rotation, it is impossible to distinguish absolute rest from absolute uniform straight line motion. So galilean relativity is not the kinematic relativity of Buridan! Kinematic relativity is really a kinematical statement, i. e. it is about how motion can be described in general, while galilean relativity is a dynamical statement, as it is about how motion can be predicted, and it claims that the dynamical laws take the same form within the class of uniform straight line motions.

To put it in a nutshell, Newton's theory of motion is a sandwich of absolutism between two layers of relativity/relationality.

1. *Kinematic relativity*: motion is motion with respect to something.
2. *Absolute space*: there is a preferred reference system, space itself, and motion with respect to it is called absolute.
3. *Galilean relativity*: absolute uniform straight line motion is undetectable.

The layer of 'absolute space' is crucial to distinguish kinematic relativity from galilean relativity.

This warming-up has shown that already at the level of a theory that we believe to know well, the relational aspects of it are delicate to grasp. We can now move on to more modern theories.

## 18.2 GENERAL RELATIVITY

In GR, the common view in many textbooks is that space-time is a differentiable manifold over which is defined a metric. The picture that people have in mind then, is that of a grid which is curved by masses (see figure 53). The problem with this image is that it lets people think that space-time *is* the continuum of points of the underlying topological manifold, which would be basic entities of the world, and on top of which would come an additional metric structure to measure the distance between the points. This view is false, because it overlooks a central aspect of GR, which is its relational novelty with respect to classical mechanics.

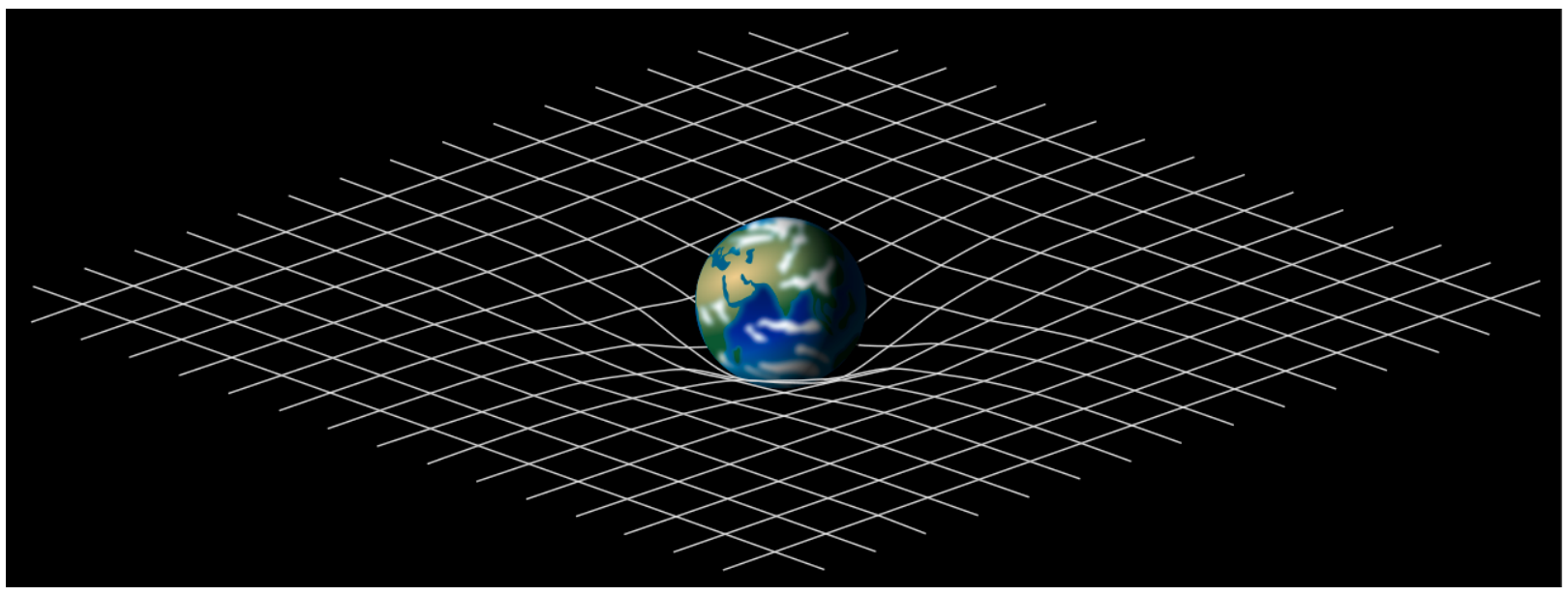

Figure 53: A manifold curved by masses.

The three layers structure of relationality in classical mechanics might be transposable to GR. The first layer (kinematic relativity) is still there: it is the possibility to choose any grid to attribute numbers to points. It's called *general covariance*, induced by *passive* diffeomorphisms. The third layer (galilean relativity) is extended to accelerated motion (from a ship to a lift), and is then called *diffeomorphism invariance*, induced by *active* diffeomorphisms. General covariance and diffeomorphism invariance are conceptually different, but formally the same, which may convey the impression that there is not much room for any notion of absolute spacetime in between.

Indeed, there is no possible physical existence to be attributed to the points of the space-time manifold. This is the consequence of the so-called *hole argument*. Imagine a space-time with a given distribution of matter and a compact empty region $H$ (the hole), see figure 54. Consider two points $A$ and $B$ in $H$. Although $H$ is empty, a metric $g_{ab}$,

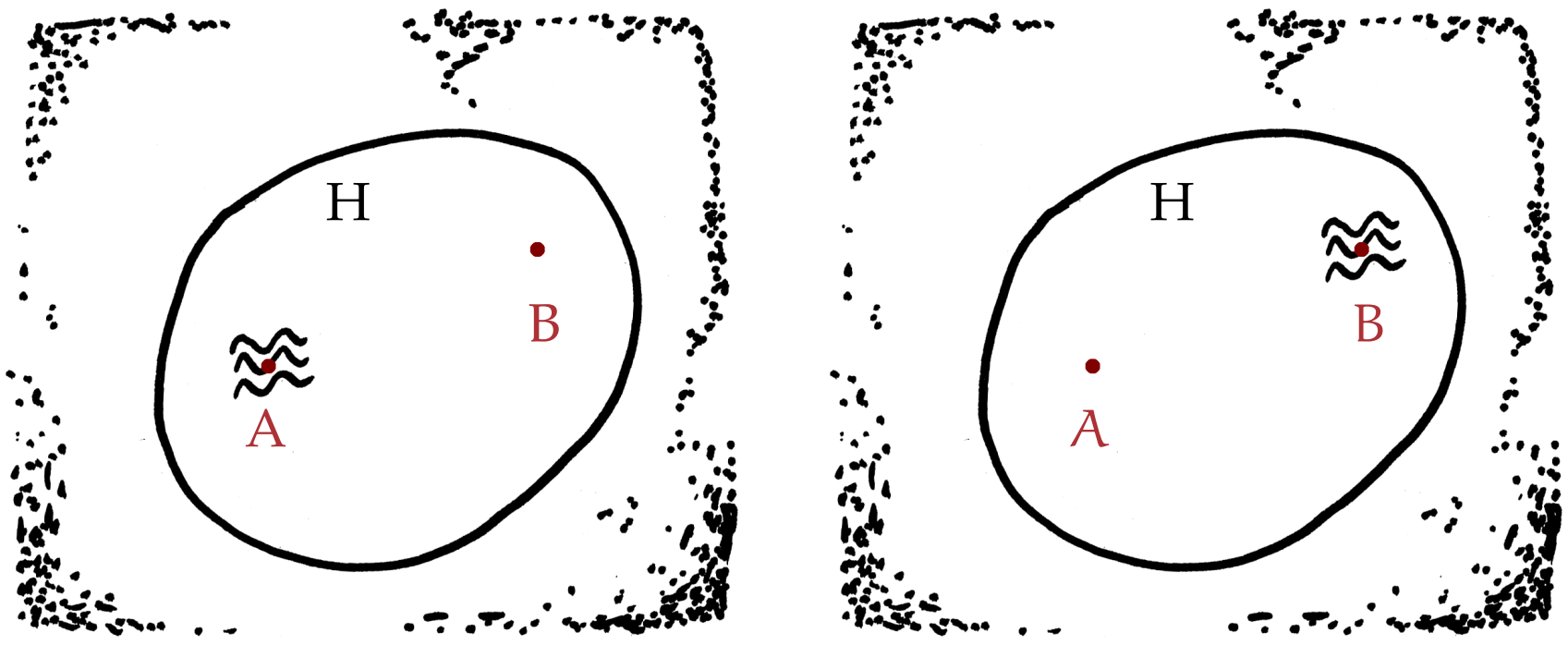

(a) Space-time with metric $g_{ab}$. (b) Space-time with metric $\tilde{g}_{ab}$.

Figure 54: The hole argument.

solution of Einstein's equation, may describe a curved geometry in $A$, and a flat one in $B$ (figure 54a). There exist active diffeomorphisms, which leaves the metric invariant outside the hole, but transform it inside to a new one $\tilde{g}_{ab}$, which is flat in $A$ and curved in $B$ (figure 54b). According to the diffeomorphism invariance of the theory, both $g_{ab}$ and $\tilde{g}_{ab}$ describe the same physical situation, so that we must conclude that the statement 'space-time is curved in $A$' is not physically meaningful. The manifold picture, made of points like $A$, is an absolute representation of space-time, in the sense that it conveys the impression that there is a preferred substratum with respect to which I can define a notion of absolute motion, like in Newton's theory. The hole argument shows that this view is not viable.

The conclusion is hard to admit, because it breaks the widespread metaphysical postulate of realism. Realism is the stance that there is a correspondence from the mathematical entities of a theory to the physically existing objects. The relational aspects of GR comes to challenge this view. The breakdown of realism does not mean that it is impossible to describe physics with maths, but it makes the correspondence indirect: it is harder to say what is observable and what is not. In Newton's theory, the coordinates $\vec{x}$ have an operational meaning, that is a concrete procedure to measure its value, while in GR, the coordinates $x^a$ do not bear this interpretation. Instead, the observables of GR are independent of the coordinates $x^a$.

So the manifold cannot be seen as an absolute space-time like in Newton's theory. However, it is hard to say precisely the positive content of this fact: if space-time is not a continuum of points, what is space-time?

An easy answer –but false– would be to say that there is no space-time, there are only distances between material points. This is false because Einstein equations admit vacuum solutions, that is non-trivial solutions which describe an empty space-time. You can have gravitational waves for instance, which may carry enough energy to make you fall.

There is a way to catch the remaining absolute core of GR, saying that physical space-time is an equivalence class of solutions of Einstein equations, two representatives being related by a diffeomorphism. This view challenges our mental representations, far from the naive image of figure 53, because the familiar notion of locality gets drowned into the abstract and global notion of equivalence class. In this context, space-time is relational, because the values taken by a metric field $g_{ab}$, a representative of an equivalent class, are only meaningful with respect to some other (matter) field. This is the standard way to attribute meaning to the coordinates, by considering four scalar matter fields, or four satellites [158]. For instance, the Ricci scalar $R(x)$ is not an observable per se, unless $x$ is attributed a physically grounded meaning, like the place where some matter field $\phi$ takes some value. Finally, the role of the topological manifold is only to provide a common ground where to compare various fields 'at the same point'.

## 18.3 RELATIONAL QUANTUM MECHANICS

Now let's turn to QM. The fact that QM exhibits relational aspects is not universally acknowledged. One reason may be that it is not explicit in the name, like in relativity theory. What is explicit in the name of 'quantum mechanics' is the idea of discreteness. The relational aspect of it was only recognised long after its discovery.

It starts in 1957 with an article by Hugh Everett entitled *'Relative State' formulation of Quantum Mechanics* [62]. He remarks this fact that has been much discussed later on under the name of Wigner's friend[1]: consider a first observer B that carries on a measurement over a system C, and consider a second observer A that describes the overall situation without interacting with it (see figure 55). Denote $\rho_C^0$ (dens-

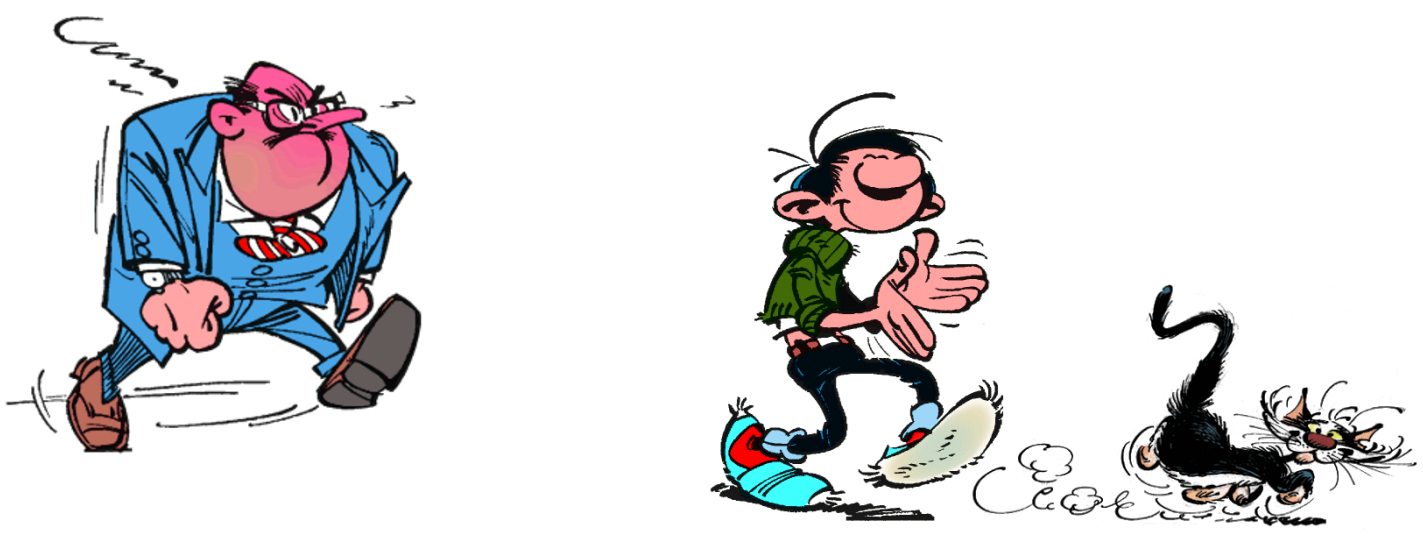

A B C

Figure 55: Wigner's friend (Belgian version).

ity matrix) the initial state of C. The two observers describe a different evolution:

A. The evolution is unitary and the state becomes

$$\rho_C^1 = \mathrm{Tr}_B \left[U(\rho_B^0 \otimes \rho_C^0)U^{-1}\right] \tag{671}$$

with U a unitary operator and $\rho_B^0$ the initial state of B.

B. The state collapses to

$$\rho_C^1 = P\rho_C^0 P^{-1}, \tag{672}$$

with P a projector on one of the eigenstates of the observable.

The two final states are different.

This thought experiment has given rise to a large amount of literature about what it means. It has notably fuelled many-world interpretations. According to them, at the moment of the measurement, the world splits in two branches, and so the second observer also splits in two, each branch corresponding to a different result. Over the years many-world interpretations have been refined. For instance, nowadays, they would say that the splitting happens only locally and then expands progressively [191]. In a sense, it has come closer to a relational interpretation.

Still, in my opinion, the many-worlds interpretation commit the same kind of mistake that was committed by Ptolemy, namely, it attributes to the world some properties which are only perspectival: Ptolemy takes epicycles as the real motion of planets, whereas it is

1 Although the article of Wigner comes later in 1961 [196].

nothing but his own motion projected in the sky. Similarly, in many-worlds, one thinks that the world splits in branches, whereas it may only indicate that the points of views are many.

At least, this is what the relational interpretation, as understood by Rovelli in 1995, would claim [156]. The two observers get different final states, and that's it. The state of a system is not something absolute, it is only a description relative to some observer. So, when an experiment is described, it is crucial to make explicit the referent observer, otherwise paradoxes are encountered. If one sticks to a given observer, no contradiction shows up. Such an interpretation is *epistemic*, as opposed to *ontic*, meaning it does not grant an objective reality to the wave-function, but understands it rather as a state of knowledge (or belief).

Is such an interpretation so surprising?

Maybe not. If we are shown a cup like in figure 56a, we see a left-handed cup... but the boy holding it sees a right-handed cup. So it is pretty obvious that different observers give a different account of the same system. To the question, 'where is the handle?', one answers 'left', the other 'right'. It is just an example of kinematic relativity, and it is not too surprising. What may be surprising is when relationality

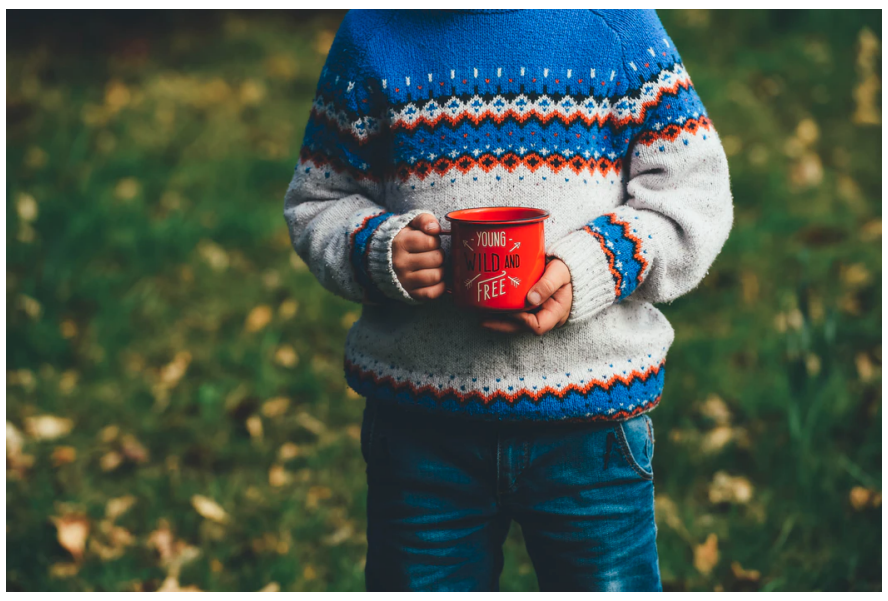

(a) Left-handed or right-handed cup?

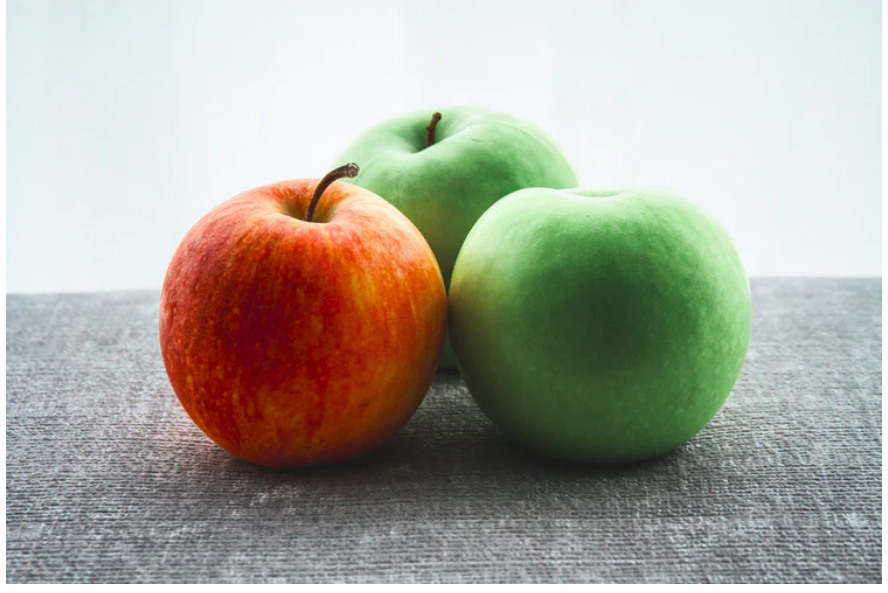

(b) Colourful apples (figuring quarks).

Figure 56: External and internal properties.

is claimed for properties which are internal, as opposed to external degrees of freedom. It seems strange to say that the colour of a quark depends, not only on the quark, but also on the observer who is looking at it (see figure 56b). But in Relational Quantum Mechanics (RQM), the distinction between internal and external properties is only superficial.

In fact, the measurement of a property establishes a correlation between the system and the observer, so that the measurement of the colour of a quark can also be regarded, symmetrically, as the measurement of the observer by the quark. Physics describes relations between systems rather that systems themselves. This is the relation core of RQM. QM is not only this (it is also probability, discreteness...), but it includes a relational lesson.

This raises a difficulty about the ontology. What is primordial, systems or relations between systems? My opinion is that the notion of system is secondary. A system is an abstract notion useful to recast past measurements whose result is assumed to remain unchanged throughout later experiments. A system is defined by constant properties which serve as a signature to track its identity. For instance, the mass of a particle identifies it as a system because it remains constant when, for instance, it goes through a Stern-Gerlach. So, the definition of a system relies on a peculiar choice of observables. This choice is contingent to the experiments we are interested in, but it is not something absolute. If there is an interaction with, for instance, an antiparticle, then the particle disappears. Or one could say the system is still the same but the mass has changed and is now zero, or the particle has become a photon.

So, a system is defined by a certain invariance. For a physicist, the ontological quest consists in finding which invariance are relevant or not. In a sense, physics aims at finding the right demarcation line between the object and the subject, the system and the observer, what is relative and what is absolute.

A cup is recognised as objective by considering that it is the same whatever the perspective it is looked from. Formally, it is tempting to say that the cup *is* the equivalence class of all the perspectives on the cup. This view is a funny way to define a cup, and not completely convincing because no-one has ever poured tea in an equivalence class. There is a gap between the formal definition and the practical use of the term. The same gap makes us sceptical with the definition of space-time as an equivalence class of solutions of the Einstein equations. But formally it is correct. This remark about the primacy of relations on systems is a general metaphysical statement, not specific to QM or GR, as the example of the cup illustrates. A given physical theory will additionally specify some mathematical structure on top of the equivalence class. For instance, in QM, the equivalence class of a system has the structure of a Hilbert space, which is not the case in the example of the (classical) cup. A canonical example is given by the definition of a particle as an irreducible representation of the Poincaré group. This group encodes all the various perspectives that an observer can have on a particle.

## 18.4 QUANTUM GRAVITY

So a priori, one can expect quantum gravity to be relational at two levels. One because of QM, one because of GR. It is tempting to say that these are the two sides of the same coin, but we probably miss an adequate theoretical formalism to say it in a precise manner.

The observables of quantum gravity are some geometric properties of space-time. Then, it could be that the relativity of localisation is just a particular instantiation of a more general principle of quantum relationality. Can we say something more precise by looking at the current theories of quantum gravity?

Let's try with the covariant formulation of LQG, introduced in chapter 16. Let's come back to the physical interpretation of spin-foams in the context of the more general formalism, called *general boundary QFT*, developed by Oeckl [135, 136].

A system is a 3D hypersurface of space-time $\Sigma$, and the observables are geometric properties of it. $\Sigma$ can be described by a quantum state of space $\xi$, which belongs to the Hilbert space of LQG. Then, there is a procedure to associate to this state, a number $\rho(\xi)$, called an *amplitude*. It is not a transition amplitude between two states, as in standard QFT, but it is the amplitude of one single state. Two questions: how is it computed? What does it mean?

Its computation has been described in section 16.2. It requires building a spin-foam whose boundary is given by this state. The amplitude of the state is approximated by the spinfoam amplitude, whose computation follows a standard procedure. A better approximation is given by a finer spinfoam catching more degrees of freedom.

What about its meaning? In standard QFT, an amplitude is a transition amplitude between an 'in' and an 'out' state. 'In' and 'out' refer to a preferred time slicing of space-time. But such a global notion of time fades away in quantum gravity. Our intuition is stuck as long as one regards the state and its amplitude as objective entities of reality, that is independent of the observer. Let's adopt instead an epistemic interpretation of the quantum states. If a state $\psi$ encodes the complete information that an observer $\mathcal{O}$ can have about $\Sigma$, a linear subspace $\mathcal{S}$ carries only partial knowledge of $\Sigma$. If $\mathcal{A}$ is a linear subspace of $\mathcal{S}$, then the conditional probability that the state is found in $\mathcal{A}$ knowing that it is in $\mathcal{S}$ is

$$P(\mathcal{A}|\mathcal{S}) = \frac{\sum_{i\in I} |\rho(\xi_i)|^2}{\sum_{j\in J} |\rho(\zeta_j)|^2}, \tag{673}$$

where $\xi_i$ (resp. $\zeta_j$) is an orthonormal basis of $\mathcal{A}$ (resp. $\mathcal{S}$). So the amplitude is understood in terms of a probability of transition between states of knowledge (or states of belief). It is a conditional probability: the probability that such is the case, knowing that such it is.

Our preliminary analysis of relationality in the context of quantum gravity highlights the virtue of an epistemic interpretation of states, such as RQM. From GR, the space-diffeomorphism invariance is directly included in the formalism, as the states $\xi$ are not states *on* space, but directly states *of* space. The time-diffeomorphism invariance is implemented in the abandonment of the concept of transition amplitude. Thus, quantum gravity does not seem to exhibit a single notion of relationality, but rather encompasses various aspects of it, coming both from GR and QM.

# 19

# THE NOTION OF LOCALITY IN RELATIONAL QUANTUM MECHANICS

A central challenge of quantum gravity is the understanding of the emergence of the notion of locality from a quantum space-time picture. Many physicists advocate the idea of a form of fundamental non-locality to solve the information-loss paradox, with late Hawking radiation correlated to early radiation (see chapter 12). These ideas are fuelled by the AdS/CFT conjecture and some related holographic ideas. It is also fuelled by the stronger result of violation of Bell's inequalities. These later experiments challenge our views about what locality really means.

There have been claims that Relational Quantum Mechanics (RQM) is local [172], but it is not clear then how it accounts for the effects that go under the usual name of quantum non-locality. In this chapter, almost entirely taken from [125], we propose a reinterpretation of the mathematical definition of locality in the light of RQM, and clarify in what sense RQM can be said to be local. We show that the failure of 'locality', in the sense of Bell, reduces, in the relational framework, to the existence of a common cause in an indeterministic context. In particular, there is no need to appeal to a mysterious space-like influence to understand it.

In section 19.1, we remind of the polysemia of 'locality' in physics. In section 19.2, we review the notion of 'local causality' and its mathematical definition introduced by Bell. In section 19.3, we see that this definition still makes sense in the relational interpretation, but its meaning changes to the point that there is nothing more surprising in non-locality than in the fundamental randomness of quantum mechanics.

## 19.1 NOTIONS OF LOCALITY

It has often been argued that non-locality is a fundamental feature of the world that quantum mechanics has unveiled [28, 175], but it is not clear what this means precisely. The term 'locality' is used in different contexts with different meanings. One encounters at least five different notions of 'locality' in the literature:

1. *No superluminal signalling*: signals cannot propagate faster than light;

2. *No superluminal causal influence*: causes and effects of events are no further away than permitted by the velocity of light;

3. *No space-like influence*: space-like separated quantum systems do not influence each other;

4. *Point-like interaction*: quantum systems (or fields of the lagrangian in QFT) interact only at the same point in spacetime;

5. *Local commutativity*: space-like separated local observables commute.

Of course, these various notions can be very close to one another. Sometimes, there are even seen as equivalent, but in the following we will adopt a more cautious strategy and consider them as prima facie distinct notions.

In the 1970s, John Bell proposed a precise mathematical formalisation of the principle that we called above 'no superluminal causal influence'. This formal definition, which originally goes under the name of 'local causality', is a crucial step toward the proof of his famous theorem [28]. When people say that EPR-type experiments highlight the fundamental non-locality of nature, they implicitly refer to that peculiar notion of locality.

In the context of the relational interpretation of quantum mechanics [156], there have been claims that quantum mechanics becomes local [172]. It is not clear, however, how this presumed 'locality' has to deal with the original definition of Bell. Surely, the relational interpretation must have a way to include the effects that go under the usual name of quantum non-locality. This has been recently pointed out as one of the open problems of the relational interpretation [119].

## 19.2 IS QUANTUM MECHANICS LOCALLY CAUSAL?

### 19.2.1 *The failure of local determinism*

For John Bell, a theory is said to be *locally deterministic* if the state of physical systems in a bounded region of space-time $A$ can be entirely deduced from the knowledge of the state of systems in another bounded region $B$ located inside the past light-cone of $A$ [28]. One can say for instance that Maxwell's theory of electromagnetism is locally deterministic, because one can predict the configuration of electromagnetic fields in a bounded regions of spacetime $A$ knowing the configuration of the fields over a time-slice of the past light-cone of $A$. A counter-example is quantum mechanics, which is not locally deterministic, because of the probabilistic aspects of the measurement.

It is important to notice that all the indeterminism of quantum mechanics lies in the measurement process. If one restricts to the unitary evolution, given by the Schrödinger equation, then the evolution is deterministic (and in a sense, even more deterministic than in classical mechanics, as pointed out by John Earman [58]). But once a measurement takes place, the future outcome is not determined by past measurements.

Nevertheless, even if the past does not completely determine the future, there might still be some sense in which one could say that 'the future is only influenced by the past (and not by spatially separated events)'. This is another way to state what we previously called the principle of 'no superluminal causal influence'. In this way, the notion of locality first appears as an attempt to preserve the intuitive notions of cause and effect in an indeterministic context. For sure, the notion of causation admits a large number of interpretations. John Bell proposed his own formalisation of this idea that he called 'local causality'. Later, it has often been abbreviated as 'locality', which may have sometimes introduce some confusion in the debate. Importantly for what follows, this definition relies on the notion of 'local beables'. As we shall argue below, quantum non-locality is strictly connected to what we count as local beable.

### 19.2.2 *Local beables*

Although the notion of beables seems to refer to a peculiar interpretation of quantum mechanics, or even a new theory (one of the first papers by Bell was entitled, maybe misleadingly, *The theory of local beables* [27]), the concept is meaningful in ordinary quantum mechanics. To put it in a nutshell, the term 'beable' is a fancy way to say 'element of reality'. John Bell has been more or less explicit about what he really meant:

> *The beables of the theory are those elements which might correspond to elements of reality, to things which exist. Their existence does not depend on 'observation'.* [27]

It was a surprise of quantum mechanics to realise that some very intuitive concepts like 'the position of a particle' may simply not be meaningful in the absence of an observation. Reality seemed to fade suddenly. The introduction of the concept of 'beable' was an attempt to bring back to the theory the primacy of 'things which really *are* in the world' over 'things which are observed' (observable). Thus, the concept of beable strongly depends on a certain form of realism, which might first seem to be too restrictive for a good interpretation of quantum mechanics. Is it indeed reasonable to assume the existence of 'elements of reality' in the quantum world?

In fact, this apparent drawback can be turned into an advantage if beables are seen as a very general concept whose actual content depends on the choice of the interpretation. The ontology of quantum mechanics is not given a priori by its mathematical formalism, but it is expected from the interpretation to supplement the mathematics and to answer the question 'what is real?' or equivalently 'what does physically exist?'. The answer to these questions constitutes the actual content of the term 'beable', which is thus a word whose precise meaning may change with the various possible ontologies of quantum mechanics.

Different interpretations may agree on some of the basic things that should be considered as 'real'. For instance, John Bell suggested that:

> *The beables must include the settings of switches and knobs on experimental equipment, the current in coils, and the reading of instruments. 'Observables' are* made, *somehow, out of beables.* [28]

In the following, we will focus on local beables, that is to say on beables that are localised in a bounded region of spacetime. For this to be meaningful, spacetime is assumed to be the classical spacetime of special relativity, upon which 'elements of reality' can live. It is the usual assumption of ordinary quantum mechanics and QFT, but one should note that it is still unclear what 'local beable' would mean in a context of quantum gravity where spacetime itself would be considered as a quantum field.

The notion of local beables, which we have just seen to be quite flexible, is the basic concept for Bell's formalisation of the principle of 'no superluminal causal influence' according to which causes and effects cannot propagate faster than light.

### 19.2.3 *Local causality*

Let us denote by $\{x|y\}$ the probability of some particular value $x$ of the beable $X$, knowing the particular value $y$ of the beable $Y$. Consider figure 57, with

| | |
|---|---|
| $A$ and $B$ | two space-like separated beables; |
| $\Lambda$ | the set of beables in their common past; |
| $N$ | the set of beables in the past of $A$ excluding $\Lambda$; |
| $M$ | the set of beables in the past of $B$ excluding $\Lambda$. |

Then the theory is said to be *locally causal* if, for all possible values $a, \lambda, n, b$ of the beables $A, \Lambda, N$ and $B$, we have

$$\{a|\lambda, n, b\} = \{a|\lambda, n\}. \tag{674}$$

In words, it means that, given the knowledge of all the information about the beables in the past cone of $A$, any information about the value of beable $B$ is superfluous for the prediction of the value of the beable $A$.

### 19.2.4 *Quantum mechanics is not locally causal.*

Bell emphasised that ordinary quantum mechanics is not locally causal [28]. To see this, he considers a radioactive particle in $\Lambda$. The radioactive decay will lead to the emission of an $\alpha$ particle. Suppose the beables $A$ and $B$ are the reading of Geiger counters which tell us whether an $\alpha$ particle is detected (value 1) or not (value 0). Suppose also that the radioactive particle can only emit at most one $\alpha$ particle. Knowing only the values $\Lambda = \lambda$ and $N = n$, it is impossible to predict if $A$ will detect an $\alpha$ particle or not: it is the randomness of the measurement process. So, whatever the value $a \in \{0, 1\}$ of the beable $A$, we have

$$\{a|\lambda, n\} = \frac{1}{2}.$$

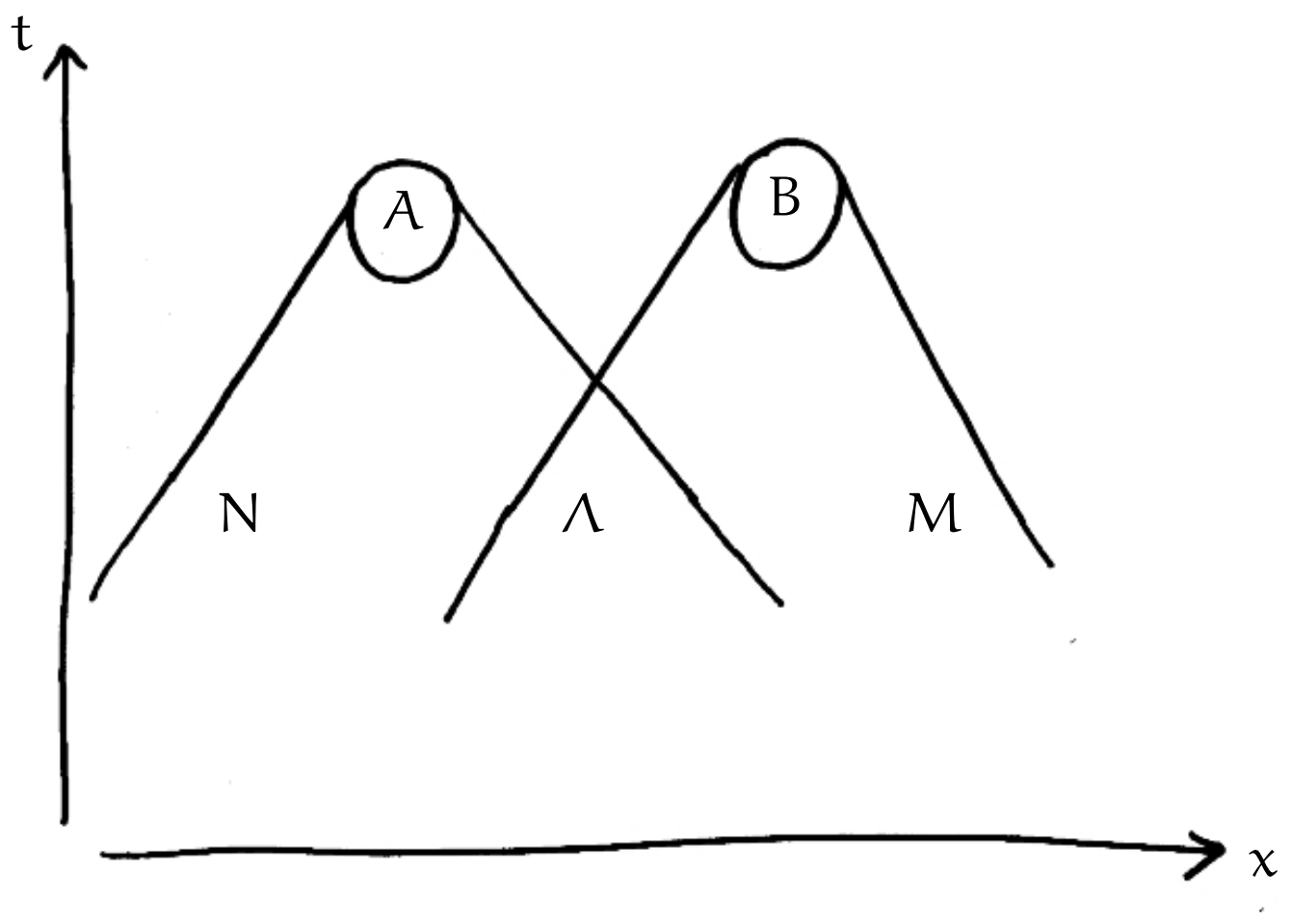

Figure 57: Space-time diagram showing the localisation of the beables A,B, Λ, N and M.

Now, if $B$ detects an $\alpha$ particle, $A$ will not, and so:

$$\{A = 1|\lambda, n, B = 1\} = 0.$$

Thus, in this scenario, we have exhibited a case where:

$$\{a|\lambda, n, b\} \neq \{a|\lambda, n\}.$$

Hence ordinary quantum mechanics is not locally causal.

Let us insist on this: in order to discuss the meaning of Bell's concept of local causality, and show that *ordinary* quantum mechanics –assumed complete– is not locally causal, it is not necessary to introduce entanglement and EPR pairs. In fact, Einstein, Podolsky and Rosen developed their argument the other way around. They started by assuming locality and then arrived to the conclusion that ordinary quantum mechanics had to be incomplete [60]. Their hope was then to embed the theory into a larger one (the so-called local hidden variables theories) which would be complete and locally causal. The importance of EPR pairs is that they have been later used to discard the possibility of such an embedding [11]. In the face of these results, quantum mechanics is acknowledged to be non-local in the sense of Bell, whether or not it is complete. The relational interpretation assumes the completeness of quantum mechanics, and therefore, discussing the meaning of non-locality in this context does not require to introduce EPR pairs; it is sufficient to consider the simpler example of radioactive decay.

### 19.2.5 *Interpreting non-locality*

The usual interpretation of non-locality is fuzzy. If 'local causality' is indeed a faithful mathematical translation of the principle of 'no superluminal causal influence', then its experimental violation should be understood as the possibility that causes and effects propagate faster than light. Furthermore, one usually stresses that this process cannot be used to transmit information, hence no violation of the principle of 'no superluminal signalling', which would have been in complete contradiction with special relativity.

However, even the simple idea of superluminal causal influence is at odds with special relativity. One often says that the outcome of a measurement in A determines a later outcome in B, but this can only be a loose way of speaking because the same ensemble of space-like separated outcomes in A and B can be equivalently interpreted as a measure in A affecting B, or a measure in B affecting A, depending on the choice of reference frame specifying a preferred time foliation. The absence of an absolute time ordering between A and B prevents us from interpreting the origin of the correlations by a causal influence from A to B or from B to A, because 'causation' is a time-oriented concept. One could argue in favour of an absolute reference frame which would justify an absolute causal orientation, but this hypothesis does not show up in the phenomenology of the experiments. So, the hypothetical non-local influence between A and B cannot be causally oriented. At best, it can be thought as a kind of mutual action at a distance that would enable 'instantaneous' space-like influence, and so would violate what was called earlier the principle of 'no space-like influence'.

Some other interpretations of non-locality argue that the mere collapse of the wave function would already be a manifestation of non-locality in ordinary quantum mechanics. This is not completely true, because the definition of Bell is a bit more subtle. In fact, it is true if one adopts an ontic interpretation of the wave-function, that is if one regards it as a beable. It is the case, for instance, in Bohmian mechanics[1] [40, 41]. But if one sees the wave-function only as a mathematical tool, then the so-called collapse of the wave function does not tell us anything deep about the local causality of the world. This kind of fake non-locality is very similar to the apparent superluminal

1 Some modern variations suggest another point of view, see [85].

propagation of the potential for Maxwell's theory in Coulomb gauge. The same thing also happens for British sovereignty since the Prince becomes the King as soon as the Queen dies, however far away in the Universe he may be. These two examples were exhibited by Bell himself. In both cases, the thing (the potential or the sovereignty) that travels faster than light is not a physical thing. This explains the necessary use of 'beables' to define a physically meaningful notion of 'locality'. By the way, this remark also discards some interpretations of non-locality claiming that EPR-type experiments would force us to choose between 'locality' and 'realism', for there is no physically meaningful notion of 'local causality' without the realism of beables.

To summarise, although local causality is mathematically well defined, its meaning is not obvious because it supervenes on that of beables, which makes it dependent on the interpretation of quantum mechanics. In the following section, we are going to see that RQM can still make sense of this definition, but it loses on the way most of its surprising features.

## 19.3 RELATIONAL LOCAL CAUSALITY

The relational interpretation of QM was proposed in 1995 [156, 160]. First of all, it is a criticism of the usual notion of 'quantum state'. The 'state' should not be taken as an absolute notion, i.e. observer independent, but rather as a book-keeping device relative to a specific observer. In this interpretation, the apparent paradox of Wigner's friend is taken as evidence that different observers may give different accounts of the same sequence of events.

This point of view might be puzzling at first sight, because it challenges the opinion that physical systems should be describable independently of any observer. This is surely a departure from strong realism where, for instance, the property 'spin up' belongs to the electron independently of any observer. Actually, the Kochen-Specker theorem has already challenged the strongly realistic point of view, showing that a complete set of properties cannot be consistently attributed to a physical system [114]. The relational interpretation challenges strong realism a bit differently though: there would be no absolute state-property of a physical system at all. However, this view is not completely at odds with realism because, from the point of view of a chosen observer, the idea of 'properties of a physical system' makes perfect sense (to the limit imposed by the Kochen-Specker theorem). Indeed, it has been recently argued that RQM would be an instantiation of 'structural realism', where 'relations', rather than 'objects', are seen to be the basic elements of the ontology [43].

### 19.3.1 *Quantum events as relational beables*

A priori, it might seem hard to fit the definition of beables as 'elements of reality that do not depend on observation' in the framework of relational quantum mechanics. John Bell certainly intended the beables to be observer-independent, but RQM does not claim the complete arbitrariness of reality neither.

We have seen earlier that the explicit content of the word 'beable' was actually given by the peculiar ontology of a chosen interpretation. In the relational interpretation, the basic elements of physical reality are the 'relational quantum events' [160]. A quantum event is an interaction between two quantum systems. In this sense, the relational interpretation gives primacy to 'relations' over 'objects'. A measurement is a special kind of interaction where one of the two systems is macroscopic. Other kinds of quantum events happen when two quantum systems entangle through local interaction, and the degrees of freedom of one become correlated with the degrees of freedom of the other. Importantly, these quantum events are themselves 'relational' because their mere existence can only be experienced through their interaction with a reference observer.

In fact, if one sticks to a particular observer, a relational beable is nothing but a quantum event. It will not be as absolute as Bell would have expected, because talking about 'quantum events' still requires us to first fix a reference observer, but a beable can still be conceived as an 'element of reality with respect to the reference observer'. Now, given a reference observer $\mathcal{O}$, *the only physically meaningful beables lie in the past cone of* $\mathcal{O}$. Indeed for $\mathcal{O}$, it is a matter of metaphysical faith to attribute an existence to events beyond the scope of its practical experience (future or space-like separated events), but it is a matter of experimental facts to attribute an existence to events in its past cone.

### 19.3.2 *Relational local causality*

So, which reference observer $\mathcal{O}$ shall we choose to reformulate the definition of local causality? In order to reasonably talk of the beables $A$ and $B$, the reference observer $\mathcal{O}$ should lie in the common future of $A$ and $B$. (see figure 58). Then, the local causality relation should be rewritten

$$\{a|\lambda, n, b\}_{\mathcal{O}} = \{a|\lambda, n\}_{\mathcal{O}}\,, \tag{675}$$

with an index $\mathcal{O}$ reminding us of the reference observer. A violation of this condition means that, from the perspective of $\mathcal{O}$, $A$ and $B$ are correlated. You can imagine $\mathcal{O}$ as an experimenter looking at two distant Geiger counters, in $A$ and $B$, equipped with a LED which lights up when they detect an $\alpha$ particle. $\mathcal{O}$ would observe that the light coming from $A$ is switched off whenever the light from $B$ is switched on, and vice versa. Now, suppose O also knows all the relevant beables in regions Λ and N. Then O knows that there can be at most one $\alpha$ particle emission, and so O can predict the correlations between A and B. What O still fails to predict is precisely which of the two Geiger counters will detect the $\alpha$ particle. This is the only observational manifestation of 'non-locality'.

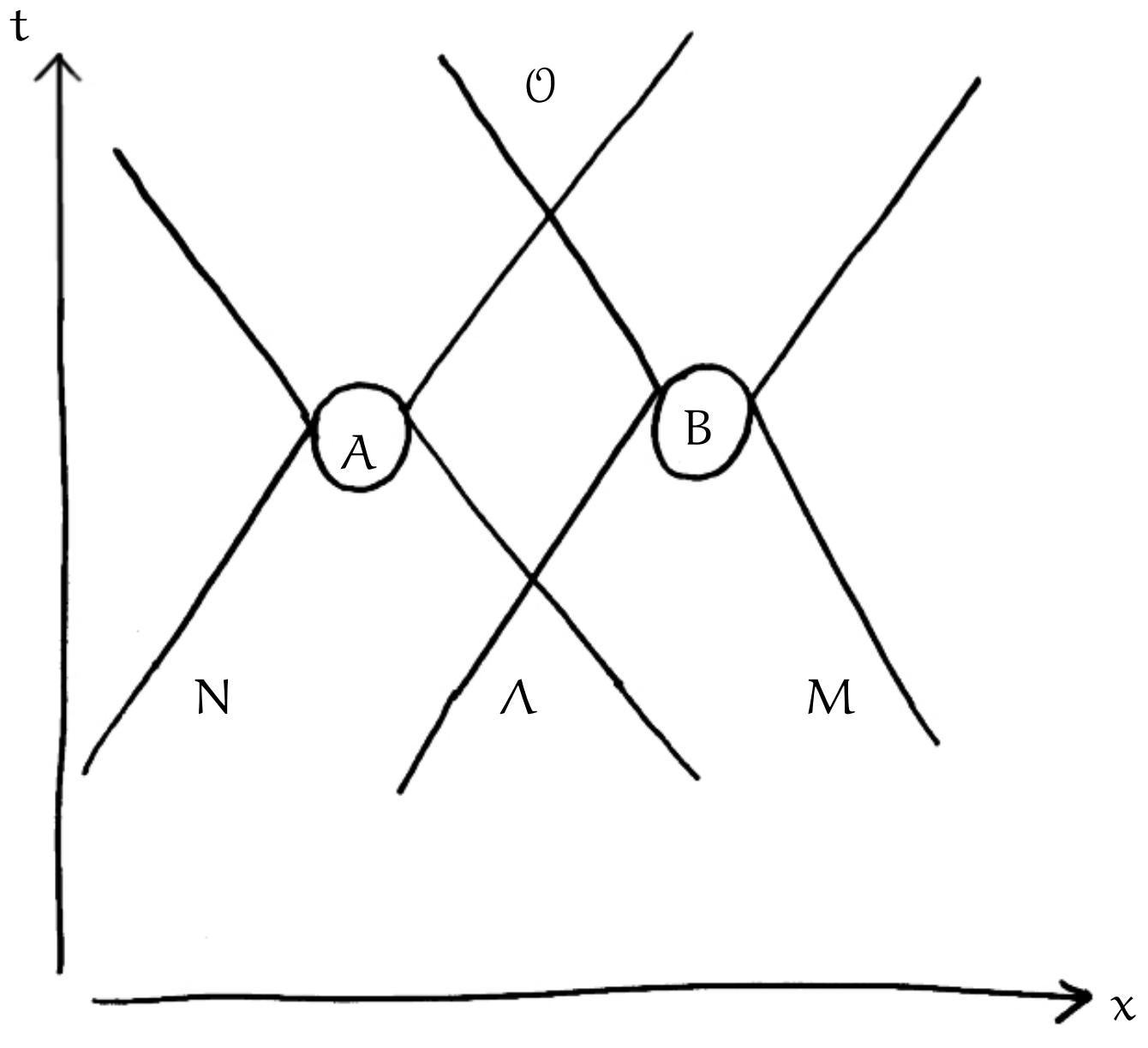

Figure 58: Space-time diagram showing the localisation of the relational beables A,B, Λ, N and M, and the observer O with respect to which they are described.

### 19.3.3 *A common cause*

Since A and B are space-like separated, a classical observer O observing correlations between past events A and B would be lead to the conclusion that there is a common cause to A and B. And indeed there is one, namely the radioactive particle in their common past Λ. Here the notion of 'common cause' does not entail a deterministic evolution for the decay of the radioactive particle. In the classical case, the notion of causation is strongly related to that of determinism, hence the idea that a complete knowledge of the past would entail a complete knowledge of the future. In quantum mechanics, there is no such a determinism, not only because there is no precise meaning of 'complete knowledge of the past' (Heisenberg uncertainty), but also because the measurement process is intrinsically probabilistic. Nevertheless, the intuitive notion of causation does not disappear from the quantum world: in our example it is still meaningful to say that the radioactive element is the 'common cause' of the correlations observed later, even if one cannot predict from the past knowledge ($\Lambda$, $N$ and $M$) whether the $\alpha$ particle will be observed in $A$ or in $B$. To be clear, we are here referring to an intuitive notion of 'common cause' and not to the formal notion introduced by Reichenbach [7]. Indeed, his attempt of formalisation is too closely tied to a deterministic context, and so, does not suit the example of a radioactive decay as was first pointed out in [200]. An interesting discussion concerning the notion of common cause in quantum mechanics can be found in [107].

## 19.4 CONCLUSION

Let us recall that the initial goal of Bell was to formalise the intuitive idea that 'causes and effects cannot go faster than light' in the context of the indeterminism of quantum mechanics. This aim was thought to be achieved with the mathematically well-defined notion of 'local causality'. With this definition at hand, EPR-type experiments have shown that quantum mechanics is fundamentally non-local. What this really means however depends on the physical content given to beables. Now we have seen that the relational interpretation of quantum mechanics forces us to reconsider the EPR-type experiments from the perspective of a future observer. As a consequence, the failure of 'local causality' in the sense of Bell can nevertheless be understood as the existence of a common cause in an indeterministic context. Surely, there is no need to appeal to a mysterious space-like separated influence to understand it.

A possible conclusion is that Bell's definition of 'local causality' does not capture finely enough the intuitive idea of an indeterministic 'no superluminal causal influence' as Bell would have liked. Interestingly, Bell himself seems to have been very conscious of the potential inadequacy of his formalisation. Just before asserting his definition of 'local causality', Bell writes very honestly:

> *Now it is precisely in cleaning up intuitive ideas for mathematics that one is likely to throw out the baby with the bathwater. So the next step should be viewed with the utmost suspicion.* [29]

With the relational interpretation, quantum mechanics is still 'non-local' in the sense of Bell, but it nonetheless remains true that 'causes and effects of events are no further away than permitted by the velocity of light'. Indeed, if the observer $\mathcal{O}$ sees a light signal from $A$, he will think the radioactive particle in the region of the past cone $\Lambda$ is the cause of the detection of an $\alpha$ particle in $A$. The same thing could be said symmetrically for $B$. Neither causes nor effects travel faster than light whatsoever. There are correlations between $A$ and $B$ because there is a common cause in their common past.

Earlier claims in [172] that RQM was local were maybe misleading: the relational interpretation is not locally causal in the sense of Bell.

However it should now be clear that this kind of non-locality cannot be interpreted as a superluminal interaction, and the relational interpretation is indeed local in all the senses listed in section 19.1.

Bell did not know about RQM, but he made very clever claims about realism in [29]. We have already recalled the enlightening examples about the gauge potential and the British sovereignty, and Bell uses them to show that conventions can always travel faster than light. Though it is not clearly stated in these terms, the idea is already almost there that the question 'what cannot go faster than light?' might be a relevant criterion to determine what should be considered as physically real and what should not. With this point of view, the failure of non-locality could have already been reinterpreted by Bell as the impossibility to attribute a relevant physical existence to $B$ from the perspective of $A$, and vice versa. This conclusion would have put him on the way to a relational interpretation, with the idea that one should rather consider a future observer $\mathcal{O}$ to talk consistently of $A$ and $B$.

With the relational interpretation, the possible weirdness of non-local experiments boils down to the weirdness of indeterminism. Surely, fundamental randomness is a characteristic feature of quantum physics; the future is not completely predictable, even in principle. Although it contradicts the prejudices of classical physics, randomness has been accepted much easier in the literature than non-locality. A reason may be that the uncertainty of any measurement already constrains classical physics to be indeterministic in practice. The shift is that classical randomness is epistemic (lack of knowledge) while quantum randomness is fundamental (irreducible indeterminism). In fact, 'non-locality' exemplifies the difficulty to grasp together 'causality' and 'indeterminism' in the same conceptual and mathematical framework.

The relational interpretation has not yet answered all the intriguing questions raised by quantum mechanics [119]. However, a lot of work has already been achieved in recent years, leading to an increasing interest in the approach [55, 108]. Indeed, we believe this interpretation is a very promising framework to think about the essential features of quantum physics, as we have shown with the example of non-locality in this chapter.

## COLOPHON

This document was typeset using the typographical look-and-feel `classicthesis` developed by André Miede, released under the GNU General Public License as published by the Free Software Foundation, and available for both LaTeX and LyX:

https://bitbucket.org/amiede/classicthesis/.

Photo of the front cover (printed version) by Guillaume Gouin on Unsplash.
Photo page xii by Adeline Guéret.
Comic strip on page xvii from *There's Treasure Everywhere –A Calvin and Hobbes Collection* by Bill Watterson copyrighted by Andrews McMeel Syndication.
Photo page xxxiv by Matteo Modica on Unsplash.
Photo page 68 by Dmitry Ratushny on Unsplash.
Photo page 164 by José Fulgencio Orenes Martínez on Unsplash.
Photo page 206 by Adrian Swancar on Unsplash.
Figure 12 inspired from [99].
Figure 16 taken from [14].
Figure 17 taken from [140].
Figure 22 taken from [71].
Figures 20 and 53 by Wikimedia Commons, licensed under Creative Commons Attribution-Share Alike 3.0 Unported.
Pictures of figure 55 inspired from the comics *Gaston*, by André Franquin.
Photo of figure 56a by Annie Spratt on Unsplash.
Photo of figure 56b by Benjamin Wong on Unsplash.

*Final Version* as of 3rd September 2020.